\documentclass[
    12pt,
    letterpaper,
    oneside,
    footnotes=multiple,
]{scrbook}

\usepackage[margin=1.0in]{geometry}
\usepackage[inline]{enumitem}
\usepackage[autooneside=false,automark]{scrlayer-scrpage}

\addtokomafont{part}{\fontsize{72}{72}\selectfont}
\addtokomafont{part}{\color{white}}
\setkomafont{chapter}{\color{SlateHeading}\bfseries\LARGE\sffamily\sansmath\sisetup{detect-all}}
\setkomafont{section}{\color{SlateHeading}\normalfont\Large\sffamily\sansmath\sisetup{detect-all}}
\setkomafont{subsection}{\color{SlateHeading}\normalfont\large\sffamily\sansmath\sisetup{detect-all}}
\setkomafont{subsubsection}{\color{SlateHeading}\normalfont\large\sffamily\sansmath\sisetup{detect-all}}
\setkomafont{paragraph}{\color{SlateHeading}\normalfont\normalsize\sffamily\sansmath\sisetup{detect-all}}
\setkomafont{captionlabel}{\color{SlateHeading}\normalfont\small\sffamily\sansmath\sisetup{detect-all}\setstretch{1.0}}
\setkomafont{caption}{\normalfont\small}
\setkomafont{title}{\bfseries\sffamily\HUGE}
\setkomafont{subtitle}{\color{DarkSlateGray}\bfseries\sffamily\LARGE}
\setcapindent{0pt}

\deffootnote[0em]{0em}{1em}{\textsuperscript{\thefootnotemark}}
\setfootnoterule[0pt]{10em}

\addtokomafont{partentry}{\Large}
\addtokomafont{chapterentry}{\mdseries\large}
\RedeclareSectionCommands[
  toclinefill=\quad,
  tocraggedpagenumber=true,
]{part,chapter,section,subsection}
\RedeclareSectionCommands[tocdynnumwidth]{chapter,section}
\addtokomafont{partentrypagenumber}{\color{SlateHeading}}
\addtokomafont{chapterentrypagenumber}{\color{SlateHeading}}

\setkomafont{pageheadfoot}{\small\sffamily\upshape\color{SlateHeading}}
\setkomafont{pagenumber}{\small\sffamily\upshape\color{SlateHeading}}
\ihead{\leftmark}
\ohead{\ifstr{\leftmark}{\rightmark}{}{\rightmark}}
\lefoot{}
\rofoot{}

\newlist{compactitem}{itemize}{1}
\setlist[compactitem]{label=\textbullet,itemsep=0.1\baselineskip,parsep=0.1\baselineskip}

\usepackage{verbatim}

\usepackage[
    usenames,
    dvipsnames,
    svgnames,
    x11names,
    table,
]{xcolor}
\usepackage{graphicx}
\usepackage{tikz}
\usepackage[outline]{contour}
\usepackage{eso-pic}
\definecolor{ElectricBlue}{rgb}{0.00, 0.15, 0.45}
\definecolor{DeepGreen}{rgb}{0.00, 0.30, 0.00}
\definecolor{SlateHeading}{rgb}{0.225, 0.275, 0.275}

\usepackage[T1]{fontenc}
\usepackage{amsmath,amssymb}
\IfFileExists{fourier.sty}
    {\usepackage{fourier}}
    {}
\IfFileExists{erewhon.sty} %
    {\usepackage[p,lining,scale=0.99]{erewhon}}
    {}
\IfFileExists{plex-sans.sty} %
    {\usepackage[scale=0.90,textmd]{plex-sans}}
    {\usepackage{helvet}}
\IfFileExists{zi4.sty} %
    {\usepackage[varl,scaled=1.0]{zi4}}
    {}
\IfFileExists{euscript.sty} %
    {\usepackage[mathcal]{euscript}}
    {}
\usepackage{setspace}
\usepackage{bbding}
\usepackage{sansmath}
\renewcommand{\alpha}{\otheralpha}
\renewcommand{\beta}{\otherbeta}
\renewcommand{\gamma}{\othergamma}
\renewcommand{\delta}{\otherdelta}
\renewcommand{\epsilon}{\otherepsilon}
\renewcommand{\varepsilon}{\othervarepsilon}
\renewcommand{\zeta}{\otherzeta}
\renewcommand{\eta}{\othereta}
\renewcommand{\theta}{\othertheta}
\renewcommand{\vartheta}{\othervartheta}
\renewcommand{\iota}{\otheriota}
\renewcommand{\kappa}{\otherkappa}
\renewcommand{\varkappa}{\othervarkappa}
\renewcommand{\lambda}{\otherlambda}
\renewcommand{\mu}{\othermu}
\renewcommand{\nu}{\othernu}
\renewcommand{\xi}{\otherxi}
\renewcommand{\pi}{\otherpi}
\renewcommand{\varpi}{\othervarpi}
\renewcommand{\rho}{\otherrho}
\renewcommand{\varrho}{\othervarrho}
\renewcommand{\sigma}{\othersigma}
\renewcommand{\varsigma}{\othervarsigma}
\renewcommand{\tau}{\othertau}
\renewcommand{\upsilon}{\otherupsilon}
\renewcommand{\phi}{\otherphi}
\renewcommand{\varphi}{\othervarphi}
\renewcommand{\chi}{\otherchi}
\renewcommand{\psi}{\otherpsi}

\usepackage{etoolbox}
\newrobustcmd{\bftab}{\fontseries{b}\selectfont}
\newrobustcmd{\ittab}{\fontshape{it}\selectfont}
\usepackage{microtype}
\usepackage[
    range-phrase=--,
    range-units=single,
    retain-explicit-plus=true,
    retain-unity-mantissa=false,
]{siunitx}
\usepackage[
    citecolor=ElectricBlue,
    colorlinks=true,
    linkcolor=ElectricBlue,
    urlcolor=DeepGreen,
]{hyperref}
\usepackage{url}
\usepackage{soul}  %

\usepackage{calc}
\usepackage{xspace}
\usepackage{tabularray}

\usepackage{subcaption}
\usepackage{float}
\usepackage{array, booktabs, rotating, makecell}
\usepackage{multirow, bigdelim}
\usepackage{wrapfig}
\usepackage{colortbl} %
\usepackage[capitalise]{cleveref}
\crefformat{chapter}{\S#2#1#3}
\crefrangeformat{chapter}{\S\S#3#1#4--#5#2#6}
\crefname{chapter}{\S}{\S\S}

\usepackage[
    backgroundcolor=LightSteelBlue4,
    framemethod=tikz,
    hidealllines=true,
    roundcorner=4pt,
]{mdframed}
\DeclareCaptionType{boxcaption}[Box]
\newcommand{\callout}[1]{{\normalfont\large\bfseries\sffamily\color{SlateHeading} #1}}

\newenvironment{boxenv}[3]
  {
    \begin{boxcaption}[#1]%
    \noindent\begin{minipage}{#3}
      \begin{mdframed}[backgroundcolor=#2]
      \normalfont\normalsize\sisetup{detect-all}%
      }    
  {\end{mdframed}\end{minipage}\end{boxcaption}}
\crefname{boxcaption}{Box}{Boxes}

\usepackage[
    autocite=superscript,
    autopunct=true,
    backref=true,
    biblabel=brackets,
    doi=false,
    eprint=true,
    giveninits=true,
    hyperref=true,
    language=australian,
    maxbibnames=3,
    minbibnames=1,
    maxcitenames=2,
    mincitenames=1,
    sorting=none,
    style=phys,
    url=false,
    urldate=long,
]{biblatex}
\DeclareSourcemap{
  \maps[datatype=bibtex]{
    \map{
      \pertype{article}
      \step[fieldsource=journal, match=\regexp{arXiv}, final]
      \step[fieldset=journal, null]
      \step[fieldset=pages, null]
      \step[fieldset=year, null]
      \step[fieldset=primaryClass, null]
      \step[fieldset=eid, null]
      }
    \map{
      \pertype{article}
      \step[fieldsource=journal, final]
      \step[fieldset=archivePrefix, null]
      \step[fieldset=eprint, null]
      \step[fieldset=primaryClass, null]
      \step[fieldset=eid, null]
      }
    }
  }

\DeclareFieldFormat{labelnumberwidth}{#1\adddot}
\DeclareFieldFormat{titlecase}{#1}
\DeclareFieldFormat[unpublished,misc]{titlecase}{#1}
\DeclareFieldFormat[techreport,report]{title}{\mkbibemph{#1}}
\DeclareFieldFormat[techreport,report]{citetitle}{\mkbibitalic{#1}}

\DeclareFieldFormat
  [
   thesis, %
   report, %
  ]{title}{\href{\thefield{url}}{\mkbibitalic{#1}}}

\DeclareFieldFormat[unpublished]{title}{\href{\thefield{url}}{#1}}

\DeclareFieldFormat[unpublished,misc]{titlecase}{#1}

\newcommand{\rmd}{\mathrm{d}}

\newcommand{\mr}[1]{\mathrm{#1}}

\newcommand{\GW}{gravitational wave}  %
\newcommand{\SNR}{signal-to-noise ratio}

\newcommand{\gwd}{gravitational-wave detector}

\newcommand{\gwds}{gravitational-wave detectors}

\newcommand{\gwo}{gravitational-wave observatory}

\def\bmt{beamtube}
\def\bmts{beamtubes}

\def\fortykm{\SI{40}{km}}
\def\twentykm{\SI{20}{km}}

\newcommand{\Msol}{\ensuremath{M_\odot}\xspace}

\newcommand{\Msun}{\ensuremath{M_\odot}\xspace}

\DeclareSIUnit{\ppm}{ppm}
\DeclareSIUnit{\torr}{torr}
\DeclareSIUnit{\clight}{\text{\ensuremath{c}}}
\DeclareSIUnit{\inch}{in}
\DeclareSIUnit{\atomicmassunit}{u}

\graphicspath{{../figures/}{./figures/}}

\definecolor{spring}{rgb}{0.7,0.9,0.7}
\definecolor{brick}{rgb}{0.7,0.2,0.1}
\definecolor{redHL}{rgb}{1.0,0.7,0.7}
\definecolor{blueHL}{rgb}{0.7,0.7,1.0}
\sethlcolor{redHL}

\newcommand{\sref}[1]{\S\ref{sec:#1}}
\newcommand{\Sref}[1]{\S\ref{sec:#1}}
\newcommand{\xlabel}[1]{\label{sec:#1}}
\crefformat{section}{\S#2#1#3}
\crefmultiformat{section}{\S\S#2#1#3}{ and~#2#1#3}{, #2#1#3}{, and~#2#1#3}

\newcommand{\fref}[1]{\cref{fig:#1}}

\newcommand{\flabel}[1]{\label{fig:#1}}

\newcommand{\tref}[1]{\cref{tab:#1}}

\newcommand{\tlabel}[1]{\label{tab:#1}}

\newcommand{\flowdown}{Table~\ref{flowdown}}

\newcommand{\simt}{\ensuremath{\sim}}

\newcommand{\hlight}[1]{}  %

\newcommand{\CE}{Cosmic Explorer\xspace}
\newcommand{\ET}{Einstein Telescope\xspace}

\definecolor{CEwET}{rgb}{0.84, 0.7, 0.93}

\definecolor{ccnone}{rgb}{0.9, 0.9, 0.9}
\definecolor{ccstart}{rgb}{0.87, 0.8 , 0.47}
\definecolor{ccgood}{rgb}{0.13, 0.53, 0.2 }
\definecolor{ccbetter}{rgb}{0.0, 0.42, 0.24}

\definecolor{colhr}{rgb}{0.86, 0.02, 0.05}
\definecolor{colmr}{rgb}{0.95, 0.58, 0.18}
\definecolor{collr}{rgb}{1.  , 0.77, 0.31}

\newlength{\FDheight}
\makeatletter
\newcommand\footnoteref[1]{\protected@xdef\@thefnmark{\ref{#1}}\@footnotemark}
\makeatother

\title{Cosmic Explorer}
\subtitle{A Horizon Study}
\author{}
\date{2021}

\begin{document}

\newcommand{\MACROCOLOR}{black}

\newcommand{\CEBBHPERYEAR}{\textcolor{\MACROCOLOR}{$\num{100000}$}}

\newcommand{\GWTCGWS}{\textcolor{\MACROCOLOR}{50}}

\newcommand{\CEVSGWTC}{\textcolor{\MACROCOLOR}{\num{2000}}}

\newcommand{\CENEARBBHPERYEAR}{\textcolor{\MACROCOLOR}{8}}
\newcommand{\CENEARBBHMEDSNR}{\textcolor{\MACROCOLOR}{600}}
\newcommand{\CENEARBBHLOUDSNR}{\textcolor{\MACROCOLOR}{2700}}

\newcommand{\CEFARBBHPERYEAR}{\textcolor{\MACROCOLOR}{\num{60000}}}
\newcommand{\CEFARBBHMEDSNR}{\textcolor{\MACROCOLOR}{20}}

\newcommand{\CEFARTHERBBHPERYEAR}{\textcolor{\MACROCOLOR}{\num{10000}}}

\newcommand{\CEBNSPERYEAR}{\textcolor{\MACROCOLOR}{\num{300000}}}
\newcommand{\CEFARBNSPERYEAR}{\textcolor{\MACROCOLOR}{$2\times 10^5$}}
\newcommand{\CEMEDBNSPERYEAR}{\textcolor{\MACROCOLOR}{\num{100000}}}
\newcommand{\CELOUDBNSPERYEAR}{\textcolor{\MACROCOLOR}{5}}

\newcommand{\INITCEBBHNEARMEDSNR}{\textcolor{\MACROCOLOR}{390}}
\newcommand{\INITCEBBHNEARLOUDSNR}{\textcolor{\MACROCOLOR}{1700}}
\newcommand{\VOYNETBBHNEARMEDSNR}{\textcolor{\MACROCOLOR}{150}}
\newcommand{\VOYNETBBHNEARLOUDSNR}{\textcolor{\MACROCOLOR}{610}}
\newcommand{\APLUSNETBBHNEARMEDSNR}{\textcolor{\MACROCOLOR}{70}}
\newcommand{\APLUSNETBBHNEARLOUDSNR}{\textcolor{\MACROCOLOR}{280}}
\newcommand{\ETTWOCEBBHNEARLOUDSNR}{\textcolor{\MACROCOLOR}{4100}}

\newcommand{\CostTwentyManagementCivil}{12}
\newcommand{\CostTwentyManagementCivilInf}{15}
\newcommand{\CostTwentyManagementVac}{17}
\newcommand{\CostTwentyManagementVacInf}{21}
\newcommand{\CostTwentyManagementDet}{18}
\newcommand{\CostTwentyManagementDetInf}{22}
\newcommand{\CostTwentyManagementTotal}{47}
\newcommand{\CostTwentyManagementTotalInf}{58}
\newcommand{\CostTwentyDesignCivil}{2.5}
\newcommand{\CostTwentyDesignCivilInf}{3.2}
\newcommand{\CostTwentyDesignVac}{0.7}
\newcommand{\CostTwentyDesignVacInf}{0.8}
\newcommand{\CostTwentyDesignDet}{0.9}
\newcommand{\CostTwentyDesignDetInf}{1.2}
\newcommand{\CostTwentyDesignTotal}{4.1}
\newcommand{\CostTwentyDesignTotalInf}{5.2}
\newcommand{\CostTwentyRealizationCivil}{118}
\newcommand{\CostTwentyRealizationCivilInf}{148}
\newcommand{\CostTwentyRealizationVac}{168}
\newcommand{\CostTwentyRealizationVacInf}{210}
\newcommand{\CostTwentyRealizationDet}{180}
\newcommand{\CostTwentyRealizationDetInf}{225}
\newcommand{\CostTwentyRealizationTotal}{466}
\newcommand{\CostTwentyRealizationTotalInf}{583}
\newcommand{\CostTwentyTotal}{517}
\newcommand{\CostTwentyTotalInf}{646}
\newcommand{\CostFortyManagementCivil}{23}
\newcommand{\CostFortyManagementCivilInf}{29}
\newcommand{\CostFortyManagementVac}{31}
\newcommand{\CostFortyManagementVacInf}{38}
\newcommand{\CostFortyManagementDet}{18}
\newcommand{\CostFortyManagementDetInf}{22}
\newcommand{\CostFortyManagementTotal}{72}
\newcommand{\CostFortyManagementTotalInf}{89}
\newcommand{\CostFortyDesignCivil}{2.5}
\newcommand{\CostFortyDesignCivilInf}{3.2}
\newcommand{\CostFortyDesignVac}{0.7}
\newcommand{\CostFortyDesignVacInf}{0.8}
\newcommand{\CostFortyDesignDet}{0.9}
\newcommand{\CostFortyDesignDetInf}{1.2}
\newcommand{\CostFortyDesignTotal}{4.1}
\newcommand{\CostFortyDesignTotalInf}{5.2}
\newcommand{\CostFortyRealizationCivil}{234}
\newcommand{\CostFortyRealizationCivilInf}{293}
\newcommand{\CostFortyRealizationVac}{306}
\newcommand{\CostFortyRealizationVacInf}{383}
\newcommand{\CostFortyRealizationDet}{180}
\newcommand{\CostFortyRealizationDetInf}{225}
\newcommand{\CostFortyRealizationTotal}{720}
\newcommand{\CostFortyRealizationTotalInf}{901}
\newcommand{\CostFortyTotal}{796}
\newcommand{\CostFortyTotalInf}{995}
\newcommand{\CostContingencyTwenty}{103}
\newcommand{\CostContingencyTwentyInf}{129}
\newcommand{\CostContingencyForty}{159}
\newcommand{\CostContingencyFortyInf}{199}
\newcommand{\CostContingencyProject}{12}
\newcommand{\CostContingencyProjectInf}{15}
\newcommand{\CostProjectProjectwideManagement}{20}
\newcommand{\CostProjectProjectwideManagementInf}{25}
\newcommand{\CostProjectProjectwideCoordination}{5}
\newcommand{\CostProjectProjectwideCoordinationInf}{6}
\newcommand{\CostProjectProjectwideComputing}{10}
\newcommand{\CostProjectProjectwideComputingInf}{13}
\newcommand{\CostProjectProjectwideTotal}{35}
\newcommand{\CostProjectProjectwideTotalInf}{44}
\newcommand{\CostProjectDesignCivil}{15}
\newcommand{\CostProjectDesignCivilInf}{19}
\newcommand{\CostProjectDesignVac}{7}
\newcommand{\CostProjectDesignVacInf}{8}
\newcommand{\CostProjectDesignDet}{5}
\newcommand{\CostProjectDesignDetInf}{6}
\newcommand{\CostProjectDesignTotal}{27}
\newcommand{\CostProjectDesignTotalInf}{33}
\newcommand{\CostProjectTotal}{62}
\newcommand{\CostProjectTotalInf}{77}
\newcommand{\CostTotal}{1649}
\newcommand{\CostTotalInf}{2061}
\newcommand{\CostTwoTwenty}{1314}
\newcommand{\CostTwoTwentyInf}{1642}
\newcommand{\CostJustForty}{1029}
\newcommand{\CostJustFortyInf}{1286}

\newcommand{\CostOverviewCivil}{422}
\newcommand{\CostOverviewCivilInf}{528}
\newcommand{\CostOverviewVac}{569}
\newcommand{\CostOverviewVacInf}{712}
\newcommand{\CostOverviewDet}{432}
\newcommand{\CostOverviewDetInf}{540}
\newcommand{\CostOverviewTotal}{1650}
\newcommand{\CostOverviewTotalInf}{2062}
\newcommand{\CostOverviewProject}{227}
\newcommand{\CostOverviewProjectInf}{283}

\newcommand{\PCTCIVn}{26}
\newcommand{\PCTVACn}{34}
\newcommand{\PCTDETn}{26}
\newcommand{\PCTMANn}{14}

\newcommand{\CostOpsFacility}{18}
\newcommand{\CostOpsVac}{7.7}
\newcommand{\CostOpsDet}{16.9}
\newcommand{\CostOpsComp}{7.2}
\newcommand{\CostOpsManagement}{5.7}
\newcommand{\CostOpsCommunity}{4.7}
\newcommand{\CostOpsFacilityInf}{22.5}
\newcommand{\CostOpsVacInf}{9.6}
\newcommand{\CostOpsDetInf}{21.1}
\newcommand{\CostOpsCompInf}{9}
\newcommand{\CostOpsManagementInf}{7.1}
\newcommand{\CostOpsCommunityInf}{5.9}
\newcommand{\CostOpsTotal}{60.2}
\newcommand{\CostOpsTotalInf}{75.3}

\newcommand{\opsPCTFACn}{30}
\newcommand{\opsPCTVACn}{13}
\newcommand{\opsPCTDETn}{28}
\newcommand{\opsPCTCMPn}{12}
\newcommand{\opsPCTMNGn}{9}
\newcommand{\opsPCTCOMn}{8}

\newcommand{\PCTCIV}{\PCTCIVn\%}
\newcommand{\PCTVAC}{\PCTVACn\%}
\newcommand{\PCTDET}{\PCTDETn\%}
\newcommand{\PCTMAN}{\PCTMANn\%}

\newcommand{\opsPCTFAC}{\opsPCTFACn\%}
\newcommand{\opsPCTVAC}{\opsPCTVACn\%}
\newcommand{\opsPCTDET}{\opsPCTDETn\%}
\newcommand{\opsPCTCMP}{\opsPCTCMPn\%}
\newcommand{\opsPCTMNG}{\opsPCTMNGn\%}
\newcommand{\opsPCTCOM}{\opsPCTCOMn\%}

\newcommand\aplusBnsHorizon{0.19}
\newcommand\aplusBbhHorizon{2.7}
\newcommand\aplusMaxMtot{760}
\newcommand\aplusBnsSNR{75} %
\newcommand\aplusEarlyWarning{4} %
\newcommand\voyBnsHorizon{0.45}
\newcommand\voyBbhHorizon{7.5}
\newcommand\voyMaxMtot{950}
\newcommand\voyBnsSNR{169} %
\newcommand\voyEarlyWarning{10.4} %
\newcommand\etBnsHorizon{3.7}
\newcommand\etBbhHorizon{57}
\newcommand\etMaxMtot{4.1e+03}
\newcommand\etBnsSNR{850} %
\newcommand\etEarlyWarning{6.6\,\si{\hour}} %
\newcommand\ceOneBnsHorizon{4.2}
\newcommand\ceOneBbhHorizon{34}
\newcommand\ceOneMaxMtot{2.2e+03}
\newcommand\ceOneBnsSNR{900} %
\newcommand\ceOneEarlyWarning{94} %
\newcommand\ceTwoGlassBnsHorizon{8.3}
\newcommand\ceTwoGlassBbhHorizon{41}
\newcommand\ceTwoGlassMaxMtot{2.4e+03}
\newcommand\ceTwoGlassBnsSNR{1.26e+03} %
\newcommand\ceTwoGlassEarlyWarning{103} %
\newcommand\ceTwoSiliconBnsHorizon{11.7}
\newcommand\ceTwoSiliconBbhHorizon{41}
\newcommand\ceTwoSiliconMaxMtot{2.4e+03}
\newcommand\ceTwoSiliconBnsSNR{1.46e+03} %
\newcommand\ceTwoSiliconEarlyWarning{103} %

\frontmatter
\newgeometry{margin=0in}
\begin{titlepage}
\pagecolor{black}
\color{white}
\thispagestyle{empty}

\begin{tikzpicture}[remember picture,overlay]
    \node[anchor=center] at (current page.center)
        {\includegraphics[width=\paperwidth]{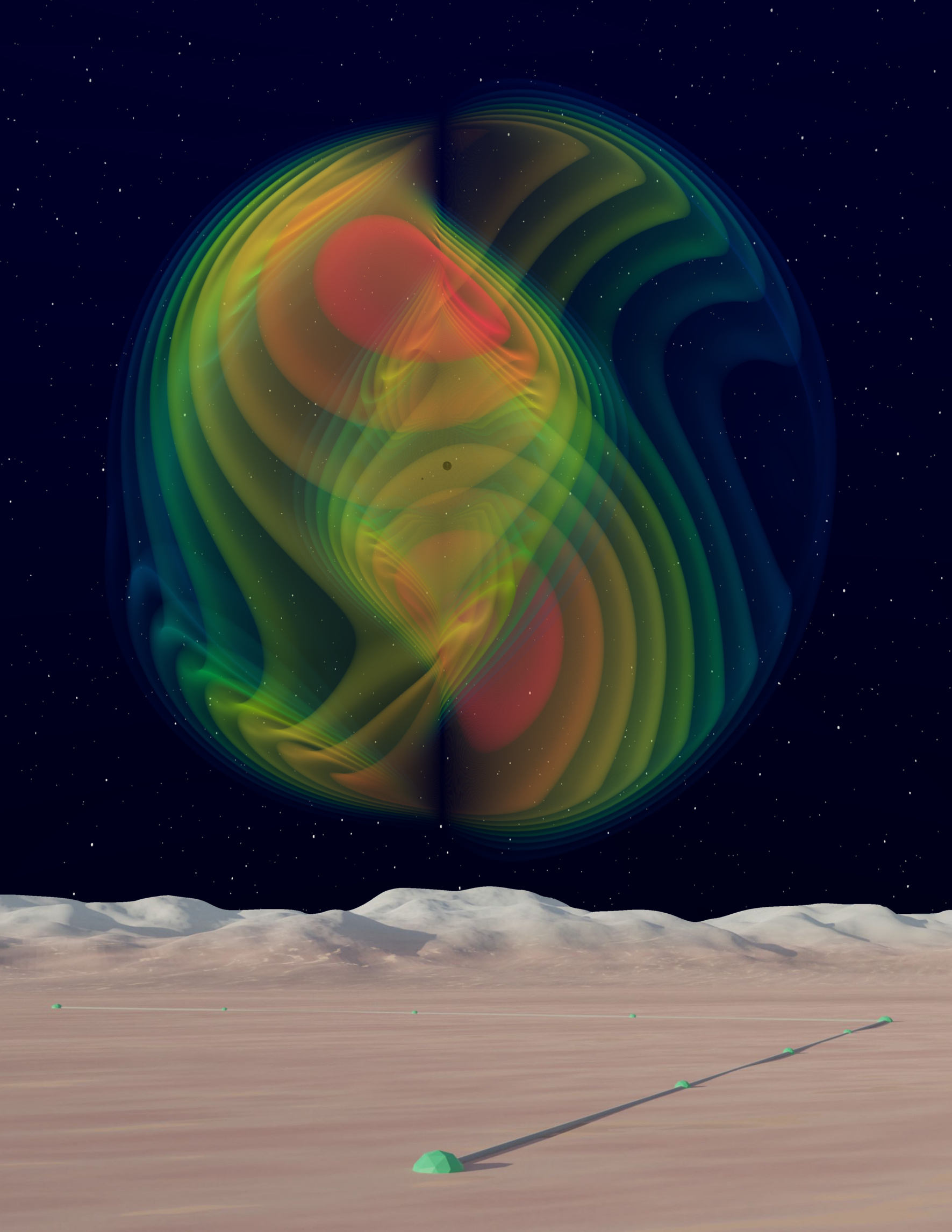}};
    \node[anchor=north west,inner sep=3mm] at (current page.north west)
        {\includegraphics[width=2.6in]{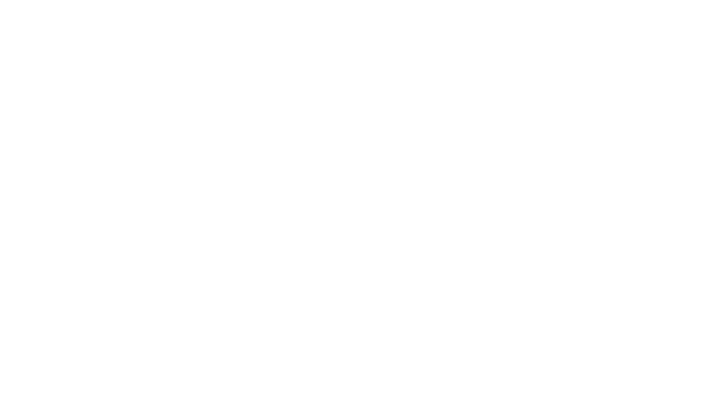}};
\end{tikzpicture}

\begin{center}
    \vphantom{=}

    \vspace{4.0in}
        
    \contourlength{0.75pt}
    \contour{black}{\color{SlateGray2}{\textbf{\textsf{\LARGE A Horizon Study for}}}}

    \vspace{0.30in} 

    \contourlength{2.0pt}
    \contour{black}{\textbf{\textsf{\fontsize{64}{72}\selectfont Cosmic Explorer}}}

    \vspace{0.30in} 
        
    \contourlength{1.0pt}
    \contour{black}{\color{SlateGray2}\textbf{\textsf{\fontsize{30}{36}\selectfont Science, Observatories, and Community}}}

    \vspace{2in}

    {}
\end{center}
\end{titlepage}
\restoregeometry

\nopagecolor
\color{black}

\thispagestyle{empty}

\newgeometry{margin=0.9in}
\noindent\textit{Cosmic Explorer Technical Report \href{https://dcc.cosmicexplorer.org/P2100003/public}{CE--P2100003--v7}} \\
\textit{October 2021}

\section*{Authors}
\vspace{-0.25\baselineskip}
\begin{minipage}{\linewidth}

\begin{flushleft}
    \mbox{Matthew~Evans,\footnote{\label{fn:LigoMIT}LIGO Laboratory, Massachusetts Institute of Technology, Cambridge, MA 02139, USA}}
	\mbox{Rana~X~Adhikari},\footnote{\label{fn:LIGOCIT}LIGO Laboratory, California Institute of Technology, Pasadena, CA 91125, USA}
    \mbox{Chaitanya~Afle,\footnote{\label{fn:Syracuse}Department of Physics, Syracuse University, Syracuse, NY 13244, USA}}
    \mbox{Stefan~W.~Ballmer,\textsuperscript{\ref{fn:Syracuse}}}
    \mbox{Sylvia~Biscoveanu,\textsuperscript{\ref{fn:LigoMIT}}}
    \mbox{Ssohrab~Borhanian,\footnote{\label{fn:PennState}Institute for Gravitation and the Cosmos, Department of Physics, Pennsylvania State University, University Park, PA 16802, USA}}
    \mbox{Duncan~A.~Brown,\textsuperscript{\ref{fn:Syracuse}}}
    \mbox{Yanbei~Chen,\footnote{\label{fn:CaltechCart}Caltech CaRT, Pasadena, CA 91125, USA}}
    \mbox{Robert~Eisenstein,\textsuperscript{\ref{fn:LigoMIT}}}
    \mbox{Alexandra~Gruson,\footnote{\label{fn:CSUFullerton}Nicholas and Lee Begovich Center for Gravitational-Wave Physics and Astronomy, California State University, Fullerton, Fullerton, CA 92831, USA}}
    \mbox{Anuradha~Gupta,\textsuperscript{\ref{fn:PennState},}\footnote{\label{fn:UMiss}Department of Physics and Astronomy, University of Mississippi, University, MS 38677, USA}}
    \mbox{Evan~D.~Hall,\textsuperscript{\ref{fn:LigoMIT}}}
    \mbox{Rachael~Huxford,\textsuperscript{\ref{fn:PennState}}}
    \mbox{Brittany~Kamai,\footnote{\label{fn:UCSantaCruz}Department of Astronomy \& Astrophysics, University of California Santa Cruz, Santa Cruz, CA 95064, USA}\textsuperscript{,}\footnote{\label{fn:CaltechEAS}Department of Mechanical Engineering, California Institute of Technology, Pasadena, CA 91125, USA}}
    \mbox{Rahul~Kashyap,\textsuperscript{\ref{fn:PennState}}}
    \mbox{Jeff~S.~Kissel,\footnote{\label{fn:LigoHanford}LIGO Hanford Observatory, Richland, WA 99352, USA}}
    \mbox{Kevin~Kuns,\textsuperscript{\ref{fn:LigoMIT}}}
    \mbox{Philippe~Landry,\textsuperscript{\ref{fn:CSUFullerton}}}
    \mbox{Amber~Lenon,\textsuperscript{\ref{fn:Syracuse}}}
    \mbox{Geoffrey~Lovelace,\textsuperscript{\ref{fn:CSUFullerton}}}
    \mbox{Lee~McCuller,\textsuperscript{\ref{fn:LigoMIT}}}
    \mbox{Ken~K.~Y.~Ng,\textsuperscript{\ref{fn:LigoMIT}}}
    \mbox{Alexander~H.~Nitz,\footnote{\label{fn:AEI}Max-Planck-Institut f\"ur Gravitationsphysik (Albert-Einstein-Institut), D-30167 Hannover, Germany}\textsuperscript{,}\footnote{\label{fn:Hannover}Leibniz Universit\"at Hannover, D-30167 Hannover, Germany}}
    \mbox{Jocelyn~Read,\textsuperscript{\ref{fn:CSUFullerton}}}
    \mbox{B.~S.~Sathyaprakash,\textsuperscript{\ref{fn:PennState},}\footnote{School of Physics and Astronomy, Cardiff University, Cardiff, UK}}
    \mbox{David~H.~Shoemaker,\textsuperscript{\ref{fn:LigoMIT}}}
    \mbox{Bram~J.~J.~Slagmolen,\footnote{\label{fn:OzGravANU}OzGrav-ANU, Centre for Gravitational Astrophysics, College of Science, The Australian National University, ACT 2601, Australia}}
    \mbox{Joshua~R.~Smith,\textsuperscript{\ref{fn:CSUFullerton}}}
    \mbox{Varun~Srivastava,\textsuperscript{\ref{fn:Syracuse}}}
    \mbox{Ling~Sun,\textsuperscript{\ref{fn:OzGravANU}}}
    \mbox{Salvatore~Vitale,\textsuperscript{\ref{fn:LigoMIT}}}
    \mbox{Rainer~Weiss\textsuperscript{\ref{fn:LigoMIT}}}
\end{flushleft}
\end{minipage}

\vspace{0.75\baselineskip}
\noindent
Correspondence: \href{mailto:ce-questions@cosmicexplorer.org}{\texttt{ce-questions@cosmicexplorer.org}}

\section*{Cover Image}
\vspace{-0.25\baselineskip}
The cover image shows an artistic rendering of Cosmic Explorer [credit: Evan Hall (MIT)] beneath a numerical-relativity simulation of a binary black hole emitting gravitational waves [credit: Nils~Fischer, Harald~Pfeiffer, Alessandra~Buonanno (Max Planck Institute for Gravitational Physics), Simulating eXtreme Spacetimes (SXS) Collaboration].

\begin{tikzpicture}[remember picture,overlay]
    \node[anchor=center] at ([yshift=1.5in]current page.south)
        {\includegraphics[width=1.5in]{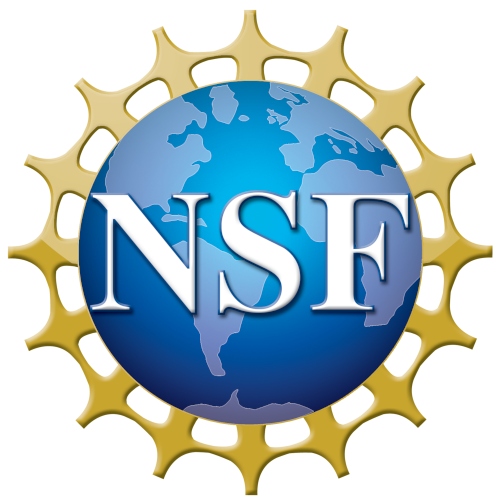}};
    \node[anchor=center] at ([yshift=0.5in]current page.south)
{This study was funded by the National Science Foundation.};
\end{tikzpicture}

\restoregeometry
\setstretch{1.00}
\tableofcontents
\setstretch{1.10}

\mainmatter

\begin{spacing}{1.03} %
\chapter{Executive Summary \hlight{[Duncan]}}
\label{sec:executive}
Gravitational-wave astronomy has revolutionized humanity's view of the universe. 
Investment in the field has rewarded the scientific community with the
first direct detection of a binary black hole merger and the multimessenger
observation of a neutron-star merger. Each of these was a watershed
moment in astronomy, made possible because gravitational waves reveal the cosmos in a way
that no other probe can.  Since the first detection of gravitational waves in 2015,
the National Science Foundation's LIGO and its partner
observatory, the European Union's Virgo, have detected over fifty binary black hole mergers
and a second neutron star merger\,---\,a rate of discovery that has amazed even
the most optimistic scientists.

\vspace*{0.5em}

\noindent
This Horizon Study describes a next-generation ground-based gravitational-wave
observatory: \textbf{Cosmic Explorer}. With ten times the sensitivity of
Advanced LIGO, Cosmic Explorer will push the reach of gravitational-wave astronomy
towards the edge of the observable universe ($z \sim 100$).  This Horizon
Study presents the science objectives for Cosmic Explorer, and describes and
evaluates its design concepts. Cosmic Explorer will continue the United
States' leadership in gravitational-wave astronomy in the international effort
to build a ``Third-Generation'' (3G) observatory network that will make
discoveries transformative across astronomy, physics, and cosmology.

\vspace*{0.5em}

\noindent
Major discoveries in astronomy are driven by three related improvements:
better sensitivity, higher precision, and opening a new
observational window. Cosmic Explorer promises all of these. 
The nature of gravity means that with a one order-of-magnitude  sensitivity improvement
 over current detectors Cosmic Explorer will see
gravitational-wave sources across the history of the universe.  With its unprecedented sensitivity, 
Cosmic Explorer will make discoveries that cannot yet be anticipated, especially since gravitational waves reach
into regions of the universe that electromagnetic observations cannot explore.
With Cosmic Explorer, scientists can use the universe
as a laboratory to test the laws of physics and study the nature of matter.  
In addition to Cosmic Explorer's extraordinary discovery potential, this Horizon Study focuses on
three key science areas in which Cosmic Explorer will make a particularly dramatic impact:

\vspace*{0.5em}

\noindent \textbf{Black Holes and Neutron Stars Throughout Cosmic
Time.} 
Understanding how the universe 
made the first black holes, and how these first
black holes grew, is one of the most important
unsolved problems in astrophysics.
Cosmic Explorer will detect gravitational waves from binary
black holes and neutron stars out to the edge of the visible universe,
providing a view of Cosmic 
Dawn complementary to that of the James Webb Space
Telescope. Cosmic Explorer will see evidence for the first
stars by detecting the mergers of
the black holes they leave behind. The millions of mergers detected by Cosmic
Explorer will map the population of compact objects across time, detect
mergers of the
first black holes that contributed to seeding the universe's structure, 
explore the physics of massive stars, and reveal the processes that create black holes and neutron stars.

\noindent \textbf{Dynamics of Dense Matter.} While a quantitative theory of
nuclei, neutron-rich matter and deconfined quark matter has begun to emerge,
understanding the nature of strongly interacting matter is an unsolved
problem in physics.  By
observing many hundreds of loud neutron star mergers and measuring the stars' radii
to $100$~m or better, Cosmic Explorer will probe the phase structure of
quantum chromodynamics, revealing the nuclear equation of state and its
phase transitions.  Cosmic Explorer's ability to detect and study the hot,
dense remnants of neutron
star mergers will provide an entirely new way
of mapping out the dense, finite-temperature region of the
quantum chromodynamics phase
space, a region that is currently unexplored.  A plethora of
multimessenger
observations will map heavy-element nucleosynthesis, explain the build-up of
the chemical elements that are the building blocks of our world, and
explore the physics of the binary-merger engine powering short gamma-ray bursts. 

\vspace*{0.5em}

\noindent \textbf{Extreme Gravity and Fundamental Physics.} Cosmic Explorer's
increased discovery aperture will allow it to observe both loud and rare
gravitational-wave events\,---\,events that will reveal physics of the most
extreme gravity in the universe as well as events from unusual and novel
objects. LIGO and Virgo are already detecting events that we do not
fully understand. With its
higher-fidelity detections Cosmic Explorer will
reveal the nature of these
mysterious sources.
Cosmic Explorer will be able to look for the effects of dark
matter in the
cores of neutron stars and probe the nature of dark energy by looking for its
imprint
in gravitational-wave signals from the cosmos.  %
Cosmic Explorer's precision observations of black holes could help develop a viable theory of quantum gravity.

\vspace*{1em}

\enlargethispage*{1000pt}
\noindent
Cosmic Explorer's order-of-magnitude sensitivity improvement will be realized 
using a dual-recycled
Fabry--P\'{e}rot Michelson interferometer, as in Advanced LIGO.
Cosmic Explorer's increased
sensitivity comes primarily from scaling up the detector's length from 4 to 40~km.
This increases the amplitude of the observed signals with
effectively no increase in the detector noise.  
From the topographical and
geological point of view, many sites exist that could accommodate a
Cosmic Explorer design with a
\fortykm\ detector at facilities in the continental United States. 
When selecting sites, partnership with the local and regional communities, and
Indigenous Peoples, will be of utmost importance to ensure that the presence of
a Cosmic Explorer observatory respects the cultural, environmental,
socio-economic, political, and other aspects of its host communities.  Hazards
including earthquakes, floods, storms, and fires must also be considered,
especially with the view of a long-lived facility and a changing environment.

\vspace*{0.5em}

\noindent
There are many design choices that could realize some or all of Cosmic Explorer's wide
range of scientific opportunity. The reference concept considered in this
Horizon Study is a 40~km detector and a 20~km detector, both located in the United States, with a total estimated cost of  \$\CostTotalInf M (2030~USD). Alternative
configurations of two 20~km detectors a single 40~km detector are estimated to cost \$\CostTwoTwentyInf M (2030~USD) and \$\CostJustFortyInf M (2030~USD), respectively. 
(Clearly, the accuracy of these estimates does not warrant the number of digits given here,
 but they are maintained to ensure consistency with the more detailed breakdowns given in \sref{cost_est}.)
If the project design stage begins in the early 2020s, then Cosmic Explorer's first observing runs could take place in the mid-2030s.
As the field moves forward, it will be
essential to engage the broadest possible set of scientific stakeholders in
Cosmic Explorer's science to define its operational parameters. This includes
defining Cosmic Explorer's scientific priorities and the best detector
technologies, network design, and operational parameters needed to deliver
this science. 

\vspace*{0.5em}

\noindent
This Horizon Study includes a preliminary study (summarized in \cref{ex:flowdown}) that shows that although
the reference concept can achieve Cosmic Explorer's science goals without
other next-generation gravitational-wave detectors, its scientific output is enhanced when operating
 as part of an international network. 
Different configurations for
Cosmic Explorer are examined, embedded in a global network
that includes the Einstein Telescope, %
a potential detector in Australia, and the
existing second-generation (2G) observatories. The impact of
downscoping Cosmic Explorer's
reference design (e.g., to a single 40~km detector) is also investigated.
The community is encouraged to use this as a launch point to engage as Cosmic Explorer's 
design parameters are developed in the coming years.

\vspace*{0.5em}

\noindent
This Horizon Study describes a Cosmic Explorer project organization that follows the model successfully
employed by LIGO: two US-based Cosmic Explorer facilities constructed in one
project, followed by a transition to an operations organization.  The project
will use well-established methods for addressing
technical, managerial and political risks.  Cosmic Explorer's timeline will have distinct stages over several decades: concept development; observatory
design and site preparation; construction and commissioning; initial
operations; operations at nominal sensitivity;
observatory upgrades; and operations.  Cosmic Explorer's facilities are
intended to be long-lived, allowing for detector upgrades
with technologies yet to be discovered.

\vspace*{0.5em}

\noindent
The operations stage will embrace daily operations, production of observation
data, and low-latency astrophysical searches to produce astronomical alerts.
Given the substantial investment and broad community support that will be
required to realize a US next-generation gravitational-wave observatory,
Cosmic Explorer is planned to be an Open Data facility in its operations
phase. In this model, the Cosmic Explorer project is responsible for detector
operations, calibration, curation of the detector data streams, and
dissemination of detector data and rapid alerts to the scientific community.
Cosmic Explorer will generate a data set that provides a unique, rich, and
deep view of the universe over its lifetime. An open data approach will
facilitate scientific collaboration, maximize the scientific community's
investment in the project, and provide opportunity for scientists from small
institutions and historically underrepresented institutions.

\vspace*{0.5em}

\noindent
Gravitational-wave astronomy is global.
As part of a multimessenger network of international gravitational-wave observatories,
astro-particle detectors, and telescopes across the electromagnetic spectrum,
Cosmic Explorer will precisely localize and study the
nature of a multitude of sources.  
Experiments probing heavy ions and rare isotopes will help Cosmic Explorer determine
the physics of dense matter. 
Gravitational waves are generated by physical processes that are vastly
different from those that generate other forms of radiation and particles, and their
detections allow us to see into regions of the universe that cannot be observed in any other way. It
would be a profound anomaly in astronomy if nothing new and interesting came from
Cosmic Explorer's vast improvement in sensitivity.

\vspace*{0.5em}
\noindent
With foundations laid by decades of National Science Foundation investment and the work 
of a large community of scientists, Cosmic Explorer
is poised to propel another revolution in our understanding of the universe.
The community is invited to join the effort to define,
shape, and realize Cosmic Explorer: the future of gravitational-wave
astronomy.
\begin{table}[h]
    \centering
    \scriptsize
\begin{tblr}{
    column{3} = {leftsep=2.5pt,rightsep=7.2pt},
    row{3-12}    = {ht=6ex},
    column{3-18} = {wd=1.2ex},
    hline{1,Z}  = {wd=1pt,fg=black},
    hline{2}    = {wd=0.5pt,fg=black},
    cell{1}{1}  = {c=2}{halign=c,font=\large},  %
    cell{2}{1}  = {halign=c},  %
    cell{2}{2}  = {halign=c},  %
    cell{1}{3}  = {valign=m,bg=Ivory3},  %
    cell{1}{4}  = {c=5}{halign=c,bg=LightSteelBlue1},  %
    cell{1}{9}  = {c=5}{halign=c,bg=CEwET},  %
    cell{1}{14} = {c=5}{halign=c,bg=Plum1},  %
    vline{3,4,9,14} = {wd=1.5pt,fg=white},
    hline{6,10,12} = {wd=1.5pt,fg=white},
    cell{3}{1} = {r=3}{valign=m},
    vline{2}   = {3-7}{wd=2pt,fg=black},
    cell{3}{3}  = {bg=ccnone},
    cell{3}{4}  = {bg=ccnone},
    cell{3}{5}  = {bg=ccnone},
    cell{3}{6}  = {bg=ccstart},
    cell{3}{7}  = {bg=ccgood},
    cell{3}{8}  = {bg=ccbetter},
    cell{3}{9}  = {bg=ccbetter},
    cell{3}{10} = {bg=ccbetter},
    cell{3}{11} = {bg=ccbetter},
    cell{3}{12} = {bg=ccbetter},
    cell{3}{13} = {bg=ccbetter},
    cell{3}{14} = {bg=ccbetter},
    cell{3}{15} = {bg=ccbetter},
    cell{3}{16} = {bg=ccbetter},
    cell{3}{17} = {bg=ccbetter},
    cell{3}{18} = {bg=ccbetter},
    cell{4}{3}  = {bg=ccnone},
    cell{4}{4}  = {bg=ccstart},
    cell{4}{5}  = {bg=ccstart},
    cell{4}{6}  = {bg=ccgood},
    cell{4}{7}  = {bg=ccgood},
    cell{4}{8}  = {bg=ccbetter},
    cell{4}{9}  = {bg=ccbetter},
    cell{4}{10} = {bg=ccbetter},
    cell{4}{11} = {bg=ccbetter},
    cell{4}{12} = {bg=ccbetter},
    cell{4}{13} = {bg=ccbetter},
    cell{4}{14} = {bg=ccbetter},
    cell{4}{15} = {bg=ccbetter},
    cell{4}{16} = {bg=ccbetter},
    cell{4}{17} = {bg=ccbetter},
    cell{4}{18} = {bg=ccbetter},
    cell{5}{3}  = {bg=ccnone},
    cell{5}{4}  = {bg=ccstart},
    cell{5}{5}  = {bg=ccgood},
    cell{5}{6}  = {bg=ccgood},
    cell{5}{7}  = {bg=ccgood},
    cell{5}{8}  = {bg=ccbetter},
    cell{5}{9}  = {bg=ccbetter},
    cell{5}{10} = {bg=ccbetter},
    cell{5}{11} = {bg=ccbetter},
    cell{5}{12} = {bg=ccbetter},
    cell{5}{13} = {bg=ccbetter},
    cell{5}{14} = {bg=ccbetter},
    cell{5}{15} = {bg=ccbetter},
    cell{5}{16} = {bg=ccbetter},
    cell{5}{17} = {bg=ccbetter},
    cell{5}{18} = {bg=ccbetter},
    cell{6}{1} = {r=4}{valign=m},
    vline{2}   = {6-9}{wd=2pt,fg=black},
    cell{6}{3}  = {bg=ccnone},
    cell{6}{4}  = {bg=ccstart},
    cell{6}{5}  = {bg=ccgood},
    cell{6}{6}  = {bg=ccgood},
    cell{6}{7}  = {bg=ccbetter},
    cell{6}{8}  = {bg=ccbetter},
    cell{6}{9}  = {bg=ccgood},
    cell{6}{10} = {bg=ccbetter},
    cell{6}{11} = {bg=ccbetter},
    cell{6}{12} = {bg=ccbetter},
    cell{6}{13} = {bg=ccbetter},
    cell{6}{14} = {bg=ccbetter},
    cell{6}{15} = {bg=ccbetter},
    cell{6}{16} = {bg=ccbetter},
    cell{6}{17} = {bg=ccbetter},
    cell{6}{18} = {bg=ccbetter},
    cell{7}{3}  = {bg=ccnone},
    cell{7}{4}  = {bg=ccgood},
    cell{7}{5}  = {bg=ccstart},
    cell{7}{6}  = {bg=ccbetter},
    cell{7}{7}  = {bg=ccbetter},
    cell{7}{8}  = {bg=ccgood},
    cell{7}{9}  = {bg=ccbetter},
    cell{7}{10} = {bg=ccgood},
    cell{7}{11} = {bg=ccbetter},
    cell{7}{12} = {bg=ccbetter},
    cell{7}{13} = {bg=ccbetter},
    cell{7}{14} = {bg=ccbetter},
    cell{7}{15} = {bg=ccbetter},
    cell{7}{16} = {bg=ccbetter},
    cell{7}{17} = {bg=ccbetter},
    cell{7}{18} = {bg=ccbetter},
    cell{8}{3}  = {bg=ccnone},
    cell{8}{4}  = {bg=ccstart},
    cell{8}{5}  = {bg=ccstart},
    cell{8}{6}  = {bg=ccgood},
    cell{8}{7}  = {bg=ccgood},
    cell{8}{8}  = {bg=ccbetter},
    cell{8}{9}  = {bg=ccgood},
    cell{8}{10} = {bg=ccbetter},
    cell{8}{11} = {bg=ccbetter},
    cell{8}{12} = {bg=ccbetter},
    cell{8}{13} = {bg=ccbetter},
    cell{8}{14} = {bg=ccbetter},
    cell{8}{15} = {bg=ccbetter},
    cell{8}{16} = {bg=ccbetter},
    cell{8}{17} = {bg=ccbetter},
    cell{8}{18} = {bg=ccbetter},
    cell{9}{3}  = {bg=ccnone},
    cell{9}{4}  = {bg=ccstart},
    cell{9}{5}  = {bg=ccstart},
    cell{9}{6}  = {bg=ccgood},
    cell{9}{7}  = {bg=ccgood},
    cell{9}{8}  = {bg=ccbetter},
    cell{9}{9}  = {bg=ccbetter},
    cell{9}{10} = {bg=ccbetter},
    cell{9}{11} = {bg=ccbetter},
    cell{9}{12} = {bg=ccbetter},
    cell{9}{13} = {bg=ccbetter},
    cell{9}{14} = {bg=ccbetter},
    cell{9}{15} = {bg=ccbetter},
    cell{9}{16} = {bg=ccbetter},
    cell{9}{17} = {bg=ccbetter},
    cell{9}{18} = {bg=ccbetter},
    row{10} = {ht=12ex},
    cell{10}{3}  = {bg=ccnone},
    cell{10}{4}  = {bg=ccnone},
    cell{10}{5}  = {bg=ccstart},
    cell{10}{6}  = {bg=ccstart},
    cell{10}{7}  = {bg=ccbetter},
    cell{10}{8}  = {bg=ccbetter},
    cell{10}{9}  = {bg=ccstart},
    cell{10}{10} = {bg=ccbetter},
    cell{10}{11} = {bg=ccgood},
    cell{10}{12} = {bg=ccbetter},
    cell{10}{13} = {bg=ccbetter},
    cell{10}{14} = {bg=ccgood},
    cell{10}{15} = {bg=ccbetter},
    cell{10}{16} = {bg=ccbetter},
    cell{10}{17} = {bg=ccbetter},
    cell{10}{18} = {bg=ccbetter},
    cell{11}{3}  = {bg=ccnone},
    cell{11}{4}  = {bg=ccnone},
    cell{11}{5}  = {bg=ccstart},
    cell{11}{6}  = {bg=ccstart},
    cell{11}{7}  = {bg=ccgood},
    cell{11}{8}  = {bg=ccbetter},
    cell{11}{9}  = {bg=ccstart},
    cell{11}{10} = {bg=ccbetter},
    cell{11}{11} = {bg=ccgood},
    cell{11}{12} = {bg=ccbetter},
    cell{11}{13} = {bg=ccbetter},
    cell{11}{14} = {bg=ccgood},
    cell{11}{15} = {bg=ccbetter},
    cell{11}{16} = {bg=ccgood},
    cell{11}{17} = {bg=ccbetter},
    cell{11}{18} = {bg=ccbetter},
    cell{12}{4}  = {bg=colhr},
    cell{12}{5}  = {bg=collr},
    cell{12}{6}  = {bg=colmr},
    cell{12}{7}  = {bg=collr},
    cell{12}{8}  = {bg=collr},
    cell{12}{9}  = {bg=colhr},
    cell{12}{10} = {bg=collr},
    cell{12}{11} = {bg=colmr},
    cell{12}{12} = {bg=collr},
    cell{12}{13} = {bg=collr},
    cell{12}{14} = {bg=colhr},
    cell{12}{15} = {bg=collr},
    cell{12}{16} = {bg=colmr},
    cell{12}{17} = {bg=collr},
    cell{12}{18} = {bg=collr}
}
    \textbf{\textsf{Science}} & &
    {\textbf{\textsf{No}} \\ \textbf{\textsf{CE}}} &
    \textbf{\textsf{CE with 2G}} & & & & &
    \textbf{\textsf{CE with ET}} & & & & &
    \textbf{\textsf{CE, ET, CE South}} & & & & \\
    \textbf{\textsf{Theme}} & \textbf{\textsf{Goals}} &
    \rotatebox{90}{\textsf{2G}} &
    \rotatebox{90}{\textsf{20}} & \rotatebox{90}{\textsf{40}} & \rotatebox{90}{\textsf{20+20}} & \rotatebox{90}{\textsf{20+40}} & \rotatebox{90}{\textsf{40+40}} &
    \rotatebox{90}{\textsf{20}} & \rotatebox{90}{\textsf{40}} & \rotatebox{90}{\textsf{20+20}} & \rotatebox{90}{\textsf{20+40}} & \rotatebox{90}{\textsf{40+40}} &
    \rotatebox{90}{\textsf{20}} & \rotatebox{90}{\textsf{40}} & \rotatebox{90}{\textsf{20+20}} & \rotatebox{90}{\textsf{20+40}} & \rotatebox{90}{\textsf{40+40}} \\
    {Black holes and \\ neutron stars \\ throughout cosmic \\ time}  & {Black holes from the \\ first stars} & & & & & & & & & & & & & & & & \\
    & Seed black holes & & & & & & & & & & & & & & & & \\
    & {Formation and evolution \\ of compact objects} & & & & & & & & & & & & & & & & \\
    {Dynamics of dense \\ matter}  & {Neutron star structure and \\ composition} & & & & & & & & & & & & & & & & \\
    & {New phases in quantum \\ chromodynamics} & & & & & & & & & & & & & & & & \\
    & {Chemical evolution of the \\ universe} & & & & & & & & & & & & & & & & \\
    & {Gamma-ray burst jet engine} & & & & & & & & & & & & & & & & \\
    {Extreme gravity and \\ fundamental physics} & & & & & & & & & & & & & & & & & \\
    Discovery potential & & & & & & & & & & & & & & & & & \\
    Technical risk & & & & & & & & & & & & & & & & & \\
\end{tblr}

    \caption{%
    \protect%
This table indicates the accessibility of astrophysical sources that can
advance key next-generation science goals. A US Cosmic Explorer consisting of
one \twentykm\  observatory, one \fortykm\  observatory, or a pair of observatories of 20
or \fortykm\  length are evaluated in the presence a background network that
includes second-generation (2G) gravitational-wave observatories, the EU
Einstein Telescope (ET), and a \twentykm\  Cosmic Explorer-like detector located in
Australia (CE South). For each goal, the colors range from gray (least
favorable, science goal not achieved) to green (good, science achievable)
and dark green (most favorable). 
The high-level conclusions of this comparison are summarized in \cref{box:ex_ts_conclusions} on
the adjacent page and
detailed descriptions of the metrics that
determine the criteria can be found in \cref{subsec:impact_on_science_goals}.
The final row, labeled ``Technical risk'', represents the risk that Cosmic Explorer's
 scientific output will be limited by technology shortfalls;
 light orange is lowest risk, and red is highest risk.
We emphasize that this study is a starting point for community input on Cosmic Explorer.

    \label{ex:flowdown}}
    \vspace*{2em}
\end{table}
\end{spacing}

\newpage

\begin{boxenv}{p}{green!10}{\textwidth}
	\caption{Impact of Different Detector Configurations on Cosmic Explorer's Science Goals.}
	\label{box:ex_ts_conclusions}
	\small
\vspace*{0.5em}
\noindent
\cref{ex:flowdown} compares the accessibility of the astrophysical sources and
observations needed to achieve Cosmic Explorer's Science Goals with different
observatories  of lengths (\twentykm\ or \fortykm) operating in different
configurations of the global detector network (existing 2G observatories,
 the Einstein Telescope and an Australian detector CE South). \\

\noindent
For each science goal, a gray box
indicates that a science goal is not achieved. Green
indicates that a science goal will be achieved, with dark green indicating the
most favorable configuration for maximizing science from that goal. Yellow
indicates that a configuration is unfavorable for achieving a science goal.\\

\noindent
Several high-level conclusions can be drawn from this comparison:
\begin{itemize}
\item Cosmic Explorer's science goals cannot be accomplished with 
 second-generation detectors. 
\item Two U.S.~Cosmic Explorer observatories  of \fortykm\ length, or one observatory of
\fortykm\ \emph{and} a second observatory of \twentykm\ length can achieve all 
science goals without relying on the construction of other next-generation
observatories.
\item If only one Cosmic Explorer observatory is constructed in the U.S.,
achieving many of Cosmic Explorer's science goals will require the construction
of the Einstein Telescope (ET) in Europe, or a second Cosmic Explorer observatory
elsewhere in the world (e.g, CE South).
This is due to reduction in the network's ability to measure the distance,
 inclination, and sky location of sources.
\item Studying new phases of quantum chromodynamics through the hot,
post-merger signatures of binary neutron stars is most economically achieved with a global network that
contains a \twentykm\ Cosmic Explorer observatory for optimizing high-frequency sensitivity. This
observatory could be in the U.S., in Australia (CE South), or elsewhere.
\item  Studies of extreme gravity and fundamental physics with gravitational waves are best done with
the sensitivity provided by a \fortykm\ Cosmic Explorer observatory.
\item In the absence of other next-generation observatories, a Cosmic Explorer
consisting of two \fortykm\  observatories is the most favorable configuration
for all goals except those involving post-merger physics of neutron stars.
\end{itemize}

\end{boxenv}

\newpage

\chapter{Purpose and Scope \hlight{[Matt]}}
\xlabel{purpose}

The LIGO observatories are continuing to extend their astrophysical reach
 into new discovery space,
 but in the coming decade they will reach the limits imposed by their facility size and lifetime.
These observatories will be replaced by a new generation of gravitational-wave observatories,
 known as third-generation (3G) or next-generation observatories,
 with longer baselines and new infrastructures.
The international community's vision for next-generation science is detailed in white papers published by the Gravitational Wave International Committee~\autocite{GWIC3GDocs},
 and plans toward a next-generation gravitational-wave observatory in Europe,
 Einstein Telescope, are well underway~\autocite{ETDesign2020}.

This Horizon Study is part of the development stage of a major-facility project~\autocite{NSFMFG}, the US-based next generation gravitational-wave observatory known as Cosmic Explorer (CE).
The purpose of this document is to provide a clear vision of the science enabled by CE,
  a reference concept for the CE instrument and its evolution,
  and initial cost estimates for its construction and operation.
It is intended to inform the scientific community, and the agencies which fund that community,
 with the goal of providing a foundation for further development of CE in those communities
 while spurring action toward CE's construction.
This document,
 together with reports from the Gravitational Wave International Committee (GWIC)~\autocite{GWIC3GSynergies},
 will form the point of departure for the process leading to
 the design stage of the Cosmic Explorer Project.\autocite{NSFMFG}

The major science themes that will be addressed by Cosmic Explorer and their associated goals are presented in \cref{sec:science_overview,ss:status,ch:keyquestions}.
\Cref{sec:overview} presents the reference concept for CE based on ground-based laser
interferometric detection technology, as well as a discussion of alternative technologies.

Given that funding for the next generation of detectors must be directed so as to maximize
 the resulting scientific output, this Horizon Study presents a preliminary trade study for CE in \sref{trade}.
This trade study documents the impact of design and funding choices on the
scientific output of CE, and in particular on the key science goals presented in \cref{ch:keyquestions}.
Of particular interest are the overall length of the CE arms, since this is the primary cost driver,
and the possibility of building multiple observatories in the US. The community is encouraged to use this trade study as a launch point to engage in the development of Cosmic Explorer's 
design parameters.

With the key scientific objectives and overview of the CE design in hand,
 the rest of the document focuses on the technology needed to achieve Cosmic Explorer.
The technical design concept, and site and infrastructure requirements are presented in \sref{design}.
A data-management plan, and the human and computational resources needed to deliver open data and multimessenger alerts are presented in \sref{computing}, along with a survey of the broader computing requirements and analysis costs.

The plans presented in this Horizon Study can only be successful with continued
 input from and strong endorsement by the scientific community,
 and early engagement with local communities, including Indigenous Peoples,
 at potential observatory locations.
The vision of Cosmic Explorer as part of diverse local and global communities,
 and its anticipated role as part of the global gravitational-wave network,
 are presented in \sref{global}.

\Sref{project}\ presents a cost estimate for CE construction and operation,
 along with a timeline and management outline for the project.
This includes a discussion of technical and project management risk,
 based on risk management strategies employed by LIGO and other large projects.
The timeline starts with ongoing research and development work,
  and lays out a path to astrophysical observation with CE.
Both Initial and Advanced  LIGO were delivered on time and on budget and CE will benefit greatly 
 from the experience earned through the LIGO project.
The management plan and cost-budget schedule presented here is based upon
 the successful Advanced LIGO management model, taking into account the lessons learned.
Finally, conclusions are reported in \sref{conclusion}.

\setpartpreamble{
    \AddToShipoutPictureBG*{
    \begin{tikzpicture}[remember picture,overlay,inner sep=0]
        \node[anchor=north] at ([yshift=0.01\paperheight]current page.north)
            {\includegraphics[height=1.05\paperheight]{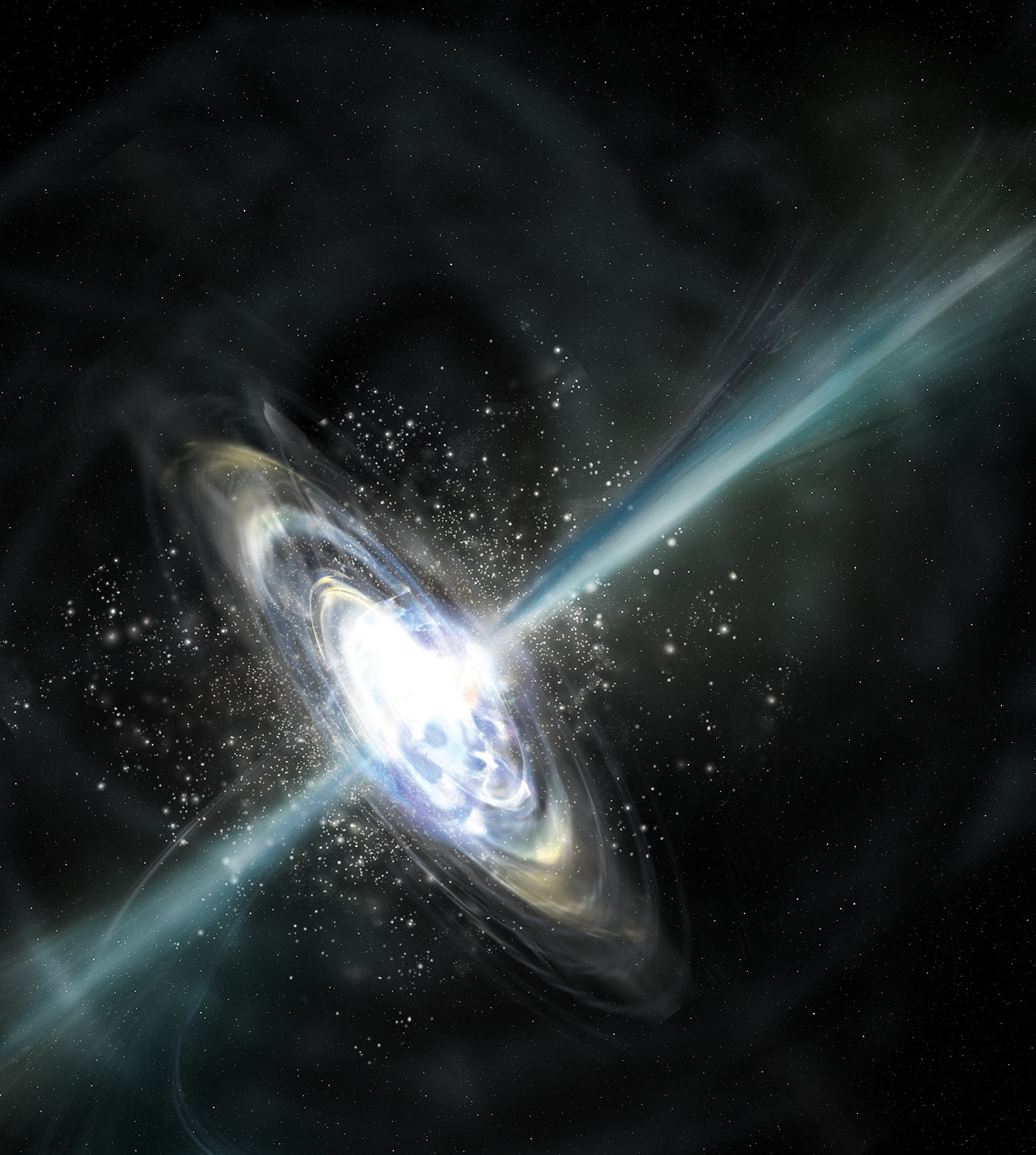}};
        \node[align=left,text=white,inner sep=6pt,anchor=south east] at (current page.south east) {Credit: Aurore Simonnet, Sonoma State University};
        \filldraw[draw=black,fill=black,opacity=0.3] (current page.south west) rectangle (current page.north east);
    \end{tikzpicture}
    }
}
\addpart{Science Objectives}

\newcommand{\NUMBNS}{10^5} %
\newcommand{\BNSHORIZON}{4} %
\newcommand{\NUMLOUDBNS}{10} %
\newcommand{\SNRLOUDBNS}{300} %
\newcommand{\NSRADIUSERR}{0.1} %
\newcommand{\NUMPMO}{100} %
\newcommand{\NUMBNSLOC}{20} %
\newcommand{\BNSLOC}{0.1} %
\newcommand{\BNSCOMPLETEZ}{1} %

\chapter{Overview \hlight{[Josh]}}
\xlabel{science_overview}

\begin{figure}[ht]
    \begin{center}
\includegraphics[width=\textwidth]{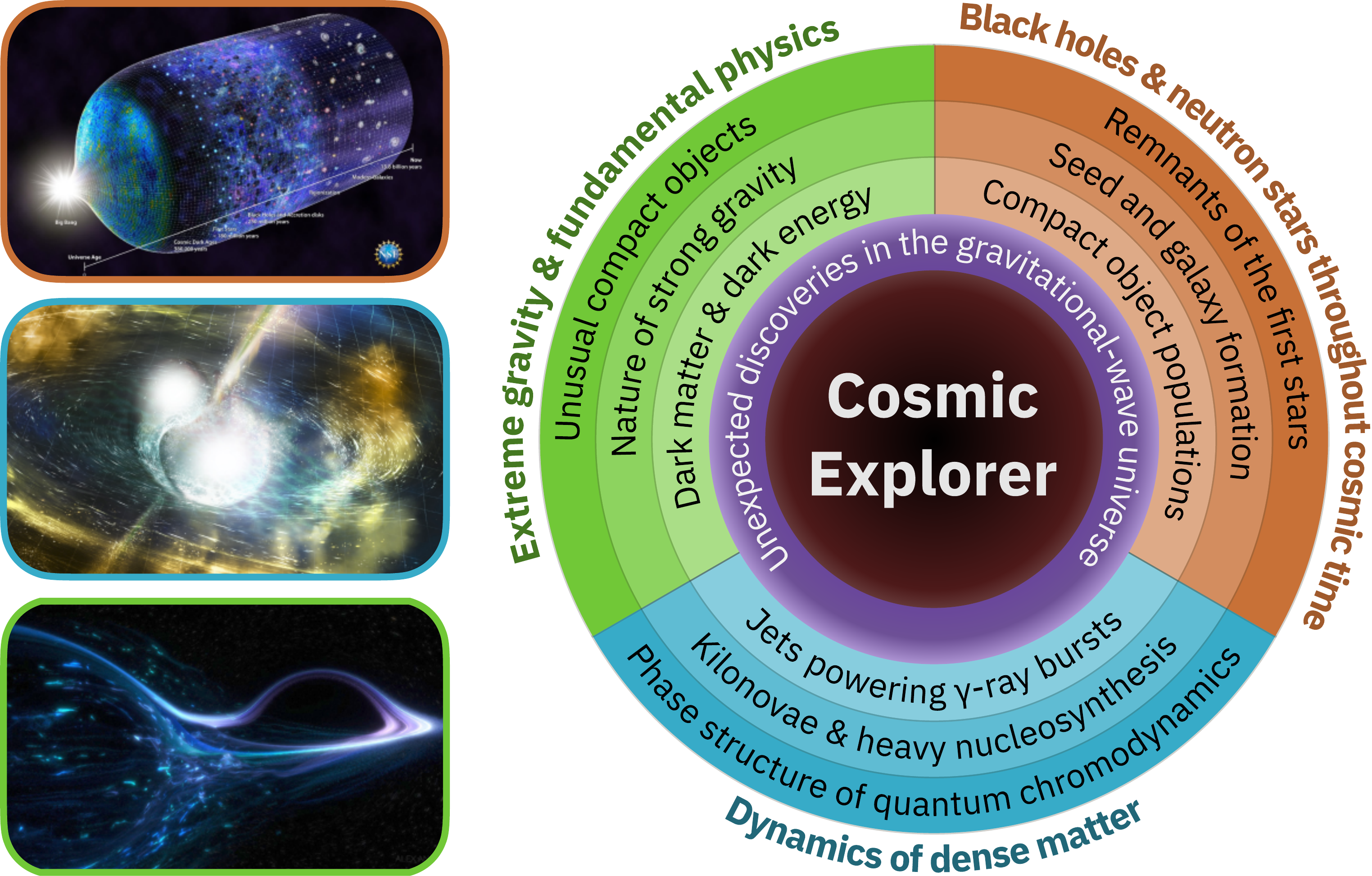}
    \end{center}
\caption{\label{fig:science_dial}Central science themes and objectives that will be addressed by Cosmic Explorer. Cosmic Explorer's greatly increased sensitivity over today's detectors provides access to significantly more sources, spread out over cosmic time, as well as high-fidelity measurements of strong, nearby sources. \cref{ch:keyquestions} provides a more detailed description of the science enabled by Cosmic Explorer. Descriptions and credits for images to left, from top to bottom: A timeline of the universe, N.R.Fuller, National Science Foundation; Merging neutron stars, Aurore Simonnet, Sonoma State University; Black hole and mystery object, Alex Andrix, independent artist and Virgo/EGO.}
\end{figure}

The gravitational-wave discoveries by Advanced LIGO and Advanced Virgo have opened a new window on the universe. There is significant international interest in and mobilization toward developing a next generation of ground-based gravitational-wave observatories capable of observing gravitational waves throughout the history of star formation and exploring the workings of gravity at its most turbulent and extreme. Broad and detailed community studies of the potential for a network of such observatories (and its synergy with other types of gravitational-wave observatories and electromagnetic and astro-particle observatories) have been organized by the Gravitational-Wave International Committee (GWIC) and summarized in a series of white papers~\autocite{GWIC3GDocs}.
The science case for a next-generation network, extensively described in the GWIC 3G Science Case Report along with a series of 2020 Astro Decadal Survey white papers\autocite{Sathyaprakash:2019rom,Kalogera:2019bdd,Kalogera:2019sui,Sathyaprakash:2019nnu,Sathyaprakash:2019yqt}, is highly compelling, and the Einstein Telescope team has independently developed a science case for their planned facility\autocite{2020JCAP...03..050M}.

Based on these reports and our interactions with the community,
 we focus in this report on \textbf{three central scientific themes}
 that Cosmic Explorer will address (illustrated in \cref{fig:science_dial}):
\begin{enumerate}
\item \textbf{Black Holes and Neutron Stars Throughout Cosmic Time}, \cref{ss:cosmictime}
\item \textbf{Dynamics of Dense Matter}, \cref{ss:densematter}
\item \textbf{Extreme Gravity and Fundamental Physics}, \cref{ss:xg}
\end{enumerate}
Additionally, we discuss the broad and deep discovery aperture of Cosmic Explorer and its potential for observing unexpected phenomena, \cref{ss:discoverypotential}.

\begin{figure}[ht]
    \centering
    \begin{tikzpicture}[remember picture,overlay,inner sep=-2pt]
        \node[anchor=north west] at ([yshift=0.05\paperheight]current page.north west) {\includegraphics[height=1.20\paperheight]{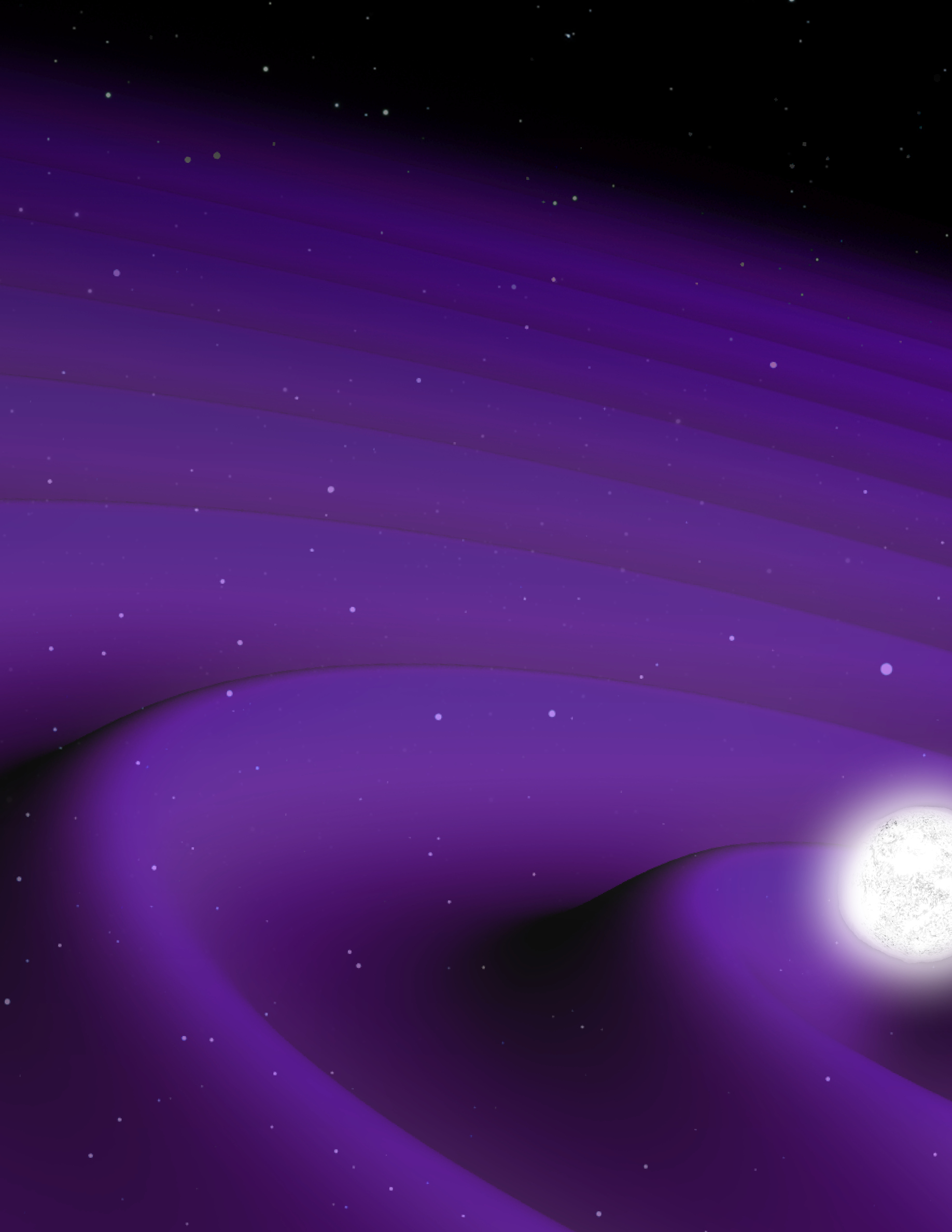}};
        \filldraw[draw=black,fill=black,opacity=0.5] (current page.south west) rectangle (current page.north east);
        \node[anchor=center] at ([yshift=0.1\paperheight]current page.center)
            {\includegraphics[width=0.95\paperwidth]{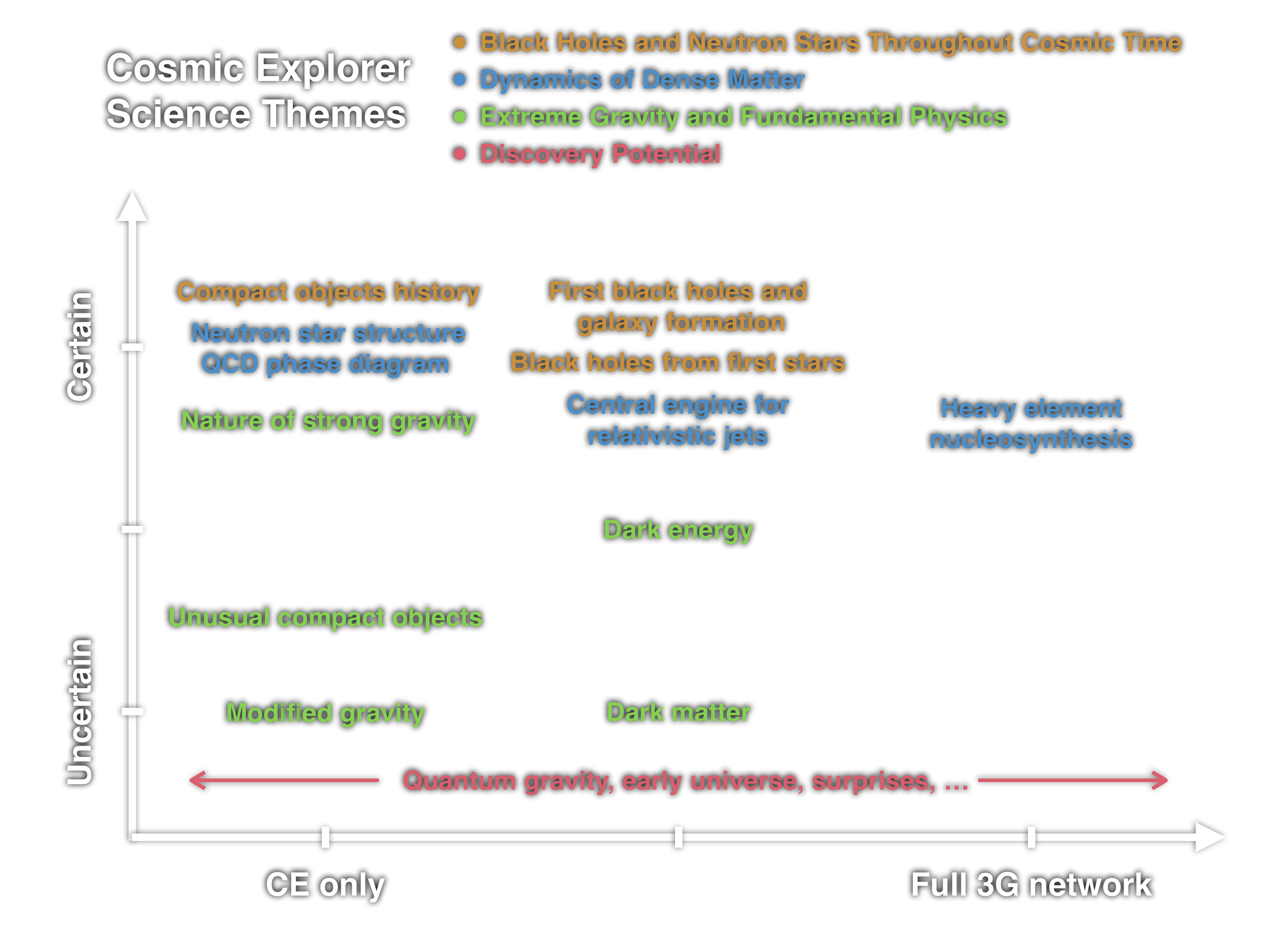}};
    \end{tikzpicture}
    \begin{minipage}{\textwidth}
        \vspace{6in}
    \end{minipage}
    \addtokomafont{captionlabel}{\color{SlateGray3}}
    \addtokomafont{caption}{\color{white}}
    \hypersetup{citecolor=SkyBlue1,linkcolor=SkyBlue1,urlcolor=DarkSeaGreen1}
    \caption{The key science objectives for Cosmic Explorer's central themes, located on axes that qualitatively assess the likelihood of observing gravitational waves related to those objectives, ranging from certain to uncertain, and the number of detectors required to achieve those objectives, ranging from Cosmic Explorer alone (two sites, as described in \cref{box:concept}), through Cosmic Explorer with a second-generation network, to Cosmic Explorer as part of a full third-generation network. The ``Discovery Potential'' objectives in pink are uncertain and span the space from alone to requiring a full network. (Background image is credit ESA, \href{https://creativecommons.org/licenses/by-sa/3.0/igo/}{CC BY-SA 3.0 IGO}, and was modified.)}
    \label{fig:opportunities}
\end{figure}

Associated with each scientific theme are a number of key objectives. Some of these are achievable by a pair of Cosmic Explorer observatories operating on their own, while others will require an additional observatory or a network to achieve. Similarly, some objectives are certain, based on Cosmic Explorer's expected performance and what we have already learned about gravitational-wave science, while others are not certain but would provide extraordinary or even revolutionary outcomes. \cref{fig:opportunities} shows the key science objectives for Cosmic Explorer arranged on axes corresponding to these considerations.

\begin{figure}[ht]
    \centering
    \hspace*{-0.53cm}\includegraphics[width=0.68\textwidth]{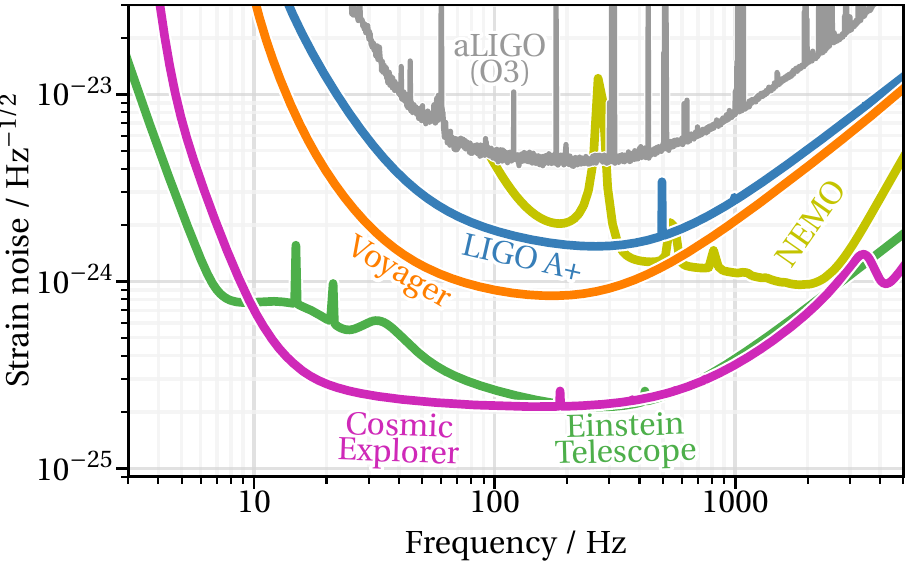}
    \includegraphics[width=0.705\textwidth]{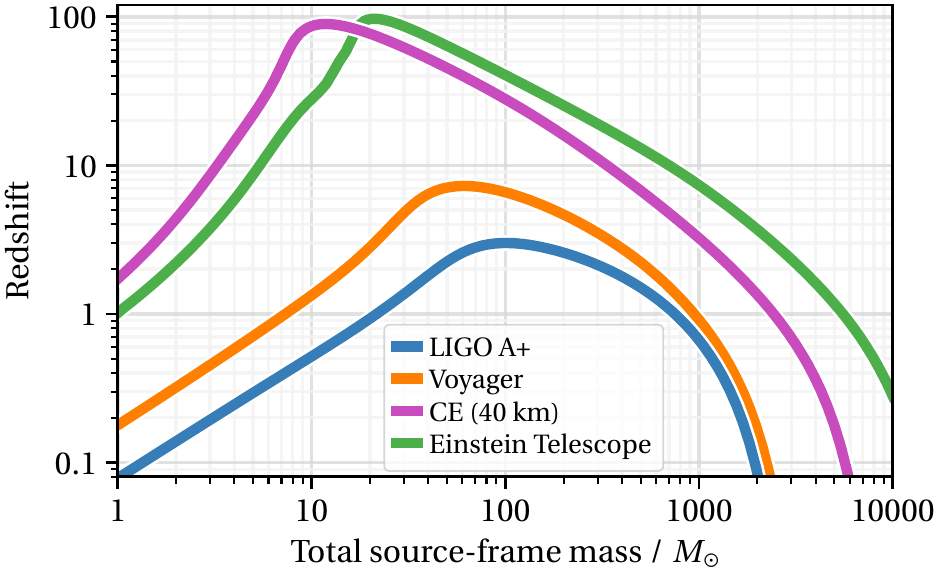}
    \caption{\emph{Top:} Amplitude spectral densities of detector noise for Cosmic Explorer (CE), the current (O3) and upgraded (A+) sensitivities of Advanced LIGO, LIGO Voyager, NEMO, and the three paired detectors of the triangular Einstein Telescope (see \cref{ss:status} for observatory descriptions). At each frequency the noise is referred to the strain produced by a source with optimal orientation and polarization. \emph{Bottom:} Maximum redshift (vertical axis) at which an equal-mass binary of given source-frame total mass (horizontal axis) can be observed with a signal-to-noise ratio of 8~\autocite{2021CQGra..38e5010C}. Different curves represent different detectors. For binary neutron stars (total mass ${\sim} 3 M_\odot$), CE will give access to redshifts larger than 1, where most of the mergers are expected to happen. For binary black holes, it will enable the exploration of redshifts of 10 and above, where mergers of black holes formed by either the first stellar population in the universe (Pop III stars) or by quantum fluctuations shortly after the Big Bang (primordial black holes) might be found.} %
    \label{fig:noise_curves}
\end{figure}

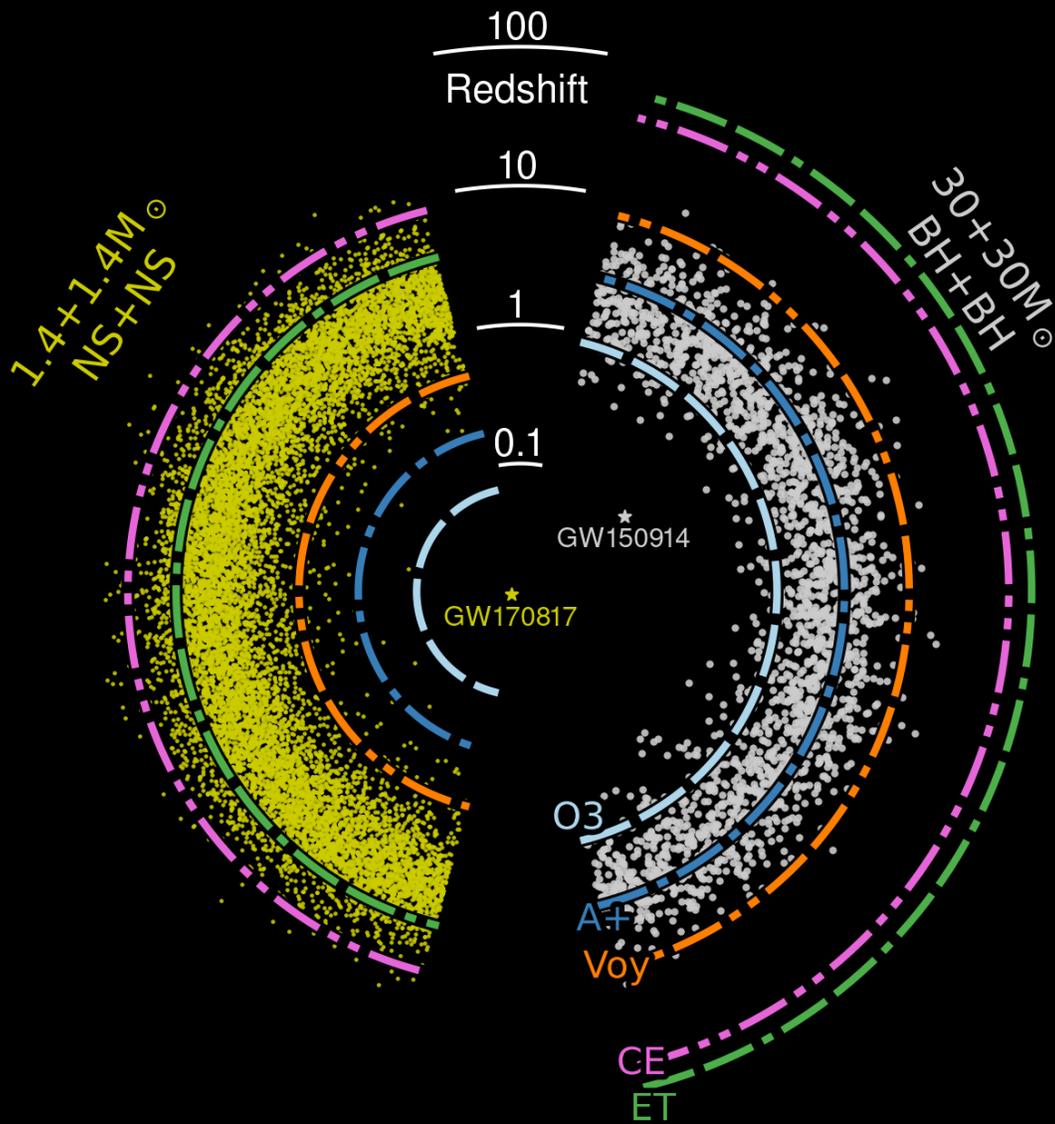
\begin{figure}[p]
    \begin{tikzpicture}[remember picture,overlay]
        \filldraw[draw=black,fill=black] (current page.south west) rectangle (current page.north east);
    \end{tikzpicture}    
    \centering
    \includegraphics[width=\textwidth]{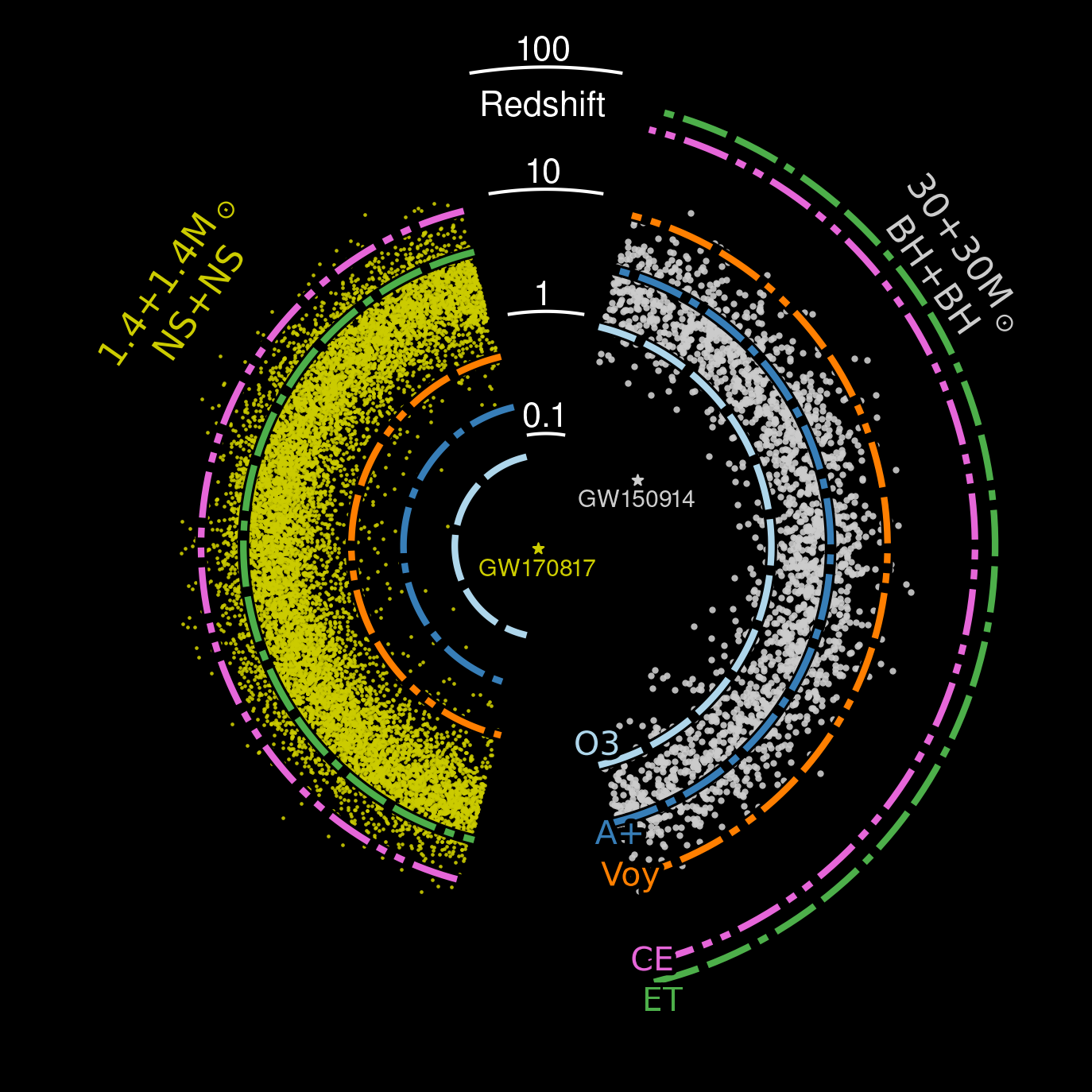}
    \addtokomafont{captionlabel}{\color{SlateGray3}}
    \addtokomafont{caption}{\color{white}}
    \hypersetup{citecolor=SkyBlue1,linkcolor=SkyBlue1,urlcolor=DarkSeaGreen1}
    \caption{
Astrophysical horizon of current and proposed future detectors for compact
binary systems. As in the bottom of \cref{fig:noise_curves}, the lines indicate
the maximum redshift at which a detection with signal-to-noise ratio 8 could be
made.  The detectors shown here are Advanced LIGO during its third observing
run (``O3''), Advanced LIGO at its anticipated sensitivity for the fifth
observing run (``A+''), a possible cryogenic upgrade of LIGO called Voyager
(``Voy''), the Einstein Telescope (``ET''), and Cosmic Explorer (``CE'',
 see \cref{ss:status} for observatory descriptions). The
yellow and white dots are for a simulated population of binary neutron star
mergers and binary black hole mergers, respectively, following
\textcite{Madau:2014bja} with a characteristic binary merger time of 100
million years.
    }
    \label{fig:donut}
\end{figure}

A ``trade study'' of how effectively different variants of Cosmic Explorer could address these science themes, along with a resulting science-driven design concept for Cosmic Explorer, are presented later in \cref{sec:trade} and \cref{sec:design}.
A science traceability matrix describes the measurements required to accomplish the science goals, and the instrument requirements needed to achieve these measurements.
This study relies on knowledge that was not available when today's detectors were designed\,---\,e.g., event rates and observations of gravitational-wave sources. In contrast, Cosmic Explorer's ultimate design will be informed by what the existing observatories have discovered, and will continue to discover, about the gravitational-wave universe.
The science traceability matrix reveals that addressing Cosmic Explorer's science goals will require detectors with a strain sensitivity of \SI{6e-25}{\big/\!\sqrt{\Hz}} at \SI{10}{\Hz} and \SI{2e-25}{\big/\!\sqrt{\Hz}} at \SI{100}{\Hz}; compared to Advanced LIGO, this an order of magnitude sensitivity improvement at \SI{100}{\Hz}, and opens a new low-frequency gravitational-wave band for observation. 
This leap in sensitivity between the 2G and 3G detectors, shown in \cref{fig:noise_curves} and \cref{fig:donut}, will expand humanity's gravitational-wave access from the first nearby discoveries to the majority of stellar-mass black hole and neutron star coalescences in the universe. The current status of the 2G detectors and path toward the next generation is outlined below in \cref{ss:status}.


\chapter{Status of Ground-Based Gravitational-Wave Observatories \hlight{[Josh]}}
\label{ss:status}

A century after Einstein's prediction of gravitational waves, Advanced LIGO~\autocite{LIGOScientific:2014pky} debuted this new ``sense'' for humanity by observing coalescing binary systems of black holes~\autocite{LIGOScientific:2016aoc,LIGOScientific:2016sjg,LIGOScientific:2016dsl} with up to tens of solar masses~\autocite{LIGOScientific:2016kwr} and enabling tests of relativity in the strong gravity regime~\autocite{LIGOScientific:2016lio}. In 2017, LIGO and its European partner Virgo~\autocite{VIRGO:2014yos} inaugurated an era of gravitational-wave multimessenger astronomy by discovering a binary neutron star merger with a gamma-ray-burst counterpart~\autocite{LIGOScientific:2017vwq}, and rapidly sharing its source location with the broader astronomical community, triggering observations of the system across the electromagnetic spectrum~\autocite{LIGOScientific:2017ync}. The scientific goldmine opened by these observations has connected binary neutron star mergers to short gamma-ray bursts, equated the speed of gravity to very precisely that of light~\autocite{LIGOScientific:2017zic}, provided a gravitational-wave measure of the local Hubble constant~\autocite{LIGOScientific:2017adf}, gave strong clues about the origin of heavy elements~\autocite{2017Sci...358.1570D}, and probed the properties of ultra-dense nuclear matter~\autocite{LIGOScientific:2017vwq,2018PhRvL.121i1102D}.

These successes of Advanced LIGO were the result of forty years of international
collaboration and research and development, from prototypes of increasing scale to the first generation LIGO detectors. These efforts delivered the technology and engineering needed to build the second-generation detectors. They also trained the early-career scientists and engineers who envisioned, built, and are operating these observatories, and fostered a community that is searching for gravitational-wave sources and extracting astrophysical information from these observations.

The LIGO and Virgo observatories have continued to increase their reach and discovery rate, revealing populations of astrophysical events~\autocite{2019PhRvX...9c1040A, 2019ApJ...872..195N,  2020ApJ...891..123N,  2020PhRvD.101h3030V} and routinely issuing alerts to the broader astronomical community~\autocite{LIGOScientific:2019gag}. At its 2020 sensitivity, this network was reporting observations of tens of black hole mergers and of order one merger involving a neutron star per year~\autocite{LIGOScientific:2020kqk}.
At the time of writing, the 2G observatory network is being strengthened by the Japanese KAGRA observatory~\autocite{KAGRA:2018plz} and by an enhancement to Advanced LIGO known as A+~\autocite{Miller:2014kma}.
A planned joint US--India detector, LIGO India~\autocite{2017CSci..113..672S}, has selected and acquired a site and is awaiting approval from the Indian government.
Additionally, progress has been made toward ``Voyager'' technology to possibly maximize the potential of the existing LIGO facilities by implementing cryogenic silicon optics and suspensions and reducing quantum and Newtonian (gravity gradient) noise~\autocite{2020CQGra..37p5003A, 2017Cryo...81...83S}.
Options for lasers, photodiodes, and electro-optics for Voyager's planned \SI{2}{\um} operating wavelength, as well as cryogenic engineering solutions to cool the suspended optics, have been identified and will soon be tested in the Caltech 40-m prototype~\autocite{Mariner} and the European ETpathfinder prototype~\autocite{etpathfinder}.
LIGO and its partners are providing dramatic and deepening insights into the populations and lifetimes of black holes and the inner workings of neutron stars. However, these \SIrange{3}{4}{\km} detectors can offer only a glimpse of the full gravitational-wave universe.

Three 3G observatory concepts have emerged. In Europe, plans have solidified for the Einstein Telescope or ET.
The first ET Design Report was published in 2011 and updated in 2020~\autocite{ETDesign2020}. It describes a single-site observatory with \SI{10}{\km}-long arms located \SIrange{200}{300}{\meter} underground to reduce seismic motion and Newtonian noise, with low-frequency detectors operated at cryogenic temperatures (\SIrange{10}{20}{\kelvin}) to reduce thermal noise and high-frequency detectors using very high laser power and frequency-dependent squeezed light to reduce the impact of the laser light's quantum noise.
In 2021, Einstein Telescope was included in the roadmap of the European Strategic Forum for Research Infrastructures~\autocite{ESFRI}.
ET will enter a preparatory phase with construction possible by 2026 and observations by 2035~\autocite{ETweb}.
An Australian detector concept, NEMO~\autocite{2020PASA...37...47A}, is a \SI{4}{\km}-baseline detector targeting excellent sensitivity in the high frequency band (\SIrange{1}{3}{\kHz}) associated with gravitational-waves from the postmerger phase of neutron stars.
Cosmic Explorer~\autocite{Reitze:2019iox, LIGOScientific:2016wof} is the planned United States contribution to the next-generation gravitational-wave observatory network and the focus of this study. 
These observatories will have strong synergy with space-based gravitational-wave observatories, astro-particle detectors, and telescopes across the electromagnetic spectrum.

With that introduction to the detectors, we now turn to a discussion of the scientific opportunities of broad interest that will be opened by Cosmic Explorer.

\chapter{Key Science Questions}
\label{ch:keyquestions}

The key science questions to be addressed by Cosmic Explorer are presented below. The gravitational-wave spectrum is rich with sources, and more scientific opportunities will be explored by the third generation network than described here. This selection was made to focus on the most compelling science that will be accessible with Cosmic Explorer.

\section{Black Holes and Neutron Stars Throughout Cosmic Time \hlight{[Salvo]}}
\label{ss:cosmictime}

\begin{boxenv}{b}{orange!10}{\textwidth}
\caption{Key Science Question 1}
\label{box:cosmictime}
\bigskip
{\centering
\callout{How have the populations of black holes and neutron stars \linebreak evolved over the history of the universe?} \\}
\bigskip
Cosmic Explorer will detect gravitational waves from black holes and neutron stars in binaries to redshifts of ${\sim}10$ and above, allowing us to:
\begin{itemize}[leftmargin=*,itemsep=1pt,topsep=2pt,parsep=1pt]
\item Shed light on Population III stars through the black holes they might have left behind;
\item Measure the properties of the first black holes and their role in forming supermassive black holes and  galaxies;
\item Characterize the populations of compact objects and their evolution.
\end{itemize}
\end{boxenv}

Cosmic Explorer can detect stellar-mass black hole mergers from when the universe was less than 500~million years old. This immense reach will reveal for the first time the complete population of stellar-mass black holes in binaries, starting from an epoch when the universe was still assembling its first stars. 
Cosmic Explorer will detect hundreds of thousands of black-hole mergers each year, measuring their distances, masses, and spins.
These observations will reveal the black-hole merger rate, the underlying star formation rate, how both have changed throughout cosmic time, and how both are correlated with galaxy evolution.

\subsection{Remnants of the First Stars}

The first stars formed when the universe was only a few hundred million years old.
With no previous generation of stars to process the primordial gases of the universe,
 these stars, known as Population III or ``Pop III'' stars,
 were almost entirely composed of hydrogen and helium~\autocite{1996ApJ...472L..63O}.
Due to their pristine composition, they are believed to have been extremely massive,
 with masses more than a hundred times that of the sun~\autocite{2002ApJ...564...23B}.
Despite intensive observational efforts, to date there are no claims of detection of Pop III stars. Since the rate at which stars burn their nuclear fuel increases dramatically with their mass, it is plausible that no Pop III stars emitted light long enough to be observed in the local universe. 
The James Webb Space Telescope~\autocite{JWST} should be able to observe the very first galaxies in the universe, containing only Pop III stars~\autocite{2012MNRAS.427.2212Z}; though it will be able to and study their role in the epoch of reionization, it will not be able to observe individual stars from this epoch~\autocite{2013MNRAS.429.3658R}.
 
 After burning their nuclear fuel, Pop III stars, like other stars, may collapse to form compact objects
 such as black holes. If the mass of Pop III stars is above \simt 230 \Msun, they should not trigger a pair instability supernova\,---\,an explosion that entirely destroys the star, leaving no compact object behind\,---\,but instead directly collapse into a black hole with minimal mass loss. Less massive stars might leave behind black holes below the pair-instability supernova mass gap, i.e., with masses up to \simt 60 \Msun~\autocite{1995ApJS..101..181W,2002A&A...382...28S,2002ApJ...567..532H,2021MNRAS.501.4514C}.
Depending on the initial mass distribution (mass function) of Pop III stars,
 we expect to find black holes in the early universe with masses of tens to hundreds of solar masses,
 possibly with a mass-gap between \simt 60 and \simt 150~\Msol due to pair instability supernovae~\autocite{2014ApJ...792...44C}. Detecting and characterizing the black holes generated by Pop III stars can thus be a powerful method to study the properties of their progenitor stars, including their masses and composition. Knowledge of the mass function of the first generation of stars in the universe could change our understanding of how galaxies formed (see next subsection).

Unfortunately, both existing X-ray telescopes, and those currently proposed for the 2030s (e.g., the Lynx~\autocite{Lynx} and Athena~\autocite{Athena} X-ray observatories) would only be able to detect and measure the mass of supermassive black holes, leaving totally unexplored a region of the mass spectrum that is likely to have been populated by Pop III stars. The LISA gravitational-wave detector~\autocite{LISA:2017pwj} will access stellar-mass black holes in extreme mass ratio inspirals up to redshifts of ${\sim} 5$~\autocite{2017PhRvD..95j3012B,2017JPhCS.840a2021G}, and  IMBHs up to redshifts of ${\gtrsim} 10$. By contrast, Cosmic Explorer can detect stellar-mass black holes to redshifts beyond $10$, if they merge in binary systems. Gravitational-wave detectors in the next decade would thus be able to probe a wide range of possible Pop III stars remnants, with Cosmic Explorer targeting the stellar-mass region. 

\subsection{Seed Black Holes and Galaxy Formation}

Supermassive black holes (SMBHs), with masses of millions to billions of solar masses, are known to exist at the center of most galaxies, significantly impacting the evolution, energetics, and dynamics of their host galaxies~\autocite{Ferrarese:2000se,Kormendy:2013dxa}. The study of galaxy formation is thus intimately related to understanding how and when the central black holes formed\,---\,an area of extremely intense research.

A key open question is: how did supermassive black holes form so early in the history of the universe? Compelling evidence shows that SMBHs of billions of solar masses already existed at redshift of ${\gtrsim} 7.6$~\autocite{2021ApJ...907L...1W}, when the universe was only 670 million years old. The relatively short time scale over which SMBHs were produced challenges our understanding of how black holes form and grow. The two main scenarios suggested to explain the presence of high-redshift SMBHs are
(1) direct collapse of hydrogen clouds into black holes, followed by gas accretion, and
(2) repeated mergers of smaller black holes through gravitational runaway processes~\autocite{2020ARA&A..58..257G}.

If the first black holes in the universe were the remnants of the first generation of stars, as mentioned above they could have masses up to a few hundred solar masses. Repeated mergers of black holes, starting from tens or hundreds of solar masses, passing through the regime of intermediate-mass black holes (IMBHs, in the mass range ${\sim}[10^2,10^5]M_\odot$), could eventually result in SMBHs of billions of solar masses. LIGO and Virgo have detected gravitational waves from the merger of two heavy black holes, GW190521, which gave birth to a black hole of \simt 150 \Msun at redshift of \simt 0.8.
The merger product might represent the first-ever detection of an IMBH, though a light one. Advanced detectors will observe more of these systems in the next few years, and start measuring their merger rate and mass distribution. Advanced detectors or their upgrades (\cref{fig:noise_curves} bottom), however, will not be able to detect a 100$-$100~\Msun IMBH at redshifts larger than 3. %
By contrast, Cosmic Explorer will be able to observe them to redshifts larger than 10. Measuring the mass function and merger rate of heavy stellar-mass and IMBHs at those redshifts would directly illuminate their role in the formation of SMBHs.
{Some heavy BBH or IMBH sources could be detected both by LISA and\,---\,months to years later\,---\,by ground-based detectors~\autocite{2016PhRvL.116w1102S}, which would significantly improve the estimates of all of the source parameters~\autocite{2016PhRvL.117e1102V}, including enabling more stringent tests of general relativity~\autocite{2016PhRvL.116x1104B}. The exciting prospects of gravitational-wave multi-banding have recently been discussed elsewhere.~\autocite{2019BAAS...51c.109C}}

\subsection{Formation and Evolution of Compact Objects}

LIGO and Virgo have detected black holes with masses as light as 2.5~\Msun (if indeed the lighter component of GW190814 is a black hole~\autocite{2020ApJ...896L..44A}) and as heavy as 150~\Msun~\autocite{2020PhRvL.125j1102A}. While most of the black hole binaries seem to have small spins (or, more precisely, a small projection of the mass-weighted total spin along the orbital angular momentum, $\chi_\mathrm{eff}$~\autocite{Damour:2001tu,Racine:2008qv}), some systems show large spins, or spins which are misaligned with the orbit~\autocite{LIGOScientific:2020ibl}. It is unlikely that this variety of parameters can be the result of a single astrophysical formation channel~\autocite{LIGOScientific:2020kqk,Zevin:2020gbd}. 
In fact, different formation scenarios are expected to result in different distributions for the 
masses and spins of the black holes (see Refs.~\cite{2018arXiv180909130M,2018arXiv180605820M,2021arXiv210600699M} for recent reviews).
The two channels which are usually expected to produce most black hole binaries are isolated formation in galactic fields~\autocite{1976IAUS...73...35V,1976IAUS...73...75P,1976ApJ...207..574S,1993MNRAS.260..675T,Dominik:2012kk}  and dynamical encounters in dense environments (e.g., globular clusters~\autocite{1993Natur.364..421K,1993Natur.364..423S,2000ApJ...528L..17P,Rodriguez:2016kxx,Rastello:2018elx}, nuclear star clusters, and the disks of active galactic nuclei~\autocite{Bartos:2016dgn,Graham:2020gwr}). Dynamical formation may produce heavier black holes than isolated binary evolution due to the possibility of repeated mergers~\autocite{Gerosa:2017kvu,Fishbach:2017dwv,Chatziioannou:2019dsz,Kimball:2020qyd}. Other parameters of the environment, e.g., the metallicity, also strongly affect the mass of the resulting compact objects~\autocite{Belczynski:2001uc}. Similarly, the magnitude and orientation of the spins in the binary will depend on where the system formed, and on details of the supernovae explosions that created the black holes~\autocite{OShaughnessy:2017eks}.
Gravitational-wave measurements of these parameters can aid in understanding where black hole binaries formed, providing precious clues about the evolution and properties of their environments, and thus of galaxies and their surroundings.

While LIGO and Virgo have opened up this new frontier of observational astrophysics, they are also intrinsically limited in their sensitivity. Even in the A+ configuration, LIGO will not be able to observe black hole binaries at redshifts larger than 3.\footnote{This is for 50$-$50~\Msun binary black holes. For lighter and heavier systems the horizon will be smaller, as seen in \cref{fig:noise_curves}.} While still providing significant understanding of the local universe, they will not be able to say much about the production of black holes or their properties earlier in the history of the universe. Precise measurements of the merger rate and mass distribution of merging binaries across a large range of redshifts will make it possible to probe environmental impacts on compact binary yields, delay times between star formation and merger, and ultimately untangle the formation channels and their uncertain physics. Searches for the stochastic signal produced by unresolved sub-threshold binaries will give hints about their high-redshift distribution, but with large uncertainties (compare, e.g., Refs.~\cite{Callister:2020arv} and~\cite{2020arXiv201209876N}).
Understanding how the rate and sites of production of black hole binaries evolved across the age of the universe will provide hints about, for example, the evolution of the star formation rate, a quantity which is hard to measure at high-redshift using photons~\autocite{Madau:2014bja}.  More importantly, advanced detectors, and the proposed upgrades such as Voyager, would entirely miss black hole mergers at redshifts larger than ${\sim}7$~\autocite{2020arXiv201209876N}. This will preclude probing the formation and merger of black holes created by Pop III stars, as described above, and other possible high-redshift channels, e.g., primordial black holes created during the inflationary epoch of the universe.  It is only with next-generation detectors like CE that we will be able to understand the efficiency and characteristics of each formation channel that generates black hole binaries, anywhere in the observable universe.
 
Similar considerations can be made for neutron stars. The discovery of GW190425~\autocite{2020ApJ...892L...3A}, a binary neutron star system much heavier than known systems in our galaxy suggests that the astrophysical properties of neutron stars might be more diverse than what has been observed in the Milky Way. Next-generation observatories will provide access to neutron star binaries all the way to redshift \simt 10, and give us a clear picture of how their parameters vary across cosmic history and galactic environments.

\section{Dynamics of Dense Matter \hlight{[Phil]}}
\label{ss:densematter}

Neutron stars are made of the densest known matter in the universe and can support incredibly large magnetic fields. The merger of two neutron stars is so cataclysmic that it can produce the brightest electromagnetic emission in the cosmos and trigger the formation of heavy elements like gold and platinum. Six decades after the discovery of neutron stars, we still do not understand how matter behaves at the extreme densities and pressures attained in their cores, or how they can generate magnetic fields a million times stronger than those created on Earth. Cosmic Explorer will detect ${\sim} \NUMBNS$ binary neutron star mergers per year out to a redshift of $\BNSHORIZON$, of which ${\sim} \NUMLOUDBNS$ are expected to have signal-to-noise ratios above $\SNRLOUDBNS$.  Cosmic Explorer
could also detect signals from the 3000 known neutron stars in our
Galaxy. The dense-matter science these observations will enable is highlighted in \cref{box:matter}. %

\begin{boxenv}{t}{blue!10}{\textwidth}
\caption{Key Science Question 2}
\label{box:matter}
\bigskip
{\centering
\callout{How does matter behave under the most extreme conditions in the universe?} \\}
\bigskip
Cosmic Explorer will measure gravitational radiation from binary neutron star coalescences and provide the precise source localizations required for multimessenger astronomy, allowing us to:
\begin{itemize}[leftmargin=*,itemsep=1pt,topsep=2pt,parsep=1pt]
\item Determine the internal structure and composition of neutron stars;
\item Explore new regions in the phase diagram of quantum chromodynamics;
\item Map heavy element nucleosynthesis in the universe through counterpart kilonovae and distant mergers;
\item Reveal the central engine for the highly relativistic jets that power short gamma-ray bursts.
\end{itemize}
\end{boxenv}

\subsection{Neutron Star Structure and Composition}

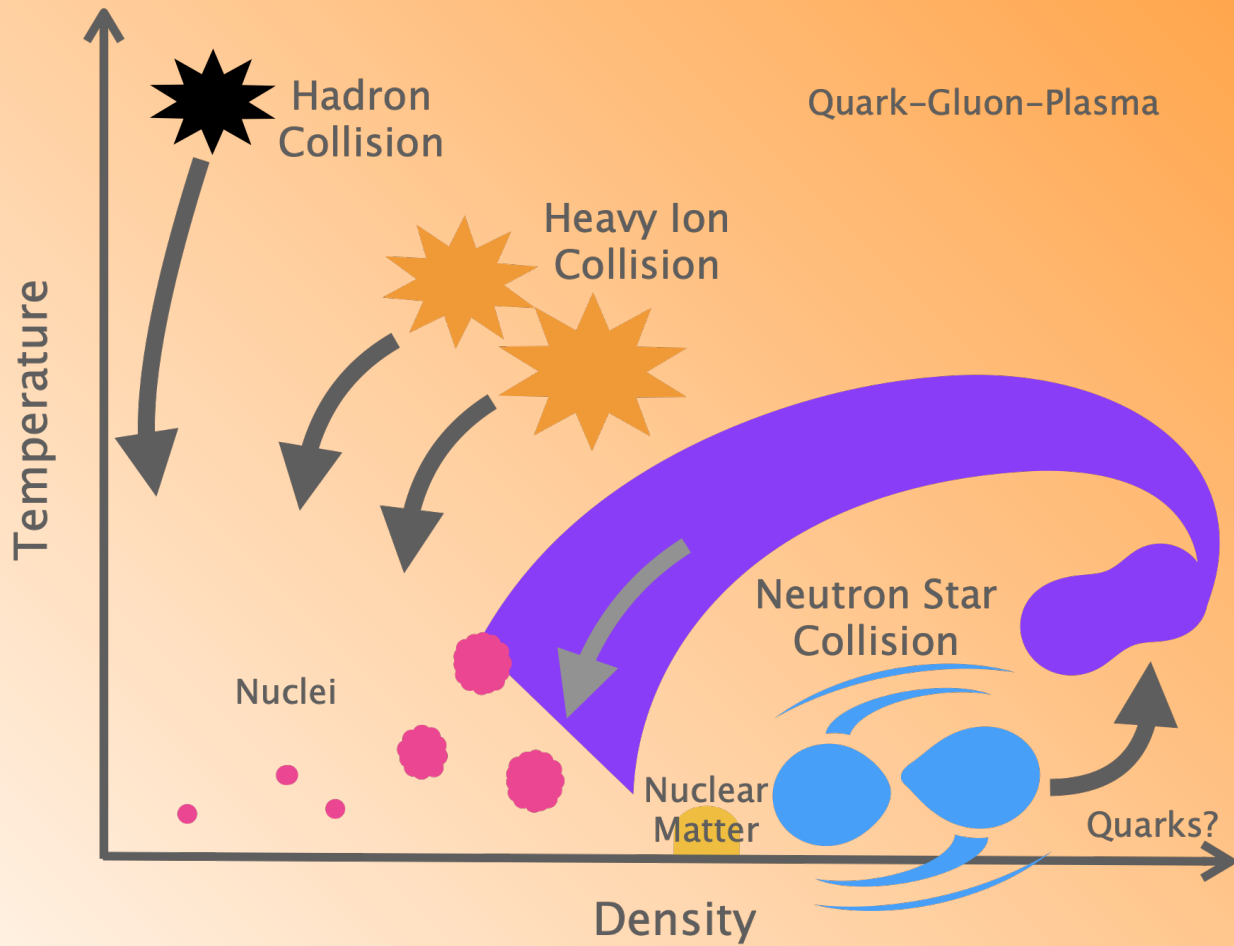
\begin{figure}[p]
    \begin{tikzpicture}[remember picture,overlay,inner sep=-2pt]
    \fill[shading=axis,rectangle,left color=white,right color=white!30!orange,shading angle=135, minimum width=\paperwidth, minimum height=\paperheight] (current page.south west) rectangle (current page.north east);
    \end{tikzpicture}
    \centering
    \includegraphics[width=\textwidth]{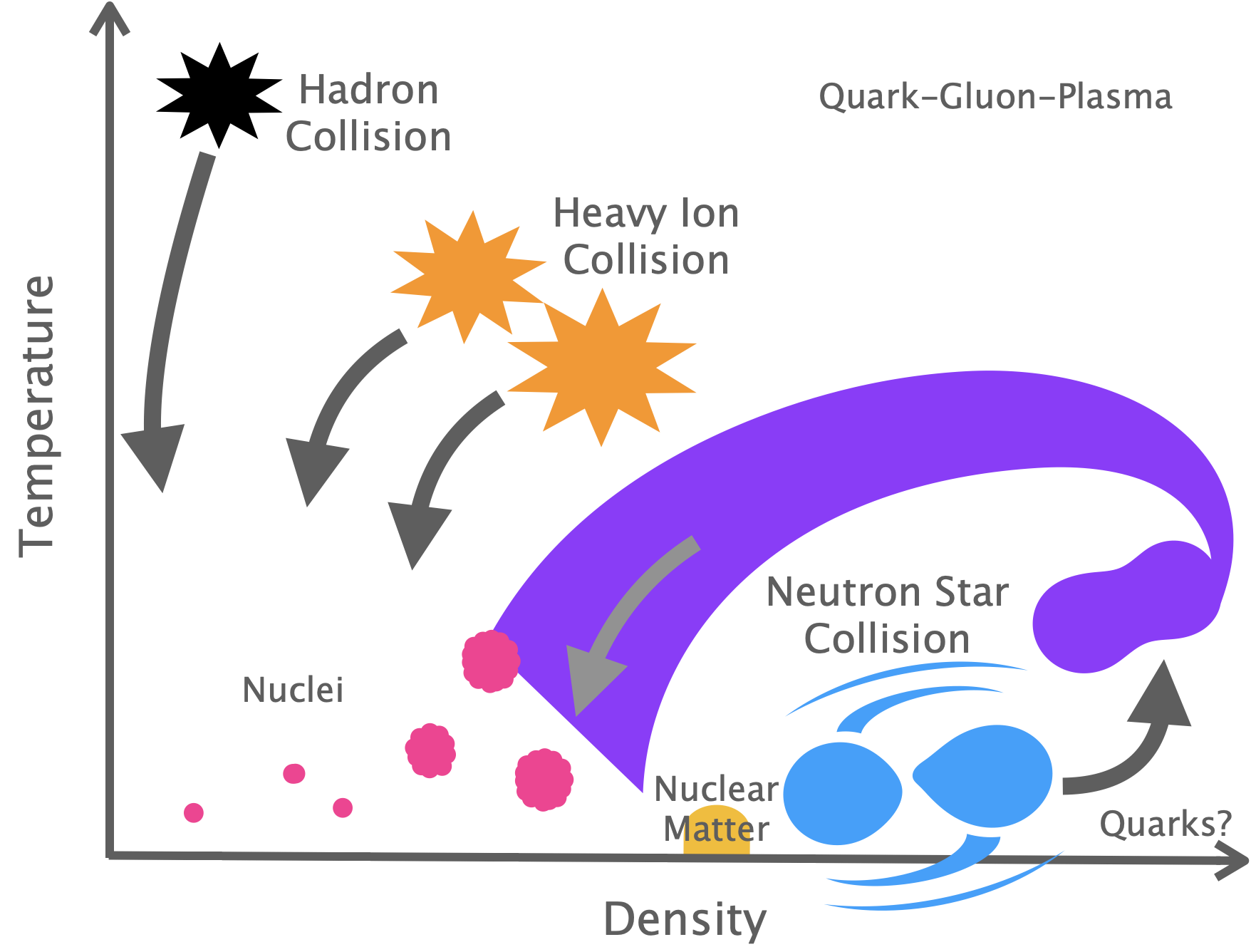}
    \caption{\label{fig:qcd_diagram}Schematic quantum chromodynamic phase diagram. At low densities and temperatures, matter is arranged into nuclei. As the density increases, the nuclei disintegrate into a sea of interacting neutrons and protons, so-called nuclear matter.  Terrestrial collider experiments probe dense nuclear matter at high temperature, as it ``melts'' into a quark-gluon plasma and recombines.
In contrast, the compressed cores of neutron stars hold supranuclear-density matter in a cold neutron-rich equilibrium. At the highest densities, phase transitions involving strange particles or quark matter may occur.  Gravitational-wave observations of neutron stars explore this low-temperature, high-density phase from the inspiral, and a unique high-temperature, high-density region of the phase diagram after the stars collide.  The kilonova counterpart to a neutron star merger traces the r-process nucleosynthesis that transforms neutron-star matter into heavy nuclei.}
\end{figure}

Neutron stars are excellent astrophysical laboratories for ultra-dense matter. Subtle signatures of the stellar interior are encoded in the gravitational waves emitted when neutron stars coalesce~\autocite{2008PhRvD..77b1502F,2010PhRvD..81l3016H,2012PhRvD..85l3007D,2020GReGr..52..109C}, allowing us to probe the fundamental properties and constituents of matter in a phase that is inaccessible to terrestrial experiments~\autocite{2017JPhG...44j4002S}. The matter in neutron star cores is so dense that it cannot be described in terms of individual nucleons. It reaches equilibrium as a neutron-rich fluid and may even transition to deconfined quarks at the highest densities~\autocite{2018RPPh...81e6902B}. The phase structure of dense matter is shown schematically in \cref{fig:qcd_diagram}.

As we enter the Cosmic Explorer era, our understanding of neutron star matter will be informed by current- and near-future observations including gravitational waves~\autocite{2018PhRvL.121p1101A,2019EPJA...55...97T,2020Sci...370.1450D,2019PhRvD.100j3009H, 2020PhRvD.101l3007L}, X-ray spectra~\autocite{2018MNRAS.476.4713S}, bursts~\autocite{2017A&A...608A..31N}, and pulsar profile modeling~\autocite{2019ApJ...887L..24M,2020ApJ...893L..21R,2020ApJ...892...55J,2019arXiv190303035R}. Together, these are expected to yield ${\sim}$\SI{5}{\percent} errors on neutron star radii for dozens of sources.

Cosmic Explorer's precision measurements of tidal deformabilities across the neutron-star mass spectrum will constrain the neutron-star radius better than $\NSRADIUSERR$ km\,---\,one part in 100\,---\,for hundreds of systems per year, revolutionizing our knowledge of the equation of state that characterizes the ultradense matter in inspiraling neutron stars. These observations will enable us to distinguish between competing first-principles models for ultra-dense matter~\autocite{2020PhRvC.102e5803E} and resolve the nature of the phase transition to quark matter~\autocite{2020PhRvD.101d4019C}, if it occurs in neutron-star cores.  Gravitational-wave astroseismology may also emerge from the dynamic excitation of oscillations in the component neutron stars, yielding new insight on their structure and composition~\autocite{1999MNRAS.308..153H,2013ApJ...769..121W,2020NatCo..11.2553P,2020PhRvD.101j4028P,Andersson:2021qdq}, including the dense nuclear physics of transport properties such as viscosities and conductivities.

Cosmic Explorer may also observe gravitational waves from spinning neutron stars; quasi-periodic signals lasting for millions of years~\autocite{Riles:2017evm}. Their emission requires nonaxisymmetry, which could be supported by elastic stresses in the crust, deformations due to magnetic fields, thermal asymmetries or unstable oscillations driven by accretion~\autocite{Andersson:1997xt, Riles:2017evm,Sieniawska:2019hmd}. Cosmic Explorer may reveal if gravitational waves explain the spin-down rates of millisecond pulsars~\autocite{2018ApJ...863L..40W}, and the corresponding ellipticities are within direct detection reach. Gravitational-wave emission could also explain the relatively low spins of accreting neutron stars in low-mass X-ray binaries \autocite{2003Natur.424...42C, 2008AIPC.1068...67C, 2017ApJ...850..106P}. Observing these long-lived gravitational waves would provide additional insight into the formation and thermal, spin and magnetic field evolution of neutron stars, as well as the
properties of the solid crust~\autocite{Sieniawska:2019hmd,2019Univ....5..217S}.

\subsection{New Phases in Quantum Chromodynamics}

After a binary neutron star merges, oscillations of the hot, extremely dense remnant produce postmerger gravitational radiation. This heretofore undetected signal probes the unexplored high-density, finite-temperature region of the quantum chromodynamic phase diagram. As indicated in \cref{fig:qcd_diagram}, this region is inaccessible to collider experiments and difficult to observe directly with electromagnetic astronomy. This is where novel forms of matter are most likely to appear~\autocite{2009PhRvD..80l3009Y,2016PhRvL.117d2501K,2016EPJA...52...50O,2016PhRvL.117d2501K,2016EPJA...52...50O,2019PhRvL.122f1101M}. Cosmic Explorer is well-suited to observing postmerger gravitational waves~\autocite{2018PhRvD..98d4044M,2018PhRvD..97b4049Y,2019PhRvD..99j2004M}: it is expected to detect ${\sim} \NUMPMO$ postmerger signals every year in a 3G network.

Measurements of the dominant postmerger gravitational-wave frequency~\autocite{2007A&A...467..395O,2010PhRvD..82h4043B,2011PhRvD..83l4008H,2012PhRvD..86f3001B,2014PhRvD..90b3002B,2015PhRvD..91f4001T,2015PhRvD..92d4045P,2015PhRvD..91f4027K,2016PhRvD..93l4051R} will reveal dense-matter dynamics with finite temperature~\autocite{2016PhR...621..127L}, rapid rotation~\autocite{1994ApJ...424..823C} and strong magnetic fields~\autocite{2001ApJ...546.1126S}. These observations will shape theoretical models describing fundamental many-body nuclear interactions and answer questions about the composition of matter at its most extreme, such as whether quark matter is realized at high densities~\autocite{2019PhRvL.122f1102B,2019PhRvL.122f1101M,2020PhRvD.102l3023B}. Direct gravitational-wave observations of postmerger remnants will also help determine the threshold mass for collapse of a rotationally supported neutron star~\autocite{2013PhRvL.111m1101B,2014PhRvD..90b3002B,2017PhRvD..96d3004H,2017MNRAS.471.4956B,2019ApJ...872L..16K}, which has implications for the neutron-star mass distribution~\autocite{2020arXiv200101747W}, compact binary formation scenarios~\autocite{2021MNRAS.500.1380M} and predictions for electromagnetic emission from neutron star mergers~\autocite{2018MNRAS.480.3871C}.

Massive stars undergoing core-collapse supernova also generate gravitational waves from the dynamics of hot, high-density matter in their central regions. Searches for supernova and various other burst-like sources are well-developed~\autocite{2019PhRvD.100b4017A, 2020PhRvD.101h4002A, LIGOScientific:2021nrg,  2020arXiv201014550T}. Cosmic Explorer will be sensitive to supernovae within the Milky Way and its satellites, which are expected to occur once every few decades~\autocite{2019PhRvD.100d3026S}. Core collapses should be common enough to have a reasonable chance of occurring during the few-decades-long lifetime of Cosmic Explorer. A core collapse supernova seen by Cosmic Explorer will have a significantly larger signal-to-noise ratio than one seen by current gravitational-wave detectors, and could be detected by a contemporaneous neutrino detector like DUNE~\autocite{2021EPJC...81..423A} giving a spectacular multimessenger event. Detection of a core-collapse event in gravitational waves would provide a unique channel for observing the explosion's central engine~\autocite{2020arXiv201004356A} and the equation of state of newly formed protoneutron star~\autocite{2018ApJ...861...10M}. Detection of a supernova would be spectacular, allowing measurement of the progenitor core's rotational energy and frequency measurements for oscillations driven by fallback onto the protoneutron star~\autocite{2021PhRvD.103b3005A,Jardine:2021fsf}.  Beyond supernova, gravitational-waves could be generated by other dynamic neutron-star processes such as: accretion, magnetar outbursts, or pulsar glitches, and by the engines of short-duration astrophysical transients such as fast radio bursts. Cosmic Explorer could provide unique insights into the engine of these events.

\subsection{Chemical Evolution of the Universe}

Stellar nucleosynthesis is responsible for the transformation of primordial hydrogen and helium into the light elements, and this elemental history is woven into the gravitational-wave observations described in Section~\cref{ss:cosmictime}. The merger rates and mass distributions of black holes and neutron stars over the history of the universe reflect the presence of these elements (`metallicity') during star formation~\autocite{2019MNRAS.490.3740N,2019MNRAS.482.5012C,2020ApJ...898..152S} and provide insight into nuclear reaction rates in massive stars~\autocite{2020ApJ...902L..36F}.

A different process must be invoked for elements heavier than iron.
The observation of GW170817 and its electromagnetic counterpart established binary neutron star mergers as a key site of heavy element nucleosynthesis~\autocite{1982ApL....22..143S}. The matter ejected during a merger is hot, dense and neutron-rich, perfectly suited for sustaining rapid neutron capture (r-process) nuclear reactions. These reactions give rise to an optical and infrared afterglow\,---\,a kilonova\,---\,that can last for days or weeks~\autocite{1998ApJ...507L..59L}. Important questions about this picture remain to be answered: is binary neutron star nucleosynthesis the sole, or merely dominant, source of heavy elements in the universe~\autocite{2018ApJ...855...99C}? How do the binary's properties affect the quantity and composition of the ejected matter?

As part of a third-generation network, Cosmic Explorer will localize ${\sim} \NUMBNSLOC$ binary neutron star mergers in the nearby universe to within ${\sim} \BNSLOC$\,deg.$^2$ every year, enabling electromagnetic follow-up to connect gravitational-wave and kilonova observations.  Distances and sky localizations for nearby neutron-star and neutron-star/black-hole mergers will identify their host galaxies, allowing the connection between compact binaries and their environment to be closely probed~\autocite{2020A&A...643A.113A,2020ApJ...905...21A}. In some cases, early warning of a system likely to produce matter outside the merger remnant can give electromagnetic observatories the advance notice required to capture the earliest moments of the kilonova~\autocite{2020arXiv200508830K}.
Cosmic Explorer will also record essentially all neutron star mergers out to redshift 1, so even poorly localized gravitational-wave events can be connected with independently identified kilonovae from surveys like the Vera C. Rubin Observatory's LSST and the Roman Space Telescope~\autocite{2021arXiv210512268C}, enabling follow-up across the electromagnetic spectrum~\autocite{2016ARNPS..66...23F,2019ApJ...880L..15M,}. Facilities such as the James Webb Space Telescope \autocite{2018ConPh..59..251K}, the Extremely Large Telescope \autocite{2016A&A...593A..24G}, the Giant Magellan Telescope~\autocite{McCarthy2018} and the Thirty Meter Telescope~\autocite{Skidmore2017}, will allow us to characterize the nature of the merger through deep imaging and spectroscopy.
Precise measurement of the source properties from the inspiral signal will break degeneracies in kilonova models, helping to pin down the rates of specific nuclear reactions. We will learn about the feedback of neutron-star merger nucleosynthesis on stellar and galactic evolution~\autocite{2015ApJ...804L..35I}, as well as the conditions under which matter is present in the circumbinary environment~\autocite{2020A&A...639A..15D}.
By establishing the rate and distribution of neutron star mergers out to cosmological distances, Cosmic Explorer will also map the history of chemical evolution in the universe beyond the reach of multimessenger astronomy.

\subsection{Gamma-Ray Burst Jet Engine}

Gamma-ray bursts are the most energetic electromagnetic phenomena in the universe. Although the connection between the short-duration subclass of these bursts and neutron-star mergers was confirmed by the multimessenger observation of GW170817~\autocite{2014ARA&A..52...43B}, the fundamental mechanism that produces this high-energy emission remains to be understood. The features of short gamma-ray burst light curves and spectra suggest that they originate in highly relativistic outflows of matter from the postmerger remnant~\autocite{2018IJMPD..2742004C}. However, the central engine powering these relativistic jets is still a matter of debate: is it an accreting black hole, or a strongly magnetized, rotating neutron star (magnetar)~\autocite{2013MNRAS.430.1061R, 2013ApJ...771L..26G}?  Cosmic Explorer will address this question by identifying the gravitational waves associated with all the observable gamma-ray bursts originating in neutron star mergers, thanks to its complete coverage of the binary neutron star population out to a redshift of $\BNSCOMPLETEZ$. A third-generation gravitational-wave detector network will measure the inclination angle of each jet, providing a comprehensive view of jet structures~\autocite{2017Sci...358.1559K,2018NatAs...2..751L,2018MNRAS.478L..18T,2018Natur.554..207M}, the time delay between merger and prompt emission~\autocite{2019FrPhy..1464402Z,2021ApJ...908..152M}, and the nature of afterglow emission~\autocite{2018NatAs...2..751L}.  Gravitational-wave information will also distinguish binary neutron star from neutron-star black-hole coalescences, revealing possible phenomenological differences in their gamma-ray emission~\autocite{2020NewA...7501306Y,2020ApJ...895...58G}.  Future gamma- and X-ray observatories, such as the Einstein Probe~\autocite{2015arXiv150607735Y}, eXTP~\autocite{eXTP:2018lyi}, ECLAIRs~\autocite{Wei:2016eox}, Athena~\autocite{Willingale:2013axa}, THESEUS~\autocite{2018AdSpR..62..191A} and TAP~\autocite{2019BAAS...51g..85C}, will be critical to increasing the reach for multimessenger follow-up of gravitational-wave sources.

The subset of Cosmic Explorer's well-localized binary neutron star mergers that produce a long-lived neutron star remnant will teach us about the origin and geometry of the ultra-strong magnetic fields supported by magnetars~\autocite{2015SSRv..191..315M}. The existence of magnetars is known from electromagnetic observations~\autocite{2005Natur.434.1107P}, but the amplification mechanism that allows their magnetic fields to grow so strong is a mystery~\autocite{2015ApJ...809...39G}. Electromagnetic follow-up of these special events will allow magnetohydrodynamic simulations of magnetic field amplification and jet creation to be put to the test. Better knowledge of the magnetic fields neutron star matter can support may shed light on a wide range of photospheric emission phenomena, including radio and gamma-ray pulses~\autocite{2008ApJ...688..499T}.

\section{Extreme Gravity and Fundamental Physics \hlight{[Geoffrey]}}
\label{ss:xg}

\begin{boxenv}{t}{green!10}{\textwidth}
\caption{Key Science Question 3}
\label{box:xg}
\bigskip
{\centering
\callout{What is the nature of the strongest gravity in the universe, and what does that nature reveal about the laws of physics?} \\}
\bigskip
Cosmic Explorer's observations of loud and rare gravitational waves will reveal the (potentially new) physics of the most extreme gravity in the universe, allowing us to:
\begin{itemize}[leftmargin=*,itemsep=1pt,topsep=2pt,parsep=1pt]
\item Probe the nature of strong gravity with unprecedented fidelity;
\item Discover unusual and (if they exist) novel compact objects impossible to detect today;
\item Probe the nature of dark matter and dark energy.
\end{itemize}
\end{boxenv}

\begin{figure}
    \centering
	\includegraphics[width=5in]{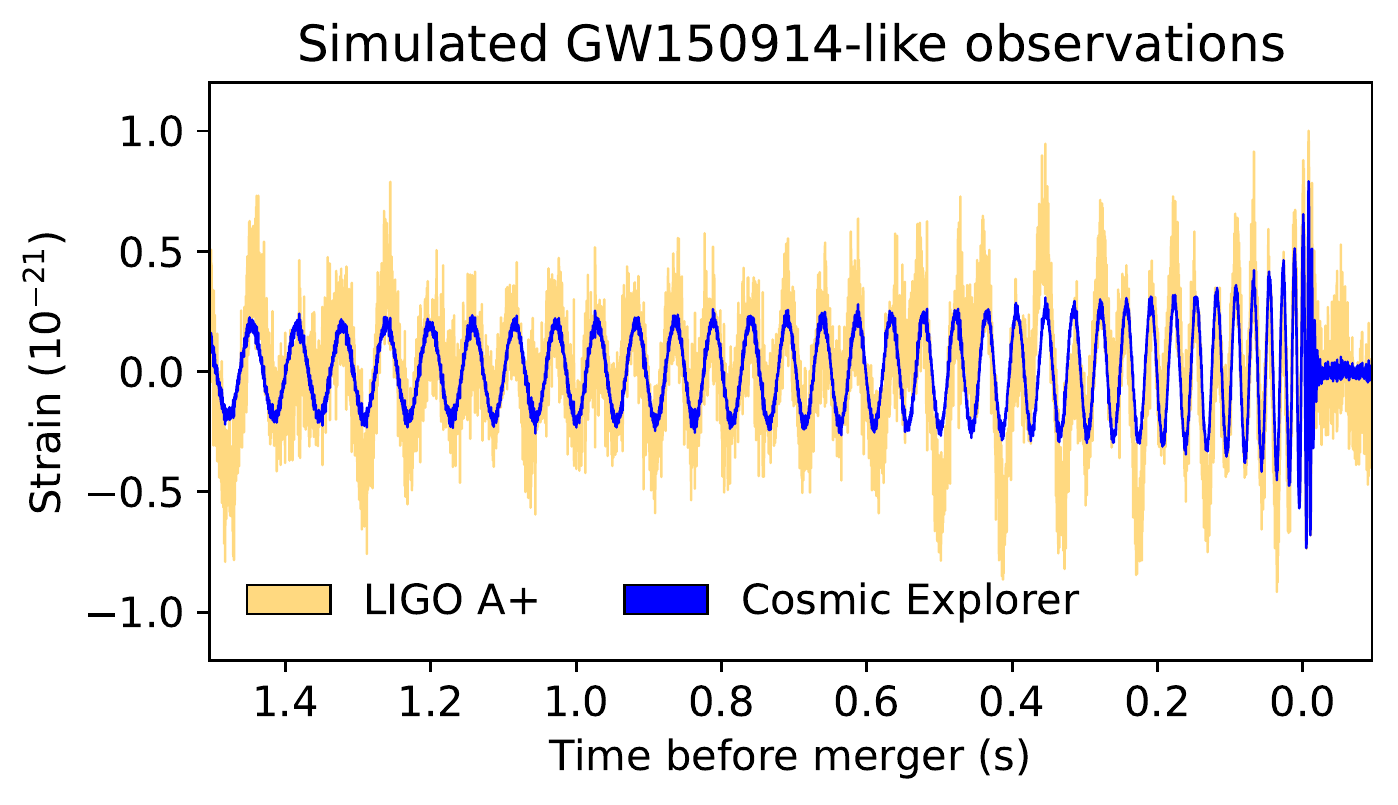}
\caption{Simulated gravitational-wave detector strain measurements for a GW150914-like signal from two merging black holes. The strain is shown as a function of time for the signal superimposed on both simulated Cosmic Explorer noise (blue) and simulated LIGO A+ noise (yellow). \label{fig:GW150914vsNoise}}
\end{figure}

Cosmic Explorer will reveal the physics of the strongest gravity in the universe in unprecedented detail, thanks to two crucial dividends from Cosmic Explorer's tremendous advance in sensitivity over current-generation gravitational-wave observatories.  First,
in three years of operation, a single \SI{40}{\km} Cosmic Explorer observatory, thanks to its increase in sensitivity, particularly at lower frequencies where CE is more than an order of magnitude more
sensitive than LIGO A+, would likely detect at least one signal from merging black holes with a signal-to-noise ratio greater than \CENEARBBHLOUDSNR~(the loudest such signal to date, GW150914, had a signal-to-noise ratio of 24). \Cref{fig:GW150914vsNoise} illustrates the impact of Cosmic Explorer's tremendous sensitivity gain by simulating the gravitational-wave strain data a GW150914-like gravitational wave would produce in Cosmic Explorer (and, for comparison, in LIGO A+).

Second, Cosmic Explorer will detect waves from sources too rare for us to observe today: in each year of its operation, Cosmic Explorer will observe approximately \CEBBHPERYEAR~binary black holes\,---\,\CEVSGWTC~times the total number of gravitational waves observed to date from any source~\autocite{2019PhRvD.100j4036A, LIGOScientific:2020ibl}.

Together, these advances will enable Cosmic Explorer to reveal the nature and nonlinear behavior of the strongest gravity in the universe with incredible clarity, perhaps revealing physics beyond general relativity whose effects are too subtle for us to recognize today. This will open a wide window on fundamental physics. Through possible effects on the gravitational waves it observes, Cosmic Explorer has the potential to shed light on longstanding mysteries in physics, including the unknown natures of dark matter and dark energy. Cosmic Explorer will also reveal the precise nature of the sources of its many observations, uncovering rare black-hole and neutron star binaries and (if they exist) novel impostors that mimic these conventional binaries.
Realizing just one of these potential discoveries would revolutionize our understanding of extreme gravity and fundamental physics.

\subsection{Nature of Strong Gravity}
General relativity describes gravity as curvature of a 4-dimensional spacetime\Autocite{2016PhRvD..94h4002Y}.
Gravitational-wave detectors are unique in their ability to probe regions of strong spacetime curvature: observations of gravitational waves, beginning with GW150914 \emph{et seq.}\autocite{LIGOScientific:2016aoc, LIGOScientific:2017vwq,2019PhRvX...9c1040A,2019PhRvD.100f4064A}, and parallel developments in accurate numerical simulations\autocite{2018Sci...361..366B} of binary black-hole coalescences, are giving us a first glimpse of strong spacetime curvature, including constraints on the nature and behavior of strong gravity~\autocite{2019PhRvL.123a1102A, 2019PhRvD.100j4036A,2019PhRvL.123l1101I,2018PhRvL.120t1102A, 2018JCAP...07..048P, 2019PhRvL.123k1102I, 2019PhRvL.123s1101N}.

Detectors on Earth are sensitive to gravitational waves from stellar-mass black holes, which, near their horizons, have stronger spacetime curvature than any other object we have observed. And because the gravitational waves emitted by coalescing binary black holes depend only on the warped vacuum spacetime surrounding the black holes' horizons, these waves present the cleanest opportunity to probe strong gravity's fundamental nature.
Making the most of this opportunity is crucial not only for understanding extreme gravity in isolation, but also for understanding its role when combined with matter and electromagnetic fields (as with neutron-star mergers~\autocite{LIGOScientific:2017vwq}).

So far, although there are many proposed alternatives, all observational and experimental tests are consistent with general relativity\autocite{will2018, 1965VA......6....1W, LIGOScientific:2016lio,2019PhRvL.123a1102A,2019PhRvD.100j4036A, 2019PhRvL.123l1101I, 2019PhRvL.123s1101N}. In particular, the loudest signal from merging black holes to date, GW150914, is in good agreement with general relativity \autocite{2016PhyOJ...9...52P}.  %
GW150914 is consistent with the ``No-Hair Theorem''\autocite{2019PhRvL.123k1102I} at about the ${\sim} \SI{10}{\percent}$ level: specifically, two inferences of the remnant mass and spin (one via recovering the fundamental ringdown gravitational-wave mode and one overtone, using data as early as the time of peak amplitude, and the other via recovering the fundamental mode using data starting 3 ms after the peak) agree to within ${\sim} \SI{10}{\percent}$\,---\,evidence to this confidence level that the remnant is a Kerr black hole.  GW170817, a binary-neutron-star coalescence\autocite{LIGOScientific:2017vwq} accompanied by electromagnetic
counterparts~\autocite{2017ApJ...848L..12A}, constrained the graviton mass to $< \SI{4.7e-23}{\eV/\clight\squared}$, provided tight constraints on possible violations of Lorentz and parity invariances, and constrained the gravitational wave speed to light speed within about one part in $10^{16}$, ruling out a number of alternative gravity theories that were invoked to explain dark energy.

Cosmic Explorer will test general relativity at unprecedented precision, with the potential to discover physical effects, either predicted by general relativity or by theories beyond general relativity, that are too subtle for current-generation instruments to measure. For instance, unlike today's detectors, next-generation detectors, such as Cosmic Explorer, will be sufficiently sensitive to detect (statistically, from many observations) the gravitational-wave memory effect, which is a change in displacement, predicted by general relativity, that remains after a gravitational wave has passed by~\autocite{2018PhRvL.121g1102Y}. As another example,
general relativity has a massless exchange boson (i.e., the graviton); with coincident detection of gravitational waves and gamma-ray bursts at redshifts of $z\sim 5$, Cosmic Explorer and its electromagnetic partner observatories would constrain the graviton mass three orders of magnitude better than current observatories.
General relativity also is Lorentz invariant, although its experimental confirmation is not as robust~\autocite{GWIC3GDocs} as the tremendous accuracy achieved in particle physics~\autocite{Mattingly:2005re}, and it is parity invariant, although some quantum-gravity theories (e.g., Ref.~\autocite{Ashtekar:1988sw, 2021arXiv210111153O}) predict parity violations. As another example, general relativity satisfies the equivalence principle and has two tensor polarizations for gravitational waves. But alternative theories that, motivated by the low-energy limit of quantum gravity, introduce additional degrees of freedom (such as a scalar field), violate the equivalence principle and lead to additional gravitational-wave polarizations\,---\,while modifying compact-binary gravitational-wave emission~\autocite{Palenzuela:2013hsa, Shibata:2013pra, Will:2014kxa}. Discovering even one such violation of general relativity, however small, would revolutionize our understanding of fundamental physics.

\subsection{Unusual and Novel Compact Objects}
Cosmic Explorer's high-fidelity observations of stellar-mass coalescing objects, together with its cosmological reach, will present an excellent opportunity for exploring the nature of merging compact objects. The
large number of detected stellar-mass black-hole and neutron-star mergers will likely include uncommon mergers too rare for even upgraded detectors in the current observatories, such as black holes with extremal spin, the inspiral of a neutron star into an intermediate-mass black hole~\autocite{Brown:2006pj,Amaro-Seoane:2018gbb}, a binary black hole with enough surrounding matter to produce an electromagnetic counterpart~\autocite{Loeb:2016fzn,2017MNRAS.464..946S,Graham:2020gwr}, or binaries with a supernova precursor~\autocite{2018ApJ...855L..12M}. Measuring the properties of these rare mergers could revolutionize our understanding of the nature of compact objects. %

Cosmic Explorer will also explore with unprecedented clarity whether some compact binaries might contain objects other than black holes and neutron stars. All observations so far are consistent with coalescing black holes and neutron stars,
but Cosmic Explorer will probe whether new types of compact binaries \autocite{2019LRR....22....4C} exist (e.g., binaries whose constituents include so-called ``great impostors''\autocite{2020PhRvL.124g1101T}, gravastars\autocite{2004PNAS..101.9545M,2016PhRvL.116q1101C, 2016PhRvD..94h4016C, 2007CQGra..24.4191C},
boson stars\autocite{2012LRR....15....6L,2017PhRvD..96f4050B}
quark stars\autocite{1965Ap......1..251I}, or
Planck stars\autocite{2014IJMPD..2342026R}), as they could have different tidal properties or quasi-normal modes of oscillation. If they do exist, it remains an open question by how much (or how little) the gravitational waves they emit differ from the waves emitted by conventional black-hole and neutron-star binaries\autocite{2017PhRvD..96b4002S,2016PhyOJ...9...52P}.
At minimum, Cosmic Explorer's enormously increased sensitivity and throughput will allow detailed tests of the %
Kerr black hole paradigm\autocite{2015CQGra..32l4006T, 2019GReGr..51..140B, 2018GReGr..50...49B, 2020PhRvD.101d4045C}, a necessary prerequisite for recognizing other kinds of novel compact objects\autocite{2019PhRvD.100l4030C,2019LRR....22....4C}. There are, however, black hole mimickers whose inspiral and quasi-normal mode spectrum might be similar to black holes in general relativity but exhibit post-merger signals such as echoes of the ringdown signal due to modified structure of the horizon. And if novel compact objects exist but are rare, Cosmic Explorer's cosmic reach will greatly increase our potential to observe them.

\subsection{Dark Matter and Dark Energy}
Dark matter\,---\,one of the major factors that governs the dynamics of the universe\,---\,has remained elusive decades after its gravitational influence on baryonic matter was discovered. And on the largest scales, the universe's expansion is driven by the invisible dark energy whose nature we do not yet comprehend and whose value appears to be too small to be consistent with vacuum energy. The astronomical community is making a substantial, ongoing effort to probe these dark sectors (i.e., dark matter~\autocite{2018PhR...761....1B} and dark energy \autocite{2012AIPC.1458..285W}).

Gravitational-wave observations present a unique opportunity for synergistic, complementary efforts to better understand dark matter and dark energy. Confirming that the universe appears the same in the gravitational-and electromagnetic-waves would reinforce our degree of belief in cosmological models (in the Bayesian sense), but any departure between the two would be tremendously consequential.

\paragraph{Dark Matter}
Approximately \SI{85}{\percent} of the mass in the universe is thought to consist of dark matter~\autocite{2020A&A...641A...6P}. Despite compelling evidence for the existence of dark matter from galaxy rotation curves, gravitational lensing, and the cosmic microwave background, its fundamental nature remains a mystery~\autocite{2005PhR...405..279B}.

To date, the only observational evidence for dark matter is via their passive gravitational influence on visible matter. Gravitational waves are an exciting new astrophysical probe of dark matter, complementing searches at high-energy colliders and underground direct-detection experiments~\autocite{2020PHYBRBK,Strategy:2019vxc}, that might reveal the nature of dark matter in several different scenarios~\autocite{Bertone:2019irm}. Cosmic Explorer's greatly improved sensitivity and cosmic reach will enable it to investigate these scenarios.
For instance, because of their strong gravitational fields and extreme densities, neutron stars might capture ambient dark matter over time through scattering off nucleons~\autocite{2010PhRvD..81l3521D,2010PhRvD..82f3531K}, or they might even produce dark matter, thanks to the exceptionally high energies achieved in binary neutron star mergers~\autocite{2019PhRvD.100h3005D}. If a neutron star were to contain dark matter, the dark matter would affect the neutron star's internal structure and hence its tidal properties~\autocite{2018PhRvD..97e5016B}. The dark matter concentration would likely depend on the neutron star's age, mass, and environment in this scenario, leading to variations in the neutron-star tidal deformability, maximum neutron-star mass throughout the population, and perhaps the implosion of neutron stars when dark matter forms mini black holes in neutron stars' cores.

There are other possibilities where dark matter might be observable with gravitational waves. Ultra-light bosonic dark matter could become self-gravitating on its own, forming a novel compact object whose properties differ from those of a neutron star~\autocite{2017PhRvD..95h4014C,2017PhRvD..96b4002S,2018PhRvD..98h3020C,2019MNRAS.483..908D}.
A significant concentration of ultra-light bosonic dark matter in the vicinity of a black-hole binary could spin down the black holes through superradiance~\autocite{2011PhRvD..83d4026A,2013PhRvD..87l4026D,2015CQGra..32m4001B,2017PhRvL.119d1101E} to spins below values that are characteristic of the black-hole and boson stars. The spin distribution of detected black holes therefore might reveal the existence of ultralight bosons~\autocite{2015PhRvD..91h4011A,2021PhRvD.103f3010N,2020arXiv201106010N}.
The cloud itself would produce continuous gravitational waves when it oscillates or a burst of gravitational waves when it collapses~\autocite{2015PhRvD..91h4011A,2017PhRvD..96f4050B,2017PhRvD..96c5019B,Bertone:2019irm}. In the former case, level transitions or annihilations in the boson cloud are predicted to emit the continuous gravitational waves monochromatically, with frequency determined by the boson and binary masses~\autocite{Bertone:2019irm}.
Searches for ultralight dark matter particles via the clouds they create around black holes only assume a coupling through through gravity; this type of search would still be viable even if dark matter does not have any type of electroweak or strong interaction with baryonic matter.
%

\paragraph{Dark Energy}
More than two thirds of the total energy in the observable universe is dark energy. The accelerating expansion of the universe reveals the ubiquitous nature of dark energy on the largest length scales, but the nature of dark energy remains one of the biggest outstanding mysteries in physics.

Since dark energy interacts only through gravitational interactions, a number of modified theories of gravity beyond general relativity have been proposed as explanations of dark energy. These theories include effects that Cosmic Explorer and its partner gravitational-wave and electromagnetic-wave observatories might detect, if they exist (e.g., Ref.~\autocite{Ezquiaga:2018btd} and the references therein, and more broadly, the discussion of the nature of gravity above). For instance, some theories of gravity beyond general relativity predict differences in the observed gravitational-wave and electromagnetic-wave luminosity distances caused by gravitational-wave damping. Cosmic Explorer's cosmic reach puts it in a strong position to search for these effects.

But whether or not Cosmic Explorer observes effects beyond general relativity, it will be well positioned to provide independent observations that complement electromagnetic observations, especially observations that are in tension. For example, gravitational-wave observations can independently address today's tension between the expansion rate of the universe determined by supernova observations and the rate deduced from the cold-dark-matter models that agree with the spectrum of the fluctuations in the cosmic microwave background~\autocite{Feeney:2018mkj}. Cosmic Explorer's gravitational-wave observations of binary black holes~\autocite{You:2020wju} at cosmological distances in particular will be standard sirens. Like standard candles (such as type Ia supernovae), gravitational waves from merging compact binaries provide a reliable measure of distance; specifically, the luminosity distance can be inferred by comparing the observed waves to theoretical model gravitational waveforms (e.g., using the technique of matched filtering~\autocite{Schutz:1986gp}). Combining the standard-siren distance with a measure of redshift (either from an electromagnetic counterpart or through statistical methods) will provide measurements of the cosmic expansion history that are independent from conventional measurements using standard candles and the other elements of the standard cosmic distance ladder~\autocite{GWIC3GDocs, Dalal:2006qt, Taylor:2012db, Cai:2016sby}, improving our understanding of the dark energy equation of state beyond what would be possible with electromagnetic observations alone (e.g., Fig.~9 of Ref.~\cite{Belgacem:2018lbp}).

\section{Discovery Potential \hlight{[Sathya]}}
\label{ss:discoverypotential}

Historically, major discoveries in astronomy have been facilitated by three related improvements in detector technology: deeper sensitivity, new bands of observation and higher precision. Improved sensitivity helps sample larger volumes and provide more complete surveys, enabling the discovery of rare events that otherwise do not make the cut, e.g., Type Ia supernovae that eventually led to the discovery of the recent accelerated expansion of the universe. Opening a new frequency window has been critical to identifying entirely new classes of sources\,---\,the cosmic microwave background, quasars and gamma-ray bursts are just the tip of the iceberg examples. Increased precision has often helped discover subtle physical effects or phenomena, e.g., the discrepancy in the Hubble constant inferred from Type Ia supernovae and the Planck mission. 

\paragraph{A deeper, wider and sharper observational window}
Cosmic Explorer will make progress at once in sensitivity, bandwidth and precision, catalyzing unprecedented discovery potential. Gravitational-wave observations will be deeply penetrating, and the signals are generated by physical processes that are vastly different from those that generate other forms of radiation and particles. It would be a profound anomaly in astronomy if nothing new and interesting came from Cosmic Explorer's vast improvement in sensitivity.

\begin{figure*}[p]
    \begin{tikzpicture}[remember picture,overlay,inner sep=-2pt]
        \node[anchor=north east] at (current page.north east) {\includegraphics[width=1.01\paperwidth]{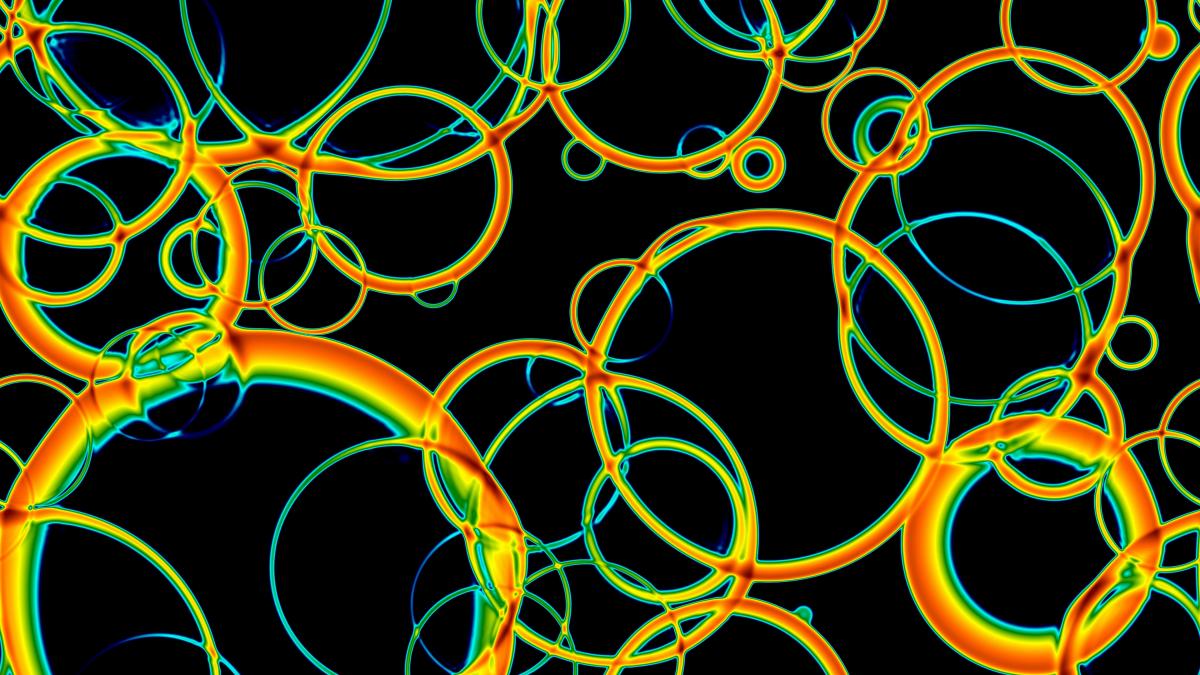}};
        \node[anchor=north east] at ([yshift=-0.40\paperheight]current page.north east) {\includegraphics[width=1.01\paperwidth]{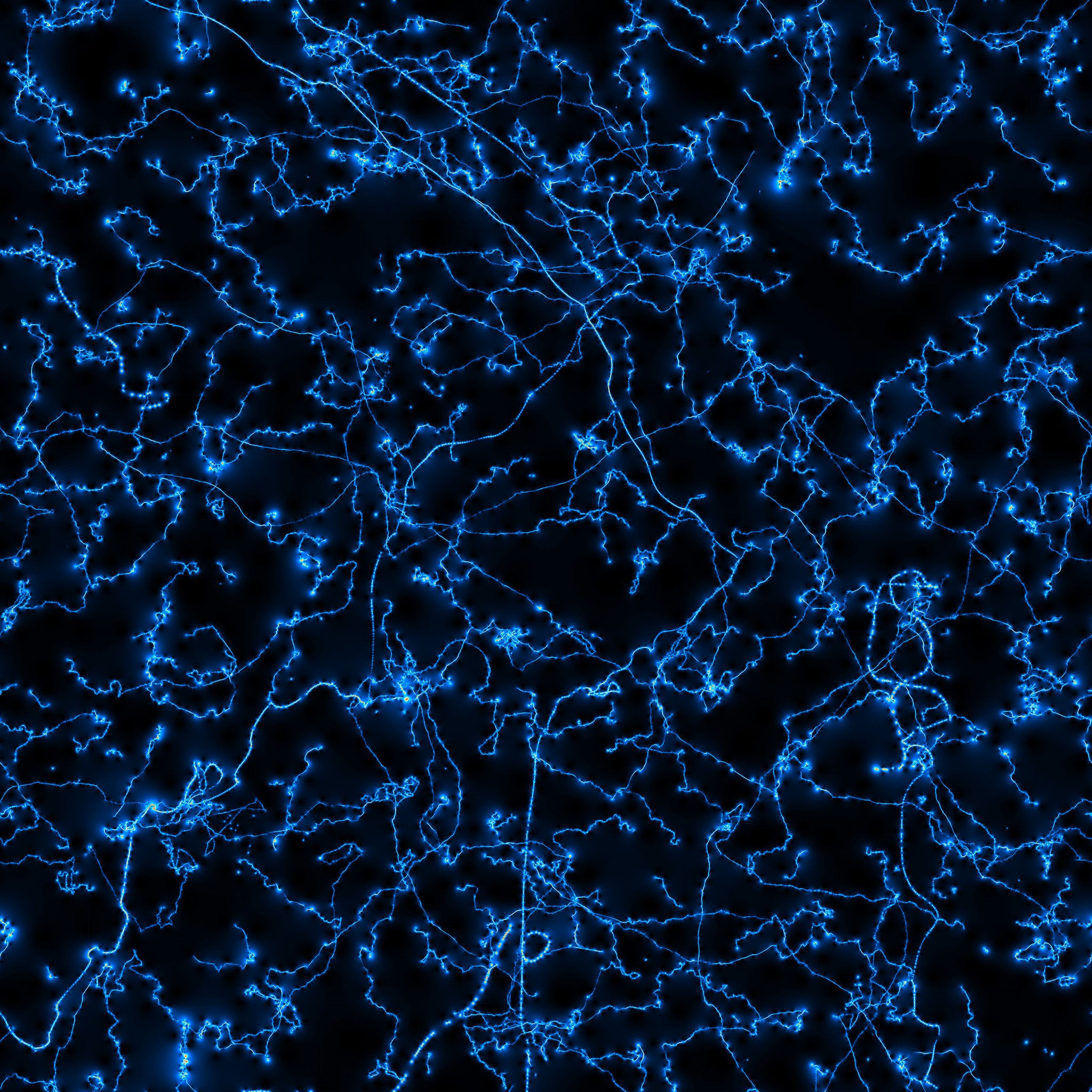}};
        \filldraw[draw=black,fill=black] (current page.south west) rectangle ([yshift=-0.80\paperheight]current page.north east);
    \end{tikzpicture}
    \begin{minipage}{\textwidth} %
        \vspace{8.125in}
    \end{minipage}
    \addtokomafont{captionlabel}{\color{SlateGray3}}
    \addtokomafont{caption}{\color{white}}
    \hypersetup{citecolor=DodgerBlue2,linkcolor=DodgerBlue2}
\caption{Cosmic Explorer's discovery potential is enabled by increased sensitivity, greater bandwidth, and high-precision measurements. The top image (credit: D. Weir, University of Helsinki) shows bubble collisions in the early universe, and the bottom image (credit: Chris Ringeval, UCLouvain) shows a visualization of cosmic strings, which are topological defects produced following inflation. Both sources could produce stochastic backgrounds detectable by a pair of Cosmic Explorer detectors.}
\label{fig:discovery potential}
\end{figure*}

Compared to current detectors, Cosmic Explorer will peer far deeper into the universe and considerably widen the observed frequency range, especially to low frequencies. Because gravitational-wave observatories are sensitive to amplitude, which falls off inversely with the distance from the source, a factor ten increase in strain sensitivity is equivalent to a factor ten increase in the \emph{diameter} of a telescope.
Cosmic Explorer will thus have dramatically greater discovery potential in a similar way to the ``discovery aperture'' opened by much larger and more advanced telescopes. Many new telescopes that have greatly expanded our view of the cosmos end up being known for a different and more dramatic discovery than what was predicted in their science cases. The phenomenon of serendipitous discovery has been discussed in articles on exploration of the unknown and serendipitous astronomy~\autocite{2009astro2010S.154K, Lang39}, some prominent examples being the cosmic microwave background, the discovery of pulsars, Cygnus X-1, and fast radio bursts.

\paragraph{Opportunities for new discoveries} Gravitational radiation results from coherent motion of bulk matter and there likely are fewer ways for it to be generated than electromagnetic radiation. However, the gravitational-wave spectrum has already proven to be source rich, with many bright emitters. The fact that we know very little about much of the universe's energy budget promises discoveries of either completely unexplored or highly speculative, but plausible, sources and phenomena. Examples include axionic clouds around black holes, dark matter in the form of subsolar primordial black hole binaries, stochastic backgrounds from early-universe phase transitions, quantum gravity signatures in the fine structure of black hole horizons or modified boundary conditions.  Moreover, gravitational waves often emanate from systems that cannot be observed with electromagnetic astronomy, allowing us to, e.g., probe the dense regions of the earliest epochs of the universe, directly observe the core-collapse supernova mechanism, and explore the nature of dense matter in the interior of neutron stars (see, \cref{fig:discovery potential}). 

\begin{boxenv}{t}{teal!10!}{\textwidth}
\caption{Key Science Question 4}
\label{box:discovery potential}
\bigskip
{\centering
\callout{What discoveries might be possible with improved sensitivity, \linebreak bandwidth and precision measurement?} \\}
\bigskip
Cosmic Explorer will have greater sensitivity and bandwidth and measure sources with exquisite precision and help in discoveries in astronomy and fundamental physics:
\begin{itemize}[leftmargin=*,itemsep=1pt,topsep=2pt,parsep=1pt]
\item Do quantum gravity effects manifest in the structure of black hole horizons?
\item What could primeval phase transitions reveal about the energy scales in the Standard Model?
\item How would gravitational wave observations help explore new particles and fields?
\end{itemize}
\end{boxenv}

\subsection{Quantum Gravity}
General relativity and quantum theory, the two founding pillars of modern physics, have both been vindicated time and again by high precision laboratory experiments, astronomical observations and cosmological measurements. Yet, there is no satisfactory theory of quantum gravity to date, but, more critically, general relativity is at odds with the fundamental principles of quantum theory that physical states obey unitary evolution. The latter is brought to bear in the bizarre behavior of black holes that are formed by collapse of matter in pure quantum states and yet when they evaporate by Hawking radiation, which is purely thermal, the observed states are mixed quantum states and information is irretrievably lost\autocite{2005PhRvD..72h4013H}. This information paradox that arises in semi-classical gravity is largely suspected to be cured by a quantum theory but every proposal for a quantum theory of gravity violates one or more of the basic tenets of general relativity, e.g., the local Lorentz violation, or its predictions, e.g., the existence of additional polarizations in the radiative field~\autocite{Will:2014kxa}. 

It is largely expected that black holes could reveal violations of general relativity in the form of failure of the no-hair theorem as a result of quantum effects near black hole horizons~\autocite{2014arXiv1409.7977S, 2015dvali}.  As another example of how Cosmic Explorer might help address open questions in quantum gravity, a recent article states that ``there has been a striking realization that physics resolving the black hole information paradox could imply postmerger gravitational wave echoes''\autocite{2018PhRvD..98d4021C}. These echoes have not yet been observed~\autocite{Cardoso:2016oxy, Mark:2017dnq, 2018PhRvD..98d4021C, Wang:2020ayy}, but Cosmic Explorer's extremely high sensitivity could reveal them, should they exist. 

The vast cosmological distances, redshifts in excess of $z\sim 20$, over which gravitational waves travel will severely constrain violation of local Lorentz invariance and the graviton mass~\autocite{Will:2014kxa}. Such violations or a non-zero graviton mass would cause dispersion in the observed waves and hence help to discover new physics predicted by certain quantum gravity theories. At the same time, propagation effects could also reveal the presence of large extra spatial dimensions that lead to different values for the luminosity distance to a source as inferred by gravitational wave and electromagnetic observations \autocite{Belgacem:2018lbp, Pardo:2018ipy} or cause birefringence of the waves predicted in certain formulations of string theory \autocite{Alexander:2009tp, Alexander:2017jmt}. 
The presence of additional polarizations predicted in certain modified theories of gravity, instead of the two degrees of freedom in general relativity, could also be explored by future detector networks \autocite{Will:2014kxa, Isi:2017fbj}.

\subsection{New Particles and Fields}
The vast horizon of Cosmic Explorer will help either discover or set stringent limits on the existence of particles and fields on a variety of different scales.  The composition of dark matter is largely unknown but it could be composed, at least in part, of ultralight bosons such as QCD axions~\autocite{Arvanitaki:2009fg}, dark photons or other light particles~\autocite{Brito:2017zvb}, spanning a wide mass range of masses \autocite{Brito:2017zvb,Arvanitaki:2009fg}, from \SI{e-33}{\eV} to \SI{e-10}{\eV}.  In particular, the Compton wavelength of ultralight bosons in the mass range \SI{e-20}{\eV} to \SI{e-10}{\eV} corresponds to the horizon size of black holes with masses from $10M_\odot$ to $10^{10}M_\odot$.  Although these ultralight fields may not interact with other Standard Model particles, the equivalence principle implies that their gravitational interaction with, for instance, black holes, could have observable consequences. For example, bosonic fields  whose Compton wavelength matches the horizon scale of an astrophysical black hole could form bound states (often called ``gravitational atoms'') around black holes and extract their rotational energy and angular momentum via the mechanism of superradiance~\autocite{Zeldovich:1971, Penrose:1971uk}.  This would result in a Bose-Einstein condensate that acts as a source of continuously emitted gravitational waves. Cosmic Explorer would have access to the higher end of the mass range from \SI{e-13}{\eV} to \SI{e-10}{\eV}, which correspond to QCD axions. 

After LIGO's first discovery of stellar black holes with unusually large masses~\autocite{Carr:2020xqk}, primordial black holes were proposed as viable candidate sources, in which case they would also constitute at least a fraction of the dark matter. Searches for subsolar mass black holes have not produced any detections so far, leading to some of the best upper limits~\autocite{LIGOScientific:2019kan} on the fraction of dark matter in black holes of mass 0.2--1.0$M_\odot$. Cosmic Explorer and partner observatories will constrain the fraction of dark matter in such black holes at the level of $10^{-5}$ of the total budget.
Moreover, since stellar evolution cannot produce black holes below about $3M_\odot$, the observation of subsolar mass black holes could indicate that they were produced in the primordial universe, or it may instead point to novel black-hole formation mechanisms driven by dark matter interactions~\autocite{Kouvaris:2018wnh,Shandera:2018xkn,Dasgupta:2020mqg}.

Aside from generating gravitational-wave signatures, new particles and fields may produce a signal in Cosmic Explorer by direct interaction with the detector itself, particularly if such a field couples to the Standard Model. Possible mechanisms include time variations in fundamental physical constants~\autocite{Stadnik:2015kia,Grote:2019uvn,Vermeulen:2021epa,Aiello:2021wlp}, fluctuating force gradients~\autocite{Pierce:2018xmy,Guo:2019ker}, or optical birefringence~\autocite{Nagano:2019rbw}.

\subsection{Stochastic Gravitational-Wave Backgrounds}
\begin{figure*}
\centering
\includegraphics[width=0.75\textwidth]{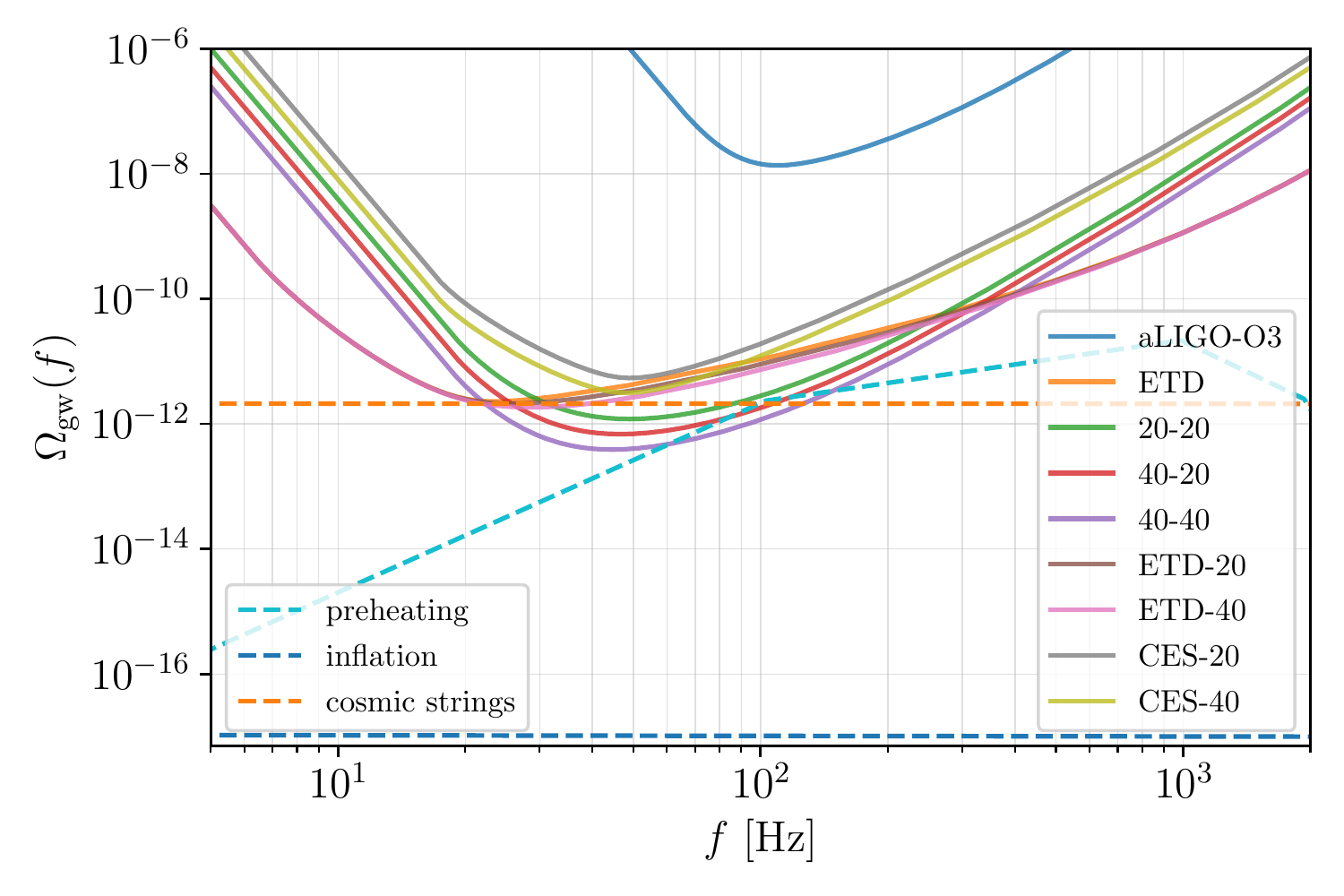}
    \caption{Sensitivity of Cosmic Explorer detector networks to stochastic gravitational-wave backgrounds compared to that of the triple-detector Einstein Telescope (ET-D). The labels 20--20, 40--20 and 40--40 correspond to two Cosmic Explorer detectors in the United States, with the numbers indicating the arm length in kilometers. ETD--20, ETD--40 correspond to the sensitivity of a triple-detector Einstein Telescope combined with a Cosmic Explorer in the United States. Likewise, CES--20 and CES--40 correspond to the sensitivity of a 20 km Cosmic Explorer in Australia and a 20- or 40-km Cosmic Explorer in the United States. A stochastic background with energy density $\Omega_{\mathrm{gw}}(f)$ calculated at a reference frequency of \SI{300}{\Hz} would be detected with a signal-to-noise ratio of 3 after one year of observation if it crosses the curve for that network~\autocite{2013PhRvD..88l4032T}. Dashed lines show the expected backgrounds for cosmic strings ($G\mu$ = \SI[retain-unity-mantissa=true]{1e-11}, with fiducial model parameters from Ref.~\cite{2012PhRvD..85l2003S}), preheating (for hybrid inflation occurring at \SI{e9}{\GeV} as calculated in Ref.~\cite{2007PhRvL..99v1301E}), and standard slow-roll inflation~\autocite{2018CQGra..35p3001C}.}
\label{fig:pi_curve}
\end{figure*}

A stochastic gravitational-wave background is expected to arise due to the superposition of individually unresolvable gravitational waves of both astrophysical and cosmological origin.
The astrophysical backgrounds of stellar-mass compact binary mergers that are the targets of current ground-based gravitational-wave detectors (e.g., Refs.~\cite{2011ApJ...739...86Z, 2012PhRvD..85j4024W, 2016PhRvL.116m1102A, 2018PhRvL.120i1101A, 2019PhRvD.100f1101A, 2021arXiv210112130T}) will be nearly entirely resolvable with the next generation observatories, but residual unresolvable signals can contaminate the measurements of much weaker cosmological backgrounds~\autocite{2017PhRvL.118o1105R, 2020PhRvD.102f3009S, 2020PhRvD.102b4051S}.
While this poses a computational challenge, recent methods~\autocite{2020PhRvL.125x1101B} demonstrate that Cosmic Explorer can provide a unique opportunity to probe the early universe with gravitational waves.
Standard slow-roll inflationary models are expected to produce a stochastic background with dimensionless energy density $\Omega_{\mathrm{gw}}\sim 10^{-17}$~\autocite{2000PhR...331..283M, 2016A&A...594A..13P}, too weak to be directly detected by all but the most ambitious space-based gravitational-wave detectors~\autocite{2005PhRvD..72h3005C, 2011CQGra..28i4011K}.
However, nonstandard inflationary and cosmological models can produce backgrounds due to processes like preheating, first-order phase transitions, and cosmic strings~\autocite{2018CQGra..35p3001C}, all with energy densities within the reach of Cosmic Explorer (see \cref{fig:pi_curve}).

\paragraph{Particle Production and Preheating} Following inflation, the universe must undergo a period of particle production during which the inflation field couples to other particle species into which it eventually decays.
If this process occurs non-perturbatively, it is called \textit{preheating} (see Refs.~\cite{2010ARNPS..60...27A, 2015IJMPD..2430003A} for reviews).
While the amplitude of the background generated during this process of particle production is expected to be independent of the temperature scale at which it occurs, standard inflationary models predict that it will peak at $f\sim 10^{7}-10^{8}~\mathrm{Hz}$, well beyond the frequency band of Cosmic Explorer~\autocite{1997PhRvD..56..653K, 2008PhRvD..77j3519E}.
However, for hybrid inflation occurring around ${\sim} 10^{9}~\mathrm{GeV}$, the background from preheating peaks in the band of ground-based detectors with an energy density of $\Omega_{\mathrm{gw}}\sim 10^{-11}$, which is detectable with Cosmic Explorer~\autocite{1998hep.ph....4205G, 2006JCAP...04..010E, 2007PhRvL..99v1301E}.
Even a non-detection of such a background could be used to constrain the physics and temperature scale of particle production in the early universe.

\paragraph{Cosmic Strings} Cosmic strings are one-dimensional topological defects produced in spontaneous symmetry breaking phase transitions following inflation~\autocite{1976JPhA....9.1387K}.
When a string folds upon itself, it produces a loop, which oscillates under its tension, emitting gravitational waves in a series of harmonic modes~\autocite{1985PhRvD..31.3052V, Shellard:1987bv, 1990PhRvD..41.1751L}.
Cusps and kinks are formed when cosmic string loops intersect~\autocite{1991PhRvD..43.3173A, 1994PhRvD..50.2496A}.
These string features emit higher frequency bursts of gravitational radiation~\autocite{2001PhRvD..64f4008D} whose superposition creates a stochastic gravitational-wave background accessible to ground-based gravitational-wave detectors~\autocite{2005PhRvD..71f3510D, 2007PhRvL..98k1101S}.
The spectrum of the background depends on the cosmic string tension, $G\mu$, and the loop model, among other parameters~\autocite{2009Natur.460..990A, 2015CQGra..32d5003H}.
The current generation of ground-based gravitational-wave detectors has placed the most stringent upper limit on the string tension to date, $G\mu \lesssim 4 \times 10^{-15}$~\autocite{LIGOScientific:2021nrg}, using a loop model based on Ref.~\cite{2019JCAP...06..015A}, distinct from the model of Ref.~\cite{2007PhRvL..99v1301E} shown in \cref{fig:pi_curve}. Cosmic Explorer will allow us to probe tensions that are several orders of magnitude smaller, offering a window into beyond-the-standard-model physics at the highest energies.

\paragraph{Phase Transitions} Phase transitions in the early universe, such as the decoupling of the electromagnetic and weak forces, can also produce a stochastic gravitational-wave background under some modifications of the Standard Model if they are strongly first order~\autocite{1994PhRvD..49.2837K, 2016JCAP...04..001C, 2017PhRvL.119n1301I, 2018RSPTA.37670126W}; i.e., if there is a discontinuity in the first derivative of the free energy during the transition.
In this scenario, gravitational waves are emitted due to the collision of bubbles of the new phase~\autocite{1992PhRvD..45.4514K, 2008JCAP...09..022H} and due to the anisotropic stresses generated by magnetohydrodynamical turbulence and discontinuities in the shocked plasma surrounding the expanding bubbles~\autocite{2008PhRvD..77l4015C,2009JCAP...12..024C, 2014PhRvL.112d1301H, 2015PhRvD..92l3009H}.
A first-order electroweak phase transition also has implications for electroweak baryogenesis, which could provide an explanation for the cosmic baryon asymmetry~\autocite{2012NJPh...14l5003M}.
The peak frequency of the stochastic background energy density spectrum depends on the energy scale of the transition,  with a transition occurring at $~10^{9}~\mathrm{GeV}$ producing a background peaking in the frequency band of ground-based gravitational-wave detectors~\autocite{2000PhR...331..283M, 2008PhRvD..78d3003K}.
Such a background is expected to have an amplitude of $\Omega_{\mathrm{gw}}\sim 10^{-12 \pm 2}$, within the range of Cosmic Explorer~\autocite{2009PhRvD..79h3519C, 2014PhRvD..90j7502G}.

A first-order phase transition in the early universe may also have occurred
from Peccei--Quinn symmetry breaking, which is responsible for producing the
QCD axion and an associated stochastic gravitational-wave background. Depending
on the temperature of the transition and the mechanism of the symmetry
breaking, the background may be of sufficient strength and of appropriate
frequency to be detected by ground-based gravitational-wave observatories,
particularly third-generation observatories~\autocite{DelleRose:2019pgi,
VonHarling:2019rgb}. Searches for this gravitational-wave background would
provide a valuable complement to dedicated axion-search experiments.

The detection of a cosmological stochastic background by Cosmic Explorer would represent the accomplishment of one of the most ambitious goals of gravitational-wave astronomy, and even a non-detection would allow for constraints on beyond-the-standard-model physics at energies orders of magnitude larger than those accessible with particle accelerators.

\setpartpreamble{
    \AddToShipoutPictureBG*{
    \begin{tikzpicture}[remember picture,overlay,inner sep=0]
        \node[anchor=north] at ([yshift=0.01\paperheight]current page.north)
            {\includegraphics[height=1.1\paperheight]{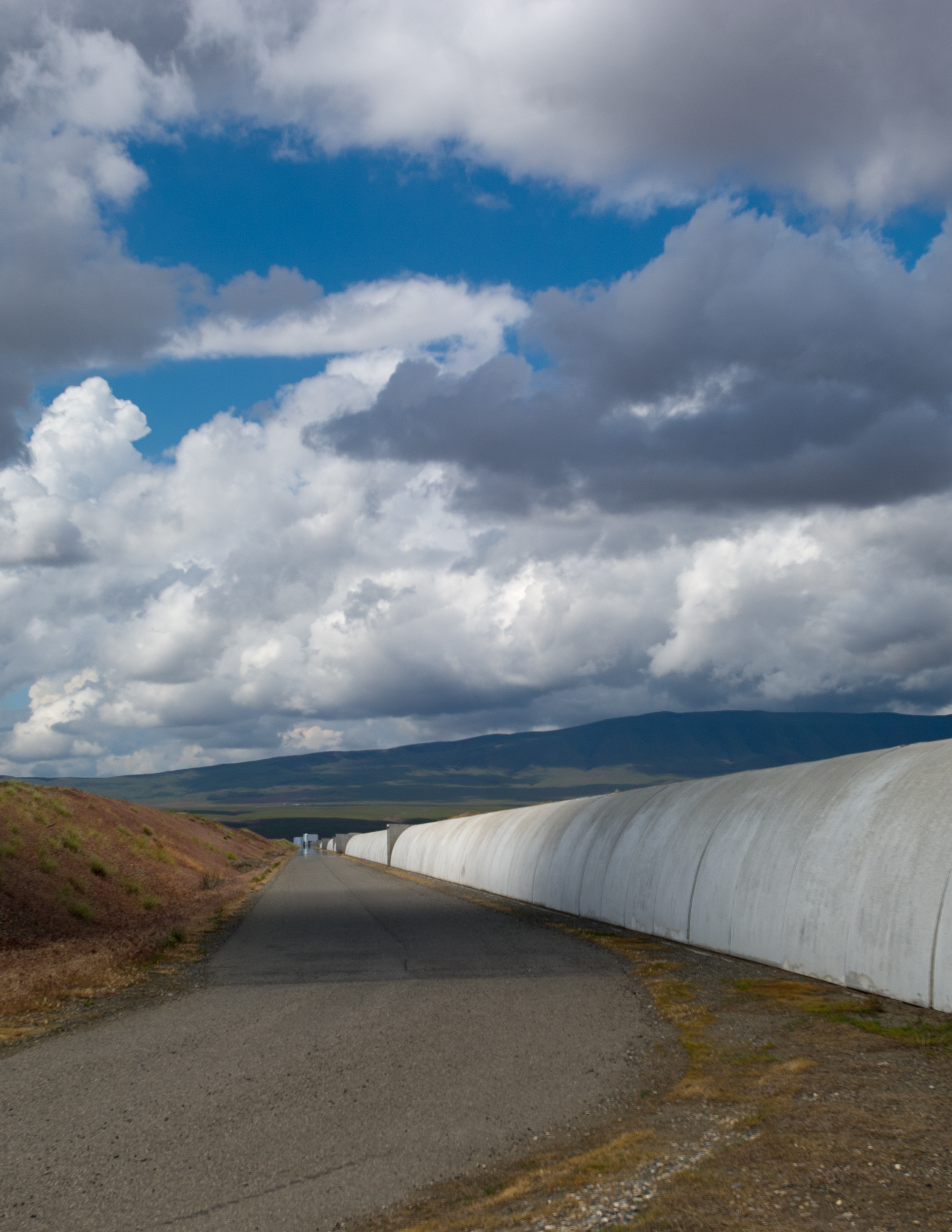}};
        \node[align=left,text=white,inner sep=6pt,anchor=south east] at (current page.south east) {LIGO Hanford Observatory. Credit: LIGO.};
        \filldraw[draw=black,fill=black,opacity=0.3] (current page.south west) rectangle (current page.north east);
    \end{tikzpicture}
    }
}
\addpart{Observatories}

\chapter{A Science-Driven Design for Cosmic Explorer \hlight{[Stefan]}}
\xlabel{overview}

\begin{figure}[ht]
\centering
\includegraphics[width=0.6\textwidth]{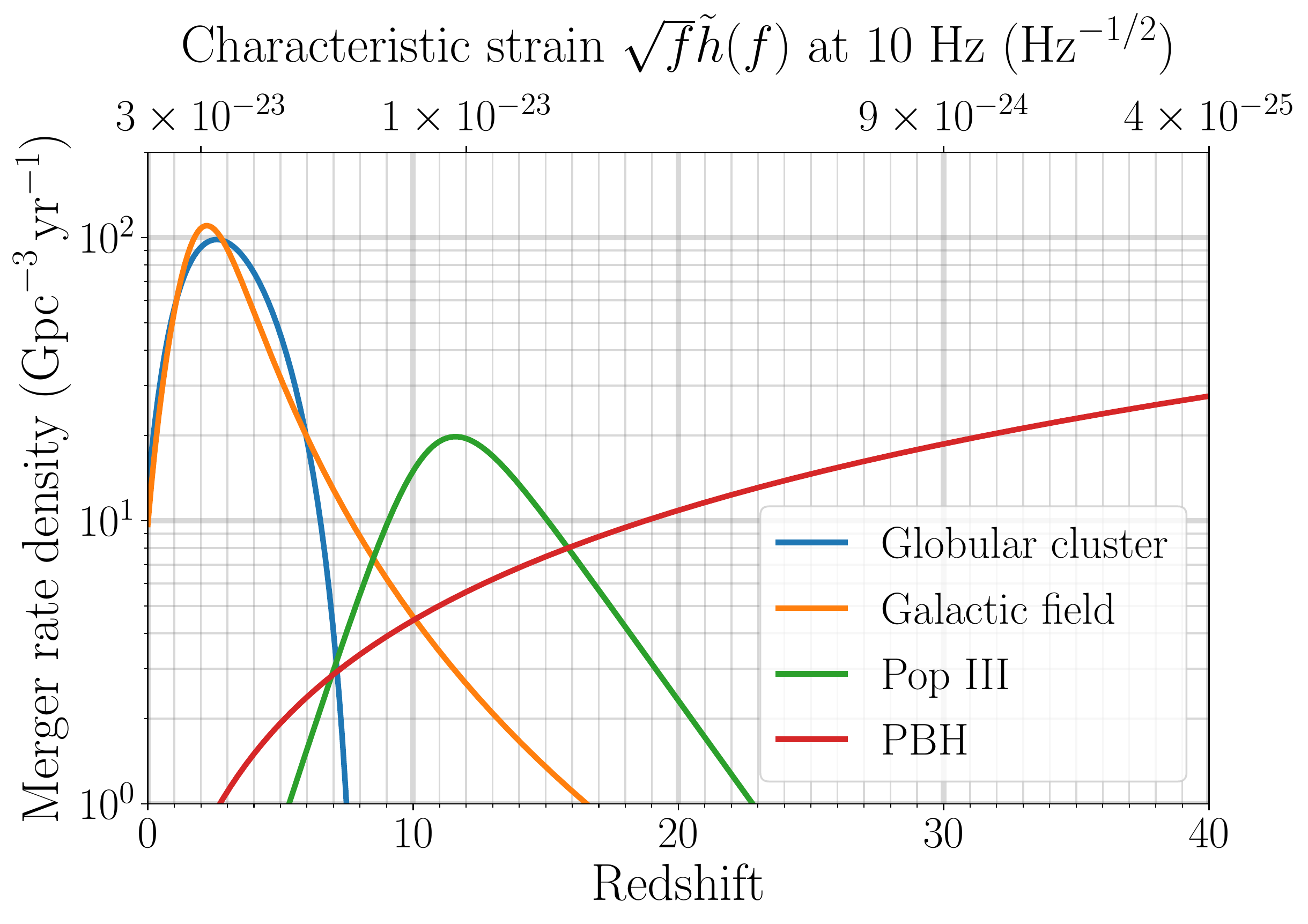}
\caption{Merger rate densities of a few representative populations of compact objects as a function of redshift. Galactic field and globular cluster formation are expected to produce both binary black holes (BBHs) and binary neutron stars, while the other channels will only produce BBHs. The curves for galactic field, globular cluster and Pop~III formation are taken from Ng et~al.~\autocite{2020arXiv201209876N}, and are based on population synthesis analyses by Refs.~\cite{2016Natur.534..512B,2017MNRAS.471.4702B,2018ApJ...866L...5R}. The primordial black hole (PBH) merger rate is taken from Refs.~\cite{2021arXiv210203809D,2019JCAP...02..018R}. 
The top axis gives the characteristic strain calculated at \SI{10}{\Hz} (as measured in the detector frame) of an optimally oriented 30--30~\Msun BBH placed at the corresponding redshift indicated on the bottom axis. Here the characteristic strain is defined as $\sqrt{f} \bigl|\widetilde{h}\bigl(f\bigr)\bigr|$, where $\widetilde{h}\bigl(f\bigr)$ is the Fourier transform of the gravitational-wave signal. The strain does not follow a simple linear trend with redshift due to (1) the non-linear relation between luminosity distance and redshift~\autocite{1999astro.ph..5116H} and (2) the fact that if the source is far enough, what will be observed at \SI{10}{\Hz} are the merger and the ringdown.}\label{fig:populations}
\end{figure}

Achieving the science goals laid out in \sref{science_overview} requires a gravitational-wave observatory capable of reliably measuring the strain from mergers in the band between a few hertz and a few kilohertz with a peak sensitivity of approximately \SI{3e-25}{\big/\sqrt{\Hz}}.  This strain sensitivity guarantees that remnant mergers from the first stars are still observable (\cref{fig:populations}, adapted from Ref.~\cite{2020arXiv201209876N}). The upper limit of the frequency band is dictated by the highest frequency signals expected from the lightest known compact objects: neutron stars.
The lower edge of the frequency band of any terrestrial detector is dictated by the seismic motion of the detector and by Newtonian noise, the coupling of seismic and atmospheric fluctuations through direct Newtonian gravity.
Since seismic and Newtonian noise are displacement noises, the low-frequency strain sensitivity can be improved by lengthening the detector arms.

Since gravitational-wave detectors are essentially antennas, the highest frequency of interest also sets the ideal scale of the antenna: a few tens of kilometers for signals at a few kilohertz, about ten times the size of existing detectors.

\section{Design Concept for Cosmic Explorer \hlight{[Stefan]}}
\label{subsec:designoverview}

The interferometric technology used in current gravitational-wave detectors such as Advanced LIGO and Virgo is the most mature and it forms the basis for the Cosmic Explorer
 detector concept.
In addition to the discussion of interferometric technology in \cref{subsubLaserIFO},
 in this section we also briefly discuss potential alternative technologies: space-based interferometers in \cref{subsubLaserSpace}, atom interferometers in \cref{subsubAtom}, and torsion pendulum detectors in \cref{subsubLaserTorsion}.

The scale of the facility required to achieve the science goals outlined in \sref{science_overview} represents a major investment. We therefore plan for this facility to have a lifespan of about 50 years and the flexibility
to host a number of iterations of detector designs.
This will allow funding agencies to capitalize on future research and development breakthroughs,
 should the operational life-span of CE be extended beyond the initial mandate
 (which is expected to be 20 years).

The technology to be installed when the Cosmic Explorer observatories are built
features the lowest possible technical risk to achieve the most readily accessible science goals.
The corresponding detector is largely a scaled-up version of current room-temperature,
 fused-silica-based interferometers, with some incremental improvements in non-critical technologies
 (see \cref{{subsec:technology_drivers}}).
This will be followed by a sequence of planned upgrades that
 incorporate currently less developed technologies as they become available
 (see \cref{subsec:reference_design}).
In addition to the planned upgrade path to achieving the CE target sensitivity and science goals,
 a second path involving \SI{2}{\mu m} lasers and cryogenic silicon mirrors (a.k.a. ``Voyager technology''
  or simply ``\SI{2}{\mu m} technology'')
 is discussed as a potential alternative should the incremental approach based on current technology
 encounter unexpected challenges.
Beyond its role as a technology alternative for the Cosmic Explorer science goals presented here,
  \SI{2}{\mu m} technology may present an opportunity  for maximizing the output
  of the Cosmic Explorer observatories in the future.
While \SI{2}{\mu m} technology is much less mature than the currently deployed  \SI{1}{\mu m} technology,
  this technology, or some other future detector technology  that has not yet been conceptualized,
  may eventually allow the CE observatories to push toward the
  fundamental physical limits of the facility (see \cref{subsec:silicon_upgrades}).

\section{Technology survey \hlight{[Stefan, Evan]}}
\label{subsec:other_technologies}

Here we survey a number of potential technologies for detecting gravitational waves: ground-based laser interferometry (which we choose as the technology for the CE reference concept), space-based interferometry, atom interferometry, and torsion bars.  We also look at the cryogenic silicon-based upgrade proposal Voyager for the currently existing LIGO observatories. A comparison of low-frequency terrestrial \GW{} detection methods was also given by Harms et~al.~\autocite{2013PhRvD..88l2003H}

\subsection[]{Ground-Based Laser Interferometry}
\label{subsubLaserIFO}

All direct detections of \GW{}s to date have been made with laser interferometers, more specifically Michelson interferometers enhanced with optical cavities in a so-called ``dual-recycled Fabry--P\'{e}rot Michelson'' (\textsc{drfpmi}) configuration.
Astrophysically sensitive laser interferometers of this type are the result of a global R\&D effort spanning four decades: whereas early laboratory prototypes in the 1980s achieved peak strain sensitivities of about \SI{e-19}{\big/\sqrt{\Hz}} at kilohertz frequencies~\autocite{1988PhRvD..38..423S}, the current kilometer-scale detectors achieve peak strain sensitivities better than \SI{e-23}{\big/\sqrt{\Hz}} down to several tens of hertz.

The Cosmic Explorer reference concept (\cref{subsec:reference_design}) adopts the \textsc{drfpmi} interferometer as the working technology.
This design builds on the success of the existing \textsc{drfpmi} research program, aiming to extend the sensitivity of this class of laser interferometers by one more order of magnitude, achieving peak strain sensitivities better than \SI{1e-24}{\big/\sqrt{\Hz}} down to \SI{5}{\Hz}.
This sensitivity improvement is due to a combination of longer interferometer arm cavities, realizable in the 2030s at new facility, and a set of technology improvements that can be achieved in the 2020s and 2030s.

Other laser interferometer topologies have been proposed for gravitational wave detection~\autocite{1990PhLA..147..251B,2002PhRvD..66l2004P}.
However, none of these topologies will achieve cosmological reach unless, as with the Cosmic Explorer \textsc{drfpmi} design, a combination of longer facilities and technology improvements is assumed.
Moreover, these interferometer topologies are still in the laboratory prototyping phase: compared to the \textsc{drfpmi} program, any program needed to realize these alternate topologies carries more risk and cannot leverage as much existing R\&D. The Cosmic Explorer facility however would be able to accommodate a corresponding upgrade should one of these topologies turn out to be beneficial. This is particularly true for an above-ground facility like Cosmic Explorer, where significant changes to the observatory's vertex and end stations are possible.

\subsection[]{Space Missions}
\label{subsubLaserSpace}

Gravitational-wave interferometry is also being pursued for implementation in
space, with a science program that is largely complementary to that of
ground-based gravitational-wave interferometers. The first anticipated space
mission, LISA~\autocite{2017arXiv170200786A}, is scheduled to launch in the
mid-2030s, and other missions include TianQin~\autocite{TianQin:2015yph,
TianQin:2020hid}, Taiji~\autocite{2021CmPhy...4...34T} and
DECIGO~\autocite{2006CQGra..23S.125K, 2020arXiv200613545K}.  Going into space
has the advantage of much longer laser path lengths (2.5 million kilometers, in
the case of LISA), as well as the absence of terrestrial force noise. On the
other hand, laser power limitations and diffraction loss limit the achievable
arm power, and hence the shot noise sensitivity of space-based laser
interferometers is less than terrestrial detectors.
These characteristics make space missions most suitable for detections in the
sub-hertz band.  Space missions will detect the mergers of intermediate-mass
and supermassive black holes, as well as extreme mass-ratio mergers of
stellar-size objects and massive black holes. They will observe stellar-mass binary systems only in
their early inspiral phase. Notably, space missions will not observe
neutron-star postmerger signals.  On the other hand, space missions will be
able to observe some early-inspiral stellar-mass systems months to years
before they are observed in terrestrial detectors. Such joint
`multi-band’ observations can potentially set tighter limits for some tests of
general relativity. The planned dates for the LISA mission, with observation
starting in the late 2030s, mesh well with Cosmic Explorer's schedule promising interesting multi-band observations.

\subsection[]{Atom Interferometry}
\label{subsubAtom}

Atom interferometers have been proposed as tools to detect gravitational waves
via gradiometric measurement~\autocite{2018NatSR...814064C,
2018arXiv181200482C, 2020CQGra..37v5017C,2019arXiv191111755B,
2019BAAS...51c.453H}.  In a typical proposed setup atom interferometers are
used as interferometric inertial references, taking the place of test masses in
conventional gravitational-wave interferometers. Two or more such atom
interferometers are separated along a baseline and interrogated by a common
laser. Pulses from that laser serve as splitter, mirrors and recombiner for the
individual atom interferometers, and additionally pick up a phase modulation
due to a \GW{} passing through the baseline. This puts challenging constraints
on the laser phase front that need to be met to achieve interesting
sensitivities. For reference, the initial sensitivity goal for the MAGIS-100 experiment, using state-of-the-art parameters for projection, is about \SI{5e-15}{\big/\sqrt{\Hz}} between \SI{0.3}{\Hz} and \SI{3}{\Hz}~\autocite{2018arXiv181200482C}.

Even with orders-of-magnitude improvements in atomic flux and with baselines
exceeding \SI{10}{\km}, the audio-band strain sensitivity of these gradiometers
is limited by atomic shot noise to a level that does not surpass the
sensitivity already achieved by laser interferometers.  Instead, the proposed
sensitivity improvement over ground-based laser interferometers occurs in the
decihertz band, where the seismic noise coupling is suppressed because
the atom clouds are in free-fall.  As such, an atomic gradiometer operating at
the shot-noise limit is sensitive primarily to compact binaries in the range
$[10^3,10^4]M_\odot$, potentially extending to redshifts of a few.

Though the direct seismic noise is suppressed, atomic gradiometers are still
sensitive to seismic and atmospheric fluctuation through Newtonian coupling in
much the same way as laser interferometers.  This Newtonian noise drives many
of the proposed experiments to assume underground operation over a long
(kilometer-scale) baseline, coupled with other techniques such as noise
subtraction with auxiliary sensors and the use of dozens of atom
interferometers to exploit the different spatial correlation properties of
gravitational waves and Newtonian noise.  Even so, mitigating Newtonian noise
at decihertz frequencies, which is a prerequisite for shot-noise-limited
operation of the instrument, will require a challenging research program due to
the greater strength of geophysical noise below \SI{1}{\Hz} and the greater
number of processes that produce it~\autocite{2020arXiv200704014C,
2019LRR....22....6H}.

\subsection[]{Torsion Pendulums}
\label{subsubLaserTorsion}

Laser interferometry can be used to search for gravitational-wave-induced fluctuations in the angle between two bars, suspended from their centers of mass as torsion pendulums. Such a torsion bar antenna design offers some cancellation of mechanical noise.
The characteristic length scale of the detector is set by the size of the bars, which in the TOBA proposal~\autocite{2010PhRvL.105p1101A, 2020IJMPD..2940003S} is \SI{10}{\meter}; this proposal additionally assumes cryogenic operation underground.
The design sensitivity is of the order \SI{e-20}{\big/\sqrt{\Hz}} above \SI{1}{Hz}, which is several orders of magnitude less sensitive than ground-based interferometric detectors. Torsion bar detectors are also affected by Newtonian noise, although the coupling geometry is slightly different. Interestingly, this might make torsion bar detectors the most promising local sensors for directly measuring Newtonian noise, potentially assisting Newtonian noise mitigation in other gravitational-wave detector designs.

\subsection[]{Voyager}

Voyager is the name for a proposed cryogenic silicon upgrade intended to maximize the reach of the existing LIGO facilities~\autocite{2020CQGra..37p5003A}. Efforts toward Voyager are currently focused on the research and development needed for cryogenic silicon and  \SI{2}{\um} technology. The implementation of Voyager in the LIGO facilities would lead to increased gravitational-wave detection rates and significantly improved astrophysics. 
However, the unproved nature of the optics needs extensive development. The \SI{4}{\km} baseline would also constrain the future. Voyager would not reach the era of first stars or achieve the full set of goals envisioned for CE (see~\cref{flowdown}). Voyager would be a demonstration of technology that could be used to upgrade Cosmic Explorer, yielding important performance information in detectors significantly more sensitive than other possible technology prototypes.

\chapter{Optimizing Design Performance Versus Cost}
\label{sec:trade}

In this section we explore a range of Cosmic Explorer designs and the impact of
design choices on the scientific output of CE and the 3G network.  The
objective of this exploration is to ensure that resources expended in
construction of the CE facility are put to good use, i.e., to optimize the
science-per-dollar spent on CE.

We start by creating a science traceability matrix, shown in
\cref{traceability}, that maps the three primary science objectives described
in \cref{ch:keyquestions} to the observations needed to realize each
objective in terms of a specific measurement and its requirements.  Measurement
requirements are then mapped to the instruments and instrument requirements. We
then identify a reference configuration for CE (see \cref{box:concept}) that
can meet all of these requirements, and discuss a number of variants of this
configuration (see \cref{subsec:reduced_cost_configurations}).  These variants
differ principally in the length and number of CE facilities in the US, since
these are the primary cost drivers.  This is followed by a presentation of the
impact of these alternatives on CE's ability to achieve its key science goals.
A summary of the results of this section is given in \flowdown.

\begin{boxenv}{!h}{blue!10}{\textwidth} \caption{Cosmic Explorer Reference Concept.}\label{box:concept}

\emph{The Cosmic Explorer concept consists of two widely-separated L-shaped observatories in
the United States\,---\,one with \fortykm\ long arms and another with \twentykm\ arms.}
\\
\\
This concept maximizes the scientific output as the \SI{40}{\km} detector can
be optimized for deep broadband sensitivity, while the \SI{20}{\km} detector is
capable of tuning its sensitivity to the physics of neutron stars after they
have merged. To enable accurate source localization and coverage and to ensure
sufficient transient noise rejection, the observatories should not be
co-located. To ensure that wave polarizations can be well distinguished, the
observatories should not be parallel.
\\
\\
Two US observatories can accomplish all of the CE science goals
independent of additional international next-generation gravitational-wave
observatories\,---\,though CE will reach its maximum potential as part of a
next-generation network.  This concept also takes advantage of
efficiencies associated with simultaneous construction (as well as commissioning
and operation) of two sites within the US, as done by LIGO.
\end{boxenv}

Facility capabilities differ more than might be expected by a simple arm-length
scaling of the signal's strength. In particular, the free spectral range of
long-arm facilities ($c/2L_\mr{arm}\approx \SI{3.7}{\kilo\Hz}$ for a
\SI{40}{\km} detector) begins to limit the flexibility of the observatories to
target high-frequency signals such as the postmerger phase of neutron-star
mergers. The frequency of postmerger gravitational waves varies substantially
within current matter uncertainties and with the masses of the merging
stars~\autocite{2016CQGra..33h5003C}. As we better understand the population of
neutron-star mergers and the properties of dense matter, we can tune the
sensitivity of a shorter \SI{20}{\km} detector for optimal postmerger
physics,\autocite{2019PhRvD..99j2004M,2020PASA...37...47A} for example by
focusing on frequencies characteristic of a hadron-quark phase
transition~\autocite{2019PhRvL.122f1102B}. To compare facilities, we include
reference tunings optimized for inspiral and postmerger observation.
\section{Alternate Configurations \hlight{[Evan]}}
\label{subsec:reduced_cost_configurations}

This section describes variants which differ somewhat from the
 reference configuration of two L-shaped observatories,
 one with \fortykm\ long arms and another with \twentykm\ arms (see \cref{box:concept}).
For the reasons discussed in the following sub-sections,
 the variants involving one or two observatories
 of either \twentykm\ or \fortykm\ are carried forward into the
 subsequent trade-study discussion in \cref{sec:trade-study}.

\subsection[]{Shorter Arms (10, 20 and 30\;km) and Optical Tunings}
\xlabel{shorter_arms}

Reducing the length of the interferometer arms is a clear means of reducing the
cost of a CE facility.  As the reference concept was chosen for maximum
scientific output, this cost reduction must come at the expense of scientific
output.

\cref{tab:noise_length_scalings} shows the scalings of fundamental noises as
the detector length $L$ is varied.  In all cases, the strain-referred noise
from the geophysical and thermal sources is the same or worse as $L$ is
reduced.  The shot noise, which is the dominant noise source near and above
\SI{100}{\Hz}, is shaped not only by the length of the detector, but also by its
optical configuration.  For a given length $L$, we identify two optical
configurations of interest.  First, we identity a ``compact-binary optimized''
configuration, where the detector's shot noise is tuned to give the best
sensitivity below \SI{1}{\kHz}, where stellar-mass binaries inspiral and merge.
Second, we identify a ``postmerger optimized'' configuration, where the
detector's shot noise is tuned to give the best sensitivity around
\SIrange{2}{3}{\kHz}, at the expense of sensitivity below \SI{1}{\kHz}; this
configuration will best capture late-time signals from the aftermath of neutron
star mergers.  It is possible to convert the detector from one configuration to
the other by replacing a small number of optical components, with no facility
modification required.

The compact-binary optimized and postmerger optimized detector configurations
are shown in \cref{fig:ce_alternate_strains} as a function of arm length.
Evidently, for $L = \SI{40}{\km}$ there is only a modest difference between the
two configurations.  However, for a \SI{20}{\km} facility the difference is
significant, showing a clear trade-off to be made between the science goals.  It
is noteworthy that the \SI{30}{\km} option appears to be ``the worst of both
worlds'' in that it cannot be tuned to high frequencies, and its sensitivity at
low frequency is not as good as a longer facility.

We stress that for a given detector length, the trade-off between these two
optimized configurations is \emph{not} built into the facility and thus not a
long-term choice.  We expect to periodically switch detector
configurations (e.g., observe in compact-binary mode for a year, then observe
in postmerger mode the following year), since this requires only minor
modifications to the detector and does not require facility modification.
Furthermore, the postmerger optimized sensitivity shown in
\fref{ce_alternate_strains} is only one of a continuum of options available: in
a \SI{20}{\km} facility, for instance, any frequency between 1 and \SI{3}{\kHz}
can be targeted, and the target frequency can be changed between observing
runs.
\begin{table}[t]
    \centering
    \begin{tabular}{l l l}
    \toprule
         {\textsf{Noise}}          &
            {\textsf{Scaling}}     &
            {\textsf{Remarks}}     \\
    \midrule
        Coating Brownian            &
            $1\big/L^{3/2}$          &
            Fixed cavity geometry   \\
        Substrate Thermo-Refractive &
            $1\big/L^2$             &
            Fixed cavity geometry   \\
        Suspension Thermal          &
            $1/L, 1$                &
            Horizontal, vertical noise \\
        Seismic                     &
            $1\big/L, 1$            &
            Horizontal, vertical noise  \\
        Newtonian                   &
            $1\big/L$               &
            {}                      \\
        Residual Gas Scattering     &
            $1\big/L^{3/4}$          &
            Fixed cavity geometry   \\
        Residual Gas Damping        &
            $1\big/L$               &
            {}                      \\
        *Quantum Shot Noise         &
            $1\big/L^{1/2}$          &
            Fixed bandwidth         \\
        *Quantum Radiation pressure &
            $1\big/L^{3/2}$          &
            Fixed bandwidth         \\
    \bottomrule
    \end{tabular}
    \caption{Scalings of fundamental noises with arm length $L$, referred to
astrophysical strain~\autocite{2017CQGra..34d4001A}.  The test mass radii of
curvature are varied to hold the arm cavity geometry fixed.  In the case of the
quantum shot and radiation-pressure noises (*), the given scalings are for a
fixed detector bandwidth, but these noises could instead be optimized in a
number of different ways\,---\,hence the ``compact-binary optimized'' and
``postmerger optimized'' curves in \cref{fig:ce_alternate_strains}.
    \tlabel{noise_length_scalings}}
\end{table}
\begin{figure}
    \centering
    \includegraphics[width=0.48\textwidth]{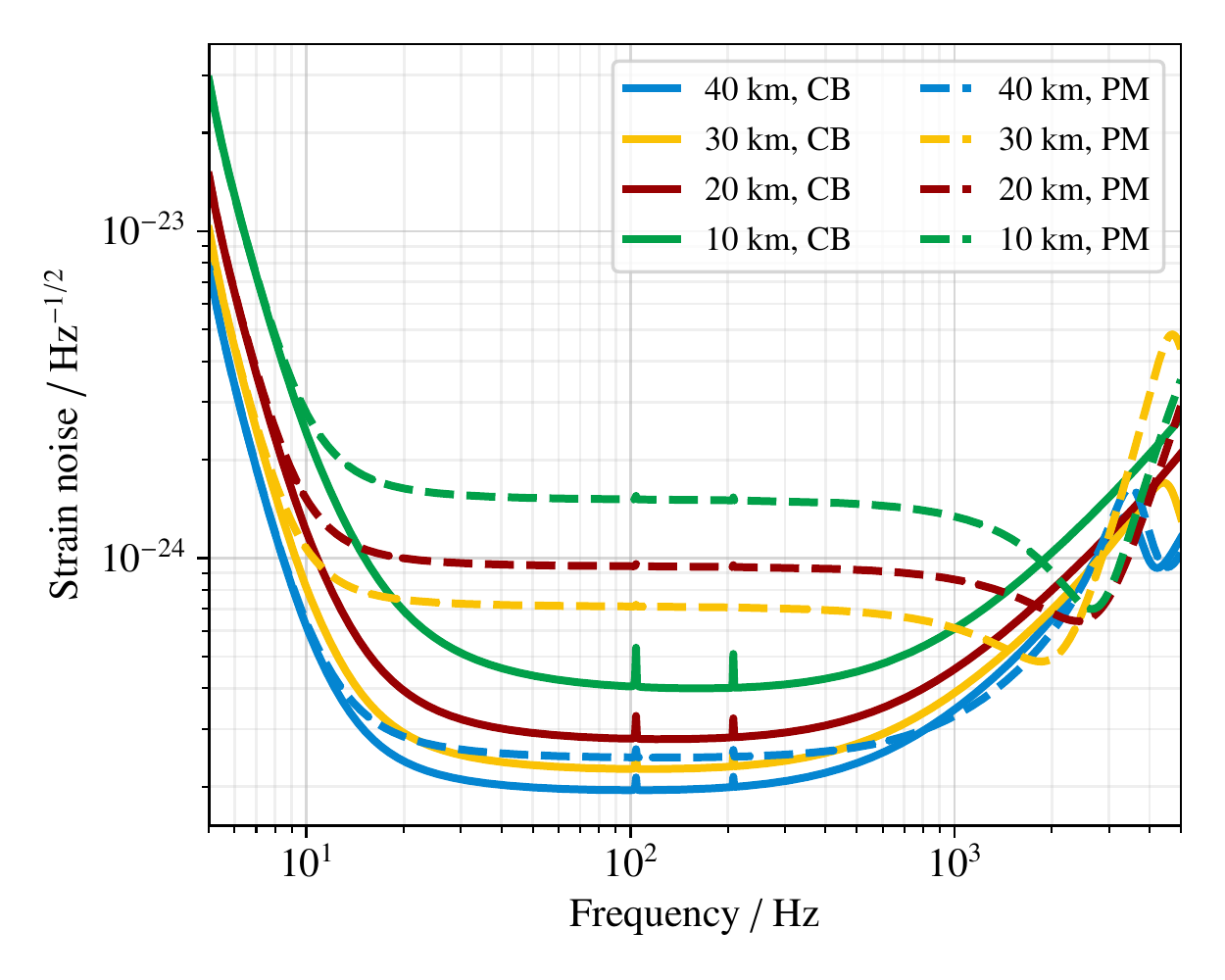}
    \includegraphics[width=0.48\textwidth]{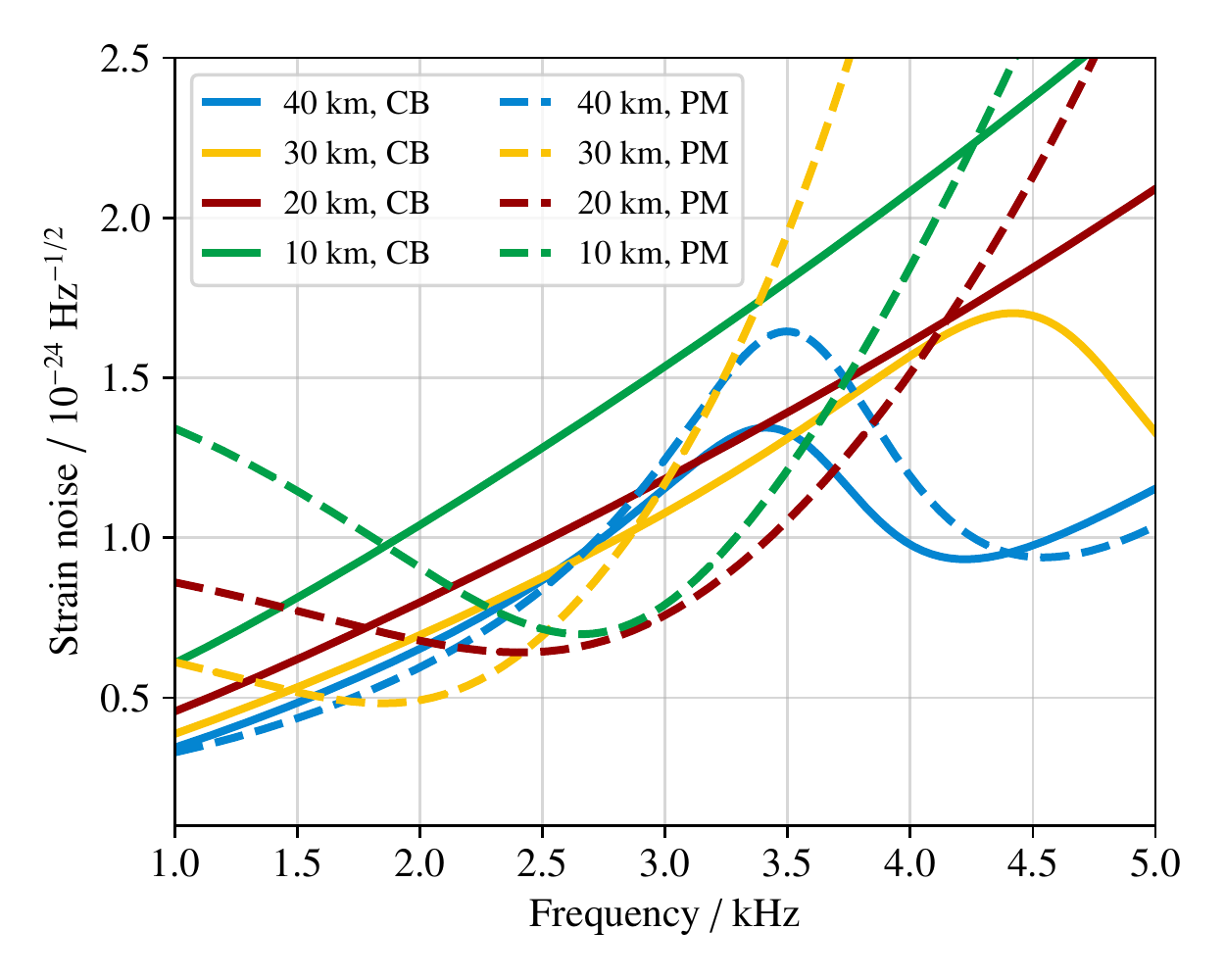}
    \caption{Amplitude spectrum of the detector noise as a function of frequency for the four Cosmic Explorer lengths considered in this comparative performance study. PM denotes the postmerger optimized configuration and CB denotes the compact-binary optimized configuration. The plot on the left shows the broadband sensitivity and the plot on the right shows the benefits of the postmerger optimized configurations at high frequencies.}
\flabel{ce_alternate_strains}
\end{figure}
\subsection[]{Multiple Interferometers}
\xlabel{Multiple Interferometers}

The current reference concept for Cosmic Explorer includes two facilities that
are geographically separated while still being located in the United States.
While the reference maximizes the scientific output of CE, alternate
configurations may involve one, two or three detectors which may or may not be
colocated.

\paragraph{One Large CE versus Two Smaller CEs}
It is apparent that a \SI{40}{km} facility will cost more than a \SI{20}{km}
facility, and one might imagine that two of the smaller facilities can be built
for a price similar to that of a single larger detector.  (This is not true;
see \cref{sec:cost_drivers}.) The merits of a longer detector lie in the
scaling of various noises with length (see \tref{noise_length_scalings}), while
those of multiple detectors lie in the freedom to sample different
gravitational-wave polarizations and to geographically separate the detectors
to improve localization of sources.  For CE, where the arm-length approaches
the wavelength of the gravitational waves of interest for some science goals,
the bandwidth of the detector is also an important consideration.  Thus, the
impact of choosing a single detector over a pair of detectors will be different
for different science goals.

As a simplified example in which high-frequency free spectral range effects are not a concern, 
consider two
\SI{10}{km} facilities versus a single \SI{20}{km} facility.  When tuned to a broadband
configuration, quantum shot noise will dominate in most of the detection band. 
If we keep the quantum noise tuned to the same fixed bandwidth, from
\tref{noise_length_scalings} we can see that these two configurations will give
similar results (i.e., both a factor of $\sqrt{2}$ more sensitive than a single
\SI{10}{km} facility).  At lower frequencies, where quantum radiation pressure,
coating Brownian noise, and Newtonian noise are dominant, the longer detector
will provide superior performance.  The same is true for mid-frequencies if we tune
the quantum noise to trade  bandwidth for
low- and mid-frequency performance. This consideration, combined with the fact
that cost is not simply proportional to arm length, drives us to consider long
detectors, and is the reason why the CE reference concept revolves around
matching the detector to the gravitational-wave wavelength relevant to our
science goals.
 
This brings us back to the original question: one large observatory, or two
smaller ones?  In \sref{trade-study} we consider combinations of
\twentykm\ and \fortykm\ observatories and the merits of each depend on the
science question being answered.  Broadly speaking, when operating without
international partners (e.g. Einstein Telescope or CE South in Australia), having two
Cosmic Explorer detectors is critical for many science goals
(see \flowdown).
However, when operating as part of a global network, the sensitivity advantage of a \fortykm\ observatory
outweighs the benefit from the additional detector in a two \twentykm\ configuration (see \flowdown) for most science goals.
As previously noted a single \fortykm\ observatory is also somewhat less
expensive than a pair of \twentykm\ observatories (see \tref{costs}).
A single \twentykm\ observatory would clearly be the worst of the considered Cosmic Explorer configurations (see \flowdown{} and \cref{sec:risk}).

\paragraph{Two Interferometers in a Single Vacuum Envelope}

Given the expense of building a CE observatory, one might expect that housing
multiple interferometers in the same vacuum envelope would allow for greater
sensitivity and flexibility with relatively little added cost.  Putting both a
\twentykm\ and a \fortykm\ CE in the same \bmts, for instance, appears to yield
an observatory capable of simultaneously optimizing compact-binary and
postmerger science.  The Initial LIGO Hanford Observatory did, in fact, operate
with two interferometers in the same \bmts, albeit for different reasons.

Unfortunately, this approach requires that the clear aperture of the \bmts\ be
large enough to put the interferometer optics side-by-side with enough
separation to avoid mechanical and optical interactions between the detectors.
This, in turn, requires that the \bmt\ diameter be roughly doubled to house two
interferometers, which leads to a number of practical and economical
challenges.

The immediate practical issue is that CE, which has a characteristic beam
diameter of several tens of centimeters along its entire length, already
requires the largest pipe diameter and vacuum hardware (e.g., gate valves)
commonly made. A special process would be required to produce many kilometers
of larger diameter pipe and special orders would be required for all \bmt\
related hardware.  The second issue is that a tube with diameter $2D$ requires
twice as much material per unit length as two tubes each with diameter $D$,
since the pipe wall thickness required to support atmospheric pressure must
increase linearly with diameter in order to maintain the same safety margin (or
else a more sophisticated manufacturing process, e.g., stiffening rings or
corrugated pipes, is required)~\autocite{TimoshenkoGere}.  These factors
combined would increase the cost of the vacuum system by more than a factor of
2, making separate vacuum systems a clearly superior approach.  Using separate
vacuum systems also avoids potential optical interactions between the
interferometers\,---\,a problem which plagued the initial LIGO Hanford
Observatory.

One approach that could work is to ``time multiplex'' the vacuum system.
That is, to build an observatory that can operate as either a  \twentykm\
\emph{or} a \fortykm\ CE by building mid-stations (\twentykm\ from the vertex)
capable of housing test-masses.  The mid-station mirrors could be installed
when dense matter science motivates the \twentykm\ postmerger-optimized
configuration.  The advantages of this configuration are relatively small, but
so is the additional expense, so it may be the best option if only one
observatory can be built.

\paragraph{Side-by-Side Interferometers}

While the previous discussion makes it clear that putting multiple
interferometers in the same vacuum envelope will not reduce cost, there remains
the possibility of placing two interferometers side-by-side in the same
observatory.  This approach could potentially save some fraction of the civil
engineering cost of CE (\PCTCIV\ of the total) relative to building two
separate observatories.  It also has the advantage of reducing the overall
project footprint, and thus its impact on the land and environment.

The disadvantages of this approach are clear: both detectors measure only one
gravitational-wave polarization, and there is no additional sky localization
information offered by the second detector.  This would result in a significant
compromise on the science goals: much like having only a single detector, CE
would be dependent on the rest of the next-generation network in order to
deliver on many science goals (see \flowdown).

\paragraph{Three-Detector Triangle}
Another option is to build a single triangular facility comprised of three
side-by-side detectors, which is the baseline design of the Einstein Telescope.
Such a facility is sensitive to both the $+$ and $\times$ polarization of
incoming gravitational waves. %
As a rough metric, we can compare the signal-to-noise performance of such a
three-detector triangular facility to two single-detector L-shaped facilities
with nonparallel arms, which jointly would be sensitive to both polarizations.

For a circularly-polarized overhead signal, a three-detector triangular
facility of length $L$ on a side collects the same \SNR{} as two L-shaped
single-detector facilities each with arm length $L' = \tfrac{9}{8}L$ and
oriented at \SI{45}{\degree} relative to one another, assuming in both cases
the detectors are shot-noise limited with the same bandwidth, and otherwise
have identical parameters.\footnote{Suppose the triangular facility has a side
length of $L$, so that each of the three detectors has an arm length $L$ and an
opening angle of \SI{60}{\degree}.  If a circularly polarized signal with
strain amplitude $h_+ = h_\times \equiv h$ is incident from directly overhead,
then the total \SNR{} of the triangular facility is $\rho_\triangle(L) =
\tfrac{3}{2}\rho(L)$, where
    \begin{equation}
        \rho(L) = \left(4\int\!\!\rmd\!{f} \frac{h^2}{S_h^{(L)}(f)}\right)^{1/2}
    \end{equation}
is the \SNR{} that would be accumulated by a single detector also of length $L$
with \SI{90}{\degree} opening angle, and $S_h^{(L)}(f)$ is the strain noise power
spectral density of such a detector.

Conversely, two L-shaped facilities of length $L'$ oriented at \SI{45}{\degree}
relative to one another would collect a total \SNR{} of
$\rho_{\llcorner\text{\rotatebox{45}{$\llcorner$}}}(L') = \sqrt{2}\rho(L')$.
Assuming that the detector parameters, including the bandwidth, are the same in
both the triangular and L-shaped cases, and assuming the signal occurs in a
frequency range dominated by shot noise, the detector noises are related by
$S_h^{(L')} = (L/L') S_h^{(L)}$~\autocite{2017CQGra..34d4001A}.  Equality of
the \SNR{s} $\rho_\triangle(L)$ and
$\rho_{\llcorner\text{\rotatebox{45}{$\llcorner$}}}(L')$ is then achieved for
$L' = \tfrac{9}{8}L$.} This means that instead of laying out a triangular
facility with total arm length $3L$, laying out two L-shaped facilities of the
same overall sensitivity would require a total arm length $4\times\tfrac{9}{8}L
= 4.5L$. However, since each detector in the triangular facility requires a
separate vacuum envelope, the triangular facility would require $6L$ of vacuum
tube, while the two L-shaped facilities would require only $4.5L$.  The
relative expense of these options will depend on the details of the sites and
local construction costs, but it is unclear that the decrease in total arm
length when building a triangle (relative to two L's) will ever be sufficient
to outweigh the added expense of manufacturing, housing, and operating a longer
total vacuum envelope and an extra interferometer.  (In CE, the vacuum system
makes up \PCTVAC\ of the cost, the detector \PCTDET, and the civil work
\PCTCIV, which suggests that a triangular facility would cost at least as much
as two L-shaped facilities with the same sensitivity.)

In light of the above, it is clear that building a three-detector triangular
facility is not advantageous relative to two co-located L-shaped facilities,
except possibly in an environment where excavation costs entirely dominate the
facility cost, which is not the case for CE.  Furthermore, with two L-shaped
facilities there is the clear advantage that the facilities can be separated by
a long baseline, as with the current LIGO facilities, and thereby achieve
better sky localization than a single triangular facility.  This option is also
favorable in that it only requires two interferometers to be built and
operated, rather than three interferometers as in the triangular design,
thereby reducing maintenance and operations costs.

\begin{sidewaystable}
\centering
\caption{Science traceability matrix for the science goals identified for Cosmic Explorer in this Horizon Study. This matrix illustrates the flow-down of Cosmic Explorer's high-level science goals, through the measurement objectives and requirements needed to accomplish these goals, to the minimum required instruments and the corresponding projected instrument performance. The instruments chosen here are the minimum set needed for Cosmic Explorer alone to realize each science goal. We emphasize that this study is a starting point for community input to the final design of Cosmic Explorer and encourage the community to provide input to expand and refine Cosmic Explorer's science traceability matrix.
\label{traceability}
 }
\scriptsize
\sisetup{retain-unity-mantissa=true}
\begin{tabular}{
m{0.15\textheight} m{0.18\textheight} 
m{0.19\textheight} m{0.145\textheight} 
m{0.235\textheight}
}

\cellcolor[HTML]{FFE1E1} &\cellcolor[HTML]{D1F3D4} &\cellcolor[HTML]{D1F3D4} &\cellcolor[HTML]{DCDDFB} &\cellcolor[HTML]{DCDDFB} \\
\cellcolor[HTML]{FFE1E1}{\large\textbf{\textsf{Science \mbox{Objectives \rule{0pt}{0.8em}}}}} &\cellcolor[HTML]{D1F3D4}{\large\textbf{\textsf{Measurement \mbox{Objectives \rule{0pt}{0.8em}}}}} &\cellcolor[HTML]{D1F3D4}{\large\textbf{\textsf{Measurement \mbox{Requirements  \rule{0pt}{0.8em}}}}} &\cellcolor[HTML]{DCDDFB}{\large\textbf{\textsf{Instruments}}} &\cellcolor[HTML]{DCDDFB}{\large\textbf{\textsf{Projected \mbox{Instrument Performance\rule{0pt}{0.8em}}}}}\\

\cellcolor[HTML]{FFE1E1} &\cellcolor[HTML]{D1F3D4} &\cellcolor[HTML]{D1F3D4} &\cellcolor[HTML]{DCDDFB} &\cellcolor[HTML]{DCDDFB} \\\arrayrulecolor{white}\hline\arrayrulecolor{white}\hline\arrayrulecolor{white}\hline\arrayrulecolor{white}\hline\arrayrulecolor{white}\hline
\cellcolor[HTML]{FFE1E1} &
\cellcolor[HTML]{D1F3D4}Observe black holes formed by Population III stars. &
\cellcolor[HTML]{D1F3D4}Detect tens of binary black hole mergers at $z>10$ and measure their redshifts better than \SI{10}{\percent}. &
\cellcolor[HTML]{DCDDFB}A \SI{40}{\km} and a \SI{20}{\km} Cosmic Explorer. &
\cellcolor[HTML]{DCDDFB} Two detectors with strain sensitivity better than \SI{1e-24}{\big/\!\sqrt{\Hz}} below \SI{10}{\Hz}, increasing to no more than \SI{2e-25}{\big/\!\sqrt{\Hz}} up to \SI{50}{\Hz}.\\
\multirow{-3}{=}[-2em]{\cellcolor[HTML]{FFE1E1}\textbf{Black holes and neutron stars throughout cosmic time.}} &
\cellcolor[HTML]{D1F3D4}Determine if supermassive black holes at $z \sim 8$ are built from hierarchical mergers of smaller black holes. &
\cellcolor[HTML]{D1F3D4}Detect thousands of binary black hole mergers at $z>8$ and measure their redshifts better than \SI{20}{\%} and source frame mass better than \SI{10}{\%}. &
\cellcolor[HTML]{DCDDFB}Two \SI{20}{\km} Cosmic Explorer, or a \SI{40}{\km} and a \SI{20}{\km} Cosmic Explorer. &
\cellcolor[HTML]{DCDDFB}Two detectors with strain sensitivity better than \SI{1e-24}{\big/\!\sqrt{\Hz}} below \SI{10}{\Hz}, increasing to no more than \SI{3e-25}{\big/\!\sqrt{\Hz}} up to \SI{500}{\Hz}.\\
\cellcolor[HTML]{FFE1E1} &
\cellcolor[HTML]{D1F3D4}Determine the formation and evolution channels of black holes and neutron stars formed at $z < 4$. &
\cellcolor[HTML]{D1F3D4}Detect tens of thousands binary black hole mergers up to redshift of 4 and measure their redshifts to better than \SI{5}{\%} and source frame mass to better than \SI{10}{\%}. &
\cellcolor[HTML]{DCDDFB}Two \SI{20}{\km} Cosmic Explorer, or a \SI{40}{\km} and a \SI{20}{\km} Cosmic Explorer. &
\cellcolor[HTML]{DCDDFB}Preferably two detectors with strain sensitivity better than \SI{5e-25}{\big/\!\sqrt{\Hz}} from \SI{10}{\Hz} to \SI{500}{\Hz}. \\\arrayrulecolor{white}\hline\arrayrulecolor{white}\hline\arrayrulecolor{white}
\hline\arrayrulecolor{white}\hline
\cellcolor[HTML]{FFE1E1} &
\cellcolor[HTML]{D1F3D4}Measure the structure and composition of neutron stars. &
\cellcolor[HTML]{D1F3D4}Measure the neutron star radius to within $< \SI{100}{\m}$ across the full mass range of neutron stars and all plausible nuclear equations of state. &
\cellcolor[HTML]{DCDDFB}A \SI{40}{\km} Cosmic Explorer, or two \SI{20}{\km} Cosmic Explorer. &
\cellcolor[HTML]{DCDDFB}Strain sensitivity better than \SI{3e-25}{\big/\!\sqrt{\Hz}} between \SI{20}{\Hz} to \SI{400}{\Hz}, rising to no more than \SI{6e-25}{\big/\!\sqrt{\Hz}} at \SI{1.5}{\kHz}. \\
\cellcolor[HTML]{FFE1E1} &
\cellcolor[HTML]{D1F3D4}Explore the high density, finite temperature phase space of quantum chromodynamics. &
\cellcolor[HTML]{D1F3D4}Detect neutron star post-merger remnants with a signal-to-noise ratio $> 8$ with measurement of merging object's masses to better than \SI{1}{\%}. &
\cellcolor[HTML]{DCDDFB}A \SI{20}{\km} Cosmic Explorer, or two \SI{40}{\km} Cosmic Explorer. &
\cellcolor[HTML]{DCDDFB}High frequency strain sensitivity better than \SI{1e-24}{\big/\!\sqrt{\Hz}} between \SIrange{2}{4}{\kHz}.\\
\cellcolor[HTML]{FFE1E1} &
\cellcolor[HTML]{D1F3D4}Trace the chemical evolution of the universe using multi messenger observations of neutron star mergers. &
\cellcolor[HTML]{D1F3D4}Detect and localize neutron star mergers to \SI{1}{deg.^2} and \SI{1}{\percent} in distance at least two minutes prior to merger. &
\cellcolor[HTML]{DCDDFB}Two \SI{20}{\km} Cosmic Explorer, or a \SI{40}{\km} and a \SI{20}{\km} Cosmic Explorer, and electromagnetic observatories. &
\cellcolor[HTML]{DCDDFB}Two detectors with strain sensitivity better than \SI{5e-25}{\big/\!\sqrt{Hz}} from \SI{20}{\Hz} to \SI{2}{\kHz}. \\
\multirow{-4}{=}[7em]{\cellcolor[HTML]{FFE1E1}\textbf{Dynamics of dense matter.}} &
\cellcolor[HTML]{D1F3D4}Connect the properties of binary neutron star mergers to the physics of the relativistic jets powering short gamma-ray bursts. &
\cellcolor[HTML]{D1F3D4}Detect mergers and measure binary inclination to within \SI{1}{\%} for neutron star mergers out to $z \sim 2$. &
\cellcolor[HTML]{DCDDFB}Two \SI{20}{\km} Cosmic Explorer, or a \SI{40}{\km} and \SI{20}{\km} Cosmic Explorer, and gamma-ray burst satellites. &
\cellcolor[HTML]{DCDDFB}Two detectors with strain sensitivity better than \SI{5e-25}{\big/\!\sqrt{Hz}} from \SI{20}{\Hz} to \SI{2}{\kHz}. \\\arrayrulecolor{white}\hline\arrayrulecolor{white}\hline\arrayrulecolor{white}\hline\arrayrulecolor{white}\hline

\cellcolor[HTML]{FFE1E1}\textbf{\mbox{Extreme gravity,} \mbox{fundamental physics, and} \mbox{discovery potential.}} &

\cellcolor[HTML]{D1F3D4}Study the nature of strong gravity, search for unusual or novel compact objects, explore the implications of quantum gravity, and probe dark matter and dark energy. &
\cellcolor[HTML]{D1F3D4}Open the largest possible discovery space in gravitational-wave strain sensitivity with a network that is sensitive to both polarizations of the gravitational-wave signal. &
\cellcolor[HTML]{DCDDFB}A \SI{40}{\km} and a \SI{20}{\km} Cosmic Explorer, preferably two \SI{40}{\km} Cosmic Explorer detectors. &
\cellcolor[HTML]{DCDDFB}At least one \SI{40}{\km} Cosmic Explorer facility with strain sensitivity better than \SI{2e-25}{\big/\!\sqrt{\Hz}} across the widest possible frequency band, tunable to \SI{1e-25}{\big/\!\sqrt{\Hz}} in a narrower band. \\
\end{tabular}
\end{sidewaystable}

\clearpage

\begin{table}[ht]
    \centering
    
    \caption{
    \protect%
This table indicates the accessibility of astrophysical sources that can
advance key next-generation science goals. A US Cosmic Explorer consisting of
one \twentykm\  observatory, one \fortykm\  observatory, or a pair of observatories of 20
or \fortykm\  length are evaluated in the presence a background network that
includes second-generation (2G) gravitational-wave observatories, the EU
Einstein Telescope (ET), and a \twentykm\  Cosmic Explorer-like detector located in
Australia (CE South). For each goal, the colors range from gray (least
favorable, science goal not achieved) to green (good, science achievable)
and dark green (most favorable).
Longer, more sensitive detectors are
generally better, and a network is required for many science goals.  For
example, studying black holes from the first stars requires a \fortykm\  detector
that can see black holes at $z\gtrsim 10$ in a network that can measure both
gravitational-wave polarizations to accurately measure the holes' redshifts.
Only observations of neutron star post-merger signatures benefit from a \twentykm\
detector. The higher bandwidth of the \twentykm\  observatory allows for better
narrow-band tuning for this particular source, although only one detector needs
to be in this configuration.
Detailed descriptions of the metrics that
determine the criteria can be found in \cref{subsec:impact_on_science_goals}.
The final row, labeled ``Technical risk'', represents the risk that Cosmic Explorer's
 scientific output will be limited by technology shortfalls;
 light orange is lowest risk, and red is highest risk.
The \twentykm\ detectors incur a higher risk rating due to the more severe sensitivity
impact from underperforming technologies (thermal noise, signal extraction cavity losses, etc.).
We emphasize that this study is a starting point for community input on Cosmic Explorer.

    \label{flowdown}}
\end{table}
\clearpage

\section{Trade-Study Outline \hlight{[Sathya]}}
\xlabel{trade-study}

The optimization of CE design in the context of a variety of potential future
global gravitational-wave detector networks is a complex task.  The process
used to perform this optimization is referred to herein as a ``trade study''
since we are looking for trade-offs which are likely to maximize the scientific
output of CE both in the near-term and integrated over the lifetime of the
facility.  This section gives a brief outline of the trade study, while leaving
a full technical description to the literature \autocite{Adhikari:2019zpy,
2019CQGra..36v5002H, Borhanian:2021abc, Borhanian:2021xyz}.

The trade study considers the performance of CE design variants both in the
context of the existing 2G detector network, and in the presence of
representative next-generation facilities.  Specifically, nine detector
locations are considered in this study: the five 2G detector sites, including
LIGO India (Hingoli, India), and four representative sites for the 3G
detectors.  Since the locations of future detectors are unknown, we choose
locations which we expect are plausible based on geophysical considerations,
knowing that the exact location of a detector has little impact on network
performance~\autocite{2019CQGra..36v5002H}.  The Einstein Telescope's reference
location is set to be the same as Virgo, while the three possible
Cosmic Explorer locations C, N, and S are set to be sites in Idaho
(USA), New Mexico (USA), and New South Wales (Australia).  The spread of these
locations around Earth is shown in \fref{locs_map}.  The right-hand plot in
\fref{locs_map} provides a graphical summary of 2G and 3G detector generations
and design concepts considered at each location.  The Cosmic Explorer
sensitivity curves used in the trade-study are shown in \cref{fig:ce_alternate_strains}.
For the sake of brevity, we include results for only 20 km and 40 km Cosmic Explorer configurations although the results are available, and will be made public online, for other arm lengths.

\begin{figure}[b!]
    \includegraphics[width=0.49\textwidth]{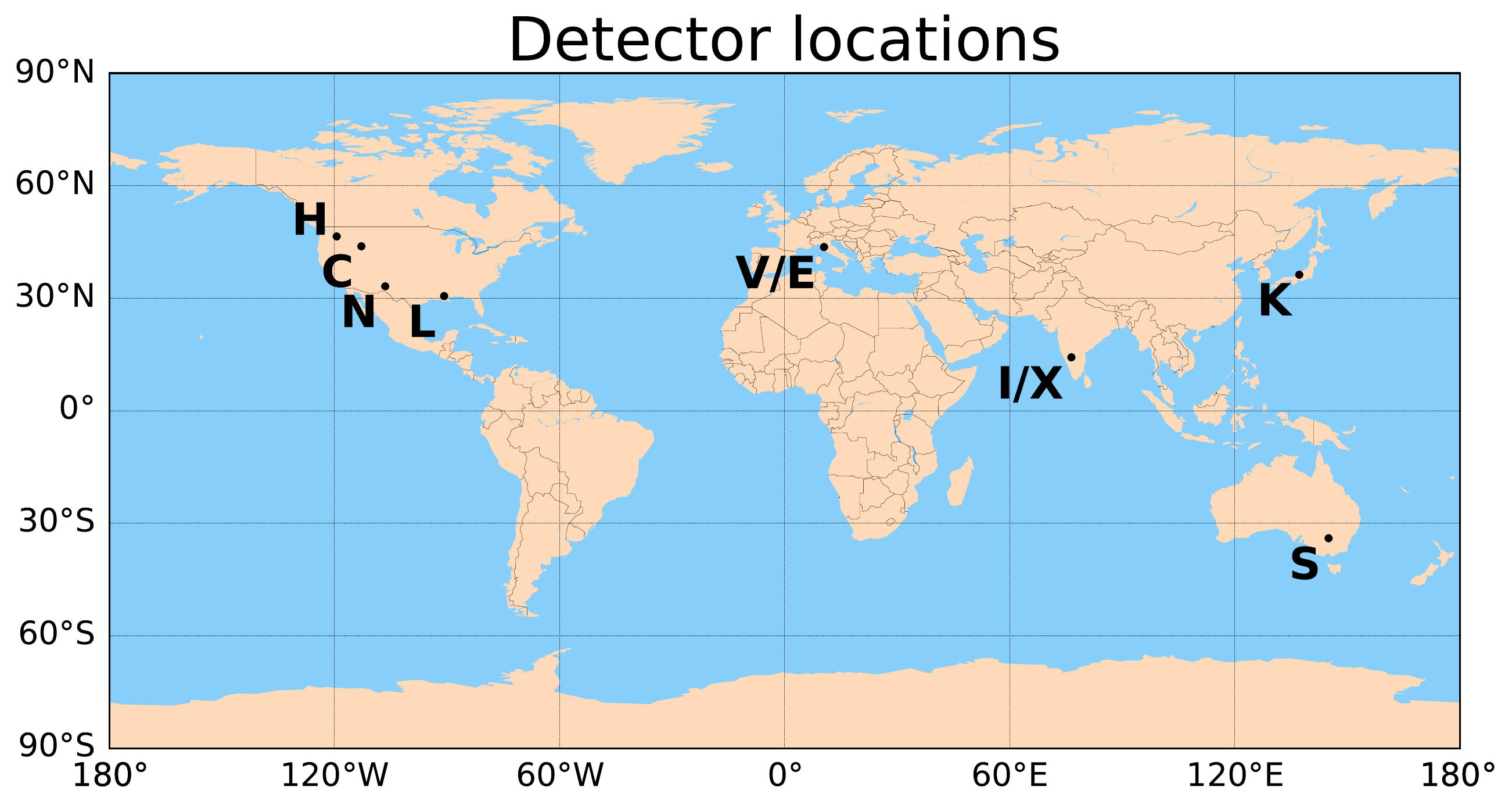}
    \includegraphics[width=0.49\textwidth]{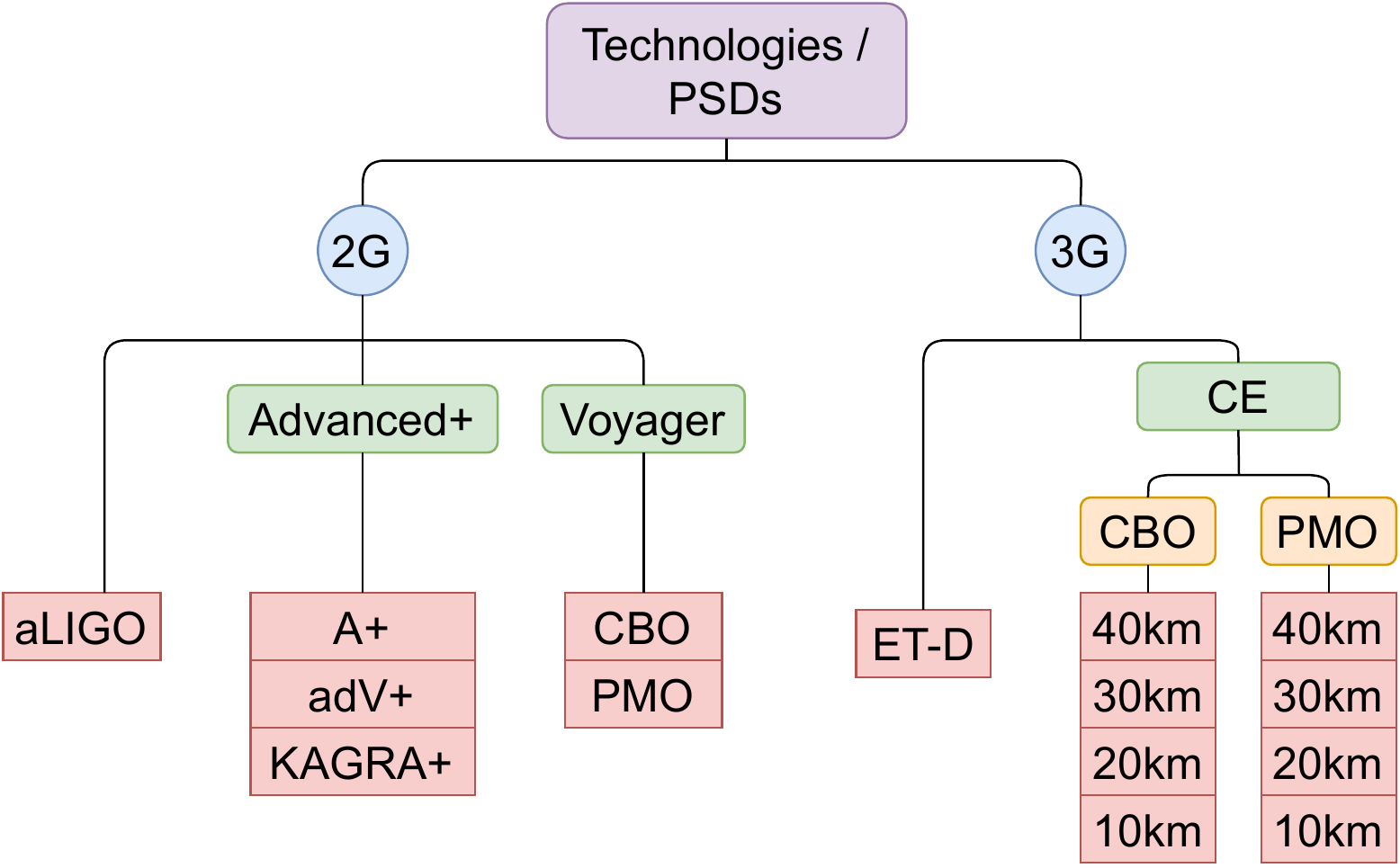}
    \caption{The left plot shows detector locations considered in this trade
study while the right plot is a graphical summary of the choices available for
installing detectors of different arm lengths and sensitivities at various
locations.  The LIGO A+, AdVirgo+, KAGRA+, and Voyager detectors are located at the 5 existing
sites (labeled H, L, V, K and I).  Since its actual future location is unknown,
but certainly in Europe, the Virgo site is used as our reference location for
the Einstein Telescope.  Cosmic Explorer facilities are considered at three possible locations, two
in the US (C and N), and one in Australia (S).  For each instance of CE in the
US, a variety of configurations are considered (encoded as different power spectral densities, PSDs, and shown on the right side, under the
green CE label).  These include arm lengths from 10 to \SI{40}{km} and compact-binary or postmerger optimizations (CBO or PMO).}
    \flabel{locs_map}
\end{figure}

\begin{figure}[p]
    \centering\includegraphics[width=0.99\textwidth]{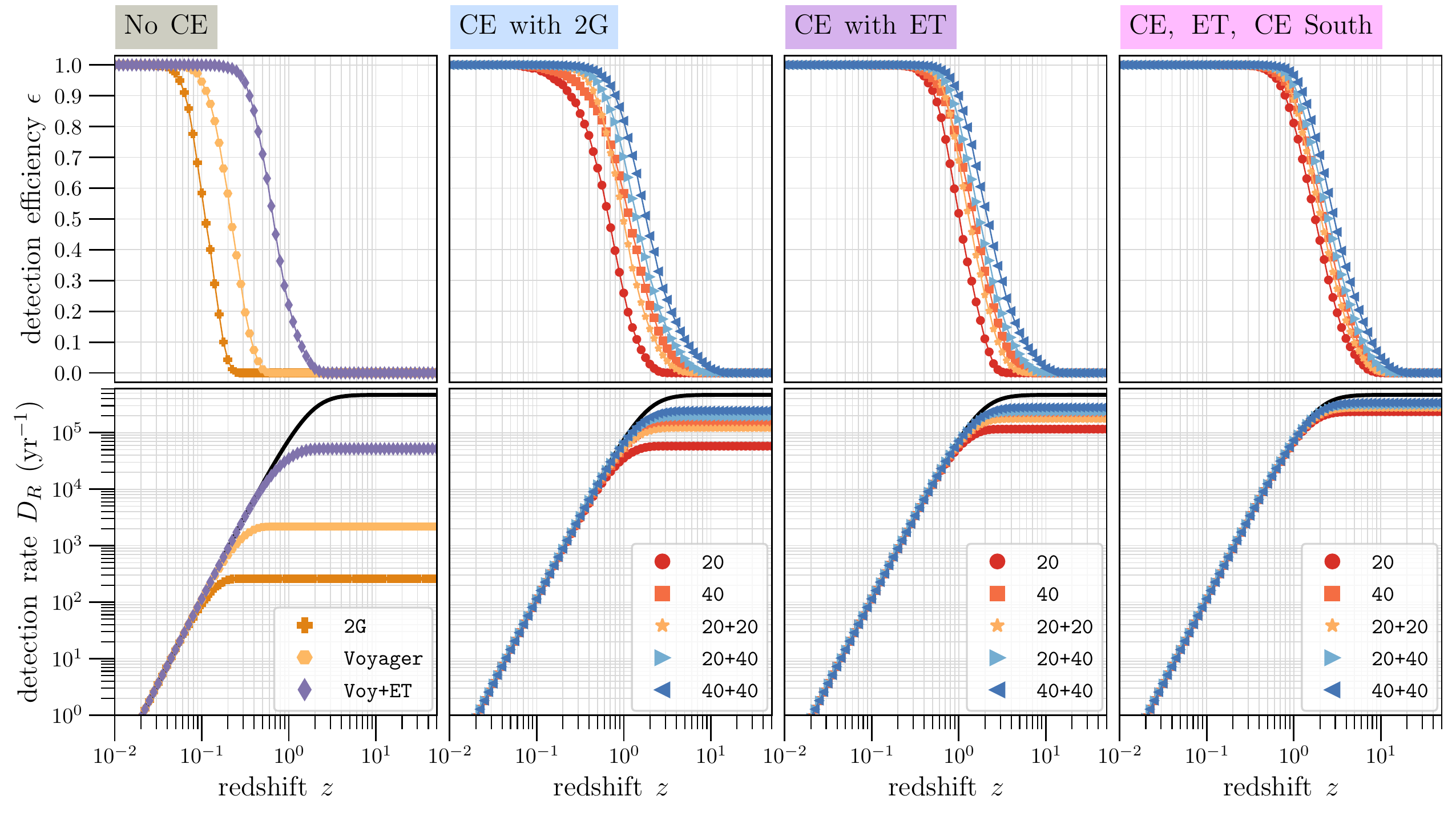}
    \centering\includegraphics[width=0.99\textwidth]{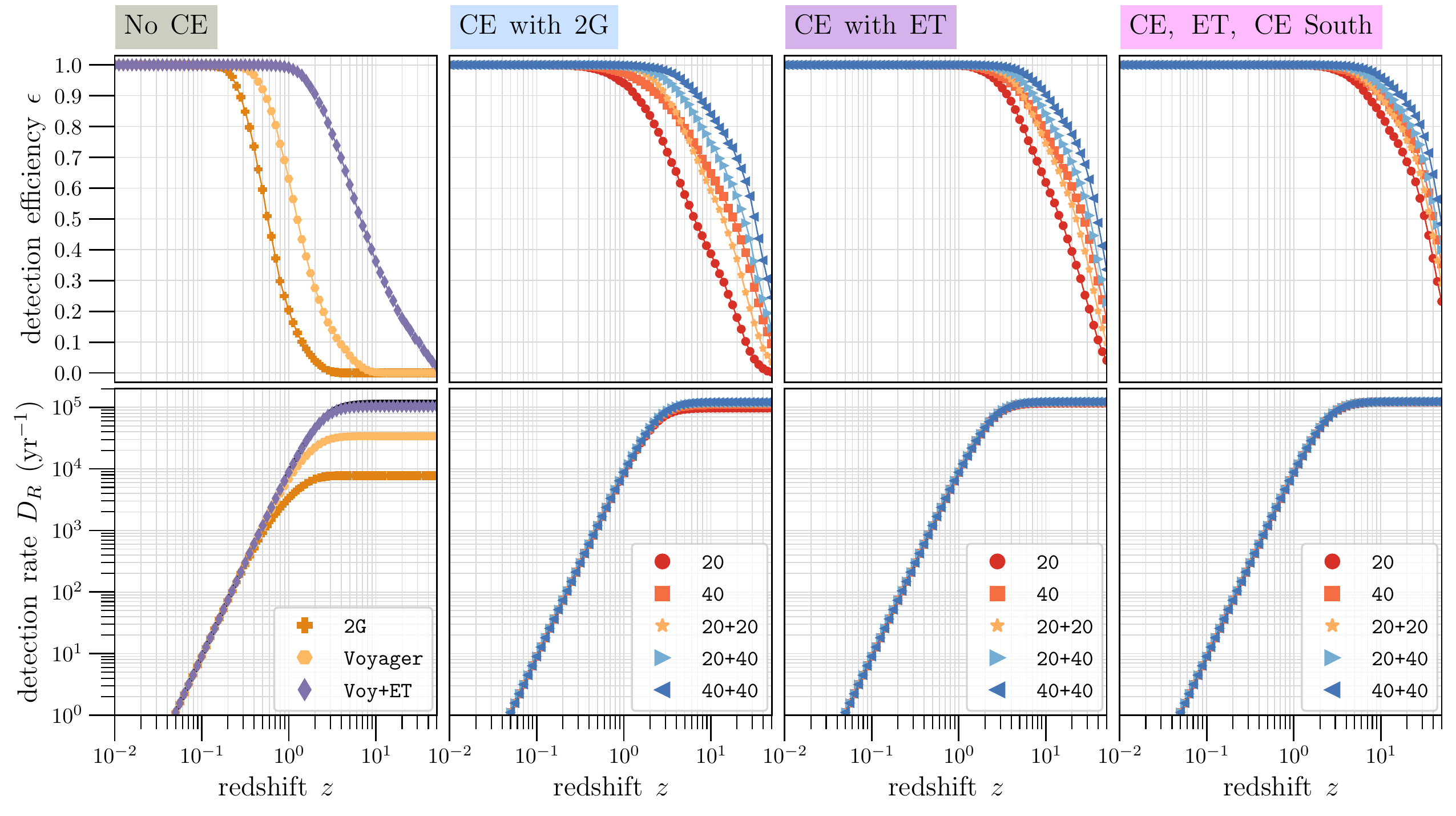}
    \caption{The detection efficiency (top and third rows) and the cumulative
detection rate from the galactic field binaries (second and fourth rows) of events with signal-to-noise ratio greater than 10 for binary neutron stars (top two panels) and binary black holes (bottom two panels).  For a given redshift $z$, the detection efficiency $\epsilon$ is defined as the ratio of the number of detected sources to the total number of sources (shown as solid, black lines in the third and bottom rows) out to that distance. 
The networks are exactly as in \flowdown.
}
    \label{fig:rates_volume}
\end{figure}

Many different performance metrics are used in the trade-study to capture
network performance and its impact on scientific output
\autocite{Adhikari:2019zpy, 2019CQGra..36v5002H, Borhanian:2021abc,
Borhanian:2021xyz}.  A key ingredient in almost all performance metrics is the
rate of events expected to be observed by different detector networks as a
function of redshift.  Based on observations so far, the local (i.e., $z\ll 1$)
merger rates inferred for the population of binary neutron stars (BNS) and
binary black holes (BBH) are
$\mathcal{R}_\text{BNS}=320\, \mathrm{Gpc^{-3}\, yr^{-1}}$ and
$\mathcal{R}_\text{BBH}=23.8\, \mathrm{Gpc^{-3}\, yr^{-1}}$~\autocite{LIGOScientific:2020kqk}, broadly consistent with expectations from multiple astrophysical formation channels~\autocite{Mandel:2021smh}.
\cref{fig:rates_volume} plots the cosmic merger rate as a function of redshift
for the two source populations. This rate model begins with a Madau--Dickinson
star-formation rate as a function of redshift~\autocite{Madau:2014bja}, and
then accounts for the characteristic time delay from binary formation to
merger, including the effects of metallicity for BBHs ~\autocite{2020arXiv201209876N}.

The component masses of the BNS population are chosen to be Gaussian
distributed with mean $1.34\,M_\odot$, standard deviation $0.15\,M_\odot$,
minimum mass $1\,M_\odot$, and maximum mass $2\,M_\odot$. The primary masses of
the BBH population are chosen to follow the so-called \textsc{power law + peak}
distribution \autocite{2021ApJ...913L...7A} with lower and upper cutoffs at
4.59 and $86.22\,M_\odot$, while the secondary mass is sampled uniformly
between the lower cutoff and the primary mass component. This deviation from
the original \textsc{power law + peak} model allows for the examination of a broader
mass ratio range with the BBH population.

\begin{boxenv}{!h}{green!10}{\textwidth}
	\caption{Questions addressed by the trade study, and their answers in brief.}
	\label{box:questions}
	\small
In addition to a general science-per-dollar optimization, we also use the trade
study as a means of answering the following frequently asked questions:
\begin{itemize}
    \item Is it better to build one large CE, or two smaller ones? \\
    \hspace{5ex}\emph{Science goals requiring excellent source localization drive the strong desire for a network
    of detectors. The significantly increased broadband sensitivity of a \fortykm\ detector is advantageous 
    for science goals that require high signal-to-noise observations. 
     Compared to one \fortykm\ observatory, two \twentykm\ observatories are somewhat more expensive (see \sref{Multiple Interferometers}).
     The \fortykm\ $\text{+}$ \fortykm\ and the \fortykm\ $\text{+}$ \twentykm\ configurations had the best performance in the study. Because of cost considerations and the tunability advantage for neutron star post-merger signal the reference configuration was chosen as the \fortykm\ $\text{+}$ \twentykm\ configuration.
     (see \flowdown).}

    \item Should a second CE be built in the US, internationally, or both?\\
    \hspace{5ex}\emph{Having a long separation between observatories is favorable
     for localizing and characterizing gravitational-wave sources, so if only two CE
     observatories are built it is best for the second to be located far away (e.g., in Australia).
     However, two observatories in the US can be separated by a sufficient distance to
      precisely localize a large number of sources, making the key science goals accessible
      (see \flowdown).}

    \item Would a triangular Einstein Telescope-like design make sense for CE? \\
    \hspace{5ex}\emph{A triangular configuration is not advantageous in places where above-ground construction is feasible (e.g., US and Australia, see \sref{Multiple Interferometers}).}

    \item To what degree are our key science goals dependent on the global detector network? \\
    \hspace{5ex}\emph{This question drives much of the complexity of our trade study,
     and the answer is graphically captured in \flowdown.
    The short answer is: the key science goals are achievable with the reference CE configuration
     (one \fortykm\ and one \twentykm\ observatory), while a network of
     three or more next-generation detectors will increase the rate at which these goals are achieved.}
\end{itemize}
\end{boxenv}
\subsection[]{Impact on Key Science Goals \hlight{[Sathya]}}
\label{subsec:impact_on_science_goals}

To assess the impact of Cosmic Explorer design choices on our capacity to
accomplish the science goals described in \cref{ch:keyquestions}, we identify a
set of performance metrics for each science goal and then evaluate the
capability of Cosmic Explorer. We perform this evaluation for several
scenarios, with Cosmic Explorer either in the presence or absence of other
detectors.  \cref{box:questions} summarizes the main conclusions of this study,
and \cref{box:rates} summarizes the key observation rates for Cosmic Explorer.

\begin{boxenv}{!h}{green!10}{\textwidth}
	\caption{Observation Rates for Compact Binaries with Cosmic Explorer.}
	\label{box:rates}

Cosmic Explorer's ability to detect\footnote{We use signal-to-noise ration greater than 10 as detection criteria.} large numbers of compact binaries out to large cosmological distances is driven by the low-frequency sensitivity of the \SI{40}{\km} detector. In one year of observations, such a detector will:
	\begin{itemize}%
		\item Observe \CEBNSPERYEAR\ binary neutron star mergers (one every 100 seconds),
		\begin{itemize}
		\item %
		including \SI{80}{\percent} of all mergers within $z=1$,
		allowing association with EM transient surveys, %
		\item of which \CELOUDBNSPERYEAR\ will have SNR $>300$, providing access to postmerger physics,
		\item for thousands will provide distance and sky localization with more than 10 minutes of early warning,
		\item and will observe half of all mergers out to $z=10$, allowing association with gamma-ray bursts and charting the history of supernovae and merger time delays; %
		\end{itemize}
		\item Observe \CEBBHPERYEAR\ binary black hole mergers from galactic field population (one every 5 minutes),
		\begin{itemize}
		\item of which \CENEARBBHPERYEAR\ will be nearby ($z<0.1$) with median SNR of \CENEARBBHMEDSNR\ and exceeding an SNR of \CENEARBBHLOUDSNR\ for the loudest source,
		\item and \CEFARBBHPERYEAR\ will be at cosmological distances $z>2$ (inaccessible to current networks whose most distant sources are $z\sim{1}$) and have median SNR of \CEFARBBHMEDSNR\ (i.e., with SNR similar to GW150914 in Advanced LIGO).
		\item and \CEFARTHERBBHPERYEAR\ will be at cosmological distances $z>4$ and have median SNR of 18.
		\end{itemize}
	\end{itemize}
\end{boxenv}
\subsection[]{Black Holes and Neutron Stars Through Cosmic Time}

\paragraph{Remnants of the First Stars}
The most sensitive astronomical telescopes (e.g., JWST) will be sensitive to
objects at a maximum redshift $z\sim 30$, some 100~Myr after the Big Bang,
while the first stars in the universe could have formed even earlier, a mere
30~Myr after the Big Bang or $z=70$. The network of ET and two CE facilities with 
at least one 40 km facility will be sensitive to such redshifts and beyond.  
Binaries of black hole remnants of first stars
could be observed by the 3G network. Decisively inferring that the observed
sources are remnants of first stars requires an accurate measurement of their
redshift. This can be done either using the whole population of detected
sources, and showing that it contains a high-redshift merger peak, or by
proving that individual sources have merged at redshifts higher than what is
expected from other astrophysical channels.

The first approach was followed by Ng et al.~\autocite{2020arXiv201209876N}
where it was shown that a network of 2 CE and one ET could reveal a peak of
mergers from Pop~III remnants at redshift of ${\sim}12$ (see also
\cref{fig:populations} above). Given the computational cost of that type of
analysis, here we use a simpler figure-of-merit based on individual sources.
Specifically, we focus on the fraction of events merging at redshifts $z\geq
10$ with fractional redshift uncertainty smaller than some threshold.  As shown
in \cref{fig:populations}, the peak of mergers from Pop.~III remnants is
expected to happen at $z\sim 12$ (though significant uncertainty exists).
Meanwhile, the main two late-universe populations, formation in galactic fields
or dynamical formation in globular clusters,\footnote{Other formation channels
have been proposed, e.g., nuclear star clusters, young star clusters and mergers
in the disk of active galactic nuclei. Here we focus on globular clusters and
galactic fields merely because they have been extensively studied in the
literature.} do not contribute significantly to the total merger rate for
redshifts above ${\sim}9$ (\cref{fig:populations}).  Our rough figure of merit
is thus the number of BBH sources for which the statistical uncertainty in redshift is
better than \SI{10}{\%}. This is an uncertainty for which the posterior
distribution for the redshift of a black hole binary that merges at the lowest
redshift we consider, i.e., 10, would exclude $z<9$ with $1\sigma$ level.  For
black holes whose true redshift is higher than 10, this criterion is
conservative in the sense that even an uncertainty larger than \SI{10}{\%}
could be sufficient to exclude $z<9$.  We find that no network without at least
two 3G detectors can satisfy our criterion. A network with only one 3G detector
(\ET or \CE) could \emph{detect} some sources at $z\geq 10$, but the associated
redshift uncertainty would be too large to definitively prove the merger did not
happen at smaller redshifts.  A network with two \SI{40}{\km}
Cosmic Explorer optimized for compact binaries
detection\footnote{\label{fn:lf_rates}For all science goals involving
populations of compact binaries, low frequency sensitivity is more important
than sensitivity above ${\sim} \SI{500}{\Hz}$.  Therefore, the
postmerger-optimized setting is not thoroughly discussed here.} would detect
roughly 200 sources per year that satisfy our criterion.  That number improves
by ten folds (${\sim} 2000$ sources) if the network is augmented to also include \ET.
This increase is due to the superior polarization resolving power of multiple detectors.
In \flowdown\ we mark in yellow networks that yield at least 10
viable sources per year, and in light (dark) green networks that give access to
at least 50 (100) viable sources per year. Details are available in 
a technical note~\autocite{FirstBBHtradestudy}.

\paragraph{Black Hole Seeds and Galaxy Formation}
If supermassive black holes at $z\sim 8$ were built from hierarchical mergers
of smaller black holes at higher redshifts we should detect lighter black hole
mergers at higher redshifts and heavier ones at lower redshifts. This requires
not only the capability of measuring the redshift of a BBH source, but also its
source-frame mass. We stress that precise measurement of the source-frame mass
does in turn require a precise measurement of the source redshift. This is
because gravitational-wave detectors measure redshifted mass parameters,
which are $(1+z)$ times larger than the astrophysically interesting
source-frame quantities~\autocite{2000PhR...331..283M}. Therefore, we expect
that networks with more detectors will do better at measuring source-frame
masses~\autocite{Vitale:2016icu,2018PhRvD..98b4029V}.

Our figure-of-merit to quantify the ability of a network to track the growth of
black holes across cosmic history will be the number of sources for which the
source-frame chirp mass and the redshift can be measured at least as well as
what has been reported for GW190521. This source is one of the most
interesting found to date in advanced detector data, being composed of two very
heavy stellar mass black holes, one of which might lie in the pair instability
mass gap (see \cref{ss:cosmictime}). Its formation pathway is not certain, but
it might be the result of previous-generation black hole
mergers~\autocite{2020ApJ...900L..13A}.

We would like our 3G network to be able to characterize similar sources with
equal or higher precision, at high redshifts, from 4 to 10. Quantitatively,
this requires a $1\sigma$ uncertainty on the estimation of the source-frame
chirp mass of ${\sim} \SI{10}{\%}$ or better, and a  $1\sigma$ uncertainty on the
estimation of the redshift of ${\sim} \SI{20}{\%}$ or better.

We find that at least two 3G detectors are needed. For example, a network of
two \SI{20}{\km} CEs (both compact-binary-optimized) will provide access to
${\sim} 1100$ viable sources per year.  Networks which can more precisely
resolve the two polarizations of gravitational-wave signals can yield
significantly higher numbers of sources that satisfy our criterion.  An
Einstein Telescope and a \SI{40}{\km} compact-binary-optimized Cosmic Explorer
would detect ${\sim} 3000$ viable sources per year, while the best network we
consider (ET, CE South and two \SI{40}{\km} compact-binary-optimized CEs, i.e.
ET plus three CE detectors) would yield nearly \num{10000} sources per year.

In \flowdown\ we mark in yellow networks that can detect at least 50
viable sources per year, and in light (dark) green networks that give access to
at least 500 (\num{2000}) viable sources per year. Details are available in 
a technical note~\autocite{FirstBBHtradestudy}.

\paragraph{Formation and Evolution of Compact Objects}
While the extremely high redshift universe will teach us about primordial black
holes and black holes from the first generation of stars, most of the black
holes in the universe are produced and merge at redshifts smaller than 4.  To
characterize the evolution and formation channels of these black holes one
needs a large number of black hole binaries with precise measurement of
redshifts and intrinsic parameters.  The figure of merit we use is 
the number of detected sources whose source-frame
chirp mass is measured to a $1\sigma$ uncertainty of \SI{10}{\%} or less, and
whose redshift is measured to a $1\sigma$ uncertainty of \SI{5}{\%} or less.
However, we only consider sources up to redshift of 4.

As one might expect, even networks without a Cosmic Explorer can yield some
viable sources per year, for example a network of 3 Voyager detectors can
detect ${\sim} 230$ viable sources per year.  While a single \SI{40}{\km}
Cosmic Explorer with 2G detectors can find ${\sim} 3000$ viable sources per
year, that number becomes \num{15000} if two \SI{20}{\km} CEs are used.  This
highlights that when the BBHs of interest are at redshifts of few, instead of
$>10$, having two \SI{20}{\km} CE detectors is more beneficial than having a
single larger detector. In turn this is due to the superior polarization
resolving power of larger networks.  The same pattern is observed even when ET
is included. A network of ET and a \SI{40}{\km} CE finds \num{27000} sources
per year, while a network of ET and two \SI{20}{\km} CEs finds \num{36000}.
Adding a CE South increases the number of viable sources by less than
\SI{10}{\%}.

In \flowdown\ we mark in yellow networks that can yield at least 250
viable sources per year, and in light (dark) green networks that give access to
at least \num{2500} (\num{25000}) viable sources per year. Details are available in 
a technical note~\autocite{FirstBBHtradestudy}.

\subsection[]{Dynamics of Dense Matter}

\paragraph{Neutron Star Structure and Composition}
Cosmic Explorer's ability to probe the structure and composition of neutron star matter is tied to the precision with which the 3G detector network can measure masses and tidal deformabilities from inspiral gravitational waves. In order to achieve a milestone in our knowledge of the neutron-star equation of state at zero temperature, masses and tidal deformabilities must be measured precisely enough to constrain the stellar radius to within $\NSRADIUSERR$ km across the full neutron-star mass spectrum. As the measurability of these parameters is essentially dictated by the inspiral SNR, this will require hundreds of observations of loud binary neutron star mergers. Adopting an SNR of 100 as the threshold above which we expect an informative tidal signature in the measured waveform, we assess the relative performance of different Cosmic Explorer configurations and networks for this science goal by comparing the yearly number of loud binary neutron star mergers they detect.

\begin{figure}[p]
	\centering
    \includegraphics[width=0.49\textwidth]{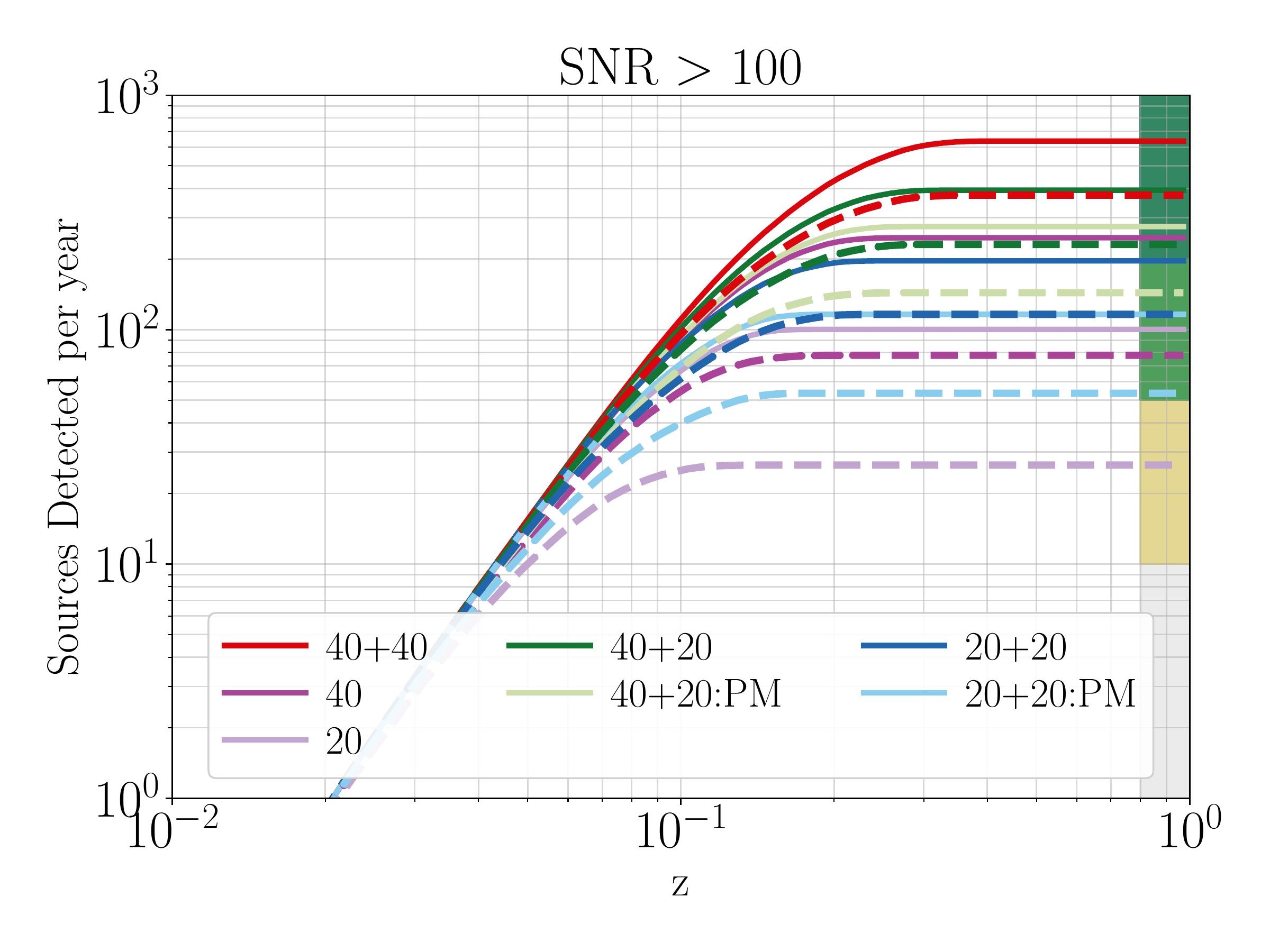}
    \includegraphics[width=0.49\textwidth]{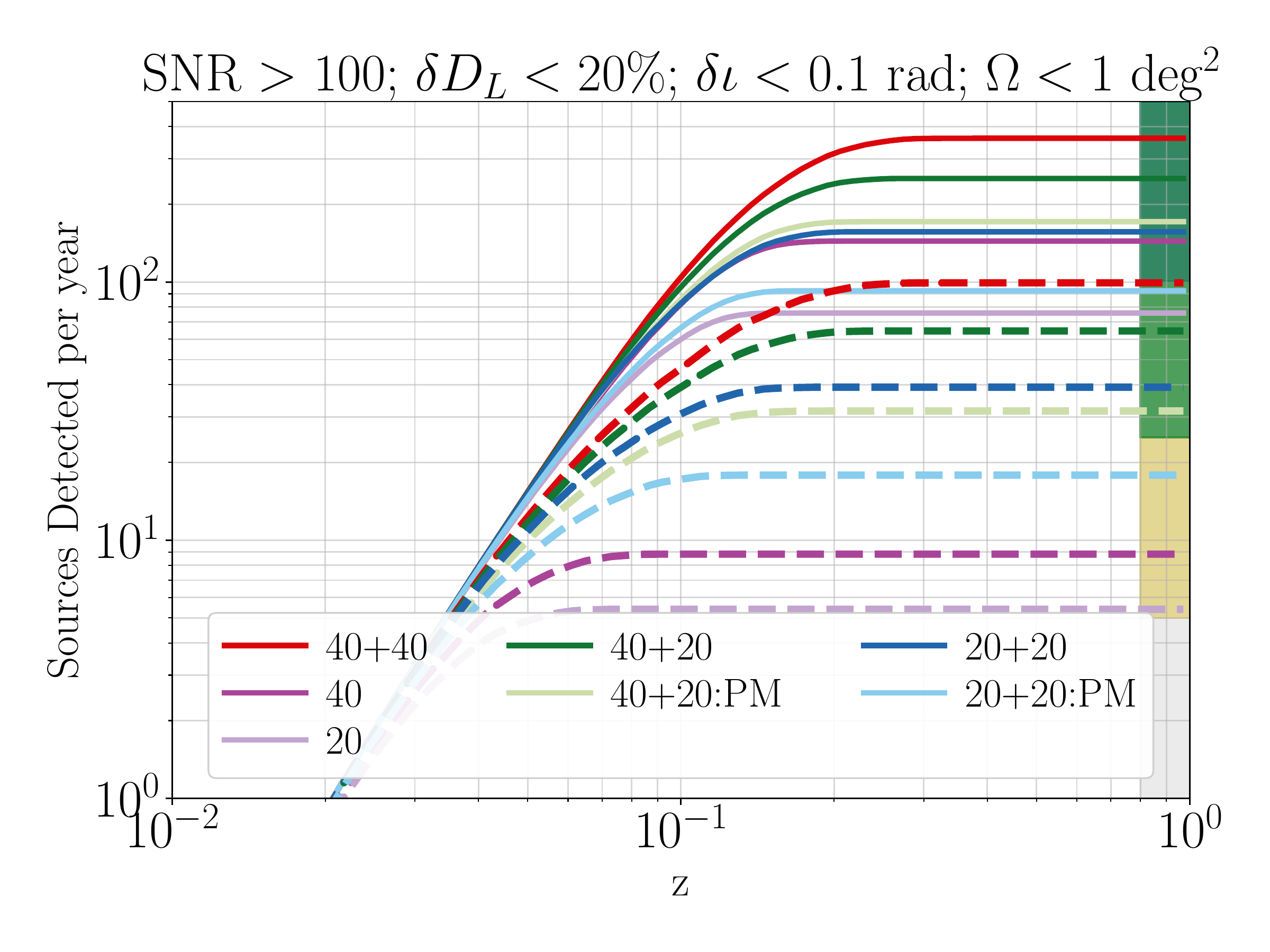}
    \includegraphics[width=0.49\textwidth]{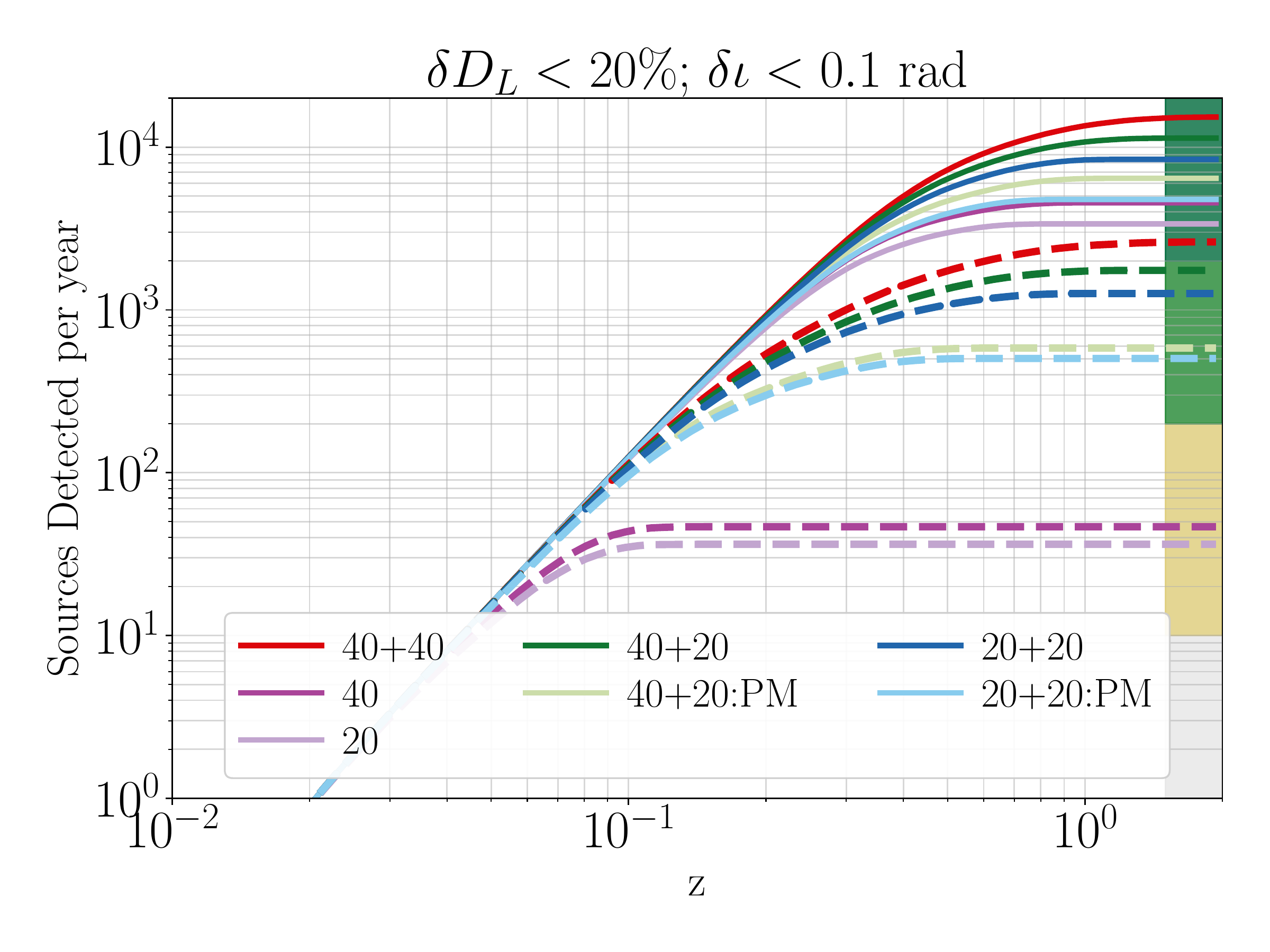}
    \caption{Cumulative number of detected BNS mergers satisfying criteria relevant to the Dynamics of Dense Matter science goals as a function of redshift, after one year of observation by different CE networks. Dashed curves refer to a 2G background network, while solid curves include ET in the network. A network's ability to achieve a given science goal is assessed based on the cumulative number of detections across all redshifts, with the colors in \flowdown\ corresponding to the colored bins shown along the right edge of the plot. \textit{Top left:} BNS mergers detected with SNR above 100, a threshold for an informative tidal signature in the signal. Networks that detect 50 (respectively, 200) such mergers per year can achieve (ensure the best return on) the Neutron Star Structure and Composition science goal. \textit{Top right:} Subset of the BNS mergers with SNR above 100 that are localized to within 1 deg$^2$ on the sky and \SI{20}{\percent} uncertainty in distance, while having their inclination $\iota$ constrained to better than $\pm 0.1$ in $\cos \iota$; these criteria allow for early warning of the merger, targeted electromagnetic follow-up and the breaking of distance and inclination degeneracies in emission models. Networks that capture 25 (respectively, 100) mergers per year according to these criteria can achieve (ensure the best return on) the Chemical Evolution of the Universe science goal. Virtually all of the sources satisfying the SNR criterion are detected within a redshift of $z = 0.5$; sources satisfying only the sky area and inclination criteria can be detected out to $z \approx 2$. \textit{Bottom:} BNS mergers with distance measured to within \SI{20}{\percent} and inclination to within $\pm 0.1$ in $\cos \iota$, such that they track the sources' redshift evolution out to $z \approx 2$ and break degeneracies in electromagnetic emission models. Networks that make 200 (2000) such detections per year can achieve (ensure the best return on) the Gamma-Ray Burst Jet Engine science goal.
}
\label{fig:bns-performance}
\end{figure}

The cumulative number of binary neutron star detections per year with SNR $> 100$ is plotted as a function of redshift in the top left panel of \cref{fig:bns-performance} for the different networks; the redshift horizon for detection of these loud sources is between $z = 0.1$ and $0.4$, depending on the network. For this science goal, the primary driver of relative performance is the number of \SI{40}{\km} Cosmic Explorer detectors in the network: all else being equal, a \SI{40}{\km} Cosmic Explorer clearly outperforms a \SI{20}{\km} one for this metric. For instance, a single \SI{40}{\km} Cosmic Explorer will observe ${\sim} 80$ binary neutron star mergers per year with SNR in excess of 100 when embedded in a 2G background network, compared to ${\sim} 30$ for a single \SI{20}{\km} detector. Similarly, a 2G network augmented with two \SI{40}{\km} Cosmic Explorers can observe five times as many loud binary neutron star mergers as a single \SI{40}{\km} detector, and four times as many as two inspiral-optimized \SI{20}{\km} detectors. (Optimization for the postmerger signal is detrimental to this science goal, as a postmerger-optimized \SI{20}{\km} Cosmic Explorer detects only about half as many loud binary neutron star mergers as the inspiral-optimized one.) The performance of a heterogeneous \SI{40}{\km}-\SI{20}{\km} Cosmic Explorer network is intermediate between the \SI{40}{\km}-\SI{40}{\km} and \SI{20}{\km}-\SI{20}{\km} pairs. The addition of a third 3G detector to the network\,---\,whether ET, or a \SI{20}{\km} Cosmic Explorer South facility\,---\,boosts the detection rate significantly: two \SI{40}{\km} Cosmic Explorers plus ET will detect ${\sim} 600$ loud binary neutron star mergers per year, while even two \SI{20}{\km} Cosmic Explorers plus ET will capture in excess of 100 such mergers per year.

Hence, the return on this science goal can be maximized with two \SI{40}{\km} Cosmic Explorer detectors. The rate of loud binary neutron star merger detections would be enhanced by the presence of ET or Cosmic Explorer South in the network, but the science target could still be fully achieved with two \SI{40}{\km} detectors (or, indeed, one \SI{40}{\km} detector and one \SI{20}{\km} detector) in a 2G background network. A single Cosmic Explorer detector of either length is also a viable choice, provided the global detector network includes ET; without ET, however, a single \SI{20}{\km} Cosmic Explorer would fail to deliver most of this science on its own. To discretize the performance assessment, in \flowdown\ we mark in light (dark) green networks that are expected to detect at least 50 (200) %
binary neutron star inspirals with signal-to-noise ratio above 100 per year. Networks marked in yellow meet this inspiral signal-to-noise ratio criterion at a rate of 1--50 per year.

%

\paragraph{New Phases in Quantum Chromodynamics}
Cosmic Explorer's capacity to map the phase structure of quantum chromodynamics
with neutron-star merger observations will depend on the 3G detector network's
sensitivity to the postmerger signal. In order to break new ground in our
understanding of the equation of state in the finite-temperature regime,
postmerger gravitational waves must be captured above a signal-to-noise
threshold of 8 for several neutron-star mergers, so as to permit measurements
of the dominant postmerger frequency.

In contrast to the inspiral signal, the postmerger signal is better captured by
a \SI{20}{\km} Cosmic Explorer detector than a \SI{40}{\km} detector because it
can be tuned for increased sensitivity in the relevant kilohertz frequency
range.  %
A postmerger-optimized \SI{20}{\km} detector will make ${\sim}75$ postmerger
observations per year, whether operating in a 2G background network or with
Einstein Telescope.  A single \SI{40}{\km} Cosmic Explorer in a 2G background
network would likely miss half of those signals. The optimal performance is
achieved when pairing the \SI{20}{\km} postmerger-optimized Cosmic Explorer
with a second Cosmic Explorer detector of either \SI{20}{\km} or \SI{40}{\km};
we can then expect ${\sim} 120$ postmerger detections per year, and ${\sim}
180$ per year if Einstein Telescope is also in the network. A network
comprising two \SI{40}{\km} Cosmic Explorer detectors returns ${\sim} 85$
postmerger observations per year.

In \flowdown\ we mark in light (dark) green networks that are expected to detect at least 50 (100) postmerger signals per year. Networks marked in yellow are expected to detect more than one postmerger signal per year. 

\paragraph{Chemical Evolution of the Universe}
The evolution of heavy elements in the universe can be traced by observing
hundreds of neutron-star mergers out to cosmological distances. The complete
picture will be built up from multimessenger observations of the sources,
particularly prompt electromagnetic observations over the entire spectrum
accessible to ground- and space-based telescopes. The key to rapid
electromagnetic follow-up is precise localization of the sources, preferably
before merger. Since the signal can last for hours in the sensitivity
window of Cosmic Explorer if it is loud enough, it will in many cases be possible to provide alerts about an imminent merger several minutes ahead of time. Electromagnetic telescopes can then slew to source and observe the
prompt emission of electromagnetic waves in the aftermath of the merger. Additionally, a precise measurement of the binary's inclination angle is important for determining whether it is being viewed face-on or edge-on. If detected, postmerger gravitational waves would also help connect the merger remnant's properties to emission models. To evaluate the ability of the various detector networks to achieve this science goal, we examine the yearly number of binary neutron star detections satisfying a combination of metrics: SNR greater than 100, for early warning of the merger; a sky localization of less than 1 deg$^2$, to enable identification of electromagnetic counterparts; and a distance uncertainty of less than \SI{20}{\percent} and an inclination uncertainty of less than $\pm 0.1$ in $\cos \iota$, to break degeneracies in emission models.

The yearly number of SNR $> 100$ binary neutron star mergers localized to
\SI{1}{deg^2} in sky area and \SI{20}{\percent} in distance, with a $\pm 0.1$ measurement of inclination $\cos{\iota}$, is plotted in the top right panel of \cref{fig:bns-performance} for the different networks. A key factor governing a network's performance is the number of 3G detectors it includes: those with a single one detect fewer than 10 signals satisfying the combined criteria per year, while those with two or three can respectively detect up to a hundred or several hundred per year. Because of ET's especially good low-frequency sensitivity, networks that include ET are particularly advantageous for localizing sources on the sky. Given that a single \SI{40}{km} Cosmic Explorer detects loud, well-localized mergers at about twice the rate of a \SI{20}{km} detector, all else being equal, networks that include \SI{40}{km} detectors outperform their \SI{20}{km} counterparts. Thus, for example, a single \SI{40}{km} Cosmic Explorer paired with ET performs as well as the three-detector network consisting of two \SI{20}{km} Cosmic Explorers plus ET, making ${\sim} 150$ such detections per year. For comparison, two \SI{40}{km} Cosmic Explorers embedded in a 2G background network detect ${\sim} 100$ equivalent mergers per year.

The signals loud enough to give early warning of the merger while satisfying the distance, inclination and sky area constraints will only be detected out to $z \approx 0.3$. However, mergers with lower SNR can be detected out to $z \approx 2$ while meeting the other criteria. The quantitative performance of the various networks without the early warning criterion is documented in~\autocite{NStradestudy}, but their relative performance is essentially the same as in the lower panel of \cref{fig:bns-performance}, which omits both the SNR and sky area targets. When emphasizing the reach of well-localized neutron-star observations out to cosmological distances, rather than the prospect of early warning, the most important factor is simply the number of 3G detectors in the network.

In terms of distance measurements, inclination constraints and sky localization, the post-merger-optimized \SI{20}{km} detectors are less effective than their inspiral-optimized counterparts. On the other hand, the inclusion of a postmerger-optimized detector in the network increases the odds of associating postmerger gravitational waves with inspiral and electromagnetic observations. This would help relate the postmerger dynamics, and the lifetime of the remnant in particular, to the features of the electromagnetic emission. For this science goal, the trade-off in converting one inspiral-optimized \SI{20}{km} Cosmic Explorer to a postmerger-optimized one is roughly a factor of two reduction in the number of detections satisfying the combined criteria.

Consequently, the optimal detector network for this science goal would involve
two \SI{40}{\km} Cosmic Explorer detectors, preferably with a third 3G detector
such as Einstein Telescope. However, any network with three 3G detectors has the ability to fully achieve the science goal (as does the ET plus \SI{40}{\km} Cosmic Explorer network). Choosing one
the three detectors to be a \SI{20}{\km} postmerger-optimized Cosmic Explorer increases the prospect of joint electromagnetic, inspiral and postmerger gravitational wave observations, and is a viable option as the overall network performance is enough to compensate for the mild loss in localization and early warning capabilities. A single Cosmic Explorer detector could not achieve this science goal on its own. %
In \flowdown, networks
colored in light (dark) green will make detections satisfying the joint criteria at least 25 (100) times per year. Yellow
networks cannot achieve these criteria at a similar rate, but nonetheless detect at least 5 such sources per year.

\paragraph{Gamma-Ray Burst Jet Engine}
The properties of the gamma-ray burst ensuing after a neutron star
merger are largely determined by the remnant, which forms the central engine
for launching and driving relativistic jets. To make associations between gamma-ray bursts and gravitational-wave events, a 3G detector network will need to capture a large fraction of the population of merging neutron stars out to the horizon $z \sim 2$ of future gamma-ray observatories, while measuring the distance to these mergers precisely, so as to provide a reasonably complete gravitational-wave catalog for identifying counterparts. Additionally, to break degeneracies in emission models, it is crucial to measure the source inclination precisely. Postmerger gravitational-wave observations are also desirable for this science goal, as they can reveal the nature of the remnant and help establish its connection to the physics of the jets. To compare the different Cosmic Explorer configurations and networks for this science goal, we investigate their ability to detect binary neutron star mergers satisfying distance and inclination criteria out to cosmological distances. As in the previous subsection, we target \SI{20}{\percent}-level uncertainty in the distance measurement, and $\pm 0.1$ uncertainty in the measurement of $\cos \iota$ for the inclination.

In \cref{fig:bns-performance}, the lower panel plots the cumulative number of binary neutron star mergers whose distance is measured to \SI{20}{\percent} and whose inclination is constrained to $\pm 0.1$ in $\cos{\iota}$ as a function of redshift for the networks we consider. The performance for this science goal is most sensitive to the background network: all of the networks that include ET outperform those that rely on a 2G background. For example, a single \SI{20}{km} Cosmic Explorer operating in tandem with ET will observe ${\sim} 3000$ mergers per year satisfying the distance and inclination criteria, compared to ${\sim} 40$ per year operating alone. The best-performing network with a 2G background, two \SI{40}{km} Cosmic Explorers, will make ${\sim} 2500$ such observations per year. The number of 3G detectors in the network also controls the redshift horizon out to which the sources' distance and inclination can be constrained precisely: for networks with two (respectively, three) 3G detectors, it is $z \sim 1$ (2), compared to $z \sim 0.1$ for a single Cosmic Explorer detector. Within a given background network, two Cosmic Explorer detectors are better than one, and the \SI{40}{km} configuration outperforms the \SI{20}{km} one for distance and inclination measurements, all else being equal.

While a network consisting of two \SI{40}{km} Cosmic Explorer detectors plus ET would detect the greatest number of sources (${\sim} 1.5\times 10^4$ per year) satisfying the distance and inclination constraints, a network that includes a postmerger-optimized \SI{20}{km} detector would have a better chance of observing postmerger gravitational waves in conjunction with a fraction of the gamma-ray bursts. This can be done without drastically compromising the distance and inclination measurement precision. For example, substituting a postmerger-optimized \SI{20}{km} Cosmic Explorer in place of one of the \SI{40}{km} detectors will still result in ${\sim} 6000$ observations with precisely determined distances and inclinations per year.

Thus, from the point of view of supplying a complete catalog of gravitational-wave observations to associate with gamma-ray bursts, two \SI{40}{\km} Cosmic Explorers operating in conjunction with ET or Cosmic Explorer South is the optimal network for this science goal. However, a heterogeneous Cosmic Explorer network involving one postmerger-optimized \SI{20}{\km} detector is an advantageous choice because of the prospect for joint postmerger, inspiral and gamma-ray observations. A global network involving at least two (and preferably three) 3G detectors is critical for this science goal: a single Cosmic Explorer cannot achieve the required coverage in redshift with a 2G background network. In \flowdown, the networks colored light (dark) green will make 200 (2000) yearly detections of binary neutron star mergers that are well-constrained in distance and inclination out to cosmological distances. %
The networks colored yellow will make between 1 and 200 such detections per year, all of which are restricted to the local universe.

\subsection[]{Extreme Gravity and Fundamental Physics}

\paragraph{The Nature of Strong Gravity}
Gravitational-wave observations of merging black holes encode the nature of the
strongest gravity in the universe\,---\,the gravity near the horizon of a
stellar-mass black hole. Observations of merging black holes with
current-generation detectors are giving us a first look at the nature and
behavior of strong gravity. So far, all of these observations are consistent
with general relativity's picture of two (initially Kerr) black holes merging
into a remnant Kerr black hole, within the experimental noise and theoretical
uncertainty of the form of the emitted gravitational waves.

But evidence of new, revolutionary physics has often been first evident in
small deviations from conventional expectations. How much we can learn about
strong gravity from a gravitational-wave observation thus depends on the
strength of the gravitational-wave signal compared to detector noise. The
higher the signal-to-noise ratio of an observation, the clearer is the
resulting view of the underlying gravitational physics, and the greater
potential for discovery.%

A Cosmic Explorer detector will give us a solid opportunity far beyond
the first look that even the best current-generation detector network could
give. For instance, in each year of its operation, a single, \SI{40}{\km}, Cosmic Explorer detector
would observe roughly \CENEARBBHPERYEAR\
merging black holes that are approximately at least as close to Earth as
GW150914 (redshift $z<0.1$), the loudest gravitational-wave observation from
merging black holes so far, with half having signal-to-noise ratios greater
than about \CENEARBBHMEDSNR\ and the top \SI{10}{\percent} having signal-to-noise ratios
greater than \CENEARBBHLOUDSNR. In contrast, half the observations with a
second-generation network using A+ (Voyager) technology would have a
signal-to-noise ratio greater than about \APLUSNETBBHNEARMEDSNR\
(\VOYNETBBHNEARMEDSNR), with the top \SI{10}{\percent} of signals having signal-to-noise
ratios greater than about \APLUSNETBBHNEARLOUDSNR\ (\VOYNETBBHNEARLOUDSNR).

A \SI{40}{\km} Cosmic Explorer detector in a network including at least one other
next-generation detector would be especially favorable for revealing the nature
of strong gravity. In a network with three
3G detectors (two \SI{40}{\km} Cosmic Explorers and one Einstein
Telescope), for instance, the top \SI{10}{\percent} of binary black holes closer than $z=0.1$ would have
signal-to-noise ratios greater than \ETTWOCEBBHNEARLOUDSNR. Also, including
more than one next-generation interferometer enables better measurement of the
gravitational waves' polarization, including any potential scalar or vector
polarizations that would indicate physics beyond general relativity. A
network of next-generation detectors that each have arm lengths no longer than
20 km would be almost as favorable but would not be quite as sensitive a
probe of the nature of gravity, because the \SI{20}{\km} detector's sensitivity
is not quite as good in the most sensitive frequency band.

\paragraph{Unusual or Novel Compact Objects}
Third-generation detectors have the potential to unmask novel objects,
including objects potentially masquerading as black holes or neutron stars, by
providing precise strain measurements that can be compared with theoretical
predictions.  Voyager would be capable of giving us a first look at rarer
compact binaries, and it also has the potential to give a first look at novel
compact objects, if they not only exist but are also sufficiently nearby to be
in Voyager's range and sufficiently distinct from black holes and neutron stars
that they do not require the signal-to-noise ratios that only detectors in
next-generation observatories can achieve. Even a single next-generation
observatory would give a much more solid picture of rare, conventional compact
binaries: one Cosmic Explorer detector would observe all of the
binary-black-hole mergers in the observable universe (about
\CEBBHPERYEAR~observations per year) and would observe nearby mergers with very
high signal-to-noise ratios, which corresponds to a greater opportunity to
recognize an observation as being unusual.

Similar considerations as for probing the nature of strong gravity apply here:
a Cosmic Explorer with \SI{40}{\km} instead of \SI{20}{\km} arms, as part
of a network with more than one next-generation detector,
would be especially favorable. The \SI{40}{\km} arm length would deliver the highest
signal-to-noise ratios, which would enable us to better recognize unusual or
novel compact binaries. More than one next-generation detector would be
especially favorable for finding unusual or novel compact binaries, because
they would increase our confidence in observations of faint unusual or novel
compact binaries, and because they would provide polarization information
not accessible to a single detector.

\paragraph{Dark Matter and Dark Energy}
The strength that any signature that dark matter or dark energy might have on
gravitational-wave observations is unknown.  Current-generation detectors have
so far not found signatures of dark matter, and they lack the cosmic reach to
provide insight into dark energy through cosmic variations in the observed
population of compact objects.

To look for such potential effects, a \SI{40}{\km} Cosmic Explorer as part of a network
with multiple next-generation detectors would be most favorable.
A \SI{40}{\km} detector would have the highest sensitivity, which would
have the best potential to reveal subtle signatures of
dark matter, dark energy, or quantum gravity. And a network with
more than one next-generation
detector will be necessary to observe some of the proposed effects, such as
additional gravitational-wave polarizations or standard sirens that rely on
multimessenger observations. While some of the proposed effects, such as
gravitational-wave echoes, would not necessarily require a network with more
than one next-generation detector, such a network would still increase our
confidence in detecting faint imprints that dark matter, dark energy,
and quantum gravity might have on gravitational-wave observations.

\chapter{Technical Overview and Design Choices \hlight{[Kevin]}}
\xlabel{design}

The LIGO and Virgo instruments have opened a new window on the universe, but they are, like Galileo's first telescope, just sensitive enough to observe the brightest sources.
Today the Advanced LIGO detectors see signals roughly weekly; when the ``A+'' upgrade is mature in 2025, it will deliver roughly ten detections per week.
The key science questions presented in \sref{science_overview} are only answerable by making observations with significantly higher fidelity over a wider frequency band, and by observing more distant sources.
As described  in \sref{overview},
 this requires new facilities with longer baselines and detectors
 with an order of magnitude greater sensitivity in the audio frequency band.
This section provides a technical overview of the Cosmic Explorer observatory that can deliver that sensitivity, including the sites, infrastructure, and vacuum systems. It also outlines the key technologies that will require research and development to enable the CE science goals. Finally, the key drivers of project costs are discussed.

\section{Reference Detector Concept \hlight{[Kevin]}}
\label{subsec:reference_design}

The Cosmic Explorer reference detector concept is a dual-recycled
Fabry--P\'{e}rot Michelson interferometer (\textsc{drfpmi}) scaled up to use
\SI{40}{\km} or \SI{20}{\km} long arms (see \cref{box:concept}). The longer arm
length will increase the amplitude of the observed signals with effectively no
increase in the noise.  Although there are areas of detector technology where
improvements will lead to increases in the sensitivity and bandwidth of the
instrument relative to the existing LIGO detectors, the dominant improvement
will come from the order-of-magnitude increase in length.

The interferometers installed in the Cosmic Explorer observatories will evolve
as the technologies and science evolve.  The first two decades of Cosmic
Explorer evolution are sketched \fref{timeline}.  Like LIGO and Advanced LIGO,
CE's sensitivity is expected to improve with time thanks to technology upgrades
and commissioning effort.  Parts of the CE nominal design may not be installed
before the CE observatories begin collecting data.  These ``planned upgrades'',
to be installed as the technology becomes available, include: low-loss readout of high-fidelity squeezed states of light
(\cref{subsubsec:high_fidelity_squeezing}), adding seismometer arrays to
subtract fluctuations in the local gravity
(\cref{subsubsec:seismic_arrays_engineering}), and improved sensors for seismic
isolation relative to what is expected to be available at the time of
construction (\sref{inertial_sensors}).

To minimize the required technical development, the initial CE detectors will use the Advanced LIGO detector design, including its A+ upgrades, scaled as needed in size, along with some advances to improve the low frequency sensitivity. This provides a straightforward approach to significant improvement using tested technology with relatively low risk. The planned upgrades will then proceed when possible given availability of new technologies and when maximally beneficial to the scientific output of the observatories. That is, the upgrades can all be performed in parallel at one or both observatories, or sequentially at one observatory at a time to avoid long
downtimes.

A second technology capable of achieving (and
possibly surpassing) the target sensitivity of CE has also been identified.
This approach uses technology currently being developed for the LIGO
Voyager detector, consisting of a \SI{2}{\um} laser and cryogenic silicon test
masses, and will only be required if a major problem is encountered with
scaling up current technology.  Research is ongoing to understand if the \SI{2}{\um} design also offers a path to higher laser power and thus sensitivity beyond the CE target (\cref{subsec:silicon_upgrades}).

\subsection[]{Overview}
\label{subsubsec:reference_design_overview}

\begin{figure}[!ht]
  \centering
  \includegraphics[height=0.6\textheight]{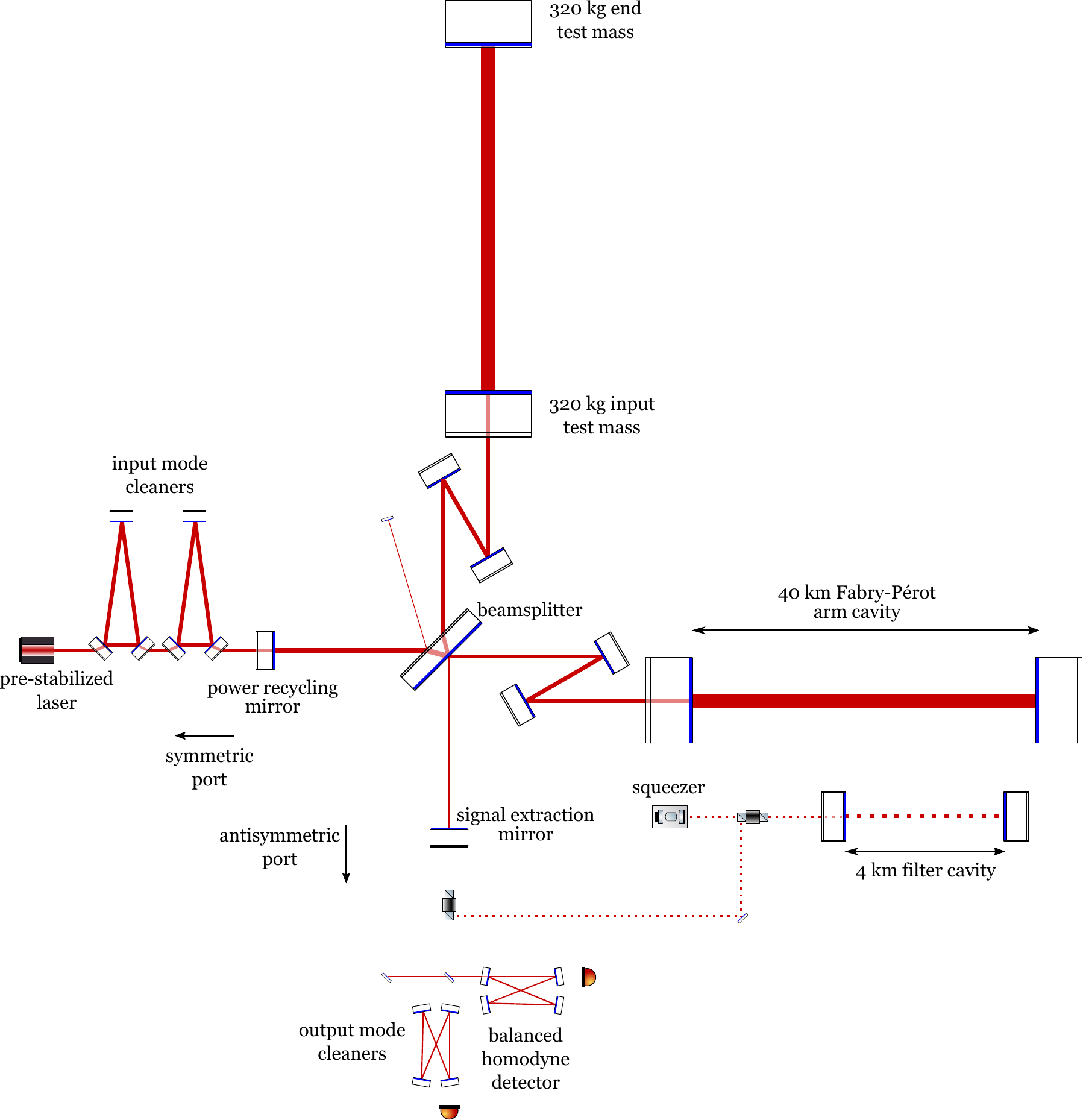}
  \caption{Simplified optical layout of the Cosmic Explorer reference detector concept for the \SI{40}{\km} implementation. The input and end test masses form the two arm cavities which, together with the beamsplitter, power recycling mirror, and signal extraction mirror, comprise the core of the dual-recycled Fabry--P\'{e}rot Michelson interferometer as described in \cref{subsubsec:reference_design_overview}. As described in \cref{subsubsec:io}, the light carrying the gravitational wave signal is spatially filtered and read out from the antisymmetric port by a balanced homodyne detector comprised of two photodiodes and output mode cleaners; a high power laser is injected into the symmetric port of the interferometer after passing through two input mode cleaners which assist in producing a frequency and intensity stabilized beam with a spatially clean mode. The squeezer generates squeezed vacuum states which are reflected off of a filter cavity and injected into the antisymmetric port to provide broadband quantum noise reduction as described in \cref{subsubsec:quantum_noise}. The beamsplitter is shown with the high-reflective surface facing the antisymmetric port rather than the laser, unlike current detectors, to minimize loss in the signal extraction cavity, but careful analysis of thermal effects is needed before finalizing the design.}
  \label{fig:reference_design}
\end{figure}

A simple Michelson interferometer measures the strain of a passing gravitational wave by measuring the difference in time for light to traverse two more-or-less perpendicular arms, at the end of which are mirrors serving as test masses. So that these test masses are free from external horizontal forces in the frequency band of interest, they are suspended from pendulums (which also isolate them from seismic noise as discussed further below). The relative phase accumulated by the light in each arm is modulated by a passing gravitational wave and the ensuing power modulation at the antisymmetric port is measured with photodetectors. Practically, the common mode rejection of a Michelson interferometer provides some suppression of laser frequency and intensity noises.

The \textsc{drfpmi}, shown in \cref{fig:reference_design}, improves on the simple Michelson interferometer in several ways. First, input test masses are added to each arm to make Fabry--P\'{e}rot cavities. This increases the power stored in the arms, which decreases the quantum shot noise. Second, a ``power recycling mirror'' added to the symmetric port further increases the power stored in the interferometer and provides passive filtering of laser noise. Finally, a ``signal extraction mirror'' added to the antisymmetric port forms a ``signal extraction cavity.'' The bandwidth of the interferometer can be tuned to enhance the sensitivity relevant for a particular science case by simply changing the reflectivity of this mirror. For example, the sensitivity to post-merger physics can be improved as shown in \cref{fig:ce_alternate_strains}, though other tunings can improve the low frequency sensitivity instead.

\begin{figure}[!h]
  \centering
  \includegraphics[width=0.8\textwidth]{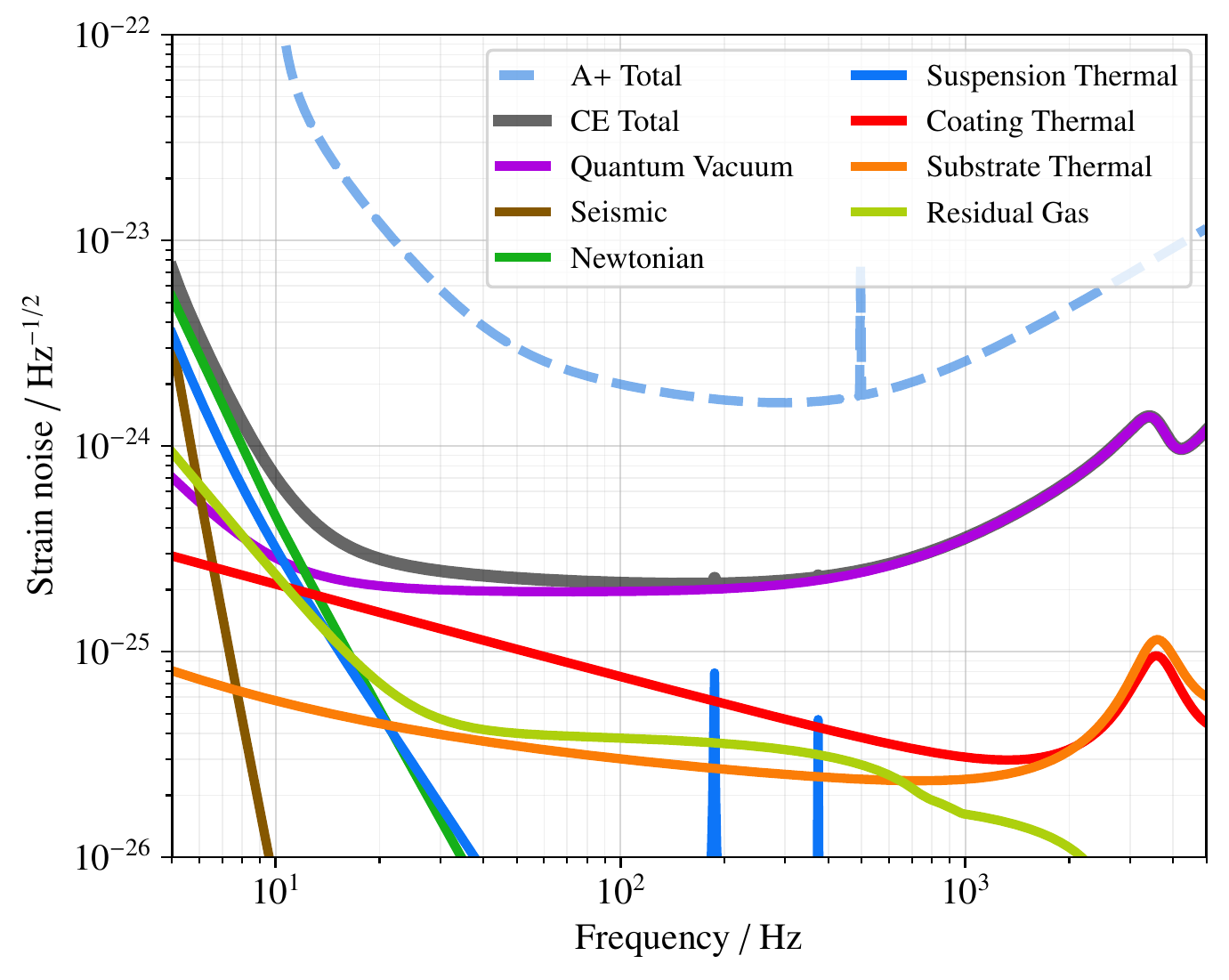}
  	\caption{Estimated spectral sensitivity (solid black) of Cosmic Explorer and the known fundamental sources of noise that contribute to this total (colored curves). The design sensitivity of LIGO A+ is also shown in dashed blue.}
  	\label{fig:nb}
\end{figure}

The other major subsystems of the interferometer are described in the following sections. \Cref{fig:reference_design} shows this reference concept including these major subsystems and \cref{tab:detector_params} presents the key design parameters. The estimated spectral sensitivity of Cosmic Explorer and the limitations imposed by the fundamental noise sources are shown in \cref{fig:nb}.

\begin{table}[t]
    \sisetup{
        scientific-notation=fixed,
    }
    \centering
    \begin{tabular}{r l S S S S}
    \toprule
         &
         {\textsf{Quantity}}      &
         {\textsf{Units}}   &
            {\textsf{LIGO A+}}       &
            {\textsf{CE}} &
            {\textsf{CE (\SI{2}{\um})}}    \\
    \midrule
        &
            Arm length  &
            \si{\km} &
            4 & 40 & 40 \\
        &
            Laser wavelength  & \si{\um} &
            1 & 1 & 2 \\
        &
            Arm power  & \si{\MW} &
            0.8 & 1.5 & 3 \\
        &
            Squeezed light & \si{\dB} &
            6 & 10 & 10 \\
        &
            Susp. point at \SI{1}{\Hz} & $\si{\pico\meter\big/\sqrt{\text{Hz}}}$ &
            10 & 0.1 & 0.1 \\

        \midrule
        Test masses &
            Material & &
            {Silica} & {Silica} & {Silicon} \\
        &
            Mass & \si{\kg} &
            40 & 320 & 320 \\
        &
            Temperature & \si{\K} &
            293 & 293 & 123 \\

        \midrule
        Suspensions &
            Total length & \si{\meter} &
            1.6 & 4 & 4 \\
        &
            Total mass & \si{\kg} &
            120 & 1500 & 1500 \\
        &
            Final stage blade & &
            {No} & {Yes} & {Yes} \\

        \midrule
        Newtonian noise &
            Rayleigh wave suppr. & \si{\dB} &
            0 & 20 & 20 \\
        &
            Body wave suppr. & \si{\dB} &
            0 & 10 & 10 \\

        \midrule
        Optical loss &
            Arm cavity (round trip) & \si{\ppm} &
            75 & 40 & 40 \\
        &
            SEC (round trip) & \si{ppm} &
            5000 & 500 & 500 \\

    \midrule
    \midrule
        & BNS horizon redshift & & 
            \aplusBnsHorizon{}        &
            \ceTwoGlassBnsHorizon{}   &
            \ceTwoSiliconBnsHorizon{} \\
        & BBH horizon redshift & &
            \aplusBbhHorizon{}        &
            \ceTwoGlassBbhHorizon{}   &
            \ceTwoSiliconBbhHorizon{} \\
        & BNS SNR, $z=0.01$  & &
            \aplusBnsSNR{}        &
            \ceTwoGlassBnsSNR{}   &
            \ceTwoSiliconBnsSNR{} \\
        & BNS warning, $z=0.01$  & \si{\minute} & 
            \aplusEarlyWarning{}        &
            \ceTwoGlassEarlyWarning{}   &
            \ceTwoSiliconEarlyWarning{} \\
    \bottomrule
    \end{tabular}
    \caption{
        Key design parameters and astrophysical performance measures for the
LIGO A+ and \SI{40}{\km} Cosmic Explorer detectors.  The astrophysical performance measures assume a 1.4--1.4$M_\odot$ binary-neutron-star (BNS) system and a
30--30$M_\odot$ binary-black-hole (BBH) system, both optimally oriented.
``BNS warning'' is the time before merger at which the event has accumulated
a signal-to-noise ratio (SNR) of 8.
    \label{tab:detector_params}}
\end{table}

\subsection[]{Quantum Noise Reduction}
\label{subsubsec:quantum_noise}

Quantum vacuum fluctuations entering the antisymmetric port of the interferometer will be the dominant limit to the sensitivity of Cosmic Explorer above \SI{20}{\Hz}. Quantum mechanics requires every mode of the electromagnetic field to have a minimum zero-point energy. These vacuum fluctuations enter any open port of the optical system.\footnote{Vacuum fluctuations entering the symmetric port contribute noise to the common mode, rather than the differential mode which carries the gravitational wave signal.} Radiation pressure noise dominates at low frequencies where fluctuations in the vacuum field amplitude quadrature beat with the main laser field to produce a fluctuating radiation pressure force on the mirrors. At higher frequencies, it is the beating of the vacuum field fluctuations in the orthogonal phase quadrature with the main laser field, shot noise,\footnote{This shot noise arising from the beating of the vacuum fluctuations entering the antisymmetric port with the main laser is a truly intrinsic phase noise of the optomechanical system. It is distinct from the related technical noise, also referred to as shot noise, where phase noise is produced by excess light incident on a photodiode beating with vacuum fluctuations.} that directly limits the accuracy with which the phase can be measured\autocite{1981PhRvD..23.1693C, Kimble:2000gu, 2002PhRvD..65d2001B}.

While Heisenberg's uncertainty relation dictates the fundamental limit below which the vacuum fluctuations cannot be reduced, fluctuations in one quadrature can be reduced at the expense of increasing the fluctuations in the orthogonal quadrature, leading to ``squeezed states.''\autocite{2019RPPh...82a6905B} As with Advanced LIGO\autocite{2019PhRvL.123w1107T}, Cosmic Explorer will use a nonlinear crystal pumped with a laser at twice the frequency of the main laser, known as a degenerate optical parametric amplifier, to produce the correlations necessary to generate such squeezed vacuum states and inject them into the antisymmetric port. In this way, radiation pressure noise can be reduced by injecting states with decreased uncertainty in the amplitude quadrature, at the expense of increased phase uncertainty and shot noise. Similarly, shot noise can be reduced by injecting states with decreased uncertainty in the phase quadrature at the expense of increased amplitude fluctuations and radiation pressure noise. However, since radiation pressure dominates at low frequencies and shot noise dominates at high frequencies, injecting squeezed vacuum with a fixed ratio of amplitude to phase uncertainty necessitates a trade-off between reducing quantum noise at high and low frequencies.

The frequency dependence required to achieve broadband quantum noise reduction can be achieved by reflecting the squeezed vacuum state generated by the parametric amplifier off of an extra optical cavity detuned from resonance, known as a filter cavity, before it is injected into the interferometer\autocite{Kimble:2000gu, 2013PhRvD..88b2002E, 2014PhRvD..90f2006K}. Cosmic Explorer will use such a \SI{4}{\km} long filter cavity to produce a frequency dependent squeezed state that rotates from amplitude squeezing at low frequencies to phase squeezing at high frequencies at the appropriate rate to achieve \SI{10}{\dB} of broadband quantum noise reduction. The net noise reduction depends on the optical losses in the system, which are expected to be reduced as upgrades are made to the detector.

\subsection[]{Optics and Coatings}
\label{subsubsec:optics_coatings}

Cosmic Explorer will use large \SI{320}{\kg} test masses made into highly reflective mirrors through the use of thin-film coatings consisting of alternating layers of high and low refractive-index materials. The more massive test masses serve to reduce quantum radiation pressure noise and, as a practical matter, they must be wide enough to accommodate the large diameter beams of a nearly diffraction limited \SI{40}{\km} long arm cavity.

The CE test masses will use the LIGO A+ technology, scaled up for their larger size. The substrate is fused silica, chosen for its low mechanical loss and correspondingly low thermal noise at room temperature, and low optical absorption at \SI{1}{\um}. The low refractive-index layers of the coatings will be silica; the high index layers are yet to be determined. These coatings must have low mechanical loss to reduce their thermal noise, yet this thermal noise is still non-negligible below ${\sim}\SI{60}{\Hz}$. The desire to reduce this noise is one factor motivating the use of cryogenics as an alternative technology as discussed in \cref{subsec:2um_alternative}.

\subsection[]{Seismic Isolation and Suspensions}
\label{subsubsec:seismic_isolation}

Each of the four Cosmic Explorer test masses will be suspended by a quadruple pendulum to isolate them from seismic disturbances\autocite{2012CQGra..29w5004A}. The suspensions provide passive $1/f^8$ filtering of seismic noise above their mechanical resonance frequencies. The suspensions themselves will be mounted on inertial seismic isolation systems which provide additional active and passive suppression of seismic noise\autocite{2015CQGra..32r5003M}.

To minimize thermal noise, the final suspension stage is fabricated monolithically from fused silica, with the test mass suspended from the penultimate mass by fused silica fibers\autocite{2012CQGra..29w5004A}. The top two masses of the suspensions are steel and are suspended by steel wires.

To reduce the vertical suspension resonances, and thus both the vertical seismic noise and thermal noise which couple into the arm length due to the Earth's curvature, each of the first three stages is suspended from steel blade springs attached to the stage above. Unlike the LIGO suspensions, however, a set of silica blade springs is added for the final stage instead of bonding the suspension fibers directly to the penultimate mass.

The inertial seismic isolation systems are similar to those of Advanced LIGO\autocite{2015CQGra..32r5003M} but with improved inertial and position sensors. It is assumed that incremental improvements will allow Cosmic Explorer to initially achieve a threefold improvement over Advanced LIGO at \SI{10}{\Hz} and a tenfold improvement at \SI{1}{\Hz}. Novel six-dimensional inertial isolators with optical readout will be used to achieve an additional threefold improvement at \SI{10}{\Hz} and tenfold improvement at \SI{1}{\Hz} to achieve the final Cosmic Explorer sensitivity. The status of this technology is described in \cref{sec:inertial_sensors}.

\subsection[]{Input and Output}
\label{subsubsec:io}

The laser source for Cosmic Explorer is similar to that of LIGO and begins with a \SI{1}{\um}, \SIrange{1}{2}{\watt} seed laser. This is amplified by a multi-stage amplifier to produce the full input power of up to \SI{200}{\W}. Together with some laser intensity and frequency stabilization and some cleaning of the spatial mode of the laser, this comprises the pre-stabilized laser\autocite{2012OExpr..2010617K}. The light from the pre-stabilized laser is then sent through two triangular cavities known as input mode cleaners to provide further laser frequency stabilization and cleaning of the spatial mode. The frequency stabilization scheme used in LIGO relies on a single mode cleaner, but the longer arms of Cosmic Explorer require a new control scheme which, while possible to implement with a single mode cleaner, greatly benefits from two.\autocite{Cahillane:2021jvt} The light exiting these mode cleaners, required to be ${\sim}\SI{140}{\watt}$ to reach the goal of \SI{1.5}{\mega\watt} arm power with the expected optical loss, is then injected into the main interferometer at the back of the power recycling mirror.

The gravitational wave signal is imprinted on the light exiting the interferometer from the signal extraction mirror. This signal is measured using a balanced homodyne detector with a local oscillator derived from a few hundred milliwatts of light extracted from the beamsplitter. The spatial mode and frequency content of the signal and local oscillator are cleaned by two bow-tie cavities known as output mode cleaners before being detected with high quantum-efficiency photodiodes.

\subsection[]{Newtonian Noise Mitigation}
\label{subsubsec:newtonian_noise}

Fluctuations in the gravitational attraction between the test masses and the environment, known as ``Newtonian noise''\autocite{2019LRR....22....6H}, are a significant low frequency noise source for Cosmic Explorer.

Seismic waves are one source of Newtonian noise. Cosmic Explorer will initially suppress noise from Rayleigh surface waves by a factor of two in amplitude. After planned upgrades, it will suppress Rayleigh waves by a factor of ten and will additionally suppress the Newtonian noise from body waves by a factor of three. A combination of several techniques may be employed to achieve these goals. One technique is to use seismometer arrays to estimate the seismic field near the test masses and to subtract its effects from the gravitational wave strain data\autocite{2012PhRvD..86j2001D}. Another approach is to directly reduce the coupling of seismic waves to the test mass by modifying the density of the material below each mass\autocite{2014CQGra..31r5011H} or intentionally deflecting or dissipating them with architected materials or seismic metamaterials\autocite{2014PhRvL.112m3901B, 2016NatSR...639356P, 2016NatSR...627717C, roux2018toward, 2019APS..APRR11006K,2020PhRvP..13c4055Z}. These techniques are discussed in detail in \cref{subsubsec:seismic_arrays_engineering}.

Cosmic Explorer is also affected by Newtonian noise due to density fluctuations in the atmosphere at infrasonic frequencies. The reference concept does not include any suppression of Newtonian infrasound noise, since it is unknown if the technology needed to do so would be mature by the 2040s. This is the dominant source of Newtonian noise after the seismic contributions have been suppressed.

\subsection[]{Computing and Controls}
\label{subsubsec:computing_controls}

Holding the detector at its stable, astrophysically sensitive operating point
requires feedback control on a large number of degrees of freedom, such as the
relative distances and angles between the suspended optics. Additionally,
bringing the detector to its operating point requires a multi-step locking
scheme.  Similar to LIGO~\autocite{2021SoftX..1300619B}, Cosmic Explorer will
use a hybrid digital and analog real-time data acquisition and controls system,
along with automated supervision of the detector locking.  The digital system
will also provide near real time calibration and astronomical alerts.

Because of their susceptibility to environmental conditions and the time required for the locking
process, current gravitational-wave interferometers have roughly \SI{75}{\%}
availability; the CE designs will strive to improve upon this. The greatest
improvement in observing time will likely come from reducing the time required
to achieve lock, e.g., with a fully deterministic locking scheme and feed-forward thermal compensation,
and from improving the instrument's robustness to environmental disturbances,
particularly from high seismicity and wind.

\subsection[]{\SI{2}{\um} Cryogenic Silicon as an Alternative Technology}
\label{subsec:2um_alternative}

Since it is possible that a major problem prevents the LIGO A+ technology from achieving the Cosmic Explorer design sensitivity, and to retain the possibility of surpassing this goal in the future, it is prudent to continue research into the technology being developed for LIGO Voyager, namely cryogenic silicon test masses and a \SI{2}{\um} laser.

One concern for the nominal \SI{1}{\um} technology is potential
roadblocks to achieving \SI{1.5}{\mega\W} power in the arm cavities.  The
presence of particulates in the test mass coatings that absorb the laser power
and thermally deform the mirrors has been an obstacle to achieving the design
power in Advanced LIGO\autocite{LIGOScientific:2021kro}. While progress has
been made in addressing this issue, which must also be solved for A+ to reach
its sensitivity goal, it is not expected to be a significant issue for the
cryogenic silicon technology due to the high thermal conductivity of silicon.
Other thermal effects may also limit the achievable arm power for the
\SI{1}{\um} technology. The power absorbed in the test mass coatings and substrates, for
instance, creates both a thermoelastic deformation of the mirror surface and a thermally induced lens.
Both effects produce wavefront distortions which must be corrected with adaptive optics~\autocite{2016ApOpt..55.8256B}. While research is still needed into
methods of doing so with the \SI{2}{\um} technology, the magnitude of these distortions should be smaller than in the \SI{1}{\um} technology.

Another motivation for research into the \SI{2}{\um} technology is the fact that coating Brownian noise is a non-negligible noise source from roughly 10 to \SI{100}{\Hz} for the \SI{1}{\um} technology. Even if the thermal effects associated with the \SI{1}{\um} technology are overcome and more power can be stored in the arm cavities, coating Brownian noise may prevent significant improvement beyond the current design. Brownian noise from promising coatings for the \SI{2}{\um} technology can be roughly a factor of 2.5 lower in comparison. Furthermore, while both technologies achieve similar performance with the long arms of Cosmic Explorer, the \SI{1}{\um} technology becomes less attractive for arms significantly shorter than \SI{20}{\km}. If, however, the more speculative crystalline GaAs/AlGaAs coatings are used instead of the amorphous A+ coatings for the \SI{1}{\um} technology, both technologies would again have similar performance for shorter arm lengths.

\section{Site and Facility \hlight{[Kevin]}}
\label{subsec:site_infra}

Several important factors must be considered when identifying sites suitable for hosting a Cosmic Explorer facility. The site must satisfy the requirements described in \cref{subsubsec:site} in order to reach design sensitivity, and sites with favorable topography can significantly decrease the cost of civil engineering work needed to accommodate the beamtubes as is discussed in \cref{subsubsec:arm_length}. Other site selection considerations are discussed in \cref{subsubsec:site_choice}.

The Cosmic Explorer building design and construction can be based upon those used for LIGO, with future research into considerations such as aerodynamic building shapes, wind fences, and other vibration reduction engineering such as berms. Operating the existing LIGO observatories has taught us the importance of designing facilities that have more immunity to environmental noise. As much as possible, equipment and personnel that cause vibration, acoustic, infrasound, and electromagnetic disturbances should be located far from the most sensitive equipment, for example in out-buildings near to the corner and end stations. In addition to the buildings housing the CE detector infrastructure, CE will require laboratories, warehouses, mechanical and electrical workshops, and offices, as well as meeting spaces for users and visitors. These buildings should be close enough to allow access to the CE site, but far enough away that they do not significantly couple anthropogenic noise into the detector. Access roads will be needed, and access to rail would be advantageous, especially for delivery of the vacuum pipe.

\subsection[]{Site and Facility Requirements}
\label{subsubsec:site}

\paragraph{Ambient seismicity}
Ground motion directly impacts the sensitivity of Cosmic Explorer, including
from seismically induced fluctuations in the local gravitational field. This
gravitational influence on the detector test masses cannot be screened or
shielded, though it can be partially ameliorated with data subtraction
techniques, by altering the properties of the ground near the test masses, and
by selecting a low-seismicity site.  The current estimate of the Cosmic
Explorer sensitivity assumes a Rayleigh-wave-dominated ground motion with
amplitude \SI{1}{\bigl(\um\big/{\second^2}\bigr)\big/\!\sqrt{\Hz}}. Based on US seismic
surveys, this ground acceleration target is not particularly onerous, and the
overall seismicity at the site is likely to be dominated by local machinery
above \SI{5}{\Hz}. A dedicated seismic survey for Cosmic Explorer must
establish both the overall ground motion amplitude above \SI{5}{\Hz}, as well
as the partitioning of the field into its bulk and surface wave components.

Additionally, experience from existing long-baseline laser interferometers
shows that high seismicity negatively impacts the controllability of the
instrument, leading to downtime and decreased sensitivity. It is therefore
important to survey the ground motion down to \SI{10}{\mHz} to understand
the requirements on the seismic isolation system.

\paragraph{Ambient infrasound}
Acoustic waves also impact the Cosmic Explorer sensitivity by generating local
gravity fluctuations. The Cosmic Explorer sensitivity model assumes an ambient
acoustic spectrum of \SI{1}{\milli\Pa\big/\!\sqrt{\Hz}}, which is in line with
the median spectrum from long-term global infrasound surveys. A dedicated
infrasound survey for Cosmic Explorer must be careful to disentangle the effect
of wind confusion noise.

\paragraph{Geotechnical issues}
Any potential site will require a geotechnical investigation to assess its
suitability and to arrive at a precise cost estimate for the construction of a
CE observatory.  In addition to standard civil-engineering aspects, this
assessment will need to evaluate the potential for seismic engineering as
discussed in \cref{subsubsec:seismic_arrays_engineering}.

\paragraph{Ambient environment}
Long-term measurements of the environment are required to determine the
variation in noise arising from the weather (for example, wind and
thunderstorms) or from anthropogenic origins (such as cars and trains).
Susceptibility to natural disasters such as flooding, earthquakes, or
hurricanes must also be evaluated.

\paragraph{Environmental stewardship}
Throughout the construction of Cosmic Explorer very careful attention will be
given to preserving the local environment\,---\,both its living ecosystems and
its non-living components.  

Possible alterations of the ecosystem might include interference with migratory
or mating patterns, the extinction of local flora and fauna, or the
introduction of damaging invasive species.  Such problems can be caused by
chemical or thermal pollution, negligent construction, or waste disposal
practices. Therefore, a thorough environmental impact study will be necessary
to identify, constrain, and remediate such effects.  This will be done with the
active participation of the local community as well as state and federal
governing agencies.

While respect for the environment is essential throughout the lifetime of the
facility, it is especially important during the initial construction phase and
during facility decommissioning.  During the operations phase, the potential
for harm is smaller but still requires careful monitoring.

\subsection[]{Vacuum System}
\xlabel{vacuum_reqs}

As with all interferometric gravitational-wave detectors, ultrahigh vacuum
(UHV) is necessary in Cosmic Explorer to reduce path-length fluctuation of the
light traveling down the arms and to reduce mechanical damping on the
detector's core optics. The vacuum system is additionally responsible for
shielding the interferometer from acoustic noise and thermal noise from the
atmosphere.  While the vacuum techniques developed for the LIGO detectors are
adequate for Cosmic Explorer, there is room for improvement and value
engineering (\sref{vac_value_eng}).  Nominal parameters of the Cosmic Explorer vacuum system
are given in \cref{tab:vacuum_parameters}.

\begin{table}[t]
  \centering
  \begin{tabular}{
  r l
}
  \toprule
  Beamtube diameter & \SI{48}{\inch} (\SI{122}{\cm}) \\
  Beamtube thickness & $\tfrac{1}{2}\,\si{\inch}$ (\SI{13}{\mm}) \\
  Beamtube material & mild steel \\
  Beamtube BRDF & \SI{1e-3}{\per\steradian}\\
  Hard close gate valves & 10, partitioning into \SI{10}{\km} sections \\
  Soft close gate valves & 32, partitioning into \SI{2}{\km} subsections \\
  \SI{2000}{\L/\s} ion pumps & 40, one for each subsection \\
  Roughing pumps & 40, one for each subsection \\
  non-evaporable getters & distributed throughout \\
  \SI{6}{\inch} pumping ports & one every \SI{250}{\m} \\
  Baffle aperture & \SI{100}{\cm} \\
  Baffle BRDF & \SI{1e-3}{\per\steradian} \\
  \bottomrule
\end{tabular}

  \caption{Reference parameters for the Cosmic Explorer vacuum system for a \SI{40}{\km} facility. \cref{fig:vacuum_system} shows a schematic of how the vacuum system is broken up into \SI{10}{\km} subsections.}
  \label{tab:vacuum_parameters}
\end{table}

The vacuum practices used with the test mass chambers of Cosmic Explorer will need to be improved relative to those for current LIGO chambers to reduce pumpdown times after in-chamber detector work. Besides increased pumping capacity, it may be necessary to heat the walls of the test mass chambers to increase the evaporation rate (known as baking), especially in the case that the mirrors are cryogenically cooled and need protection from condensation. For both the chambers and beamtubes, the hydrogen pumping speed can easily be augmented by titanium sublimation or non-evaporable getter pumps.

\paragraph{Residual gas pressure requirements}
The pressure requirements in the beamtubes and in the test mass chambers are set by the effect of two different processes. Fluctuations of the gas column density in the beamtubes induces a phase noise when the light scatters off the residual gas molecules,\autocite{1996magr.meet.1434Z} while the residual gas in the chambers exerts a force noise directly on the test masses.\autocite{2010PhLA..374.3365C,2011PhRvD..84f3007D} The gas force noise has a $1/f^2$ frequency dependence and is important at low frequencies, while the gas scattering noise is constant in frequency up to a cutoff frequency determined by the time for a molecule to cross the beam (see \cref{fig:nb}). Reaction masses are used in LIGO for electrostatic force actuation on the test masses, and the resulting squeezed film damping from their close proximity to the test masses increases the total damping noise.\autocite{2011PhRvD..84f3007D} This excess noise is not considered here since it is assumed that CE will use a different actuation scheme where this effect is negligible.

The vacuum system requirements for the partial pressures of each species are broken up into a set of requirements that must be met and a set of goals to strive for. The requirements on the beamtube pressures are that the gas scattering noise is at least a factor of five below the design sensitivity, and the requirements on the chamber pressures are that the gas damping noise is at least a factor of three below the design sensitivity. The goals on the pressures in both the beamtubes and chambers are that the residual gas noise is a factor of ten below the design sensitivity at all frequencies. In the absence of a malfunction or poor vacuum practice, such as leaks in the vacuum system or inadequate cleaning of vacuum components, achieving the low partial pressures necessary to meet these requirements will be most challenging for hydrogen, water, nitrogen, and oxygen. We thus allow these four species to saturate these requirements with each contributing one quarter to the total. Hydrocarbons could potentially make a large contribution to the noise since they are massive and have large polarizabilities. Keeping their pressures low enough to have a negligible contribution to the total noise also ensures that they do not contaminate the mirror and cause excess optical loss. These pressure requirements and goals are summarized in \cref{tab:vacuum_requirements}.

\begin{table}[ht]
  \sisetup{round-mode=places,round-precision=1,table-alignment=center,table-format=1.1e1}
  \centering
  \begin{tabular}{
    r S S S S S
}
  \toprule
  {} &
  \multicolumn{3}{c}{\textsf{Beamtubes}} &
  \multicolumn{2}{c}{\textsf{Chambers}} \\
  \textsf{Species} &
  \textsf{Req \,/\,\si{\torr}} & \textsf{Goal \,/\,\si{\torr}} &
  \textsf{LIGO Achvd \,/\,\si{\torr}} &
  \textsf{Req \,/\,\si{\torr}} & \textsf{Goal \,/\,\si{\torr}} \\
  \midrule
  He & 1.343e-09 & 3.357e-10 &  & 8.829e-10 & 7.946e-11 \\ 
$\mathrm{H}_2$ & 3.301e-10 & 8.253e-11 & 3.4e-9 & 3.107e-09 & 2.797e-10 \\ 
Ne & 1.782e-10 & 4.454e-11 & & 3.931e-10 & 3.538e-11 \\ 
$\mathrm{H}_2\mathrm{O}$ & 3.042e-11 & 7.605e-12 & 2.3e-12 & 1.04e-09 & 9.361e-11 \\ 
$\mathrm{O}_2$ & 2.11e-11 & 5.276e-12 & 2.0e-13 & 7.805e-10 & 7.025e-11 \\ 
$\mathrm{N}_2$ & 1.877e-11 & 4.692e-12 & 1.0e-13 & 8.341e-10 & 7.507e-11 \\ 
Ar & 6.672e-12 & 1.668e-12 & 9.0e-14 & 2.794e-10 & 2.515e-11 \\ 
CO & 5.78e-12 & 1.445e-12 & 2.0e-12 & 3.34e-10 & 3.006e-11 \\ 
$\mathrm{CH}_4$ & 4.836e-12 & 1.209e-12 & 2.2e-11 & 4.411e-10 & 3.97e-11 \\ 
$\mathrm{CO}_2$ & 2.779e-12 & 6.948e-13 & 4.0e-13 & 2.661e-10 & 2.395e-11 \\ 
Xe & 6.305e-13 & 1.576e-13 & & 1.541e-10 & 1.387e-11 \\ 
\SI[round-precision=0]{100}{\atomicmassunit} $\mathrm{H}_n\mathrm{C}_m$ & 8.94e-14 & 2.235e-14 &  & 1.766e-10 & 1.589e-11 \\ 
\SI[round-precision=0]{200}{\atomicmassunit} $\mathrm{H}_n\mathrm{C}_m$ & 1.682e-14 & 4.205e-15 & & 1.249e-10 & 1.124e-11 \\ 
\SI[round-precision=0]{300}{\atomicmassunit} $\mathrm{H}_n\mathrm{C}_m$ & 6.197e-15 & 1.549e-15 & & 1.019e-10 & 9.175e-12 \\ 
\SI[round-precision=0]{400}{\atomicmassunit} $\mathrm{H}_n\mathrm{C}_m$ & 3.056e-15 & 7.64e-16 & & 8.829e-11 & 7.946e-12 \\ 
\SI[round-precision=0]{500}{\atomicmassunit} $\mathrm{H}_n\mathrm{C}_m$ & 1.724e-15 & 4.31e-16 &  & 7.897e-11 & 7.107e-12 \\ 
\SI[round-precision=0]{600}{\atomicmassunit} $\mathrm{H}_n\mathrm{C}_m$ & 1.109e-15 & 2.773e-16 & & 7.209e-11 & 6.488e-12 \\ 
\bottomrule

  \end{tabular}
  \caption{Cosmic Explorer residual gas requirements and goals. The requirements are that the total gas scattering noise is a factor of five below the design sensitivity and that the total gas damping noises are a factor of three below the design sensitivity. The goals are that the total residual gas noise is a factor of ten below the design sensitivity everywhere. See text for details. The pressures achieved in the Advanced LIGO beamtube are also shown for comparison.\autocite{C982529} The $\text{H}_2$ pumping speed can easily be augmented by titanium sublimation or non-evaporable getter pumps to reach the required pressures in both the chambers and beamtubes.}
  \label{tab:vacuum_requirements}
\end{table}

\begin{figure}
  \centering
  \includegraphics[width=\textwidth]{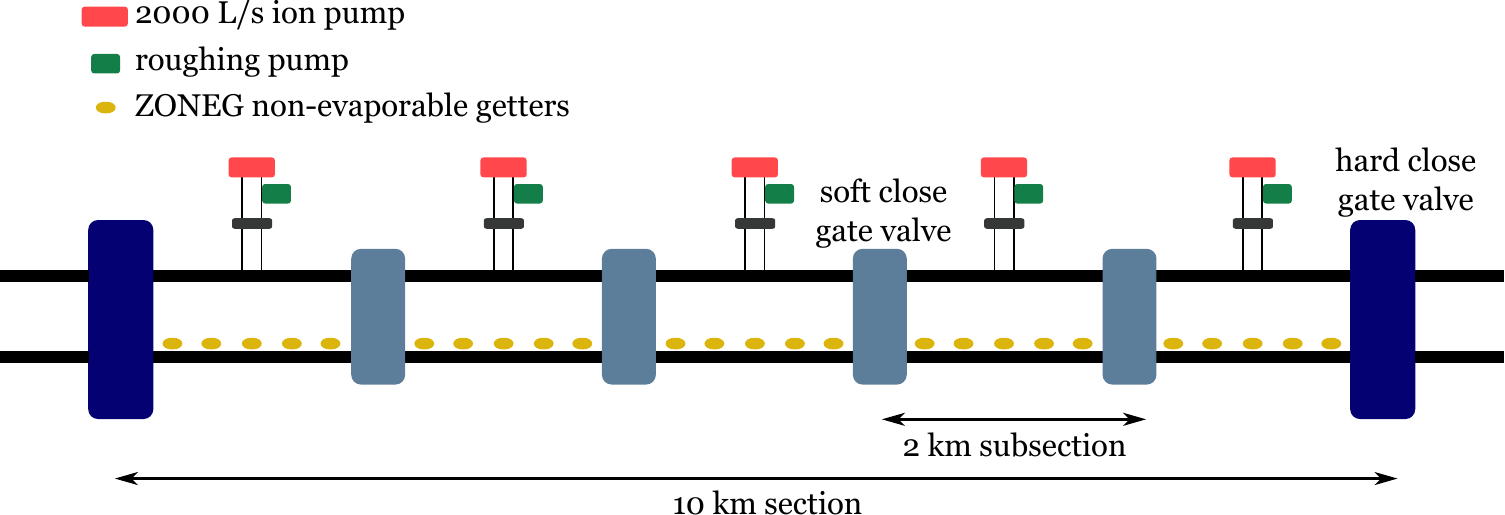}
  \caption{Schematic of a \SI{10}{\km} beamtube section. Each \SI{10}{\km} section can be pumped and serviced independently. The ends of the section are determined by commercial \SI{48}{\inch} (\SI{122}{\cm}) gate valves with elastomer O-rings that can withstand an atmospheric pressure from either side. The \SI{2}{\km} subsections are separated by soft-close gate valves used for separating regions with small pressure differences during some bake operations and for diagnostics. \SI{6}{\inch} pumping ports (not shown) are located every \SI{250}{\m} and can be used for leak hunting and diagnostics or while pumping down.}
  \label{fig:vacuum_system}
\end{figure}

\paragraph{Ultrahigh-vacuum beamtubes}
The vacuum tubing for Cosmic Explorer will be separated into \SI{10}{\km} sections, which are independently pumped. Each section is further divided into \SI{2}{\km} subsections for outgassing and leak hunting as shown in \cref{fig:vacuum_system}. The ends of the \SI{10}{\km} sections will require fully capable gate valves but the \SI{2}{\km} subsections need only the equivalent of light weight shutters to aid in the initial commissioning and operational leak checking. These shutters can be allowed to leak between subsections by as much as \SI{e-3}{\liter/\second} and do not have to bear atmospheric loads. We call this a ``soft'' close valve which should be significantly lower in cost than the ``hard'' close valves used at the ends of the \SI{10}{\km} sections. The soft close valves could be magnetically actuated and will not require penetrations in the vacuum envelope.

\paragraph{Test mass chambers}
Each test mass suspension will be enclosed in a UHV chamber. An option that will be considered is to separate the bottom two stages of the suspension (consisting of the test mass and penultimate mass) from the upper stages. This will enable the stages to be separately accessed, and can shield the test masses from materials with high outgassing. The feasibility of doing this in a manner that both withstands atmospheric loads and maintains the necessary vibration isolation needs to be investigated. Even if this can be accomplished, achieving the goal pressures in the chambers will be more challenging than in the beamtubes since high temperature bakes are not possible in order to protect the components housed in the chambers and since they will need to be opened periodically to make modifications to the detector. The other core optics, such as the beamsplitter, will also be housed in UHV chambers though the requirements on these are not as stringent.

\paragraph{Requirements due to scattered light}
Light scattered out of the interferometer can scatter off the beamtube and
reenter the interferometer.  In this way, motion of the beamtube can be
converted to noise on the light circulating in the interferometer and thus
noise at the gravitational-wave readout. Calculations show that
\SI{120}{\cm} diameter beamtubes with \SI{100}{\cm} baffle apertures are
sufficient to keep noise due to scattered light a factor of ten or more below
the nominal sensitivity of Cosmic Explorer\autocite{BackscatterTechnote, Hall:2020dps}.  See \sref{scatter} for a broader discussion of the scattered light mitigation research required for CE.

\section{Enabling Technologies \hlight{[Stefan, Josh]}}
\label{subsec:technology_drivers}

\newcommand{\RDna}{\cellcolor{ccnone}} %
\newcommand{\RDready}{\cellcolor{Green}} %
\newcommand{\RDnear}{\cellcolor{Yellow}} %
\newcommand{\RDmid}{\cellcolor{Orange}} %
\newcommand{\RDfar}{\cellcolor{Red}} %

\begin{table}[!h]
  \centering
  \small
  \sisetup{detect-all}
  \begin{tabular}{p{20em} c c c c c}
    \toprule
    \textsf{Technology}    &
    \rotatebox{90}{\textsf{2G}}     &
    \rotatebox{90}{\textsf{CE facility}}     &
    \rotatebox{90}{\textsf{initial CE}}      &
    \rotatebox{90}{\textsf{CE}}  &
    \rotatebox{90}{\textsf{CE (\SI{2}{\um})}}  \\
    \midrule

Ultrahigh-Vacuum Systems	&
    \RDready & \RDnear & \RDna  & \RDna  & \RDna \\
Large, High-Purity Mirror Substrates	&
    \RDready & \RDna    & \RDmid   & \RDmid   & \RDfar   \\
Low-Loss Mirror Coatings	&
    \RDnear  & \RDna    & \RDnear   & \RDnear   & \RDmid  \\
Optical Wavefront Control	&
    \RDnear  & \RDna    & \RDnear  & \RDnear  & \RDnear   \\
High-Fidelity Squeezed States of Light	&
    \RDnear  & \RDna   & \RDnear  & \RDmid   & \RDfar    \\
High-Power Ultrastable Laser	&
    \RDready & \RDna    & \RDnear  & \RDnear  & \RDmid    \\
Low-Noise Suspensions	&
    \RDready & \RDna    & \RDmid   & \RDmid   & \RDfar    \\
Inertial and Position Sensors	&
    \RDready & \RDna    & \RDmid   & \RDfar   & \RDfar    \\
Seismic Arrays and Engineering	&
    \RDready  &  \RDnear   & \RDnear &  \RDmid & \RDmid  \\
Environmental Monitoring	&
    \RDready & \RDna    & \RDnear & \RDnear & \RDnear  \\
Low-Noise Cryogenics	&
    \RDna    & \RDna    & \RDna   & \RDna   & \RDmid   \\
Low-Noise Control Systems	&
    \RDnear   & \RDna   & \RDmid   & \RDfar   & \RDfar    \\
Calibration Techniques	&
    \RDnear  & \RDna    & \RDmid  & \RDmid  & \RDmid   \\
Scattered Light Mitigation &
    \RDready & \RDna    & \RDnear & \RDnear & \RDmid \\
    \bottomrule
  \end{tabular}
  \caption{Summary of required research and development activities for Cosmic Explorer.
    The columns in the table indicate whether the activity involves primarily the near-future
    second generation (2G) detectors, i.e., A+, the CE facility, the initial Cosmic Explorer
    sensitivity (initial CE), the target Cosmic Explorer sensitivity (CE),
    or a realization of the target sensitivity using the \SI{2}{\um} technology (CE (\SI{2}{\um})).
   Technology readiness is indicated by green (ready), yellow (nearly ready), orange (requiring modest research), and red (requiring significant research).
  \tlabel{research}}
\end{table}

The reference detector concept for Cosmic Explorer is largely based on
 the evolution of technology currently deployed in LIGO and other gravitational-wave detectors.
Clearly, this evolution of technology will not happen without direction, effort and funding.
This section identified areas where research and development are required to realize
 the target Cosmic Explorer sensitivity.
 
 The timeline for Cosmic Explorer, which includes a collection of planned upgrades,
  is constructed so as to follow the expected technology development over the next two decades.
 Development areas are enumerated in \tref{research} and discussed in the following sections,
  along with the research required to advance them.
To ensure the continuity of gravitational-wave astronomy,
 it is critical that these development efforts are well supported
 while the current generation of observatories are still operational.
In addition to ensuring that CE will achieve its full potential,
many of the technologies required for CE may also be used to enhance existing observatories.
 This research will take place in collaboration with other projects like ETpathfinder~\autocite{etpathfinder} and the Caltech \SI{40}{\m} prototype~\autocite{Mariner}.

\subsection[]{Research on Reducing the Cost of Ultrahigh Vacuum Systems}
\xlabel{vac_value_eng}

Roughly \PCTVAC{} of the cost of CE resides in the ultrahigh vacuum (UHV)
system needed for the \bmts\ and vacuum chambers.  While the vacuum
technology used in LIGO could be used to meet the goals for CE, ongoing
research indicates that significant cost savings may be available, and as such
LIGO vacuum technology serves as a backup strategy should new techniques not be
realized.  In particular, the research described below aims to develop
technology that could meet the CE requirements and reduce the cost of UHV
systems from the estimate of around \$520~million for duplication of the LIGO
approach to less than \$340~million.

\paragraph{Techniques to eliminate high temperature bakes}
The outgassing of adsorbed water is a well established problem in UHV technology. The binding energy of water to the surface has a broad distribution with a peak around \SI{1}{\electronvolt} (\SI{e4}{\K}). The time it takes a water molecule to evaporate from the surface at a fixed temperature is exponentially dependent on the binding energy. The molecules with low binding energy evaporate quickly while the tightly bound ones can take years (even centuries) to evaporate. The distribution of binding energies leads to a $1/t$ dependence in the outgassing rate of gas species at a fixed temperature as a function of time $t$. Gases that do not adsorb, such as nitrogen and oxygen, usually pump out of a system exponentially while water remains as a long term contributor to the residual gas pressure. A standard method to remove water from a system is to heat the walls (``bake'') to increase the evaporation rate. In LIGO the \bmts\ were heated to \SI{150}{\degreeCelsius} for three weeks while the water was pumped out of the system with \SI{6000}{\liter/\second} cryopumps placed every \SI{250}{\m} along the tube. The overall bakeout costs were \$9~million in 1994 dollars for \SI{16}{\km} of tubing and, as discussed in \cref{sec:cost_drivers}, would conservatively cost \$62M (2021 USD) for the \SI{80}{\km} of tubing required for a \SI{40}{\km} CE and nearly \$100M project-wide for the 2-observatory reference concept. This cost was dominated by the electricity needed for the bake and motivates research to establish if there are more economical methods.

One method is to use lower temperature bakes with modest pumping capacity for longer durations. Modeling suggests that this could reduce the water outgassing to levels that meet the CE requirements in the beamtube. Another technique is to use a moving external heater with dry flush gas to remove water from a tube. The process would take place before the tube is evacuated and involves passing the dry gas through the tube while heating the tube to between 145 to \SI{200}{\degreeCelsius} with a movable external ring oven about a meter long. The gas density and flow rate are adjusted to keep the water entrapped in the gas from diffusing back as the oven is slowly moved from one end of the tube to the other in the direction of the gas flow. The temperature of the tube in the short region under the ring heater can be significantly higher than in a full bake which reduces the emission time of the tightly bound water and allows shorter dwell time for the ring heater at each point along the tube. It would take about 14 days to complete the bake for one \SI{2}{\km} subsystem. CERN is looking into a system using inductive rather than radiative heating.

A test planned for LIGO (and important for both LIGO and CE) is the ability to backfill a large vacuum system that has been under UHV conditions with a dry gas and recover UHV conditions without a bake\,---\,a situation that might occur after an accident or necessary repair is made. Modeling indicates that with dry gas filters now available it should be achievable.

\paragraph{Mild steel instead of stainless steel \bmts}
Low carbon steel is a quarter to a third the cost of stainless steel with comparable mechanical properties. For many years it has not been considered a satisfactory material for UHV due to some faulty measurements in the 1950s. Standard production techniques now produce carbon steel with \SIrange{0.1}{0.3}{\percent} of the entrapped hydrogen and comparable water outgassing than most stainless steels\autocite{2016JVSTA..34b1601P}. Recent preliminary measurements at CERN and NIST have provided additional evidence for the low hydrogen outgassing but more research is needed to verify these findings and to develop the practical techniques necessary to use mild steel in UHV.
The cost estimates presented in \sref{cost_est} assume the use of mild-steel \bmts.

\paragraph{Beamtube coatings}
Coating the interior of the \bmts\ with a material that has a low binding
energy for water would reduce the time needed to bake the \bmts.  Ongoing
research by metallurgists indicates that the dark oxide that forms on carbon
steel (magnetite, $\mr{Fe}_3\mr{O}_4$) may have a lower binding energy for
water, but this needs to be tested.  This naturally forming oxide, similar to ``gun bluing'',
could be generated on both the internal and external surfaces of the \bmts\ to
both lower binding energies and prevent oxidation (rust).  It may also be
worth looking into using hydrophobic silicon coatings for the internal surface,
and considering the oil-industry standard epoxy coating for the external surface.

\paragraph{Soft-close gate valves for leak hunting}
Each \bmt\ is broken up into \SI{10}{\km} long sections separated by hard close
gate valves, and each of these is further broken up into five \SI{2}{\km}
subsections separated by soft close gate valves. \SI{48}{\inch} (\SI{122}{\cm})
diameter gate valves for UHV service are commercially available from VAT and
other sources. They cost \$\num{150000} to \$\num{200000} and are able to seal a
vacuum system from atmospheric leakage to better than \SI{e-7}{\liter/\second}
as well as to withstand the atmospheric pressure load.  However, in the event
that a leak develops in the a \bmt, an uninterrupted \SI{10}{\km} segment will
make finding the problem challenging.  Experience from LIGO shows that
\SI{2}{\km} sections are manageable for leak-hunting, and development of less
expensive soft-close valves which are used only in the event of a leak could
result in significant cost savings for the CE project.
The cost estimates presented in \sref{cost_est} assume the availability of such
 soft-close valves at \SI{40}{\%} of the cost of a hard-close valve for a
  project-wide savings of roughly \$4M.

\paragraph{Nested system}
In the event that the water outgassing cannot be reduced sufficiently to meet
the CE requirements it becomes useful to investigate another option. There are
engineering and operational advantages to separating the functions of
maintaining a space against the atmospheric pressure load from that of reducing the
residual gas to the levels of UHV. One promising concept uses a nested vacuum
system with an outer shell of carbon steel tubing and an inner tube of thin
wall aluminum~\autocite{ThirdGenBeamtubeStudyTechnote}.  If the research shows
the nested system is favored, it will be useful to develop an annular soft
close valve that separates the inner and outer systems for outgassing and UHV
operations and opens quickly in the event of a pressure increase in the outer
system.

\subsection[]{Large, High-Purity Mirror Substrates}
\label{subsubsec:large_substrates}

Cosmic Explorer requires large and high-quality optical substrates for the
 main interferometer mirrors (a.k.a.\ test masses).
The Cosmic Explorer test masses will be much heavier than the current Advanced LIGO
 mirrors (\SI{320}{\kg} instead of \SI{40}{\kg}) in order to reduce quantum radiation pressure noise,
 suspension thermal noise, and all technical force noises.
Furthermore, due to the larger diffraction-limited beam size, the Cosmic Explorer mirrors
must have a diameter of more than \SI{50}{\cm} in order to hold round-trip
optical losses from aperture effects to the parts-per-million
level~\autocite{2017CQGra..34d4001A}.

\paragraph{Large silica optics}
The fused silica test masses used for CE will be roughly twice the diameter of those in
 Advanced LIGO (\SI{70}{\cm} instead of \SI{34}{\cm}), again to reduce diffraction loss from the large beams. Such large volume masses are thought to be achievable with excellent optical properties.
However, a careful engineering design, specification, and
characterization of the optics will need to be carried out as with Advanced
LIGO~\autocite{billingsley2017characterization}.
The GWIC 3G R\&D report~\autocite{GWIC3GDocs} calls out the importance of the homogeneity of the
 index of refraction for silica optics, which may be a significant challenge for the CE beamsplitters
 and input test masses due to their large diameter.

\paragraph{Large silicon optics}
The material of choice for the \SI{2}{\um} technology test mass substrates is silicon for the reasons outlined in Ref.~\cite{2020CQGra..37p5003A}. Silicon has low Brownian noise at cryogenic temperatures (unlike fused silica), its thermoelastic noise vanishes at \SI{123}{\K} where its coefficient of thermal expansion goes to zero, and this temperature is compatible with high optical power.
The diameter of the silicon mirrors for CE will be at least to \SI{80}{\cm} since \SI{2}{\um} diffraction limited beams are larger than \SI{1}{\um} beams.
There is currently no production process which can produce substrates of this size
 of sufficient purity for CE (see \cref{subsec:silicon_upgrades}).
Furthermore, there is some evidence that polishing silicon surfaces can increase their optical absorption~\autocite{Bell:2017hdb}.
Further work is required to develop a production process compatible with CE requirements
 and to test whether a silicon surface can be polished to the specifications
 without resulting in surface absorption.

\paragraph{Polishing and surface figure}
Scattered light continues to be a significant source of noise for the current generation of gravitational-wave detectors. Continued improvements in surface polishing of large optics would help to reduce the level of scattered light from the optical surfaces. Estimates of the noise from scattered light for CE suggest that the required surface polish is comparable to that already achieved in Advanced LIGO for spatial scales below a few \si{\cm}~\autocite{Hall:2020dps}. However, due to the larger CE beam size, it is necessary to ensure that this level of surface uniformity can be achieved up to spatial scales of several tens of \si{\cm}.
See \sref{scatter} for a broader discussion of the scattered light mitigation research required for CE.

\subsection[]{Low-Loss Mirror Coatings}
\label{subsubsec:low_loss_coatings}

\paragraph{1\,\textmu{}m Coatings} Current Advanced LIGO coatings are alternating layers of $\text{SiO}_2$ and Ti:$\text{Ta}_2\text{O}_5$ (nearly quarter wavelength) Bragg stacks, deposited via ion-beam sputtering. Research is underway to improve upon the mechanical loss and optical properties of these for A+. The Cosmic Explorer \SI{1}{\um} technology will require A+-like low-mechanical-loss coatings, scaled to larger substrates, and with excellent purity to minimize scattering and absorption. Cosmic Explorer will benefit from current research aimed at improving the A+ coatings. Because the room-temperature mechanical loss of the current low-index material ($\mr{SiO}_2$) is quite low, significant effort is focused on identifying a high-index material with lower mechanical loss. Doped germanium is a promising coating for A+, potentially offering a factor of 2 reduction in coating thermal noise~\autocite{Vajente:2021qru}.
While much more speculative than amorphous coatings,
 crystalline GaAs/AlGaAs~\autocite{2019OExpr..2736731K}
 could provide even lower coating thermal noise if they can be scaled to the size of Cosmic Explorer optics.
Because coating thermal noise will be a contributing noise source for Cosmic Explorer, additional gains beyond the state of the art in the coming years will be of great benefit.

Anomalous absorption from small defects in the Advanced LIGO test mass coatings has proven to be a major challenge and limitation the sensitivity of the instruments via degradation of the cavity buildup~\autocite{LIGOScientific:2021kro}. Research to solve this problem is currently a major focus for LIGO coatings, and clearly it must be solved for Cosmic Explorer as well.

\paragraph{2\,\textmu{}m Coatings} The \SI{2}{\um} coating technology will build off research and development toward LIGO Voyager and shows promise to offer improved thermal noise performance over the \SI{1}{\um} technology. The baseline assumed is ``Voyager'' coatings, scaled to \SI{80}{\cm} diameter and operated at \SI{123}{\K}. While it is expected that this area will benefit from significant further research, promising candidates for such coatings have already been identified, including amorphous $\text{SiO}_2$/Si~\autocite{2015PhRvD..92f2001M} and crystalline GaAs/AlGaAs~\autocite{2019OExpr..2736731K}. Development efforts are also needed toward scaling up from the current state-of-the-art of \SI{10}{\cm} and \SI{34}{\cm} for crystalline GaAs/AlGaAs~\autocite{2019OExpr..2736731K} and amorphous coatings, respectively.

\paragraph{Conductive Coatings}
Silica is an insulator, and hence charge can build up on its surface.  This has
been an issue in Advanced LIGO, so it will be important to limit the amount of
charge that is able to build up on the Cosmic Explorer
optics~\autocite{2020PhRvD.102f2003B}.  With R\&D underway on slightly
conductive overcoatings and with the experience of charge control in LIGO to
capitalize on, charge is not expected to be a major issue for CE.

\subsection[]{Optical Wavefront Control}
Even at the sub-part-per-million level of optical absorption achieved in the current Advanced LIGO mirror coatings, thermal lensing and thermal expansion induced by the high laser power circulating in the arm cavities (\SIrange{1.5}{3}{\mega\watt} for Cosmic Explorer) will lead to significant changes of the optical mode in the interferometer.
In addition, any anomalous absorption from small defects (see \cref{subsubsec:low_loss_coatings}) will make the problem significantly worse. If not corrected, the resulting wavefront distortions lead to excess scattering in the arm cavities, limit the power build-up, and degrade the interferometer contrast at the readout port.
Furthermore, the distorted optical modes will no longer match the output mode cleaner cavity, nor the mode of the non-classical squeezed vacuum injected from the dark port to reduce the interferometer quantum noise (see \cref{subsubsec:high_fidelity_squeezing})\,---\,unsqueezed quantum vacuum will leak into the interferometer, reducing the sensitivity gain from squeezing the quantum noise.

In order to achieve the mode-matching requirements and the high levels of frequency dependent squeezing in CE, active wavefront control actuators will be indispensable for the readout optics leading to the output mode cleaner, for the squeezed vacuum injection optics and filter cavity, as well as for the test masses themselves. New technologies are under development for active wavefront control and mode-matching for the A+ upgrade. These have direct applicability to CE with its even stricter optical loss requirements.

\subsection[]{High-Fidelity Squeezed States of Light}
\label{subsubsec:high_fidelity_squeezing}

Squeezed states of light are used to reduce quantum noise in current
gravitational-wave detectors and this technology will also be used in Cosmic
Explorer, initially at modest levels of noise reduction with planned upgrades
expected to bring CE to its target sensitivity (see
\cref{subsubsec:quantum_noise}).  Frequency-independent
squeezing is employed in the current Advanced
LIGO~\autocite{2019PhRvL.123w1107T} and Advanced
Virgo~\autocite{2019PhRvL.123w1108A} detectors, and has reached \SI{6}{\dB} of
noise reduction in GEO600\autocite{2013PhRvL.110r1101G}.  Frequency-dependent
squeezing has been achieved by reflecting squeezed light off of filter cavities
as described above~\autocite{2016PhRvL.116d1102O, 2020PhRvL.124q1102M}, and a
\SI{300}{\m} filter cavity is now being installed at the LIGO sites for the A+
upgrade. These two demonstrated technologies set the baseline \SI{6}{\dB} of
frequency-dependent squeezing for the initial CE target. Research and
development is needed to reach the required \SI{10}{\dB} of frequency-dependent
squeezing, down to about \SI{10}{\Hz}, for the final CE design and to
demonstrate this also at \SI{2}{\um} laser wavelength. 

The level of noise reduction by squeezing, in decibels, is limited to
$10\log_{10}(2\theta_{\text{rms}} + \Lambda)$, where $\theta_\text{rms}$ is the
rms phase noise of the squeezed field, and $\Lambda$ is the total
effective optical loss experienced by the field, in both cases accounting for
all mechanisms starting from the generation of the field, its propagation
through all optical elements of the detector, and its conversion to current in
the photodiode. \SI{10}{\dB} of noise reduction can be achieved assuming
baseline values of $\theta_{\text{rms}}\lesssim \SI{10}{\milli\radian}$ and
$\Lambda\lesssim \SI{8}{\%}$. The effective squeezing losses have some
broadband contributions as well as some frequency dependent contributions due
to the optical cavities of the interferometer and quantum filter cavity. Here
it is assumed that the contributions to the total loss are mostly broadband:
\SI{1}{\%} from the squeezer, \SI{2}{\%} from the Faraday isolators, \SI{1}{\%}
from the photodiodes, \SI{1}{\%} from the output mode cleaner, \SI{1}{\%} from
the pickoff mirrors, and \SI{1}{\%} from mode mismatch; finally,
frequency-dependent cavity effects contribute an additional \SI{1}{\%}, thereby
totaling \SI{8}{\%} total loss.  The following paragraphs provide an overview
of the R\&D needed to achieve these loss levels.

\paragraph{Squeezer design}
The existing design of in-vacuum squeezers used by Advanced
LIGO~\autocite{2015NatSR...518052W, 2016Optic...3..682O}, which yields
\SI{2}{\%} internal cavity losses, is nearly sufficient for the Cosmic Explorer
requirements, and only incremental improvements are needed
 to achieve the CE loss target of \SI{1}{\%}.

Research to demonstrate squeezing at \SI{2}{\um},
 where similar design constraints apply,  is
ongoing~\autocite{2018PhRvL.120t3603M}.
The main differences are related to the eventual coating
performance and crystal losses achievable at the longer wavelength.

\paragraph{Low-loss Faraday isolators}

Low-loss Faraday isolators are now used in Advanced
Virgo~\autocite{2018ApOpt..57.9705G} and Advanced LIGO. These demonstrate
\SIrange{0.5}{1}{\%} loss per pass of the squeezed light.  The current design
of a filter cavity requires a minimum of four total passes through Faraday
isolators. An additional fifth pass is used in Advanced LIGO to add optical
isolation, but can be avoided with improved isolation and scattered light
mitigation. Minor improvement is required to bring the isolator losses down to
a total of \SI{2}{\%} given the four required isolator passes.

\paragraph{High Quantum-Efficiency Photodetection}
Achieving \SI{1}{\%} loss from photodiodes requires photodiode quantum
efficiencies ${\gtrsim}\SI{99}{\%}$.
While this level of quantum efficiency has already been achieved for
\SI{1}{\um} light, it is a serious challenge for \SI{2}{\um}
light requiring significant R\&D, but there are no known fundamental obstacles
to achieving this performance~\autocite{2020CQGra..37p5003A}. Promising
candidate technologies are described in Ref.~\cite{2020CQGra..37p5003A} and
include extending the InGaAs detectors used for \SI{1}{\um} light as well as
developing HgCdTe and InAsSb detectors.

An alternative method is to add parametric amplification, with strict
requirements on its efficiency, to the output of the interferometer, possibly
by adding another squeezer
unit~\autocite{1981PhRvD..23.1693C,2019OExpr..27.7868K,2020PhRvA.102b3507B}.
This amplifies both the optical signal and quantum noise, preserving the
signal-to-noise ratio, and lower quantum efficiency photodiodes may then be
used without penalty. The integration requirements of such an optical
parametric amplifier, and its pump light, have so far not been fully studied.
However, the balanced homodyne readout of the A+ upgrade to LIGO shares a
number of requirements that are likely to translate to the parametric amplified
readout scheme.

\paragraph{Filter cavity}

Low-frequency design and high-mass mirrors give Cosmic Explorer a very low
``standard quantum limit'' (SQL) crossover frequency where the quantum noise
has equal contributions from photon shot noise and quantum radiation pressure
noise. It is necessary for the squeezed state to rotate from amplitude squeezing to phase squeezing at this frequency in order to achieve broadband quantum noise reduction, and this will be accomplished with a \SI{4}{\km} filter cavity with a finesse of
${\sim}$4000, based on the \SIrange{60}{70}{\ppm} losses achieved in the
Advanced LIGO arm cavities.  A \SI{4}{\km} filter cavity length is necessary to
prevent the loss-induced dephasing of the cavity from limiting the allowable
injected squeezing\autocite{2021arXiv210512052M}.

The additional design constraints for controlling squeezing with a \SI{40}{\km}
interferometer can be alleviated given more research into the scheme
proposed in Ref.~\cite{2020CQGra..37r5014K}, which uses the filter cavity
itself as a low-noise phase reference for the squeezed light.

\paragraph{Constraints due to the Integration of Subsystems}
In Advanced LIGO, some of the interferometer alignment signals are sensed at
the antisymmetric port using a \SI{1}{\%} transmissive mirror. Additionally, the
filter cavity also samples \SI{1}{\%} of the power for alignment control. For Cosmic
Explorer, these transmissivities should ideally be avoided or reduced, which will
impact the overall controls design of the instrument. This will require
research given that the alignment sensing and controls need to achieve the low
frequency sensitivity goals of Cosmic Explorer.

Additionally, there are integrative constraints for the squeezer and filter
cavity involving the total optical isolation, the length and alignment noises
which must be suppressed, and the sensing noise injected by the control systems.
Improved seismic isolation will reduce the required control bandwidth and
offset the requirements imposed by the lower frequency sensitivity of the CE
detectors. A more detailed design study is required.

\paragraph{Core Optics}
Optical losses in an interferometer limit the enhancement due to squeezing,
independent of other aspects of the system \autocite{2019PhRvX...9a1053M}.  In
Cosmic Explorer, the loss in the signal extraction cavity (see
\fref{reference_design}) directly limits the high-frequency
sensitivity of the detector and thus the postmerger science it can perform.
Above the detector bandwidth, the low transmission of the CE signal extraction
mirror causes a large enhancement of losses within the signal extraction
cavity, approximately $4\epsilon_{\text{SEC}}/T_{\text{SEM}}$ where
$T_{\text{SEM}}$ is the transmissivity of the signal extraction mirror and
$\epsilon_{\text{SEC}}$ is the loss in the signal extraction cavity. While
these resonantly-enhanced losses reach their maximum above the instrument
bandwidth, they are sufficiently large that the losses within the signal
extraction cavity must be minimized, so they do not degrade the detector
performance within the signal band. This includes all of the anti-reflection
coatings of vertex optics such as the input test masses, compensation plates,
and beamsplitter. Additionally, there is loss in the high-reflection coatings
of the telescope optics within the signal extraction cavity. While compensated
in other optics, the refractive index inhomogeneities in the substrate of the
beamsplitter cannot currently be compensated using surface polishing, due to
the non-normal incidence of the beam, and must be analyzed to consider its
contribution to the squeezing losses.

At high frequencies, the resonant ``dips'' in the \SI{20}{\km} instrument noise
at \SIrange{2}{4}{\kilo\Hz}, which depend on the arm length, signal extraction
cavity length, and signal extraction mirror transmissivity, depend on the
internal cavity losses. Full optimization of the squeezing level for postmerger
signals requires a careful analysis not only of the instrument response, but
also of the losses to squeezing, as the two mostly, but not entirely, align at
those high frequencies. Furthermore, it is the loss in the signal extraction
cavity that limits both the depth of these dips and the frequencies to which
they can be tuned, so it is especially important to the \SI{20}{\km} science
case that these losses be kept low.

\paragraph{Mode Matching}

The squeezed states will interact with a sequence of optical cavities: the
squeezer's optical parametric amplifier, the filter cavity, the interferometer,
then finally the output mode cleaner.
Achieving loss of $<\SI{2}{\%}$ in squeezing due specifically to mismatch
 (at all frequencies~\autocite{2021arXiv210512052M})
will require the design of the
output optics and squeezer to be extremely mindful not only of curvature
mismatch, but astigmatism and higher order aberrations as well. The overall
curvature mismatch can be corrected using active wavefront
control~\autocite{2020OExpr..2838480C,2020ApOpt..59.2784C}, as is planned for
LIGO. Research is needed into control and mitigation of astigmatism in the
telescope designs,
and the ability to measure and quantify wavefront errors in-situ using
auxiliary beams and detectors should also be improved.

The need for higher-order aberration correction may prove necessary to match
the squeezer beam to the interferometer, given that the interferometer has
contrast defect and distortion from heating. More modeling is necessary to
establish those needs.

\subsection[]{High-Power Ultrastable Laser}

Cosmic Explorer will use a similar laser source as that of LIGO, known as the pre-stabilized laser, consisting of a seed laser, laser amplifier, some frequency and intensity stabilization, and some reduction of higher order mode content.\autocite{2012OExpr..2010617K} The CE requirements on laser frequency noise incident on the interferometer are estimated to be \SI{7e-7}{\Hz\big/\!\sqrt{\Hz}}. The frequency stabilization scheme currently employed by LIGO relies on using the common mode arm cavity as the ultimate frequency reference; however, this scheme cannot achieve the frequency noise requirements for CE across the entire detection band due to its ten times longer arms. A new scheme using two suspended modecleaners can achieve these requirements without relying on the arms as a reference\autocite{Cahillane:2021jvt} and is used as the reference concept for CE. While the second mode cleaner is not strictly necessary to meet the frequency stabilization requirements, it simplifies the intensity stabilization, can reduce the complexity of the pre-stabilized laser, and reduces other technical noises.

After passing through the mode cleaners, CE will require ${\sim}\SI{140}{\watt}$ to be injected into the interferometer to reach the nominal \SI{1.5}{\mega\W} arm power given the expected optical loss. This level of output power has already been demonstrated in stabilized continuous-wave lasers~\autocite{2006JPhCS..32..270W}. If the alternative \SI{2}{\um} technology were employed, a significantly higher power of ${\sim}\SI{280}{\watt}$ would be needed to reach the nominal \SI{3}{\mega\W} of arm power. The technology available at \SI{2}{\um}  is much less advanced than \SI{1}{\um} technology, and considerable research and development toward ultrastable \SI{2}{\um} laser sources\autocite{2020OExpr..28.3280K} and their high power amplification is needed.

\subsection[]{Low-Noise Suspensions}
\label{subsec:low_noise_suspensions}

As described in \cref{subsubsec:seismic_isolation}, each of the four test
masses will be suspended by quadruple pendulum suspensions similar to those
currently employed by LIGO.  The reference concept for both technologies is a
\SI{4}{\m} long suspension chain of total mass \SI{1500}{\kg}. It is believed
that such suspensions can be achieved and supported by scaling up current
Advanced LIGO systems.

Increasing the mechanical compliance of the suspensions\,---\,in all six degrees of freedom\,---\,is key to reduce seismic and thermal noises. More compliant suspensions produce lower frequency mechanical resonances, and displacement noises are passively filtered above these frequencies. The mechanical loss of the suspension material determines the magnitude of the suspension thermal noise.

Developing highly stressed suspensions is critical for two reasons. First, this provides the opportunity to increase the suspension compliance. Second, it increases the fundamental and harmonic frequencies of the high-$Q$ transverse vibrational (``violin'') modes of the suspensions, which degrade the sensitivity in a narrow (${\sim}1/Q$) band around the mode frequencies, thus reducing the number of modes in the detection band. The development status of highly stressed materials for the two technologies is discussed below.

\paragraph{Silica}
The gravitational-wave community has much experience with manufacturing highly stressed fused silica suspension fibers using a fiber pulling technique~\autocite{2011RScI...82a1301H}. Tapered fibers are used in order to reduce thermoelastic noise at the ends of the fiber where the most bending, and therefore the most loss, occurs. The end radius is chosen to cancel the two contributions to thermoelastic loss\,---\,one from thermal expansion and one from the temperature dependence of the Young modulus. A smaller radius is chosen to maximize the stress along the length of the fiber. The maximum stress in the Advanced LIGO silica fibers is \SI{800}{\mega\pascal}\autocite{2012CQGra..29w5004A}, which provides a safety factor of about six for the breaking stress of fibers realized at the time the LIGO suspensions were designed~\autocite{2012JNCS..358.1699T}. Recent advances in these fabrication techniques allow for fibers to be manufactured with working stresses of \SI{1.2}{\giga\pascal}, which provides a safety factor of about three~\autocite{2019CQGra..36r5018L}. While these fiber fabrication techniques are mature, no fused silica blade springs have been manufactured to date, and this is a critical area of R\&D necessary to meet CE's low frequency sensitivity goals. The current design for both stages of the silica technology calls for \SI{1.2}{\giga\pascal} fiber stress and \SI{800}{\mega\pascal} blade spring stress~\autocite{Hall:2020dps}.

\paragraph{Silicon} Since the alternative silicon realization of CE operates at the zero-crossing of the thermal expansion coefficient, it is not possible to cancel the thermoelastic noise as is done for the fused silica fibers. Therefore, silicon suspensions would use silicon ribbons with dimensions chosen to maximize the stress along the entire length of the ribbon. Silicon ribbon fabrication is not well developed and the experiments most relevant to manufacturing suspensions find that the tensile strength of ribbons depends on the surface treatment and edge quality with average breaking stresses measured ranging from \num{100} to \SI{400}{\mega\pascal}, and individual samples observed as high as \SI{700}{\mega\pascal}\autocite{2017CQGra..34w5012B,2014CQGra..31b5017C}. Silicon blade springs have yet to be developed. While larger stresses have been observed in other applications, the alternative silicon based concept for CE assumes \SI{400}{\mega\pascal} of stress in both the fibers and blade springs~\autocite{Hall:2020dps}.

\subsection[]{Inertial and Position Sensors}
\xlabel{inertial_sensors}

Cosmic Explorer will benefit greatly from research and development into
low-noise inertial and position sensors.  Such sensors will enable improvements
in seismic isolation and suspension control, both of which enhance the
low-frequency sensitivity of Cosmic Explorer, and thereby improve its
localization, early warning, and/or high-redshift detection capabilities for
compact binaries.

As described in \cref{subsubsec:seismic_isolation} above, the seismic isolation for the initial CE target assumes a moderate improvement over current technology: \SI{0.1}{\pico\meter\big/\!\sqrt{\Hz}} at \SI{10}{\Hz}, which is threefold better isolation than Advanced LIGO, and \SI{1}{\pico\meter\big/\!\sqrt{\Hz}} at \SI{1}{\Hz}, which is tenfold better than Advanced LIGO~\autocite{Hall:2020dps}.
Promising technologies to achieve this include combining the mechanics of a conventional geophone (GS13) with an interferometric proof mass readout~\autocite{2018CQGra..35i5007C}.
The noise below \SI{1}{\Hz} is residual ground motion that comes from the inclusion of a position sensor signal to lock the suspension point to the ground on long timescales (a technique known as ``blending'').
Additionally, the horizontal inertial sensing is susceptible to contamination from ground tilt, and should therefore be paired with low-noise tiltmeters~\autocite{2014RScI...85a5005V}.
This is motivated by studies at LIGO Hanford that have shown that ground tilt couples significantly to the strain readout of the interferometer even after active seismic isolation~\autocite{2018PhRvL.121v1104C}. Lowering the tilt coupling, along with mitigating Newtonian noise fluctuations from the atmosphere, is an important motivator for carefully designed buildings~\autocite{2020CQGra..37r5018R}. Further research in all of these areas is warranted. 

Subsequent upgrades are planned to achieve a more significant isolation improvement: an additional threefold improvement at \SI{10}{\Hz} and tenfold improvement at \SI{1}{\Hz}.
A number of efforts are underway to meet this challenge
 worldwide~\autocite{vanHeijningen2018, 2020JInst..15P6034V, 2019CQGra..36x5006M, 2018PhRvL.120n1102Y}.
Moreover, improved low-frequency noise of the inertial sensors leads to less reliance on the low-frequency position sensor signals, thereby lessening the contamination from residual ground motion.

\subsection[]{Seismometer Arrays and Seismic Engineering}
\label{subsubsec:seismic_arrays_engineering}

Without mitigation, Newtonian noise from seismic waves would limit the
low-frequency sensitivity of Cosmic Explorer.  As stated in
\cref{subsubsec:newtonian_noise}, Newtonian noise produced from surface seismic
waves will need to be suppressed by a factor of ten and noise produced from body waves by a factor of three through a combination of careful facility design and sensor based noise cancellation.  A
combination of several techniques can be employed to achieve these goals,
outlined in Ref.~\cite{Hall:2020dps}, and briefly discussed below.

\paragraph{Low-density materials}
Newtonian noise may be reduced by lowering the overall material density near
the test mass.\autocite{2014CQGra..31r5011H} This is achieved in the
facility design by having recesses (e.g., a basement) or low-density building
materials (e.g., Geofoam) near/under the test masses, and/or by locating the
test masses well above the ground level (e.g., on the second floor).
\paragraph{Seismic metamaterials and architected structures}
Seismic waves can be deflected or dissipated before they reach the test masses
with intentionally designed structures. Seismic metamaterials are architected
structures that can reduce surface wave propagation, using above-ground
resonators, buried resonators, inclusions, and/or exclusions.
\autocite{2014PhRvL.112m3901B, 2016NatSR...639356P, 2016NatSR...627717C,
roux2018toward, 2019APS..APRR11006K,2020PhRvP..13c4055Z} While more detailed
studies will be needed on the feasibility of employing this technology, the CE
facility is expected to incorporate at least the simplest of these techniques
into its design.

\paragraph{Seismometer array subtraction}
Newtonian noise can be subtracted from the detector's strain data by estimating
the local seismic field with an array of
seismometers.\autocite{2012PhRvD..86j2001D} A recent proof of principle
experiment demonstrated a tenfold suppression in the range
\SIrange{10}{20}{\Hz}\autocite{2018PhRvL.121v1104C}, though more research is
needed to meet the Cosmic Explorer low frequency requirements down to \SI{5}{\Hz}.

\vspace{0.5\baselineskip}

\noindent The above techniques largely rely on modeling wave propagation through homogeneous media. Future studies on wave propagation through inhomogeneous media, such as stratified soil, natural topological structures, etc., will need to be conducted to ensure that these techniques are capable of reaching design sensitivity in a given seismic noise environment.

\subsection[]{Environmental Monitoring}
Monitoring of non-gravitational-wave disturbances from the environment, using
an array of instruments located at the sites as well as information from global
monitoring, has been critical for ground-based gravitational-wave observatories
to date. The main purposes of environmental monitoring are localizing and
mitigating sources of noise, assuring that the contribution of ambient
environmental noise is kept below the background noise of the detectors or
subtracted from the detector strain data, and validating candidate
gravitational-wave signals by ruling out potential sources of terrestrial
origin~\autocite{AdvLIGO:2021oxw}.

For Cosmic Explorer, Newtonian noise will place more stringent requirements on
acoustics, particularly infrasound (sound at frequency less than
\SI{20}{\Hz})~\autocite{Hall:2020dps}. CE will thus require a careful
measurement strategy, and potentially a reduction strategy, for infrasound.
Ideally, Cosmic Explorer will employ sensors that measure the atmospheric
pressure down to \SI{0.1}{\milli\pascal\big/\sqrt{\Hz}} at \SI{10}{\Hz}, and
can disentangle the acoustic field from turbulent pressure fluctuations.
Subtraction of infrasonic Newtonian noise with sensor arrays faces a number of
uncertainties~\autocite{2019LRR....22....6H}, and as such the Cosmic Explorer
sites should be selected, and facilities constructed, so as to minimize
infrasonic noise.

Requirements for the facility magnetic spectrum and coupling to the CE
electronics and sensitive equipment have not yet been set. However, magnetic
coupling in current detectors has required significant monitoring and
mitigation, so we expect this to also be the case for CE. For example, Schumann
resonances, electromagnetic resonances between the Earth's surface and the
ionosphere that are excited by lightning, will require careful measurement and
perhaps subtraction~\autocite{2018PhRvD..97j2007C}.

\subsection[]{Low-Noise Cryogenics}
A centerpiece of the alternative \SI{2}{\um} Voyager technology, as well as Einstein
Telescope, is the use of crystalline silicon at cryogenic
temperatures for the test masses and the lowest stages of the suspensions.  The
Japanese detector KAGRA, currently under construction, will use cryogenics and
sapphire test masses to achieve similar objectives.  KAGRA and ET will operate
at \SI{20}{\K}, while Voyager operates at \SI{123}{\K}.  The latter was
chosen because it not only offers some reduction of thermal noise, but is also
a zero crossing value for the thermal expansion coefficient of crystalline
silicon~\autocite{2015PhRvB..92q4113M} and thus eliminates two major
limitations to performance: thermoelastic noise and thermal aberrations from
light absorption within the optics.

A challenge for the cryogenic operation of Cosmic Explorer will be to accurately ($\SI{\pm 2}{\K}$)~\autocite{Hall:2020dps} hold the optics and suspensions at their target temperature without introducing additional noise through vibrations, acoustics, or Newtonian noise coupling. Techniques under consideration, but requiring more study, to control deviations from the target temperature include: using the vibrational eigenmodes of the test masses as temperature sensors~\autocite{2021PhRvD.103b2003B}; and injecting small localized heat modulation to the test masses and minimizing the induced displacement. For both, thermal radiation could likely be used to actuate the temperature.

It is likely that a radiative cooling approach such as that envisioned for Voyager~\autocite{2017Cryo...81...83S} would be used for CE. This involves a movable heat link that comes into contact with the test mass only during initial cool down, to speed up heat extraction, and then radiative cooling only to maintain the test mass and suspension temperature during operations. However, there are open questions such as: What are the key drivers of the overall heat budget?; What are the requirements on temperature, length (tens of meters?), and material/coating properties for the radiation shields that stop the cold optics from seeing the warmer tube and environment?; Will each core optic and suspension require a dedicated cryostat? What is the most promising cryogenic technology (e.g., pulse-tube cryocoolers in phase opposition (such as in ET) or Gifford-McMahon)?; How will vibrations be mitigated?; and what techniques (such as helium gas cool-down or retractable contacts) will be used to provide high cool-down speeds? These questions will be addressed in more detail in the design phase
of the CE project, and in light of Voyager technology research developments.

\subsection[]{Calibration Techniques}

Like current gravitational-wave detectors, Cosmic Explorer will need a calibration apparatus to turn the raw detector data into an estimated strain time series.
The greater sensitivity of Cosmic Explorer compared to today's detectors, coupled with the expected improvement in theoretical waveform accuracy (see \cref{sec:computing}), will result in calibration requirements more stringent than those of Advanced LIGO and Virgo.
In today's detectors, the systematic error in the calibration is a few percent in amplitude and a few degrees in phase across the sensitive band; the \SI{68}{\%} uncertainty on the error estimate is at most \SI{10}{\%} in amplitude and \SI{10}{\degree} in phase, and a few tens of microseconds in intersite timing~\autocite{2020CQGra..37v5008S, 2021arXiv210700129S, 2021arXiv210703294V, Virgo:2018gxa, 2010CQGra..27h4025B, Asali2019}.
The error and uncertainty are small enough that astrophysical parameter estimation is limited by noise in the detectors, rather than the calibration~\autocite{2009PhRvD..80d2005L,2014PhRvD..89h2001A,2012PhRvD..85f4034V,2019CQGra..36t5006H,2021PhRvD.103f3016V,2020PhRvD.102l2004P}.
 
The exact calibration requirements for Cosmic Explorer are not known, but are likely to be significantly below \SI{1}{\%} accuracy so as to not limit the astrophysical output of CE.
Such accuracy in calibration will enable an estimate of the Hubble constant with a \SI{0.2}{\%} uncertainty over five years via multimessenger observations with telescopes such as the Vera Rubin Observatory~\autocite{2021ApJ...908L...4C}. 
This calibration accuracy will also extend tests of general relativity, reducing the chance of calibration errors being mistaken for false-positive deviations of the observed signal waveforms from theory~\autocite{2019CQGra..36t5006H}. 
Pushing these levels of accuracy to frequencies of a few kilohertz will allow tighter constraints on the equations of state of neutron stars through observations of their postmerger phase~\autocite{2019PhRvD.100j3009H}.
Extending the accuracy to below \SI{20}{\Hz} will improve CE's ability to probe the crust and magnetic fields of rapidly rotating isolated neutron stars~\autocite{2018RSPTA.37670279S}.
Continued research is required to assess whether improvement beyond these bounds, and in particular frequency regions, may be required for other sources.

Research and development is required to extend the calibration techniques of today to reach these stringent requirements.
Currently, the level of systematic error is estimated using offline measurements; these measurements encompass the detector's response to displacement of its test masses, the test masses' response to control forces, and the detector's sensor and actuator signal processing electronics~\autocite{2020CQGra..37v5008S, 2021arXiv210700129S}.
The offline measurements are paired with online measurements of a few slowly-varying time-dependent parameters~\autocite{2017CQGra..34a5002T, 2018CQGra..35i5015V}.
These measurements are compared with a frequency-dependent model of the detector and its control system, and this model is used to produce the strain time series.
The systematic error in this calibration is limited by theoretical accuracy of the model, the residual error between the measured and modeled frequency-dependent components, and the uncertainty in the detector's absolute displacement reference or ``calibration standard'', used to measure some of the model components~\autocite{2021arXiv210700129S}. 
Each of these areas need to be improved to ensure the required systematic error. 

The primary calibration standards for the current detectors are so-called photon calibrators, which use power-modulated laser light from auxiliary lasers to drive the test masses with radiation pressure~\autocite{2016RScI...87k4503K, 2021CQGra..38g5007E}. 
The reflected light is measured by photodiodes traceable to the National Institute of Standards and Technology (NIST), thereby enabling highly accurate estimates of the radiation force on the test masses.
These systems already produce displacement with amplitude uncertainty of $\SI{0.5}{\%}$~\autocite{Bhattacharjee:2020yxe}.
Improving the photon calibrator systems to the accuracy level required for CE (perhaps $<\SI{0.1}{\%}$) should be possible with anticipated improvements in laser power standards from the global network of national metrology institutes (including NIST)~\autocite{spidell2021bilateral, pcalNIST} and the reduction of other practical limiting systematics to the systems~\autocite{Bhattacharjee:2020yxe}.

In addition to photon calibrators, other supplemental calibration standards are being considered.
Options include using existing subsystems of the detectors themselves, such as frequency modulation of the primary laser~\autocite{Goetz:2010gn} or the use of auxiliary lasers~\autocite{2017PhRvD..95f2003A}, or measurements based on the laser wavelength with the detector temporarily configured as a simple Michelson interferometer~\autocite{LIGOScientific:2010weo, 2011CQGra..28b5005A,2017PhRvD..95f2003A}.
The use of gravitational-wave sources themselves as calibration standards (sometimes referred to as astrophysical calibration) has been studied and demonstrated; however, the accuracy of these methods will not be competitive with photon calibrators, even for 3G detectors~\autocite{2019CQGra..36l5002E, Schutz:2020hyz}.
Other direct force options such as spinning gravitational calibrators (Newtonian calibrators) have demonstrated promising accuracy in early prototypes~\autocite{2021arXiv210700141R, Estevez:2020ulq} and recent work suggests that a combination of photon and gravitational calibrators could achieve an absolute accuracy of \SI{0.17}{\%}~\autocite{2018PhRvD..98b2005I}.

Work will also be needed to improve the accuracy of the detector's modeled response to gravitational-wave strain. For example, (1)~developing methods to characterize the full dual-recycled Fabry--P\'{e}rot Michelson response, with losses, to surpass the accuracy of past approximations~\autocite{2002PhLA..305..239R, Rakhmanov:2008is, Ward:2010qda, izumi2016_T1600278, hall2016_G1601599, LIGOScientific:2017aaj, 2019CQGra..36t5006H}, (2)~taking into account the multi-input multi-output complexity of the auxiliary control systems to better account for cross-coupling~\autocite{2020PhRvD.102f2003B, 2020CQGra..37v5008S}, and (3)~going beyond long-standing assumptions for how gravitational waves interact with the detectors, such as the long-wavelength approximation~\autocite{2002PhLA..305..239R, Rakhmanov:2008is}.

Producing the estimate of the systematic error function at a given time, from offline measurements, is time-consuming and inherently more uncertain than directly measuring the error by comparing displacement made by an absolute reference in real time.
To date, the desired observational duty factor has trumped characterizing the detector's calibration error, as high accuracy and precision were not of paramount importance. 
As such, many months are spent reconstructing the error estimate for all time in post-processing, after the strain data is recorded, and follow-up characterization measurements can be made. 
Research is underway to find methods of continual characterization  while retaining observation time; this includes leveraging noise subtraction techniques already demonstrated to improve the detector sensitivity~\autocite{2019CQGra..36e5011D, 2020PhRvD.101d2003V}. 
In this way, the $\SI{68}{\%}$ confidence interval on the systematic error will then only be limited by the uncertainty in the absolute reference(s) used to create the characterization signal and data integration time. 
Identifying and reducing the systematic error itself may still be time consuming.
However, newly developed data analysis techniques may sufficiently marginalize over them if armed with precise quantified knowledge of these errors~\autocite{2020PhRvD.102l2004P,2021PhRvD.103f3016V}, and other calibration pipeline development may yet yield full frequency-dependent correction of the well-characterized systematic error function in low-latency.

While all of this research and development is a challenge, the path appears clear to deliver the required precision in a timely manner for Cosmic Explorer.

\subsection[]{Scattered Light Mitigation}
\xlabel{scatter}

Noise from light that is scattered out of the main laser beams has been a persistent issue for current gravitational-wave detectors~\autocite{vajente2014stray, 2012OExpr..20.8329O}. While scattered light reduction was included in the design of the advanced detectors, much effort has been devoted to diagnosing and fixing light scattering issues after installation~\autocite{AdvLIGO:2021oxw, LIGO:2020zwl, 2021CQGra..38g5020W}. Scattered light effects, which are often driven by seismic motion, will be particularly important to address in order to ensure the excellent low-frequency performance of Cosmic Explorer. It is thus important that a scattered light mitigation design that incorporates the best practices and lessons learned from the 2G experience be developed for Cosmic Explorer. Three main areas of concern are discussed below. 

Light scattered from the coated test masses can reflect off of moving elements, such as the \bmt\ walls and baffles, and recombine with the main cavity laser mode giving rise to additional phase noise in the readout~\autocite{flanagan_thorne_bs}. For this reason, baffles were installed along the \bmts\ of the 2G detectors to deflect and absorb scattered light. Light scattering from the test masses has two main sources: roughness of the coated surfaces, which is responsible for most of the scattering at small angles; and point defects on the surfaces that show up as ``bright spots'' and are responsible for most of the scattering at large angles. The noise contribution to Cosmic Explorer surface roughness scattering can be estimated following Sec.~2.2 of Ref.~\cite{BackscatterTechnote}. Setting a noise requirement that is a factor of ten below the CE design sensitivity at all frequencies, and assuming a \bmt\ diameter of \SI{120}{\cm} (as discussed in \cref{subsec:site_infra}) 
point scattering into the baffled arms was found to be an insignificant noise source for Cosmic Explorer~\autocite{Hall:2020dps}.

A related and potentially important consideration is light from the main cavity mode being clipped by the moving baffles in the \bmts, leading to modulated diffraction that causes phase noise in the readout. Several approaches to this problem are being investigated, including modal simulations and analytic solutions.
While progress is being made, and preliminary results indicate that this coupling mechanism
 will not be problematic for CE, no firm estimates have yet been produced~\autocite{SmithGWADW2021}.  

Additionally, an evolution and front-loading of the stray light control work done for the 2G detectors will be needed for Cosmic Explorer. In Advanced LIGO, for example, it was necessary to incorporate dozens of baffles and beam dumps specifically designed, using ray-tracing software, to intercept significant stray light~\autocite{AnanyevaGWADW2021}. Some baffles also required better seismic isolation and damping of resonant motion~\autocite{AdvLIGO:2021oxw}. Significant design work is called for to identify and eliminate scatter and stray beams associated with the many optical and mechanical components of CE, along with continued R\&D into low-scatter materials and coatings that could reduce the levels of stray light.

\section{Silicon Upgrades}
\label{subsec:silicon_upgrades}

The prospect of using the \SI{2}{\um} cryogenic silicon technology with higher
arm powers than in the nominal design is a strong motivation for continued R\&D
into this technology.  Here we briefly examine the limits to which this
technology can be pushed, and then show that it could be used to approach the
sensitivity limits imposed by the Cosmic Explorer facility.

Since the cryogenic Cosmic Explorer test masses would be radiatively cooled, a strict limit on the achievable arm power is imposed by the requirement that the power absorbed in the test masses does not exceed the radiative cooling power.\autocite{2020CQGra..37p5003A} The main sources of heat are the power absorbed in the high-reflectivity (HR) optical coatings and the power absorbed in the input test mass substrates, which in turn depends on the absorption of the silicon crystal.

There are two methods of producing silicon crystals\autocite{2020CQGra..37p5003A} and, as discussed in \cref{subsubsec:large_substrates}, producing the large \SI{80}{\cm} diameter silicon CE test masses is one of the most challenging tasks for the \SI{2}{\um} technology. The magnetically stabilized Czochralski method can produce the largest crystals\,---\,up to \SI{45}{\cm} in diameter today. This is the baseline design for Voyager. The float zone technique has only realized crystals up to \SI{20}{\cm} in diameter, but they are more pure than magnetically stabilized Czochralski crystals and therefore have lower absorption. Since in either case it will likely be necessary to bond multiple silicon crystals together to make the CE test masses, it is reasonable to consider float-zone silicon. Silicon absorption of \SI{5}{\ppm/\cm} has been measured at \SI{1550}{\nano\m} wavelength\autocite{2013OptL...38.2047D} and can be taken as a starting point for the substrate absorption of a float zone CE test mass.

The emissivity of the test mass barrel can likely be made nearly 1 either by coating it with a high emissivity coating or with a broadband anti-reflective coating. The emissivities of the HR and antireflective (AR) optical coatings are unknown at this time. Assuming emissivities of \num{0.95}, \num{0.75}, and \num{0.9} for the barrel, HR, and AR surfaces, respectively, a Cosmic Explorer using float-zone silicon could possibly achieve \SI{12}{\mega\W} arm power while satisfying the heat budget, assuming the coatings have \SI{1}{\ppm} absorption. Thermal lensing in the input test mass substrates\autocite{2020PhRvD.102l2003E} would likely not be an issue at this power.

\begin{figure}[!h]
  \centering
  \includegraphics[width=0.8\textwidth]{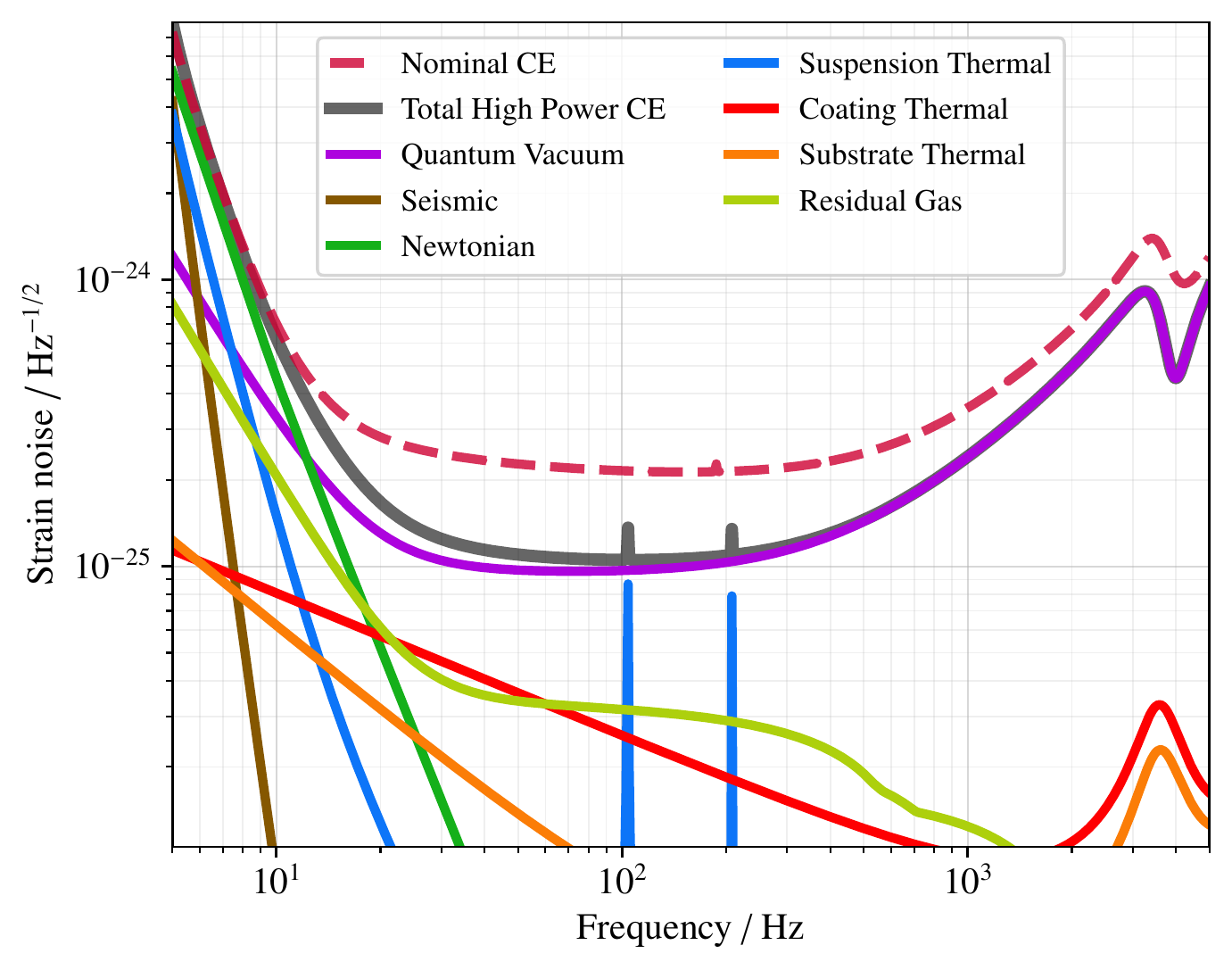}
  \caption{Noise budget for a high power silicon Cosmic Explorer with \SI{12}{\mega\W} arm power.}
  \label{fig:high_power_silicon}
\end{figure}

Such an interferometer would be roughly a factor of two more sensitive than the nominal design from about 30 to \SI{300}{\Hz}. The noise budget for such an interferometer is shown in \cref{fig:high_power_silicon}. The high frequency sensitivity is not significantly improved by going to higher power because it is dominated by loss in the signal extraction cavity (SEC), which is enhanced by the arm cavity finesse $\mathcal{F}$. The finesse is increased in proportion to the arm power so as to satisfy the heat budget by reducing the power absorbed in the input test mass substrates. Quantum noise at high frequencies is proportional to $P_\text{arm}^{-1/2}$. Since the noise due to SEC loss is proportional to $\sqrt{\mathcal{F}/P_\text{arm}}$, it is effectively constant when the power is increased in this manner, while the other quantum noises, dominant below ${\sim}\SI{300}{\Hz}$, are reduced as $P_\text{arm}^{-1/2}$.

The ultimate limits to the achievable sensitivity of any future detector that a given facility can support are usually taken to be the sum of the residual gas and Newtonian noises. If we add to this list \SI{500}{\ppm} SEC loss, the broadband sensitivity of a high power interferometer shown in \cref{fig:high_power_silicon} reaches this limit below ${\sim}\SI{10}{\Hz}$, where it is limited by Newtonian noise, and above ${\sim}\SI{600}{\Hz}$, where it is limited by SEC loss. It is possible to find a ``mid-frequency'' tuning which reaches a peak sensitivity of \SI{7e-26}{\big/\!\sqrt{\Hz}} around \SI{100}{\Hz} at the expense of sensitivity above about \SI{400}{\Hz}. It is not possible to find tunings with a resonant dip at higher frequencies due to SEC loss. Therefore, improving the high frequency sensitivity (${\gtrsim}\SI{600}{\Hz}$) beyond that shown in \cref{fig:high_power_silicon} requires some combination of reducing SEC loss below \SI{500}{\ppm}, reducing the silicon substrate absorption, reducing the coating absorption (which may become necessary), and increasing the emissivities of the test masses.

\section{Cost Drivers \hlight{[Kevin, Matt]}}
\xlabel{cost_drivers}

The initial cost estimate for CE presented in \sref{cost_est}
is based on actual costs from LIGO construction, the Advanced LIGO upgrade,
and the work of professional civil engineering and metallurgy consultants.
The exercise of developing this cost estimate brought to the fore a set
of cost-drivers which impact the technical design and scientific output
of a Cosmic Explorer observatory.
The following sections describe the primary cost-drivers
and their relationship to CE performance:
arm length, beamtube material and diameter, and observatory location.
Notably, the cost of the detectors installed in the Observatories is not a major cost driver.

\subsection[]{Arm Length}
\label{subsubsec:arm_length}

The length of an observatory's arms is the most fundamental feature in
determining its potential scientific output (see \tref{noise_length_scalings}
and \cref{box:concept}).  As such, arm length is generally increased in the
measure possible to the optimal length dictated by the science goals and thus
automatically becomes the principal cost-driver for any \gwo.

Many of the costs associated with arm length are simply proportional to the
length.  Examples of this are: the road which goes along the beamline and provides
access to the beamtube, the electrical utilities which run alongside the beamtube, the
slab which supports the beamtube, the beamtube enclosure, and the beamtube itself.  All
of these civil engineering costs are largely location independent (generally
within \SI{10}{\%} of the national average, and often a few percent lower than
the average for the reference sites considered for CE).  The sum of all costs
which are simply proportional to length is \SI{55}{\%} of the cost of a \fortykm\
facility and \SI{43}{\%} of the cost of a \twentykm\ facility.

The cost of excavation and transportation is not included in the above list of
civil engineering costs because it is highly location dependent, and generally
not proportional to the length of the facility.  As a concrete example,
consider a large dry lake bed (e.g., the Bonneville Salt Flats along interstate
80 west of Salt Lake City, UT).  The surface at such a location follows the
geoid almost perfectly: meaning that it follows the curvature of the Earth and
has constant altitude.  The arms of a \gwo\ must, however, be straight lines
since laser beams do not curve with the Earth's geoid.  The curvature of the
Earth is such that the elevation at the center of a \fortykm\ long straight
line is \SI{30}{\meter} lower than the ends.  Preparing such a site would require
excavating almost 10 million cubic meters of soil, and transporting it more
than \SI{10}{\km} on average (i.e., from the center to the ends), at a cost of
very roughly \$100 million (highly dependent on geology).  For a ``flat'' site
like this, the volume of excavation required grows with arm length squared, and
the transportation distances grow with length, such that the total cost grows
with arm length to the third power (i.e., a \fortykm\ facility would cost
\emph{8 times} that of a \twentykm\ facility).

The flat-site example drove significant interest in finding sites which minimize excavation
and transportation costs\autocite{SiteSearchTechnote}.
Such sites are slightly bowl-shaped with an elevation profile roughly \SI{30}{m} higher at
the ends than in the middle.
There are a number of wide ``valleys'' that fit this description in the western states,
and picking a location and orientation well can vastly reduce excavation
and transportation costs.
However, this search for topographically favorable sites clearly showed that the number
of advantageous and available sites decreases rapidly with arm length,
meaning that excavation costs for a \twentykm\ observatory may be less
than a \SI{5}{\%} of
the total observatory cost, while for a \fortykm\ observatory they are likely
to remain near \SI{10}{\%}
of the total simply because there are fewer \fortykm\ sites to choose from.

\subsection[]{Beamtube Material and Diameter}

The laser light propagating along the arms of a \gwd\ must travel in ultrahigh vacuum to avoid the noise associated with polarizable atoms and molecules traversing the laser beams
(see \cref{box:concept}). This fundamental performance driver is the reason that the LIGO facilities are among the largest ultrahigh vacuum systems ever built. The cost of this vacuum system, utilizing the research described in \cref{sec:vac_value_eng}, will be roughly \PCTVAC\ of the cost of a CE facility, and approximately one third of that is required to produce the beamtubes, making the choice of beamtube material and size an important cost driver.

As discussed in \sref{vacuum_reqs}, a wide variety of factors come into play
when designing a vacuum system.
While CE could use the LIGO vacuum system design with only minor modifications,
ongoing research into the vacuum properties of steel suggest that mild carbon steel
will provide superior performance at lower-cost\autocite{HenkelBeamtubeReport}.
This is the material used by the oil industry for pipelines, so it is well characterized and
readily available in a range of diameters up to \SI{48}{\inch} (\SI{122}{\cm}).
The standard pipeline wall-thickness of $\tfrac{1}{2}\,\si{\inch}$ (\SI{13}{\mm}) is sufficient to support the atmospheric load
without stiffening rings (required in the current LIGO design), making manufacturing relatively simple.
(Note: Shifting to thinner walls to save material actually increases cost, since this requires
a non-standard process at the foundry.)

Scattered light is a potentially limiting technical noise source for \gwds,
as observed in LIGO and Virgo.
Initial estimates of scattering in the CE arms indicate that \SI{48}{\inch} beamtubes are sufficient
to ensure that this technical noise will not be limiting for CE.
Calculations of scattering in the CE arms are ongoing with the objective of developing
a detailed baffling strategy for CE, based on the baffles currently used in LIGO
(see \sref{scatter} for more detail).

\subsection[]{Choice of Site}
\label{subsubsec:site_choice}

As mentioned above, the location of an observatory can have a significant impact on its cost. In addition to topography and geology, which determine the type of excavation (i.e., digging, or blasting) and the amount of transportation required, there are a variety of other factors that must be considered. A few prominent features are listed below, informed by the experience with LIGO-US and LIGO-India site identification.

\paragraph{Local Community}
A CE observatory will inevitably have a significant impact on the landscape,
environment and the local community.
As such  \emph{the local community must be included in any site selection process from the beginning}.
This issue is of paramount importance for CE and \sref{local} is dedicated to it.

\paragraph{Environmental Impact}
The long L-shaped footprint of a CE facility may have environmental impacts on
resident and migratory animal populations.
When searching for potential CE sites one should expect that many topographically favorable
locations will be eliminated by environmental constraints (e.g., sage grouse leks in Idaho),
and that some construction costs will be expended to accommodate animal populations
(e.g., wildlife bridges).

\paragraph{Land Acquisition}
Sites which are favorable for CE are vast open spaces with relatively flat terrain,
and as such they tend to be either very remote, already in use
(as national parks, military facilities, etc.), or both.
In some cases, this can facilitate land acquisition for CE (e.g., if the land is federally owned),
or make it unfeasible (e.g., if the space is a national monument).
Land acquisition may also be difficult in areas that are mostly private land
due to the length of the CE arms, which may cross many individual plots.

A potential site could be unsuitable if the land cannot be acquired or its
acquisition would greatly increase the cost of the project.  Though Cosmic
Explorer will be a surface facility, attention must be paid to, e.g., severed
mineral rights, to avoid underground activity that negatively
impacts the instrument performance.

\paragraph{Natural Hazards}
Certain potential sites could be disqualified due to unacceptably high
probability of catastrophic natural disaster (flood, fire, etc.).
In addition to consulting the historical record (e.g., the 100-year flood level),
the potential impact of climate change on the suitability of a location must
also be considered.

\paragraph{Surrounding Infrastructure.}
Some otherwise promising sites may be disqualified if they are hard to access
or frequently rendered inaccessible (e.g., due to inclement weather). The absence of anthropogenic noise sources (e.g., industry, wind farms) must be assured for the lifetime of CE. Not all potential sites will be located close enough to critical infrastructure (e.g., roads, utilities) required to construct and operate the facility, and building
such infrastructure may be prohibitively expensive.

\paragraph{Proximity to cities}
In addition, while some candidate sites tend to be remote, they must be sufficiently close to social infrastructure (hospitals, schools, etc.) to sustain an effective workforce.

\chapter{Data Management, Analysis, and Computing \hlight{[Duncan]}}
\xlabel{computing}

\section{Data Management Plan \hlight{[Duncan]}}
\xlabel{data_management}

Given the substantial United States Federal Government investment and broad
community support that will be required to realize a U.S.\ third-generation
gravitational-wave detector, Cosmic Explorer is planned to be an Open Data
facility. To realize this goal, both the construction and operations budgets
must contain sufficient funding to support personnel and computing for the
rapid release of high-quality, calibrated gravitational-wave strain data and
alerts for events of interest. Development of Cosmic Explorer will
include the creation of a Data Management Plan that will describe the technical
implementation of user access mechanisms throughout the life cycle of Cosmic
Explorer, guidelines for the use of data products, and publishing rights. The
full data management plan will be developed in consultation with input from
U.S.\ and international stakeholder funding agencies, the instrument
development team, and the broader scientific community.  Although there are
significant differences in the type and scale of the data sets, the data
policies of the NSF/DOE Vera C.~Rubin Observatory and NASA's Fermi Gamma-Ray
Space Telescope provides an excellent baseline for Cosmic Explorer's data
management plan.  This Horizon Study outlines the following principles
for Cosmic Explorer's data management plan:
\begin{enumerate}
\item The data management plan should maximize the science output of the
funding agencies' and science community's investment in the project.
\item There should be no reserved science; all types of scientific endeavors
are open to all individuals and membership of any group or collaboration does
not convey exclusive rights to any particular area of research.
\item Open data facilitates scientific collaboration, enriches research and
advances analytical capacity to inform decisions.
\item Open data supports and ensures access for junior scientists.
\item Open data supports scientists from small institutions and historically
underrepresented institutions.
\end{enumerate}
To achieve these goals, the Cosmic Explorer data set will be released as open
data as quickly as possible (i.e., as close to real time as possible) once
construction ends and operations begin.

Cosmic Explorer will generate a data set that provides a unique, rich, and
deep view of the universe over its lifetime. However, in comparison to
e.g., the Rubin Observatory data set, the volume of Cosmic Explorer's data is remarkably
small considering its scientific potential.  Almost all of the scientific
information from an interferometric gravitational-wave detector is contained
in a one-dimensional time series, with calibration information and additional
metadata that describes these data.  Whereas the Rubin Observatory is expected
to generate ${\sim} 20$~terabytes of data each night, the size of Cosmic
Explorer's primary data set will be ${\sim} 20$~gigabytes per day. An additional
${\sim} 2$~terabytes of control and monitoring data will be recorded each day
for a single interferometer. The size of alert data packets containing timing,
sky location, and source properties (e.g., masses and spins) for detected
compact-object mergers is less than 1~gigabyte per day, assuming of order one
hundred alerts per day. We anticipate no significant technical challenges to
releasing these data to the user community in real time.

The challenges of realizing open data for Cosmic Explorer are (1)~ensuring
that project construction costs have sufficient funding for the human resources
needed to develop the infrastructure to deliver open data to the community, and
(2)~ensuring that the operation budget has sufficient funds to calibrate, clean
and release the gravitational-wave time series with appropriate detector
metadata and as quickly as possible. Funding will be needed for personnel to
develop, deploy, and manage the generation of astronomical alerts for
compact-object mergers, and to provide support to the user community.  Given
the critical nature of these tasks, they should not be subject to a separate
entity supported by third-party funding or separate grants; they should be
included in the construction and operation budgets, as appropriate.

Open release of Cosmic Explorer's data will allow the broadest possible use of
the Cosmic Explorer facility, while keeping the scope of the project at a
reasonable level.  All data will be released with a liberal open license.
Reuse, redistribution, and the dissemination of derivative data products will be allowed and
encouraged.  To aid in the use of Cosmic Explorer data by the public, all
software developed for and data produced by Cosmic Explorer will be publicly
released and thoroughly and clearly documented. All data will use standard,
open formats whenever such formats exist. \cref{sec:scope_comp} and
\cref{sec:outofscope_comp} discuss the requirements for computing expected to
be with the scope of the project, as well as the broader scope of computing
that will be pursued by the wider community.

\section{Requirements for Open-Data and Analysis}
\xlabel{scope_comp}

To deliver the science goals described in \cref{ss:cosmictime}--\cref{ss:xg}, the Cosmic
Explorer project will need to provide:
\begin{enumerate}
\item Management and curation of the detector data throughout the life cycle
of the project.
\item Near-real time production of a calibrated, cleaned gravitational-wave
strain data set with metadata describing the quality of these data.
\item Production of low-latency alerts for the merger of compact-object
binaries.
\item Operation and support of a Cosmic Explorer Data Access Center for
dissemination and support of open data.
\item Periodic publication of catalogs describing the events observed in a
given period.
\end{enumerate}
Data management and the production of low-latency analysis required to deliver
alerts and prepare the data product releases for the community will be in scope for the project\,---\,neither of these tasks present significant
computational challenges, as described below.  Approximately ten FTEs will be needed to perform
tasks related to data preparation, alert generation, data curation, and user
community support during the operations phase of the project.

The bandwidth of the control systems required to operate Cosmic Explorer does
not differ significantly from that of Advanced LIGO. As in Advanced LIGO,
the cost of the digital detector control systems is not expected to be a
significant fraction of the cost of the instrument. Since the number of Cosmic
Explorer control and data channels will be similar to that of Advanced LIGO,
we expect data rates of 2~Tb per day of detector operation.  Storage and
dissemination of data of this scale is a solved problem with current
technology.

The sensitivity of Cosmic Explorer will be an order of magnitude
better than that of Advanced LIGO at 100~Hz and two orders of magnitude better
at 10~Hz. Cosmic Explorer's detector noise increases rapidly below 10~Hz and at
3~Hz the detector is five orders of magnitude less sensitive than it is at
10~Hz. For a broad-band source such as inspiraling compact objects, there is
very little signal-to-noise below 7~Hz; over \SI{99.5}{\percent} of the signal-to-noise
lies above 7~Hz. A binary neutron star waveform starting at this frequency
lasts 77 minutes from the time that it enters the detector's sensitive band to
coalescence. For a waveform of this length, the Doppler frequency modulation
due to the diurnal and orbital motion is $(\Delta f / f ) \sim 10^{-8}$ and
can be neglected in search algorithms~\autocite{1996A&A...306..317J}. Several
search algorithms already exist that can search for waveforms of this length
in a computationally efficient
manner~\autocite{2016CQGra..33q5012A,2019arXiv190108580S,2021SoftX..1400680C}.
The number of templates required in a matched filter search scales as a
function of the bandwidth of the detector and not the overall strain
sensitivity~\autocite{1999PhRvD..60b2002O}. Consequently, a
matched filter search for binary neutron stars with component masses between
$1$ and $3\,M_\odot$ using current data-analysis algorithms only requires a
factor of three times more template in its bank than required by Advanced
LIGO~\autocite{2021arXiv210314088L}.  Again, this is a scale of computing that
is accessible by current technology (using either CPUs or GPUs); implementing searches for the rapid
identification of compact-object merger events will be straightforward a
decade from now.

The computational cost of parameter measurement scales with the number of
sources observed. While this is expected to increase by three orders of
magnitude with respect to Advanced LIGO, there already exist algorithms that
can measure the parameters of, e.g., a binary neutron star (sky location,
masses, and spins) within 20 minutes of detection for Advanced LIGO using 32 cores of a
current
processor~\autocite{2010arXiv1007.4820C, 2018arXiv180608792Z, 2020ApJ...905L...9F}.
Although Cosmic Explorer will detect events at a significantly higher rate
than Advanced LIGO, the time-frequency volume of these events is relatively
sparse in the data set and so Cosmic Explorer data analysis will not be confusion
limited (as is the case, e.g., for LISA white dwarf sources).  
Consequently, there are no major obstacles
to measuring binary parameters for the generation of alerts~\autocite{2021arXiv210207692P}.  Increasing the speed of parameter
estimation from detection to time-to-release for catalogs over present-day
analysis primarily requires development of data-analysis pipelines to
automate tasks currently performed by humans, e.g., hand-checking of
convergence for detected events.

While the computational challenges are straightforward and the necessary algorithms are
already being explored, there is still significant algorithm, code, and
infrastructure development needed to realize the scientific potential of
Cosmic Explorer; these developments will allow a richer and more efficient
approach to low-latency analysis in the Cosmic Explorer era.  Adopting the
open data paradigm for Cosmic Explorer will: allow greater access to data;
opening possibilities to build upon and create new research from publicly
accessible data and alerts; facilitate research across disciplines and
foster new collaborations; help to ensure universal participation in Cosmic
Explorer's science without barriers that can prevent the participation of
underrepresented groups; and ensure compliance with funding agency mandates.

\section{Additional Computational Resources}
\xlabel{outofscope_comp}

An open data model for Cosmic Explorer leaves the community free to pursue a
wide range of science goals using human and computational resources that they
obtain through the normal process of obtaining research funding and
computational resources. These projects include but are not limited to:
searches for sources of gravitational waves beyond compact-object mergers;
low-latency analysis for gravitational waves e.g., from core collapse supernovae;
re-analysis of events or data using new waveform models; analysis of
populations of events; and comparison of signals to numerical models of sources.

Searches for unmodeled transient sources with Cosmic Explorer are likely to
be challenging but will not require significant computational resources.  The
most computationally challenging analysis will be the all-sky search for
continuous waves from pulsars. The scale is set by the need to
search over pulsar frequency, the pulsar's spin down rate, and the sky location
of the source (due to the Doppler modulation induced by the Earth's motion over
the integration period). Although the computational cost of this search does
not depend on the detector sensitivity, broad parameter space searches for
continuous gravitational waves are likely to remain bound by computational
power due to the extremely fine grid needed to search the target signal space.
At fixed computational power, the development of more sensitive search
algorithms and the use of new computing hardware will be pivotal to fully
exploit the improved reach of Cosmic Explorer.

A significant effort will be needed to develop waveforms that will be accurate
enough for third-generation gravitational wave detectors.  The accuracy
necessary to avoid biasing interpretation of an observation scales with the
square of the observation's signal-to-noise-ratio.  While the loudest
observations of merging black holes and neutron stars to date have
signal-to-noise ratios of $\approx 30$, Cosmic Explorer's observations will
include detections with signal-to-noise ratios in the thousands. These loud signals are
among the most critical for realizing our science objectives, and interpreting
them without bias will require inspiral-merger-ringdown waveform models
substantially more accurate than today's state of the
art~\autocite{2020PhRvR...2b3151P, 2020arXiv200604272F}.

Additionally, most current neutron star merger simulations are still not carried out with
realistic microphysics, and full explorations of the parameter space of neutron
star mergers are extremely challenging computationally. These issues are illustrated, for
example, in the disagreement between multimessenger constraints on the lower
limit of the neutron star radius that use different numerical simulations of
neutron stars merging in GW170817. Also note that only a handful of
high-accuracy neutron star merger waveforms exist today\,---\,and these are
not accurate enough for Cosmic Explorer~\autocite{2016PhRvD..93d4064B}.

Significant progress needs to be made over the next decade to ensure that
waveforms of sufficient accuracy and that span a large enough parameter space
are in hand to deliver the promise of Cosmic Explorer's science case. For
instance, achieving sufficiently accurate gravitational waveforms  will likely require next-generation numerical-relativity codes that
will take full advantage of the high-performance computing facilities that will
be available in the 2030s.

\setpartpreamble{
    \AddToShipoutPictureBG*{
    \begin{tikzpicture}[remember picture,overlay,inner sep=0]
        \node[anchor=north] at (current page.north)
            {\includegraphics[height=1.0\paperheight]{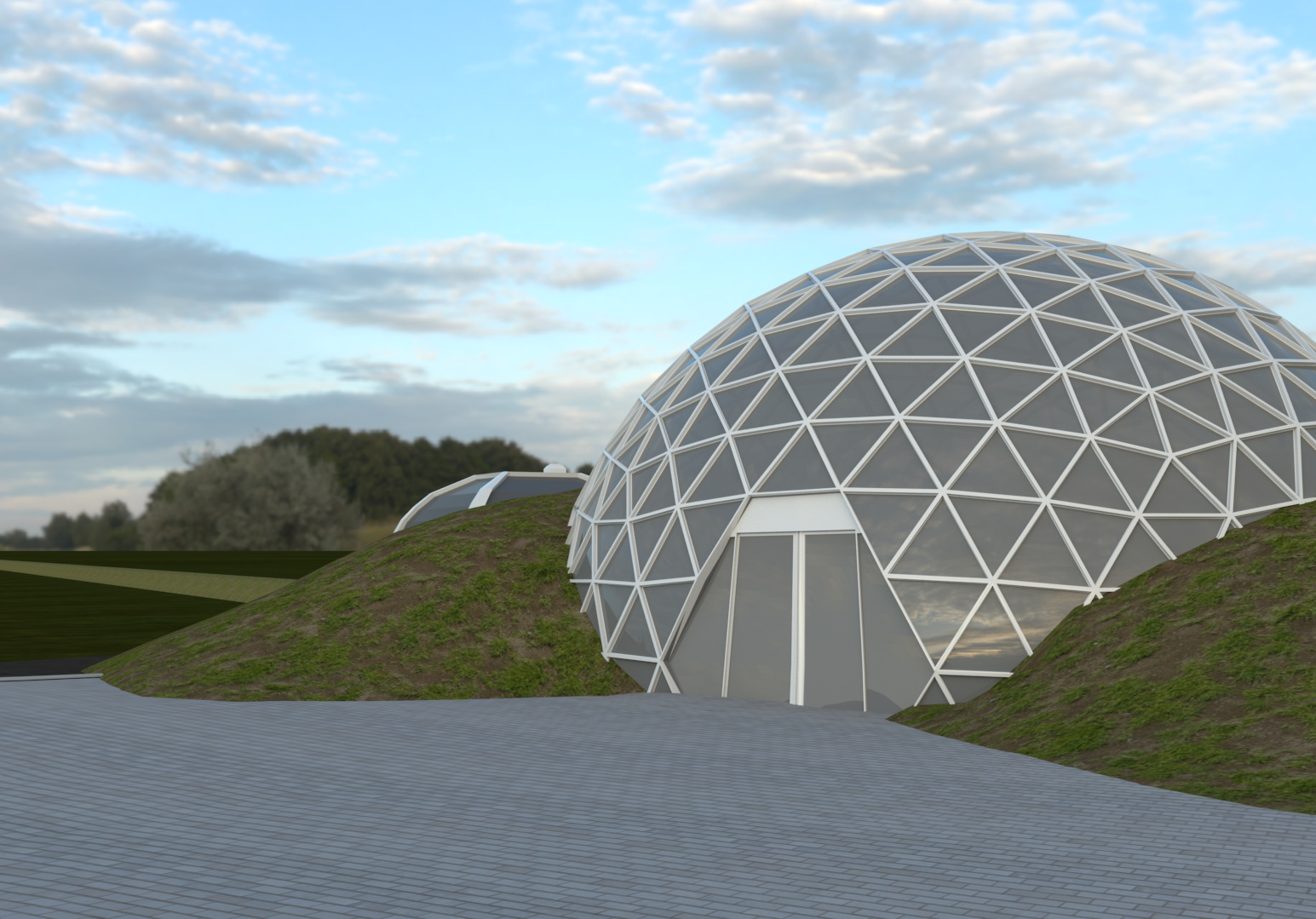}};
        \node[align=left,text=white,inner sep=6pt,anchor=south east] at (current page.south east) {Artistic visualization of a CE end station by Cal State Fullerton undergraduate Vikki Brown.};
        \filldraw[draw=white,fill=white,opacity=0.1] (current page.south west) rectangle (current page.north east);
    \end{tikzpicture}
    }
}
\addtokomafont{part}{\color{black}}
\addpart{Community, Organization, and Planning}

\chapter[CE at the Local and Global Scales]{Cosmic Explorer at the Local and Global Scales \hlight{[Evan]}}
\xlabel{global}

The Cosmic Explorer project will develop observatory designs with a
multi-dimensional approach that creates synergy with its respective local,
scientific, and global communities.  This includes designing the physical and
virtual infrastructure so that it will serve the broader goals of community
integration and engagement by developing interpersonal relationships among
members of these communities.  Early and ongoing engagement with communities
connected with Cosmic Explorer, from local to global, will be crucial to the
project's success.

In 2020 the GWIC 3G Community Networking Subcommittee published a report that
identifies potentially interested scientific communities for third-generation
gravitational-wave projects~\autocite{GWIC3GSynergies}.  The GWIC report also
outlines a communication and outreach plan for engaging the relevant
communities, and delivers concrete recommendations for next-generation
gravitational-wave projects.  This section describes specific actions already
taken toward these recommendations, and plans for realizing others in the
future.

The report identifies engagement with the public as key to the success of the
coming observatories, just as it has been for the existing observatories.
However, third-generation gravitational-wave observatories come at a time of growing
public awareness of the social impact of large scientific projects and
facilities, and Cosmic Explorer must consider how to engage the
public in this era.  Crucially, Cosmic Explorer must identify and connect with
all communities that have a potential interest in the observatory, particularly
focusing on local communities who will be impacted by its presence.  This step
lays the foundation for the critically important need to build positive and
mutually beneficial relationships with those communities.

\section{Community Integration and Engagement \hlight{[Evan]}}
\label{sec:communityfacility}

As Cosmic Explorer is a new astrophysical observatory, there is an opportunity
to reimagine the human-focused portion of the observatory, by designing the
facility to strengthen the interaction between everyone who uses and visits it
while highlighting the contribution of the local community to the Cosmic
Explorer effort.  The model currently used is to implement community engagement
plans~\autocite{T1900780}; some include the construction of science education
centers, as in Louisiana and Western Australia~\autocite{LIGOSEC,
GravityCentre}. These efforts have drawn significant local and global public
interest. Cosmic Explorer will build upon this model by facilitating an even
tighter integration between scientists and the public.

In designing a facility that brings scientists and the public together, Cosmic
Explorer can look to other scientific installations, such as Fermilab's Wilson
Hall, which combines staff offices with public gathering spaces. Beyond purely
scientific outreach activities, Wilson Hall hosts cultural activities, such as art
exhibitions, thus embedding itself in the fabric of the local community.
Additionally, the architecture of the Fermilab facility itself is designed
intentionally to conserve and restore the surrounding
environment~\autocite{FNALHistory}.
Another example of a successful meeting and exhibition space is the `Imiloa
Astronomy Center in Hawaii,\footnote{\url{imiloahawaii.org}} which presents the
astronomical knowledge of the local Native Hawaiian community alongside the
astronomical observatories in Hawaii, and emphasizes the community presence by
(for example) translating all materials into the local Indigenous
language~\autocite{APSGazetteFall2020,2019BAAS...51g.135K,Malalo}. By looking
to these examples, the Cosmic Explorer facility can be a place for scientific
workshops, community activities, teach-ins (such as the work of Karletta Chief
in the Navajo Nation~\autocite{ChiefTeachIn}) and exhibitions happening under one roof
in a welcoming environment.

\section{Building Strong Relationships with the Local Community \hlight{[Brittany]}}
\xlabel{local}

The Cosmic Explorer concept is based on observatories on the grandest of
scales. These observatories' activities will not happen in a vacuum:
they will impact the landscape in which they are built and will change the
lives of people in nearby communities.  The Cosmic Explorer project will
address issues surrounding observatory impact on the land, environment, and
host community carefully, intentionally, and as an opportunity to build a
mutually beneficial long-term relationships with its host communities.

Cosmic Explorer cannot be built and operated without ongoing local consent,
which could disintegrate if the project does not actively work to maintain a
positive relationship with the local communities. Cosmic Explorer will first
need to work to identify all relevant communities, including Indigenous
communities, to ensure that the consent is comprehensive and meaningful. Having
identified these communities, Cosmic Explorer will integrate Indigenous
leadership (elders and community leaders) and local community leaders at large
into the leadership structure of Cosmic Explorer, to ensure there is continuous
engagement around future directions of the project.  This integration could,
for example, take the form of a community-based oversight committee,
representative of the project's hosts, neighbors, and other local communities,
with decision making power (e.g., the ability to enter into binding
arbitration) over any project decision that impacts the land, environment or
community. Importantly, this leadership structure must truly represent the
perspectives of the community and cannot serve as a substitute for it.  The
Cosmic Explorer project will request funding from private partners and federal,
state, and local agencies to engage with the community and Indigenous peoples.
Such inclusion in the project and the broader scientific community is crucial
for maintaining broad consent for the project.  This support might include, for
example, funding scholarships for undergraduate students and fellowships for
graduate students and postdoctoral scholars drawn from local demographic and
Indigenous populations.  Ultimately the specific form of support will depend on
the community's needs, which can only be ascertained after proper relationships
are established.

These actions reflect Cosmic Explorer's need to prioritize community
involvement in a fundamentally new way compared to previous physics and
astronomy projects.  Having multiple members from local communities serving on
the leadership boards of the project\,---\,including Indigenous leadership,
local community groups, and others\,---\,will be necessary during all stages of
the project. The Cosmic Explorer project must start by building respectful and
meaningful relationships from the outset, evolving into permission for land
use, continuing integration and collaboration during commissioning and
operations, and importantly ensuring accountability for agreements throughout
the lifetime and decommissioning of the observatories.

\subsection[]{Indigenous Communities}

All lands within the United States are the ancestral home lands of Indigenous
Peoples~\autocite{DunbarOrtiz}. Establishing mutually beneficial relationships
with these communities is important to the project's success. Failing to
establish meaningful relationships with Indigenous communities has led to
friction, delays, and public backlash for several astronomical projects,
including the Mount Graham International Observatory, the Kitt Peak National
Observatory, and the Thirty-Meter Telescope~\autocite{Swanner2015Contested,
2020arXiv200100970K}.  The contentious relationships between these projects and
Indigenous communities has a negative impact on the communities themselves, who
often are working from previous negative experiences with academic, scientific,
and technical projects~\autocite{Nash2019Entangled, Lee2020LandGrab,
Hodge2012No}.  As the process to find a site for Cosmic Explorer begins, the
Cosmic Explorer project will first learn the history of each potential site.
The project will then connect with networks of tribal councils and leaders to
learn the most respectful ways to engage. If there is a desire for ongoing
engagement, the project will work to build and maintain a relationship with the
community.

Building these relationships will be a core driver in the way the project
engages with the local community~\autocite{2021BAAS...53d.471G}. A community's
willingness to host an observatory will depend critically on the relationships
built and the competency the project demonstrates around the community's
heritage, ancestry, values, and culture. The Cosmic Explorer project schedule
will include the time to learn together about how an observatory could achieve
mutually beneficial relationships consistent with the community's priorities.
The Cosmic Explorer project management and site search teams will make a
conscious effort to build these relationships so that dialog and consent may
follow.  To gain experience, the project management and site search teams will
study previous examples of the impact of large astronomical facilities, and
large government and industrial facilities more broadly, on Indigenous
communities.  For example, the teams will work to understand the full spectrum
of views around the local impact of the Thirty Meter Telescope, including those
of the astrophysics community, the local Hawaiian community, and those in both
communities, in order to understand how the disconnect between the communities
arose~\autocite{2019Sci...365..960C, 2020arXiv200100970K}.  The Cosmic Explorer
team will invest in internal and external development with respect to
Indigenous Peoples. Where appropriate, this investment will be enabled by
partnerships with federal, state, local, and private agencies, and involve
existing institutions such as local universities.  We envision that the
development work will include, but not be limited to, the following:

\begin{enumerate}

\item The Cosmic Explorer project will demonstrate to the Indigenous community
its commitment to understanding their culture and cultural practices.

\item The project will conduct cultural impact studies that involve the local
Indigenous community as part of any site-selection process, with the goal of
ensuring that the project is aware of and respectful toward locations of
cultural significance at the earliest possible stage of the process.

\item Indigenous communities often have protocols\,---\,e.g., practices of
respect and ceremony at physical locations. As the Cosmic Explorer project
will be a guest in local host communities, the project will learn about these
practices and create space for their perpetuation.

\item As part of developing lasting and mutually beneficial relationships, the
Cosmic Explorer project will seek out appropriate opportunities to integrate
Indigenous wisdom into its research plans, such as in the case of environmental
monitoring.  Indigenous cultures have millennia of history and
experience specific to their ancestral land and the environment, making this a
clear opportunity for working together to build mutual respect and
trust~\autocite{Barbu2021Where}.

\item Cosmic Explorer will invite the involvement of its Indigenous hosts more
broadly in order to highlight their presence at the observatory. The exact
nature of this involvement will depend on the host community’s needs and values
but may include, for example, language preservation\,---\,e.g., using Indigenous
names for parts of the observatory and notable discoveries, and translating
descriptions of discoveries into Indigenous
languages~\autocite{GreenfieldboyceYellowfly, GW150914Siksika,
2019BAAS...51g.135K, PerseveranceNavajo}. This genre of activity in particular
is an opportunity for federal, state, local, and private partners to involve
themselves.

\item Cosmic Explorer will work to eradicate anti-Indigenous rhetoric from the
gravitational-wave community and ensure that it is not creating a hostile
environment for anyone in the local community seeking to engage or for
Indigenous scholars who join the gravitational-wave astrophysics community.

\item Cosmic Explorer will elevate anti-racist work around Indigenous
communities in the gravitational-wave community, including improving cultural
competency, researching how to utilize language that is respectful of
Indigenous communities and their relationship to land, and building on
relationships and partnerships that are already in place. These actions should
lead to a collaborative environment that is intentionally structured to avoid
the perpetuation of racist practices.

\item As a commitment to the stewards of the lands, Cosmic Explorer must be in
continuous dialogue about what procedures match the practices of the host
Indigenous community. For example, during the construction phase, how and where
to put the land once it is moved must be mutually agreed
upon~\autocite{Barbu2021Where}. During the planning and decommissioning phases,
dialogue on what returning the land looks like to the host community must be
discussed, budgeted, and fully achieved, and will be codified in Cosmic
Explorer's long-term facility plans (as required by, for example, the NSF MREFC
guide).  The Cosmic Explorer project must ensure that nothing is abandoned and
everything is accounted for.
\end{enumerate}

Taken together, the above points show the necessity of the Cosmic Explorer
community developing positive relationships with local communities with
historically longstanding land tenure. This comes at a time when funding
agencies such as the NSF are increasingly recognizing the importance of
strengthening such relationships~\autocite{NSFTribalActionPlan}.

\subsection[]{The Local Community at Large}

The Cosmic Explorer project will work to cultivate a positive relationship with
additional groups in local communities, importantly starting before a site
is chosen, and continuing through the lifetime of the project. The success of
the project will rely on the members of the community being invested and
involved in CE. Members of the community will be integral to the project as
members of the CE staff, collaborators on local projects, educational partners,
and colleagues in local governance. Cosmic Explorer will engage with local
communities at forums including public libraries, local government offices,
schools, colleges and universities.

Throughout the site search process, Cosmic Explorer will reach out to local
educational institutions and develop partnerships to integrate participation of
their students, scientists, and educators with CE science and outreach. This
partnership may include science education research, sharing science and
technology development with the public, and outreach to a broad audience in the
surrounding regions, following examples such as the Southern University/LIGO
partnership in Louisiana~\autocite{SULIGO}.

In parallel, Cosmic Explorer will initiate conversations at other local hubs
such as public libraries and community centers, starting during the site
identification process and continuing during the project and observatory
lifetime, to build and maintain local relationships.

\section[CE as Part of the Scientific Community]{Cosmic Explorer as Part of the Scientific Community \hlight{[Josh, DHS]}}
\label{sec:collaboration}

The Cosmic Explorer project will only succeed with wide and explicit engagement and support by all levels of the scientific community. As the project moves forward, it will work to ramp up the frequency and depth of this engagement and support.

\subsection[]{In the Gravitational-Wave Community}
\label{sec:InTheGWCommunity}

In the context of completing this Horizon Study, preliminary steps toward integrating Cosmic Explorer into the gravitational-wave community are already underway, including the following:

\begin{itemize}
\item Cosmic Explorer's membership in the Gravitational-Wave International Committee (GWIC)~\autocite{GWIC}. This membership enables Cosmic Explorer to participate in the coordination of projects covering ground interferometers, space interferometers, and pulsar timing arrays. Membership also enables Cosmic Explorer to provide updates to GWIC and to the Gravitational-Wave Agencies Correspondents (GWAC) to inform the funding agencies covering the broad scope of gravitational-wave research~\autocite{GWAC}.
\item Presentations and outreach to communities, including those at meetings of the LIGO--Virgo--KAGRA Collaboration, ET Collaboration, Pulsar Timing Array, and LISA Consortium;
\item Organization of a one-day meeting in Summer 2020 between the teams associated with Einstein Telescope, NEMO and Cosmic Explorer;
\item Organization of the First Cosmic Explorer Meeting, a five-day remote conference that was held in October 2020 with broad community participation to discuss the technical design and the science case for CE;
\item Formation of the Cosmic Explorer Consortium~\autocite{CEConsortium} in October 2020 to provide an open and efficient way for members of the broader physics and astronomy communities to contribute to the conceptualization, design, and future use of CE. Already the consortium has more than 300 members and has begun two monthly remote meetings, one devoted to instrumental research and development and another devoted to astrophysics; and
\item Participation in discussions, plenaries, and technical talks at international Dawn Meetings. %
\end{itemize}

\subsection[]{In the Physics Community}

The Horizon Study team has worked to raise the profile of Cosmic Explorer within the wider physics community through the following:
\begin{itemize}
\item Participation in discussions, plenaries, and technical talks at American Physical Society meetings~\autocite{2019APS..APRC02003H,smith2020cosmic};
\item Invitations through the APS Division of Gravity to join the Cosmic Explorer Consortium and research meetings;
\item Participation in the DOE Snowmass2021 effort through committee leadership (Adhikari and Sathyaprakash) and through a Letter of Interest:
    \begin{enumerate}
    \item ``Cosmic Explorer: The Next-Generation U.S. Gravitational-Wave Detector'', S. Ballmer, P. Fritschel, Cosmic Explorer, LIGO Laboratory~\autocite{CESnomass}
    \end{enumerate}
\end{itemize}

\subsection[]{In the Astronomy Community}
The Horizon Study team has worked to raise the profile of Cosmic Explorer within the wider astronomy community through the following:
\begin{itemize}
\item Participation in the Astro2020 Decadal Survey through two white papers:
    \begin{enumerate}
    \item \fullcite{Reitze:2019iox}%
    \item \fullcite{Reitze:2019dyk}%

     \end{enumerate}
\end{itemize}

\subsection[]{In the Scientific Community}
The Horizon Study team has worked to raise the profile of Cosmic Explorer within the wider scientific community through the following:

\begin{itemize}
\item Launching a Cosmic Explorer website\footnote{\url{https://cosmicexplorer.org/}} to increase visibility and opportunities for engagement in the CE community;
\item Communicating Cosmic Explorer science with the public through social media about upcoming meetings, science goals, and opportunities;
\item Discussing Cosmic Explorer at SACNAS conferences, which includes scientists across all STEM disciplines; 
\item Amplifying the Cosmic Explorer Plans in Science Magazine: \fullcite{2021Sci...371.1089C}.
\end{itemize}

\subsection[]{Current and Future Work}

This Horizon Study document serves as a reference to communicate plans and gather
input and feedback. Through the current NSF funding supporting this study, the Cosmic Explorer team has already initiated efforts toward education and public outreach. All of the PIs have made Cosmic Explorer a focal point in their presentations to the public, and many of them have begun incorporating Cosmic Explorer technology and science into their classes. The project has strong engagement by graduate students and has been engaging undergraduate researchers in small but increasing numbers. These students have presented their research at their universities and in public settings.

Beyond the initial phases described above, this document will be employed to re-engage and expand discussions with members of the gravitational-wave, physics, astronomy, scientific affinity groups, and Indigenous leadership communities, including the LIGO, Virgo, and KAGRA Scientific Collaborations, Einstein Telescope, LISA, DECIGO, NanoGrav, CMBS4, the APS Division of Gravitational Physics, AAS, the DOE Snowmass, the APS Forum on Diversity \& Inclusion, AAS Committee on the Status of Minorities in Astronomy, SACNAS, AISES, to name a few. Building on the preliminary phases of community engagement described above, the Cosmic Explorer project will begin to operationalize recommendations from the Horizon Study, expecting that engagement with the broader communities will increase steadily as the Cosmic Explorer Project moves forward.

\section{Developing a Global Gravitational-Wave Network \hlight{[DHS]}}

The discussion in \cref{sec:science_overview,ss:status,ch:keyquestions} of the
observational science that is possible with the next generation of
gravitational-wave observatories illustrates the great value added were these
detectors to operate in concert as a global network.

The ground-based gravitational-wave community has recognized the imperative to
form a globally coherent effort, and has made some progress toward that
outcome.  One of the deliverables of this Horizon Study has been to contribute
to this goal. A series of NSF-supported meetings~\autocite{DawnI, DawnII,
DawnIII, DawnIV, DawnV} began in 2015 to start planning for the future. These
meetings have helped guide the development of the LIGO interferometers,
specifically establishing ``A+'' as an upgrade to the \SI{4}{\km} detectors.
They have served as valuable forums for discussion of further improvements to
the present \SI{4}{\km} baseline LIGO observatories, and to discuss CE and other
next-generation observatories, and have helped cultivate ideas for an Australian
detector that would focus mostly on studies of the coalescence phase of neutron
stars (NEMO~\autocite{2020PASA...37...47A}).

Presently, the Einstein Telescope and Cosmic Explorer are the candidate
next-generation observatories in Europe and the United States.  Assuming that
CE and ET are realized, they could naturally form the basis of the 3G network,
as Advanced LIGO and Advanced Virgo do now for the 2G network.  Should other
facilities of suitable capability come online, they too could participate.  An
effective 3G network would be a coordinated partnership that seeks to leverage
the investments in each independent observatory to create great value added.
Optimally, it would consist of three (or more) 3G detectors, geographically
distributed on the globe to provide good localization of sources on the sky.

When envisioning international partnerships for Cosmic Explorer, we will learn
from the success of the LIGO Scientific Collaboration (LSC). The LSC has data
sharing and data analysis agreements, through memoranda of understanding, with
Virgo and KAGRA. Also, the GEO collaboration, with its GEO600 detector, is a
member of the LSC, as is OzGrav. Through this mechanism, leaders from the LIGO,
Virgo, KAGRA, and GEO600 detector groups participate in the Joint Run Planning
Committee to coordinate observing runs. These partnerships have been crucial
for enabling and extracting the most information possible from the first
observations of gravitational waves and especially the sky localization
enabling multimessenger astronomy.  

In parallel, the Gravitational-Wave International Committee
(GWIC)~\autocite{GWIC} chartered a subcommittee to study detector astrophysical
and instrumental science~\autocite{GWIC3G}. Hundreds of scientists worldwide
participated in providing in-depth analyses of the observational science and
discussions of the instrumental opportunities and challenges. The documents
were improved by reviews from both experts in the field and by members of the
Gravitational-Wave Agencies Correspondents (GWAC) and are publicly available on
the GWIC web pages~\autocite{GWAC,GWIC3GDocs}. This GWIC 3G endeavor enriched
and updated the CE and ET science cases, and helped prioritize and focus the
instrumental development. GWIC remains actively engaged in coordinating the
world-wide effort and will sponsor further collaborative activities.

There remains a strong incentive for each of the major projects to form
consortia which are focused internally. Einstein Telescope has formed a
Consortium~\autocite{ETConsortium} which is providing technical and scientific
support for its proposal, and the Cosmic Explorer Consortium is also active
(\cref{sec:InTheGWCommunity}). However, the experience with the current
observatories demonstrates the synergy, efficiency, and scientific value of
close coordination at all levels. Recognizing this, the CE and ET efforts are
coordinated through exchange of members in their organizing committees, and via
informal interactions at meetings in our field: Dawn, GWADW, Amaldi, GWPAW.

The CE team greatly values closer links between the CE and ET projects, and as
early as is feasible. This will allow common technical developments, minimize
independent parallel work, and could possibly lead to economies of scale. In
the longer term, it is clear that the greatest scientific return will come from
joint planning for running and upgrades and from joint analyses of data. The
nature of this global governance is yet to be determined, but value in
significant coordination is clear.

Historically, the gravitational wave community has built strong partnerships
across many institutions, mainly in North America, Europe, Australia, Japan and
India. Into the 2030s, we will expect to see a shift into a more developed and
connected global scientific community. Cosmic Explorer will facilitate
opportunities for broader geographic participation so that scientists across
Africa, the Americas, Asian, and Pacific Island nations can be welcomed into
the global gravitational-wave community.

\section[A Respectful, Healthy, and Thriving Scientific Community]{Cultivating a Respectful, Healthy, and Thriving Scientific Community \hlight{[Brittany]}}

Innovation excels when diverse minds across many axes of identity can thrive.
The National Science Foundation has identified gender identity, race, color,
ethnicity, (dis-)ability, socio-economic status, sexual orientation, language,
nationality, age, religion, veteran status, and family structure as some of the
attributes of a diverse and high-performing workforce that will best advance
science~\autocite{NSFDiversity2012}. The Cosmic Explorer project will
continuously and comprehensively work to address all axes of diversity. As the
project builds capacity around each axis of diversity, it will understand how
people may intersect multiple groups.  Cosmic Explorer will build competency on
what impacts each demographic and work to ensure that each population has access
to participation and is represented and respected for their contributions in
every aspect of the project's work. This requires comprehensive yearly planning,
continuous engagement and regular assessment on whether those goals are
achieved.

As Cosmic Explorer continues to ensure fruitful engagement with its global
partners, it will continue to invest in learning respectful cultural practices
and designing our workflows and schedules based on this. Beyond seeking input
from the physics and astronomy communities, Cosmic Explorer will seek to learn
best practices from different organizations that have set up thriving diverse
global institutions. The project will also ensure support exists to engage with
outside consultants and facilitators to help grow our awareness around the most
effective ways to do this.

\chapter{Cosmic Explorer Project \hlight{[Matt]}}
\xlabel{project}

This section presents cost estimates, a project timeline, an operations model,
and an outline of the project management that would be brought to bear in the
project phase of Cosmic Explorer.

\section{Cost Estimates}
\xlabel{cost_est}

The cost estimate for CE presented here is based on actual costs from LIGO
construction and the Advanced LIGO upgrade.
Since the CE observatories are significantly longer than LIGO,
we have revised the estimated civil engineering costs with the guidance of
a professional civil engineering consultant (Eric Riegel, TruE Consulting),
and we have engaged a professional metallurgy consultant (Dan Henkel, Rimkus)
to help us find approaches to \bmt\ construction that simultaneously
reduce cost and increase performance.
The technical impact of design-choice cost-drivers is discussed in \sref{cost_drivers}.

Adapting the historical LIGO construction and upgrade costs to CE required a few extrapolations:
costs which depend on the length of the observatory must be scaled appropriately,
design changes due to lessons learned and research that
has been done since LIGO was built must be incorporated,
and inflation and shifts in the market prices of materials must be accounted for.
The CE cost estimate was broken down into four top-level categories:
civil engineering, vacuum system, detector, and project costs.
Each of these estimates include materials and labor,
as well as management and design costs
(see \tref{costs}).

The civil engineering costs were estimated \emph{ab initio} by our
civil engineering consultant, Eric Riegel, in consultation with LIGO engineers
(including Fred Asiri, the civil engineer in charge of the original LIGO construction).
These estimates were cross-checked against scaled LIGO costs.
Civil engineering accounts for \PCTCIV{} of the total estimated project cost
(see \tref{costfractions}).

The vacuum system cost estimate is based on recent work done by Rainer Weiss
as part of the \emph{NSF Workshop on Large Ultrahigh Vacuum Systems for Frontier
Scientific Research  Instrumentation}\autocite{VacuumWorkshopTechnote} (NSF award 1846124).
This estimate was, in turn, based on extrapolation from LIGO costs, along with
updates for current material prices and new technologies.
Vacuum systems account for \PCTVAC{} of the total estimated project cost.

Experience from the Advanced LIGO upgrade was used to estimate the cost of
design, construction, and installation of a CE detector.
Adjustments were made for sub-systems which will be significantly different,
such as the mirror suspension system, and for systems which were not part
of the Advanced LIGO upgrade (e.g., the squeezed light source).
The detectors account for \PCTDET{} of the total estimated project cost.

The project-level costs, including management, were estimated based on the
Advanced LIGO experience.
This includes separate management estimates for each of the other cost
categories, as well as project-level costs such as computing and
communication, travel and shipping between observatory locations (collectively labeled ``Coordination'' in  \tref{costfractions}).
The project-level and management costs account for \PCTMAN{} of the total estimated project cost.

Some significant uncertainties remain to be resolved in this cost estimate.
An essentially irreducible uncertainty of roughly \SI{20}{\percent} results from changes in the market
 prices of raw materials, especially steel for the vacuum system.
Furthermore, a more complete design of the facility, vacuum system and detector
 will be required to improve the cost estimates for these components.
Finally, about half of the cost of civil engineering is in excavation and site
 dependent costs that introduce some uncertainty in that
 portion of the estimate.
These uncertainties will be addressed as the Cosmic Explorer timeline becomes clearer,
 the design phase progresses, and the site selection process converges.

\begin{table}[!t]
    \centering

\begin{minipage}{1.0\linewidth}
	\centering
	\begin{tabular}{c}
		\Large \textbf{Cosmic Explorer Cost Estimates, \$(M) 2030 USD}\\[2mm]
	\end{tabular}
\end{minipage}

\resizebox{0.9\linewidth}{!}{ %

\begin{tabular}{l S S}
    \rowcolor{LightSteelBlue1}
    \textbf{Observatory Costs}    & {\textbf{20\,km}} & {\textbf{40\,km}} \\

\rowcolor{Honeydew1}
\textit{Management}	&  & \\
\rowcolor{Honeydew1}
\hspace{3mm}Civil Engineering	&~ \CostTwentyManagementCivilInf &~ \CostFortyManagementCivilInf \\ 
\rowcolor{Honeydew1}
\hspace{3mm}Vacuum System	&~ \CostTwentyManagementVacInf &~ \CostFortyManagementVacInf \\ 
\rowcolor{Honeydew1}
\hspace{3mm}Detector	&~ \CostTwentyManagementDetInf &~ \CostFortyManagementDetInf \\
\rowcolor{Honeydew2}
\hspace{3mm}\textit{Total} & \ittab \CostTwentyManagementTotalInf & \ittab \CostFortyManagementTotalInf \\

\rowcolor{LightYellow1}
\textit{Site Specific Design}	& & \\
\rowcolor{LightYellow1}
\hspace{3mm}Civil Engineering	&~ \CostTwentyDesignCivilInf &~   \CostFortyDesignCivilInf \\
\rowcolor{LightYellow1}
\hspace{3mm}Vacuum System	&~ \CostTwentyDesignVacInf &~ \CostFortyDesignVacInf \\
\rowcolor{LightYellow1}
\hspace{3mm}Detector	&~ \CostTwentyDesignDetInf &~ \CostFortyDesignDetInf \\
\rowcolor{LightYellow2}
\hspace{3mm}\textit{Total}	& \ittab \CostTwentyDesignTotalInf &  \ittab \CostFortyDesignTotalInf \\

\rowcolor{AntiqueWhite1}
\textit{Realization}	&  & \\
\rowcolor{AntiqueWhite1}
\hspace{3mm}Civil Engineering	&~ \CostTwentyRealizationCivilInf &~ \CostFortyRealizationCivilInf \\  
\rowcolor{AntiqueWhite1}
\hspace{3mm}Vacuum System	&~ \CostTwentyRealizationVacInf &~ \CostFortyRealizationVacInf \\ 
\rowcolor{AntiqueWhite1}
\hspace{3mm}Detector	&~ \CostTwentyRealizationDetInf &~ \CostFortyRealizationDetInf \\
\rowcolor{AntiqueWhite2}
\hspace{3mm}\textit{Total}	 & \ittab \CostTwentyRealizationTotalInf & \ittab \CostFortyRealizationTotalInf \\ 

\rowcolor{LightSteelBlue1}
\textbf{Observatory Total}    &  \bftab \CostTwentyTotalInf & \bftab \CostFortyTotalInf \\
\\[-2mm]

\end{tabular}

\hspace{30pt}

 \begin{tabular}{l S}
\rowcolor{Lavender}
\textbf{Project Level Costs}    &   \\

\rowcolor{Honeydew1}
\textit{Project Wide}	& \\
\rowcolor{Honeydew1}
\hspace{3mm}Management 	&~ \CostProjectProjectwideManagementInf \\
\rowcolor{Honeydew1}
\hspace{3mm}Coordination	&~ \CostProjectProjectwideCoordinationInf \\
\rowcolor{Honeydew1}
\hspace{3mm}Computing 		&~ \CostProjectProjectwideComputingInf \\
\rowcolor{Honeydew2}
\hspace{3mm} \textit{Total}	& \ittab \CostProjectProjectwideTotalInf \\ %

\rowcolor{LightYellow1}
\textit{Common Design}	&   \\
\rowcolor{LightYellow1}
\hspace{3mm}Civil Engineering	&~ \CostProjectDesignCivilInf \\
\rowcolor{LightYellow1}
\hspace{3mm}Vacuum System	&~ \CostProjectDesignVacInf \\
\rowcolor{LightYellow1}
\hspace{3mm}Detector	&~ \CostProjectDesignDetInf \\
\rowcolor{LightYellow2}
\hspace{3mm}\textit{Total}	 & \ittab \CostProjectDesignTotalInf \\
\rowcolor{Lavender}
\textbf{Project Level Total} &\textbf{\CostProjectTotalInf}\\ %
 & \\
\rowcolor{AntiqueWhite2}
\textbf{Contingency}	& \\
\rowcolor{AntiqueWhite1}
\hspace{3mm}20\,km Observatory &~ \CostContingencyTwentyInf \\ 
\rowcolor{AntiqueWhite1}
\hspace{3mm}40\,km Observatory &~ \CostContingencyFortyInf \\ 
\rowcolor{AntiqueWhite1}
\hspace{3mm}Project Level &~ \CostContingencyProjectInf \\
\\[-2mm]

  \end{tabular}%
} %
  
\begin{minipage}{1.0\linewidth}
	\centering
	\begin{tabular}{p{25em} S}
		\rowcolor{Thistle1}
		\textbf{Grand Total for Reference Concept (2 Observatories)} & \bftab \CostTotalInf \\
	\end{tabular}
\end{minipage}  
  
  \caption{Cost estimate for the Cosmic Explorer Project reference concept (one \SI{40}{\km} and one \SI{20}{\km} observatory), in millions of 2030 US dollars (see \sref{cost_2021} for 2021 USD).
  The cost estimate includes design, materials, construction, installation and project management for the civil engineering (buildings, roads, etc.), the vacuum system, and the detector.
The cost of alternate configurations can be estimated by adding the associated observatory costs to the project-level costs (e.g., \$\CostTwoTwentyInf\,M 2030 USD for two \twentykm\ observatories,
 or \$\CostJustFortyInf\,M 2030 USD for a single \fortykm\ observatory).
  \label{tab:costs}}
\end{table}

\begin{table}[b]
  \centering
  \begin{tabular}{p{20em} S S}
    \rowcolor{LightSteelBlue1}
    \textbf{Top-Level Costs}    &  {\textbf{\$(M) 2030 USD}} & {\textbf{Percent}}\\
   	\rowcolor{LightYellow1}
	Civil Engineering	 & \CostOverviewCivilInf & \PCTCIVn  \\
	\rowcolor{LightYellow1}
	Vacuum System & \CostOverviewVacInf & \PCTVACn  \\
	\rowcolor{LightYellow1}
	Detector & \CostOverviewDetInf & \PCTDETn  \\
	\rowcolor{LightYellow1}
	Management, Design, Project    &  \CostOverviewProjectInf & \PCTMANn \\
	\rowcolor{Thistle1}
	\textbf{Grand Total (2 Observatories)}    &  \bftab \CostOverviewTotalInf & 100 \\
  \end{tabular}%
  \caption{Top-level cost breakdown for Cosmic Explorer, excluding operating costs, in millions of 2030 US dollars and including \SI{20}{\percent} contingency.  Inflation is computed in then-year USD for a project starting in 2027 and completing in 2035 with a typical ramp up and ramp down, which is numerically equivalent to 2030 USD
  (see \sref{cost_2021} for more info and for estimates in 2021 USD).
  \label{tab:costfractions}}
\end{table}

\section{Timeline}
\xlabel{schedule}

The Cosmic Explorer timeline spans multiple decades
and takes place in distinct stages:
development; observatory design and site preparation;
construction and commissioning; initial operations;
planned upgrades; operations at nominal sensitivity;
future observatory upgrades and operations.

\paragraph{Development}

The development stage for Cosmic Explorer began in 2013,
 and has resulted in many relevant publications within the gravitational-wave community.
This stage will continue after the completion of this document and its endorsement
by the scientific community that Cosmic Explorer will serve.
This phase is one in which 
 the community engagement work must expand in scope as described in \sref{global},
our understanding of the key science goals discussed in \cref{ch:keyquestions} will deepen,
and the enabling technology discussed in \cref{subsec:technology_drivers}
can be further developed by the instrument community.

As noted in \sref{global}, building competence around,
and relationships with, Indigenous Peoples is a long-term endeavor that
will be critical to the success of CE and must commence as soon as possible.
This work can, and must, begin before a site is selected and may begin even before
 any specific sites are considered by making contact with national Indigenous Peoples
 organizations (e.g., SACNAS).
By opening the conversation with national Indigenous Peoples organizations,
 the relationship building process can expand to include learning about Indigenous communities,
 and eventually reaching out to specific community leaders and seeking permission to engage
 in a dialog about potential locations for a Cosmic Explorer observatory.

The technical development described \cref{subsec:technology_drivers} is also important
 to ensuring that the investment in Cosmic Explorer facilities is most effectively utilized,
 and the CE science goals are achieved.
Development of the technologies summarized in \tref{research} will require planning and funding,
 roughly at the level of \$15\,M over 4 years~\autocite{Reitze:2019iox}.
This effort may overlap with the Conceptual Design phase,
 depending on the relative timelines of CE project funding and the various research efforts.

\paragraph{Observatory Design and Site Preparation}

The project begins with dedicated funding for Cosmic Explorer design
and passes through all phases of the MREFC process.\autocite{NSFMFG}\footnote{\url{https://www.nsf.gov/bfa/lfo/}}
In addition to the design phases for the CE observatories (conceptual, preliminary, final),
 this preparatory stage will include 
 prototype construction for the CE vacuum system and
 a nationwide search for and research about potential observatory sites.
This will result in the selection of observatory construction locations (``site selection''):
 a process that we expect will be led by the relevant funding agencies.
Two or more years will be required to build relationships with local communities
and to obtain the necessary permits for construction, making it imperative that this work
be done in parallel with technical and civil design efforts.
The total time estimated for this phase of the project is 7 years.

\paragraph{Construction and Commissioning}

While some overlap between design and construction is possible, the vast majority of the
technical and civil designs will need to have finished their final design phases before
funds can be appropriated for construction.
The civil works required for a Cosmic Explorer observatory will require at least three years,
and potentially more depending on the particulars of the site.
Installation of the detector and subsystem commissioning can occur to some
degree in parallel with civil works
(i.e., as soon as the corner and end buildings are finished).
However, commissioning of each detector to the point of acceptance (i.e., transition to operations)
 will require at least one year after construction is complete,
 making the estimated time for this phase of the project 6 years.

\paragraph{First Operations Phase}

Once the CE detectors are operational,
the project will transition to the Operations Stage.
This will follow the successful model developed by LIGO:
interleaved commissioning and observation, with observation periods growing in duration
as the detector matures.
Observational campaigns will also be coordinated with other gravitational-wave observatories,
 potentially including Einstein Telescope and/or Cosmic Explorer South.
In parallel with detector operations, preparations for the planned upgrades will be underway,
 as described in \cref{subsec:reference_design}.
The duration of this phase depends on commissioning progress, upgrade readiness,
 and the success of the observation campaigns, and as such is somewhat flexible:
5 years is a plausible duration.
By this phase, we expect that the community-focused aspect of the facility (\cref{sec:communityfacility}) will also be operational; depending on the nature of the facility, this could include the arrangement of exhibits, workshops or other programs of community interest.

\paragraph{Upgrade Phase}

The upgrade of the CE detectors to their nominal configuration will bring increased sensitivity
and full access to the key science goals described in \cref{ch:keyquestions}.
Some upgrades (see \cref{subsec:reference_design})
  can be performed with relatively little disruption of observation
 (e.g., seismometer array installation), while other may require significant down-time
 (e.g., upgrading seismic isolation systems). 
To allow for installation and commissioning of the new sub-systems, interleaved with observation,
 this phase may last as long as 4 years.
The upgrades of the CE observatories may, depending on the needs of the
 scientific community, happen simultaneously or sequentially.

\paragraph{Second Operations Phase}

The second operations phase envisioned in this Horizon Study will presumably occur
in the presence of other next-generation gravitational-wave detectors,
and as such will result in ground-breaking high-fidelity access to gravitational-wave
sources from throughout the universe.
The duration of this phase is not specified here.

\paragraph{Future Work}

Though the commitment to fund Cosmic Explorer will likely be based on a 20-year duration,
 the Cosmic Explorer facilities are intended to be long-lived, with a nominal 50-year lifetime.
While this is too far into the future for meaningful speculation about particular technologies,
the CE facility design allows for improvements in fundamental noises, such as quantum
and thermal noise, well beyond the CE target sensitivity (see \cref{fig:high_power_silicon}).
As such, the CE observatories will accommodate continuous improvements to detector
technology and scientific output.

Eventually the CE facility will reach the later stages of its life cycle and be divested. It is thus important to begin planning from the outset, as recommended for large facilities~\autocite{NSFMFG}, to engage the scientific and local communities in divestment decisions and to anticipate costs. Some questions to be considered are whether the facilities could live on with new stewardship or be dismantled and cleaned up and how these decisions could strengthen the relationships fostered with the local community.

\begin{figure}[t]
    \centering
    \includegraphics[width=\textwidth]{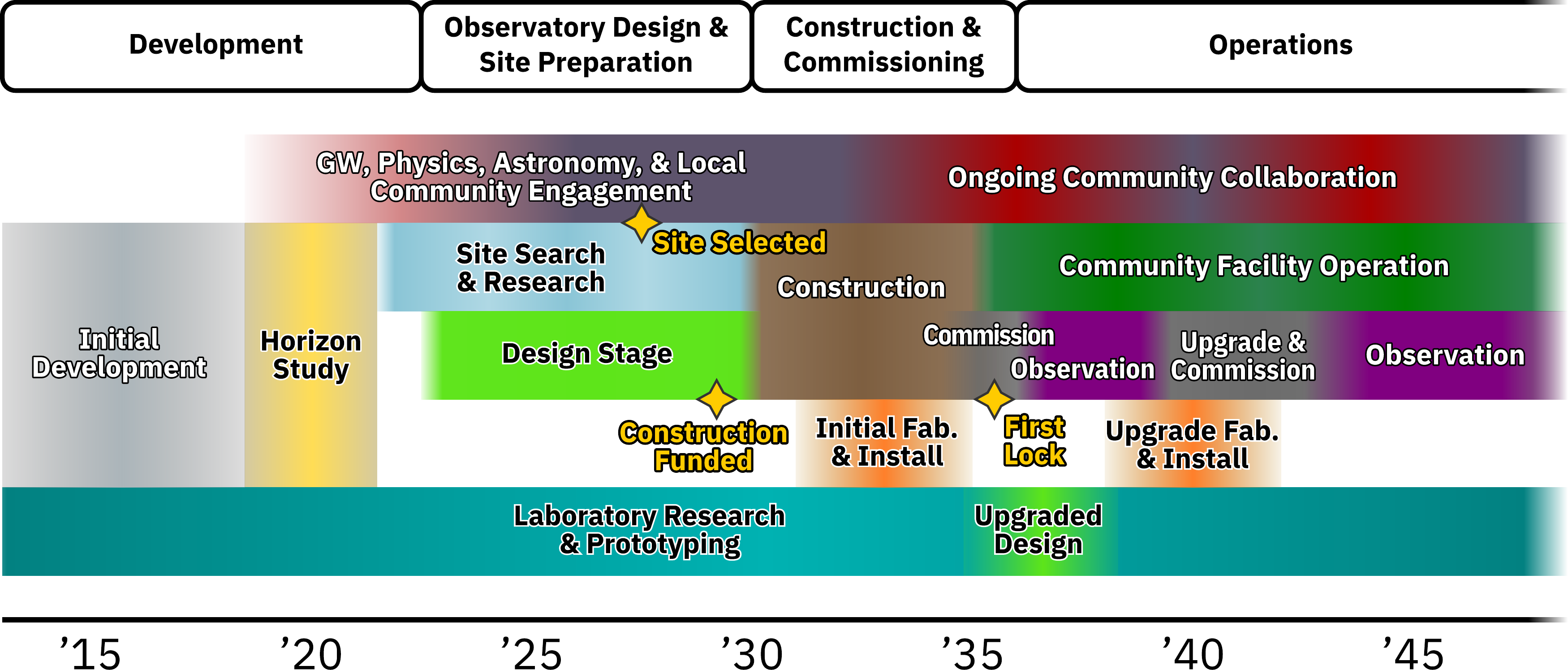}
    \caption{A top-level timeline showing a phased approach to Cosmic Explorer, as described in \sref{schedule}. The eventual divestment from the facility is not indicated.
    \flabel{timeline}}
\end{figure}

\section{Operations Model}
\xlabel{ops}

Following the successful example of LIGO,
the Cosmic Explorer construction project will have a well-defined scope which leads into the commissioning and then operations phases. The hand-off could be defined as the point at which all installation at the Observatory is complete, and acceptance of the subsystems following a successful stand-alone test campaign has been completed. The first goal of the operations and maintenance phase would be to commission the instruments to reach a useful initial astrophysical sensitivity and sufficiently robust operations, along with the ability to produce high-quality and well-calibrated astrophysical data and alerts to the broader astronomical community. Achievement of this goal leads into a phase of alternating periods of observations and detector improvements, following the successful model employed by the current gravitational-wave network.

We anticipate an open data model for Cosmic Explorer in which strain data and astronomical triggers would be released immediately to the public (see \sref{data_management}). The data distribution and associated computing infrastructure will thus be key aspects of the operations model.

The organizational and staffing model for accomplishing the Cosmic Explorer Operation and Maintenance is provided below, based on the LIGO Laboratory Operations for the Advanced LIGO Phase.
There will be persons who play roles in several groups, and many will change their focus according to the phase of activity (repairs, upgrades, commissioning, observing).

\begin{enumerate}
  \item A \textit{detector group}, consisting of engineers and scientists who specialize in various detector subsystems and electronics. This group will be responsible for maintaining, testing, documenting, and repairing controlled detector configurations with a focus on optimizing data quality and uptime. This group will also be responsible for operating the detectors.
  \item A \textit{commissioning group}, largely scientists, will commission and test the instruments and establish new detector configurations, with a focus on improving their performance.
  \item A \textit{systems engineering group}, largely engineers, will set technical standards and approve changes in controlled configurations, with a focus on system trades.
  \item A \textit{calibration and data quality group} of scientists and engineers will ensure the data are ready for astrophysical interpretation. These individuals will work with commissioners to find and resolve sources of instrumental and environmental noise, and will vet the data to correct, mark, and/or edit data as needed to allow the subsequent analysis to be made by the scientific community.
  \item An \textit{observation coordination group} of scientists will plan and coordinate observations with the gravitational-wave and astronomical network and interface with the broader scientific community on issues related to observations.
  \item An \textit{analysis, data, and computing group}, focused on the cybersecurity, computing, low-latency analysis for astronomical alerts, maintaining catalogs of observations, curating and disseminating open data, and running a ``Help Desk'' to facilitate the use of CE by the broader scientific community.
  \item A \textit{facilities group}, responsible for maintaining the physical infrastructure of the observatories. Staffing will include significant vacuum expertise, and civil, electrical, and grounds engineers and technical support.
  \item A \textit{management group}. Each site will have a lightweight management and business group, enabling safety, procurements, shipping, human resources support, and top-level direction. For the target configuration of two Cosmic Explorer Observatories, one of the two observatory sites will carry management common to the two sites to minimize costs and maximize synergy between the sites.
  \item An \textit{community engagement and integration group} responsible for
promoting integration of the CE observatories into the local community,
arranging programs and exhibits at the community-focused facility, and with
engaging the public at large.  This includes building and maintaining
synergistic relationships with local Indigenous Peoples, publishing broadly
accessible versions of high-impact scientific results which inspire community
engagement, and translating these materials into Indigenous languages.
\end{enumerate}

External advisory committees will be established to (1) aid in technical management and evolution of the observatories, (2) coordinate with the greater scientific community and help guide the timing and trades of observation vs.\ commissioning, and (3) ensure that each observatory maintains healthy relationships with local communities, including Indigenous Peoples.

The scope for the operations of Cosmic Explorer can be estimated based on
 extrapolation from the LIGO Laboratory operations.
 As described above, significant staff will be required to properly operate, maintain, and incrementally commission the detectors. The two-detector configuration of CE would profit, as LIGO does, from some economy of scale for technical and management staff; the single detector approach would allow some reduction but not a factor of two. The vertex and end stations are expected to be similar in size and complexity to the LIGO buildings. The first detector to be installed will be comparable in electronic, mechanical, and optical systems, and will require staff comparable to LIGO's to maintain it. The vertex and end-station vacuum systems for the initial detector will also be similar to LIGO; if the \SI{2}{\um} technology requiring cryogenics is used, there will be some increased operating complexity. The vacuum system will be 5 to 10 times greater in length and volume, but a great majority is passive once installed. There will be greater maintenance needs to inspect and maintain the tube, foundation, and protective cover.
There will be a significant increase\,---\,roughly a factor three\,---\,in the staffing and operating expenses for this larger vacuum system and civil infrastructure, bringing them to roughly \SI{40}{\percent} of the total operations cost (see \tref{costops}).

A significant difference in project scope for Cosmic Explorer compared to
Advanced LIGO is the staffing associated with delivering Cosmic Explorer data
and alerts to the scientific community. In an Open Data model, data will
need to be calibrated and conditioned to the point that it can be interpreted
for observational science without expert knowledge of the instrument and the
data imperfections, which must be documented. Operations staff will also have
responsibility for the production and dissemination of low-latency alerts for
known multimessenger sources. These activities will require a dedicated group
of people whose sole job is preparing data and performing initial analysis. 

It is assumed that the research and development of new detectors will be supported by proposals to funding agencies; the staff at the CE observatories would be members of groups proposing for upgrades and new detectors, complemented by many in the greater scientific community.

\begin{table}[t]
  \centering
  \begin{tabular}{p{20em} S S S}
    \rowcolor{LightSteelBlue1}
    \textbf{Yearly Operations Cost Estimates}    &  \textbf{\$(M) 2030 USD} & \textbf{Percent}\\
   	\rowcolor{LightYellow1}
	Facilities	 				& \CostOpsFacilityInf & \opsPCTFACn  \\
	\rowcolor{LightYellow1}
	Vacuum Systems 			& \CostOpsVacInf & \opsPCTVACn  \\
	\rowcolor{LightYellow1}
	Detector 					& \CostOpsDetInf & \opsPCTDETn  \\
	\rowcolor{LightYellow1}
	Analysis, Data, and Computing & \CostOpsCompInf & \opsPCTCMPn  \\
	\rowcolor{LightYellow1}
	Management  				&  \CostOpsManagementInf & \opsPCTMNGn \\
	\rowcolor{LightYellow1}
	Community Engagement  		&  \CostOpsCommunityInf & \opsPCTCOMn \\
	\rowcolor{Thistle1}
	\textbf{Grand Total  (2 Observatories)}    &  \bftab \CostOpsTotalInf & 100 \\
  \end{tabular}%
  \caption{Estimated yearly operations costs for Cosmic Explorer with two observatories,
   based on Advanced LIGO and scaled for CE facility sizes, in millions of 2030 US dollars
   (see \sref{cost_2021} for 2021 USD).
  Operations costs for a CE project with a single observatory would be roughly \SI{30}{\percent} less.
  This estimate is for observatory operations \emph{only} and does not include research
  for instrument upgrades.
  \tlabel{costops}}
\end{table}

\section{Risk Management}
\label{sec:risk}

Successful risk management starts with a careful examination of the project requirements and the construction and engineering responses in place to meet them. 
This examination leads to a good understanding of potential risk factors and their impact on the project. It is a common practice among large projects to establish a ``risk registry'' for perceived problem areas so that potential major issues are identified early and resolved, and that the project remains on schedule.  For example, the Risk Registry and Risk Management Plan for Advanced LIGO will serve as a guide to Cosmic Explorer~\autocite{M080359, M060045}.
In the next two subsections we present the main technical and management risk factors that will form the starting point of Cosmic Explorer's risk management plan.

\subsection[]{Technical Risk \hlight{[Stefan]}}
\label{subsubsec:techrisk}
Unproven technologies (cf. \cref{subsec:technology_drivers}) and identification of appropriate sites constitute the largest technical risk factors for the project. Cosmic Explorer will rely on the proven room temperature fused-silica-optic technology of Advanced LIGO operating at a wavelength of \SI{1064}{\nm}; fortunately, the success of Advanced LIGO establishes that this technology is extremely mature. As with LIGO, the Cosmic Explorer sensitivity will continue to improve through a series of planned upgrades.

The main technical risk factors are fairly limited. The ones we have identified so far are:
\begin{enumerate}
    \item \emph{Risk of acquiring site(s) with adequate space and no excess noise that could compromise interferometer sensitivity and performance} (cf. \sref{vacuum_reqs}). Finding appropriate site(s) that have not only adequate space for Cosmic Explorer, but also provide a low seismic noise background for the observatory is key to the Cosmic Explorer design. Initial surveys of North America suggest that it is possible to find adequate sites, and a detailed site survey as part of a design study can guarantee that the current seismicity of candidate sites is acceptable. But acquiring the required continuous piece of land from potentially numerous previous owners can be difficult and poses an obvious project risk. An excellent relationship with Indigenous communities and neighboring land owners is essential.  Thus the land acquisition will have to be managed carefully. Possible urban encroachment on the site(s) over the lifespan of the observatory also will have to be managed.

    \item \emph{Risk associated with the vacuum system} (cf. \cref{sec:vacuum_reqs}). An adequate vacuum system is clearly technologically feasible. However, the significant increase in required vacuum volume compared to the Advanced LIGO detectors makes the vacuum system the driving cost factor for the project. A more cost-effective vacuum system construction is desired, and requires additional research.

To mitigate the project risk ahead of construction it is essential that a vacuum system prototype be built at the engineering design stage. Ideally this prototype would be constructed by the company that will get the contract for the vacuum system. It would also be useful to prototype a test mass chamber due to its complexity. Prototypes similar in spirit were constructed as part of initial LIGO.

    \item \emph{Risk associated with larger test masses not achieving design specifications.} While almost all technology from Advanced LIGO could be directly installed in a larger facility, the longer arm lengths do require an increase in optics size compared to Advanced LIGO, beyond the capability of current coating facilities. Together with issues related to small absorbers in the test-mass mirror coatings encountered in Advanced LIGO, this puts the manufacturing capability of optics for Cosmic Explorer at a critical spot. Achieving the optics and coatings design specifications is essential for reaching the design circulating laser power (\SI{1.5}{\mega\W}) and thermal noise.  Addressing this risk will require a significant investment in proof-of-principle optics (see \cref{subsec:technology_drivers}).

	\item \emph{Risks associated with a \SI{20}{\km} detector.}
Losses in the signal extraction cavity limit the high frequency sensitivity of a \SI{20}{\km} CE observatory; specifically, they reduce the depth of the resonant dip in the post-merger tuned configuration.
Since the high frequency sensitivity is a key science driver for a \SI{20}{\km} Cosmic Explorer detector, there is also added technical risk for this \SI{20}{\km} post-merger tuning configuration arising from any excess signal extraction cavity loss. The low frequency optimized tuning is not limited by the excess loss in the signal extraction cavity. Compared with a \SI{40}{\km} detector, many technical noise sources are also closer to limiting the sensitivity for a \SI{20}{\km}. Thus, we realize that the overall risk of having a single \SI{20}{\km} detector is significantly higher than a single \SI{40}{\km} detector.
The technological challenges and the corresponding risk to CE detectors are discussed in
(cf. \cref{subsec:technology_drivers}).

    \item \emph{Risk due to an inadequate number of electromagnetic follow-up observatories.} Part of Cosmic Explorer's scientific promise is to provide sky localization and early warning for neutron star merger events. Thus the availability of a sufficient number of well-performing satellites, observatories and telescopes (X-ray, optical, radio, neutrino) is critical for reaping the full scientific benefit from Cosmic Explorer. Scheduling observation runs to maximize overlap with followup assets can mitigate that risk.
    
    \item \emph{Environmental risks:} As a facility designed for a 20+ year life span, rare but violent events such as floods, violent storms and earthquakes pose a significant but hard to predict risk for the project. The site selection, facility and instrument design will have to accommodate these risk factors.

    \item \emph{Malicious risks:}  Sufficient site security, in the form of video monitoring, digital surveillance and physical barriers (locked entrances and fencing), will be needed to protect against both unintentional accidents and intentional sabotage.   Infrastructure such as power lines, the vacuum system and the computer grid may be tempting targets for bad actors.
    \end{enumerate}

We have also identified the alternative technology being developed for the LIGO Voyager detector consisting of cryogenic silicon optics operating at a wavelength of \SI{2}{\um} which could be installed after the initial observing phase if a major problem is found which prevents the Advanced LIGO technology from reaching the sensitivity goals (cf.~\cref{subsec:2um_alternative}). This technology could also be used to surpass the nominal CE sensitivity in the future (cf.~\cref{subsec:silicon_upgrades}).

\subsection[]{Management Risk}
\label{subsubsec:manrisk}
The management risk for a large project like Cosmic Explorer is to a large extent related to the deployment and organization of human resources to address the technical and engineering challenges of the project. This includes adequate financial backing and coordination to ensure that resources and scheduling are adequate to the task.

As described in \cref{subsec:planning}, Cosmic Explorer project management will be accomplished using standard project management practices. A team of project management professionals will be tightly integrated into the project engineering and systems integration group. The use of monitoring software (e.g., Primavera, used by Advanced LIGO) and a resource-loaded schedule will help to identify areas which may have significant risk. The risk registry described in \cref{sec:risk} will be critical in managing problem areas and their potential effects on project costs and schedule.

For Cosmic Explorer, there is potential for significant management risk in at least three areas: 
(1) publicity risk associated with site acquisition; 
(2) where new technology is required to achieve performance goals
(cf. \cref{subsec:technology_drivers}); 
and (3) international partnerships that may be established as part of the core Cosmic Explorer project and the wider 3G network.

\begin{enumerate}

\item The construction of Cosmic Explorer requires the acquisition of a large, continuous. L-shaped piece of land. While for most of this land the construction impact will be limited to allowing a path for the vacuum system, this land acquisition will impact local land owners and Indigenous communities. Thus, especially in the age of social media, a meaningful and genuine relationship with local communities is absolutely essential for the success of the project. The ground work for these relationships needs to be laid early on in the project, well ahead of any attempt to acquire permission to build at a particular location. 

\item Potential loss of the expertise required to design, build and commission Cosmic Explorer.
Through the lifespan of Initial and Advanced LIGO the NSF has invested in building up that technical, scientific and engineering expertise in the form of the LIGO laboratory staff scientists and engineers, as well as associated research groups across the country. That pool expertise forms a national asset, the loss of which would set back Cosmic Explorer significantly. It is thus of particular importance to sustain this expertise during the transition from current detectors (Advanced LIGO, A+) to Cosmic Explorer.

\item International partnerships can present difficult complexities in several ways, including: the length of time needed to put them into place; the probable need for negotiations between high levels of government; differing costing protocols between countries (making cost assessments difficult to compare); ensuring that a single management structure has adequate authority; and differing work rules.  Similar issues can arise also between states and funding agencies within the U.S., though they are usually more easily managed.
\end{enumerate}

\section{Synergies with Programs at U.S. Funding Agencies}
\xlabel{otherprograms}

Cosmic Explorer's unique capabilities to explore extreme gravity and to search for new physics complement the priorities and planned missions of several funding agencies in the United States. (A comprehensive description of the ``European Strategy for Particle Physics'' is available in the 2020 Physics Briefing Book\autocite{2020PHYBRBK}.)

\begin{enumerate}
\item NSF:  The Divisions of Astronomy\autocite{20aaac_gaume} (AST, \$250M 2019 Current Plan) and Physics\autocite{20aaac_caldwell} (PHY, \$285M 2019 Actual), and the Office of Polar Programs\autocite{20aaac_Papitashvilli} (OPP, \$398M 2019 Actual) all make large investments in the study of black holes, stellar evolution, nuclear physics, dark energy and dark matter.  Instruments such as LIGO, IceCube, the Event Horizon Telescope, and optical and radio telescopes are located at many sites around the world, including the South Pole.
 
\item DOE: The DOE program\autocite{20aaac_turner} in High Energy Physics (HEP, \$1.01B 2019 request) comprises: the Energy Frontier, the Intensity Frontier and the Cosmic Frontier. The intellectual basis for the program is described in the 2014 ``P5'' report\autocite{2014PPPPP}. The overall science focus is on the Higgs boson, neutrino mass, dark matter, cosmic acceleration and ``exploring the unknown.''  The ``Cosmic Frontier'' program (\$75.5M 2019 request) supports the ongoing Fermi/GLAST, AMS, HAWC, DES and eBOSS experiments.  It is constructing the Rubin Observatory and DESI for research into dark energy, and LZ and SuperCDMS-SNOLab for dark matter searches. Via the SPT-3G it explores the CMB to study cosmic acceleration and neutrino properties (with NSF). There is also an extensive program in cosmic-ray and gamma-ray research (AMS, HAWC). Nuclear Physics programs of the Office of Science, such as the Facility for Rare Isotope Beams (FRIB), actively develop connections to nuclear astrophysics\autocite{NAP13438}. 

\item NASA:  The Astrophysics Program\autocite{20aaac_hertz} (\$1.496B 2019 request) plays the lead role in eight operating missions (e.g., Hubble, FGST, NICER, TESS),  and five in development (e.g., JWST, WFIRST).  It is a partner in the development of Euclid and LISA.  The NASA Physics of the Cosmos program focuses on on dark matter and dark energy, the evolution of galaxies and stars, and matter and energy in extreme environments.  The LISA program\autocite{LISA} in GW research is closely tied to LIGO/Virgo, and possibly to Cosmic Explorer and the European Einstein Telescope.\autocite{ETBook}
\end{enumerate}

\section{Cosmic Explorer Project Roadmap}
\label{subsec:planning}

In proceeding with the Cosmic Explorer Project, we will draw on the successful experience and expertise of the LIGO Lab and the lessons learned during LIGO's planning, construction, and operation.
The LIGO Lab now has a long history of delivering on time and on budget, has been well operated and managed, and has delivered important scientific discoveries. The LIGO Lab has consistently received high ratings for its leadership and management.  As a result of these attributes, morale within the LIGO Lab has generally been very high.  We recognize this quality as one of the greatest assets to any project and one which must be preserved for Cosmic Explorer.

We also recognize that the CE Project is significantly larger in scale and will require greater sophistication in the Project organization, and a larger breadth of participation and support in the project. 

In planning the next steps for the Project, three resources have been of particular value:
\begin{itemize}
  \item The NSF Major Facilities Guide (MFG)~\autocite{NSFMFG}.
  \item The book chapter on ``Planning, managing, and executing the design and construction of Advanced LIGO''~\autocite{shoemaker2019planning} by David Shoemaker, who led the Advanced LIGO Project. This chapter was written with future projects in mind to provide experience, generalizations, and lessons learned from both Initial and Advanced LIGO.
  \item Advice and experience shared by the management and staff of LIGO, other gravitational-wave observatories world-wide, and other large-scale scientific facilities.
\end{itemize}

The Cosmic Explorer Project will proceed along the guidelines provided in the Major Facilities Guide.  In broad outline, these are to:
\begin{itemize}
  \item collect feedback from the broad scientific community on the match of the Cosmic Explorer concept and capability with their needs and interests, using the Horizon Study as a basis for discussion. In this process the parties (individuals and institutions) interested in engaging substantively in the next Project phases can be identified. Partnership arrangements and international participation will be informally explored. 
  
  \item collect feedback from the NSF, and address any specific shortcomings to ensure that the NSF can correctly consider the Project; feedback on next steps will be welcome. 
  
  \item cast the Horizon study into the form of a Project Execution Plan, and start to address those elements most in need of refinement. Several of these elements follow.
  
  \item establish the core of a Project Office, and within it a system engineering activity.  
  
  \item establish a plan to create an accurate, detailed, costed baseline project description that provides the project performance goals, the technical aspects of the facility, its estimated cost and the time required for completion.  This will become the \emph{Reference Design} for the project.  This essential document provides the point of departure for measuring progress accurately and for assessing cost, schedule and technical performances.

  \item establish an orderly process for implementing project changes even at an early conceptual phase, and maintain an accurate record of them as they occur.

  \item identify potential critical and near-critical paths through the schedule. Ensure that early effort is allocated to assess these activities to firm up estimates, and explore mitigation where possible.

  \item develop a plan for identifying and managing risk (\emph{cf.} \cref{sec:risk})

  \item research and document ``lessons learned'' during the construction of LIGO, Advanced LIGO and A+, as well as from observatories constructed outside the U.S. (GEO, KAGRA, LIGO India, Einstein Telescope)
        and other large facilities (e.g., the Thirty Meter Telescope).

   \item draft a Scope Management Plan and explore scope contingency responses to anticipate means to explore savings from potential de-scoping options, and find decision points for exercising options.

  \item establish robust means of communication with the external physics and astronomy communities and the public.

  \item collaborate with other relevant Projects (e.g., Einstein Telescope) to leverage technical and scientific opportunities whenever possible.

  \item establish and maintain a strong community engagement and integration program with the objective of building synergistic relationships with local communities at potential Observatory sites, including Indigenous Peoples.

\end{itemize}

These activities will be focused (and iteratively tuned) with a target of creating a technical development roadmap with estimates of funding and more detail on the magnitude of the challenges associated with technical development work. A goal will be to support a critical review by mid-decade to enable a detailed engineering design study.

The pace and character of this followup activity will depend on the funding available to the Project. Seeking that funding will be one of the first activities to follow the completion of the Horizon Study.

\chapter{Conclusion \hlight{[Jocelyn]}}
\label{sec:conclusion}

This Horizon Study has described a path forward for realizing Cosmic
Explorer, a next-generation, ground-based gravitational-wave observatory in the
United States. By drawing on two decades of international effort to scale up
the proven technology that has enabled humanity's first observations of
gravitational waves, Cosmic Explorer will extend our gravitational-wave vision
to the farthest reaches of the universe.

In this study, we have described a science-driven design for Cosmic Explorer
and have considered how to optimize the design performance versus the cost. We
presented a technical overview of the detectors and a roadmap to the research
and development required to achieve them. We have further examined the
organization, planning, and community engagement that will be necessary to
design, build, and operate Cosmic Explorer.

During the next few years, we will welcome feedback from the National Science
Foundation, the National Research Council, and from the gravitational-wave
community; their guidance and endorsement will be critical to the success of
the next generation of gravitational-wave science in the United States. We
aspire for this study to prepare the way, through the next two decades, for the
ultimate design, construction, and operation of Cosmic Explorer.

Continued funding to develop enabling technologies, grow the CE community and
build relationships with potential observatory host communities is crucial to
preparing the CE Project for critical review in the mid-2020s.  If CE is
determined to be technically ready, of interest and timely, we expect that a
thorough design study will begin, leading to a complete design and construction
plan that will be funded in the late-2020s.

Once operational, its cosmic reach and exquisite sensitivity will enable Cosmic
Explorer to revolutionize our understanding of the universe while continuing
the United States' leadership in gravitational-wave science. Cosmic Explorer
will observe black holes and neutron stars throughout cosmic time, probe the
nature and behavior of the densest matter in the universe, and explore the
universe's most extreme gravity and open questions in fundamental physics. As
part of an international next-generation gravitational-wave network, Cosmic
Explorer would couple these advances in gravitational-wave astronomy with the
future of electromagnetic and particle astronomy.

\backmatter

\chapter{Acknowledgements \hlight{[Josh]}}
\xlabel{acknowledgements}

We would like to thank the National Science Foundation for supporting this study through the collaborative awards NSF--1836814, NSF--1836779, NSF--1836702, NSF--1836809, and NSF--1836734. We also thank the Australian Research Council, who provided funding through grant CE170100004, as well as the MathWorks, Inc., and the Heising-Simons foundation.
We would also like to thank the members of this study's advisory panel,
Patrick Brady, Kathryne Daniel, Gabriela Gonz\'{a}lez, Vicky Kalogera, Harald L\"{u}ck, David Reitze, Sheila Rowan, and Jim Yeck, for their advice and helpful suggestions throughout the study.
Furthermore, we are very grateful to the gravitational-wave community and especially the Gravitational-Wave International Committee, both for their  work toward the next generation of gravitational-wave observatories and for their feedback on this study.

The authors, as a group and individually, acknowledge the ancestral homelands we were on during the creation of this work.\footnote{We are grateful for the work of Native Land Digital (\url{https://native-land.ca}) for the resources they have provided.}
The authors acknowledge Indigenous Peoples as the traditional stewards of the land, and the enduring relationship that exists between them and their traditional territories.
We acknowledge the longer history of these lands and our place in that history, and we pay our respects to the Indigenous land caretakers past, present, and emerging.
The MIT authors acknowledge our presence on the traditional, ancestral, and unceded territory of the Wampanoag and Pawtucket Nations.
The Syracuse University authors acknowledge with respect the Onondaga Nation, firekeepers of the Haudenosaunee, the Indigenous peoples on whose ancestral lands Syracuse University now stands.
The Pennsylvania State University campuses are located on the original homelands of the Erie, Haudenosaunee (Seneca, Cayuga, Onondaga, Oneida, Mohawk, and Tuscarora), Lenape (Delaware Nation, Delaware Tribe, Stockbridge-Munsee), Shawnee (Absentee, Eastern, and Oklahoma), Susquehannock, and Wahzhazhe (Osage) Nations.  The Penn State authors acknowledge and honor the traditional caretakers of these lands
 and strive to understand and model their responsible stewardship.
The University of Mississippi authors recognize and acknowledge that their university is on the historic Homeland of the Chickasaw Nation.
The Caltech and Cal State Fullerton authors acknowledge and offer our respect to past and present Gabrielino-Tongva
 people and their unceded ancestral lands, including the Los Angeles Basin where Caltech and Cal State Fullerton stand today.
The U.C.~Santa Cruz authors acknowledge that the land on which the university stands is the unceded territory of the Awaswas-speaking Uypi Tribe.

Finally, the PIs on the grants that funded this work are deeply grateful to the graduate student and postdoctoral scholar co-authors who worked tirelessly to make this Horizon Study a reality.

\chapter{Cost and Inflation Estimates}
\xlabel{cost_2021}

This appendix describes the inflation estimates that went into the cost tables presented in
 \sref{project}.
This is done in two parts:
 first, the cost tables are presented in 2021 USD,
 and then the process used to compute the ``2030 USD'' estimates is described.

\vspace{1em}

\section[]{Cost Estimates in 2021 USD}

The following cost tables represent the same content as the ones presented in \sref{cost_est} and \sref{ops}.
The only difference is that they are presented in 2021 USD, i.e. without any attempt to estimate future inflation rates.

\begin{table}[h]
  \centering
  \begin{tabular}{p{20em} S S}
    \rowcolor{LightSteelBlue1}
    \textbf{Top-Level Costs}    &  {\textbf{\$(M) 2021 USD}} & {\textbf{Percent}}\\
   	\rowcolor{LightYellow1}
	Civil Engineering	 & \CostOverviewCivil & \PCTCIVn  \\
	\rowcolor{LightYellow1}
	Vacuum System & \CostOverviewVac & \PCTVACn  \\
	\rowcolor{LightYellow1}
	Detector & \CostOverviewDet & \PCTDETn  \\
	\rowcolor{LightYellow1}
	Management, Design, Project    &  \CostOverviewProject & \PCTMANn \\
	\rowcolor{Thistle1}
	\textbf{Grand Total (2 Observatories)}    &  \bftab \CostOverviewTotal & 100 \\
  \end{tabular}%
  \caption{Top-level cost breakdown for Cosmic Explorer including \SI{20}{\percent} contingency, but excluding operating costs, in millions of 2021 US dollars.  The content of this table is the same as \tref{costfractions}, but with no attempt to estimate future inflation.
  \tlabel{costfractions_2021}}
\end{table}

\begin{table}[h]
  \centering
  \begin{tabular}{p{20em} S S S}
    \rowcolor{LightSteelBlue1}
    \textbf{Yearly Operations Cost Estimates}    &  \textbf{\$(M) 2021 USD} & \textbf{Percent}\\
   	\rowcolor{LightYellow1}
	Facilities	 & \CostOpsFacility & \opsPCTFACn  \\
	\rowcolor{LightYellow1}
	Vacuum Systems & \CostOpsVac & \opsPCTVACn  \\
	\rowcolor{LightYellow1}
	Detector & \CostOpsDet & \opsPCTDETn  \\
	\rowcolor{LightYellow1}
	Analysis, Data, and Computing & \CostOpsComp & \opsPCTCMPn  \\
	\rowcolor{LightYellow1}
	Management  &  \CostOpsManagement & \opsPCTMNGn \\
	\rowcolor{LightYellow1}
	Community Engagement  &  \CostOpsCommunity & \opsPCTCOMn \\
	\rowcolor{Thistle1}
	\textbf{Grand Total  (2 Observatories)}    &  \bftab \CostOpsTotal & 100 \\
  \end{tabular}%
  \caption{Estimated yearly operations costs for Cosmic Explorer with two observatories,
   based on Advanced LIGO and scaled for CE facility sizes, in millions of 2021 US dollars.
The content of this table is the same as \tref{costops}, but with no attempt to estimate future inflation.
  \tlabel{costops_2021}}
\end{table}

\begin{table}[t]
    \centering

\begin{minipage}{1.0\linewidth}
	\centering
	\begin{tabular}{c}
		\Large \textbf{Cosmic Explorer Cost Estimates, \$(M) 2021 USD}\\[2mm]
	\end{tabular}
\end{minipage}

\resizebox{0.9\linewidth}{!}{ %

\begin{tabular}{l S S}
    \rowcolor{LightSteelBlue1}
    \textbf{Observatory Costs}    & {\textbf{20\,km}} & {\textbf{40\,km}}   \\

\rowcolor{Honeydew1}
\textit{Management}	&  &   \\
\rowcolor{Honeydew1}
\hspace{3mm}Civil Engineering	&~ \CostTwentyManagementCivil &~ \CostFortyManagementCivil   \\ 
\rowcolor{Honeydew1}
\hspace{3mm}Vacuum System	&~ \CostTwentyManagementVac &~ \CostFortyManagementVac  \\ 
\rowcolor{Honeydew1}
\hspace{3mm}Detector	&~ \CostTwentyManagementDet &~ \CostFortyManagementDet  \\
\rowcolor{Honeydew2}
\hspace{3mm}\textit{Total} & \ittab \CostTwentyManagementTotal & \ittab \CostFortyManagementTotal  \\

\rowcolor{LightYellow1}
\textit{Site Specific Design}	& &  \\
\rowcolor{LightYellow1}
\hspace{3mm}Civil Engineering	&~ \CostTwentyDesignCivil &~   \CostFortyDesignCivil \\
\rowcolor{LightYellow1}
\hspace{3mm}Vacuum System	&~ \CostTwentyDesignVac &~ \CostFortyDesignVac \\
\rowcolor{LightYellow1}
\hspace{3mm}Detector	&~ \CostTwentyDesignDet &~ \CostFortyDesignDet  \\
\rowcolor{LightYellow2}
\hspace{3mm}\textit{Total}	& \ittab \CostTwentyDesignTotal &  \ittab \CostFortyDesignTotal \\

\rowcolor{AntiqueWhite1}
\textit{Realization}	&  &  \\
\rowcolor{AntiqueWhite1}
\hspace{3mm}Civil Engineering	&~ \CostTwentyRealizationCivil &~ \CostFortyRealizationCivil  \\  
\rowcolor{AntiqueWhite1}
\hspace{3mm}Vacuum System	&~ \CostTwentyRealizationVac  &~ \CostFortyRealizationVac  \\ 
\rowcolor{AntiqueWhite1}
\hspace{3mm}Detector	&~ \CostTwentyRealizationDet &~ \CostFortyRealizationDet   \\
\rowcolor{AntiqueWhite2}
\hspace{3mm}\textit{Total}	 & \ittab \CostTwentyRealizationTotal & \ittab \CostFortyRealizationTotal \\ 

\rowcolor{LightSteelBlue1}
\textbf{Observatory Total}    &  \bftab \CostTwentyTotal & \bftab \CostFortyTotal \\
\\[-2mm]

\end{tabular}

\hspace{30pt}

 \begin{tabular}{l S}
\rowcolor{Lavender}
\textbf{Project Level Costs}    &      \\

\rowcolor{Honeydew1}
\textit{Project Wide}	&    \\
\rowcolor{Honeydew1}
\hspace{3mm}Management 	&~ \CostProjectProjectwideManagement    \\
\rowcolor{Honeydew1}
\hspace{3mm}Coordination	&~ \CostProjectProjectwideCoordination    \\
\rowcolor{Honeydew1}
\hspace{3mm}Computing 		&~ \CostProjectProjectwideComputing    \\
\rowcolor{Honeydew2}
\hspace{3mm} \textit{Total}	& \ittab \CostProjectProjectwideTotal   \\ %

\rowcolor{LightYellow1}
\textit{Common Design}	&      \\
\rowcolor{LightYellow1}
\hspace{3mm}Civil Engineering	&~ \CostProjectDesignCivil  \\
\rowcolor{LightYellow1}
\hspace{3mm}Vacuum System	&~ \CostProjectDesignVac  \\
\rowcolor{LightYellow1}
\hspace{3mm}Detector	&~ \CostProjectDesignDet   \\
\rowcolor{LightYellow2}
\hspace{3mm}\textit{Total}	 & \ittab \CostProjectDesignTotal  \\
\rowcolor{Lavender}
\textbf{Project Level Total} &\textbf{\CostProjectTotal}\\ %
 & \\
\rowcolor{AntiqueWhite2}
\textbf{Contingency}	&   \\
\rowcolor{AntiqueWhite1}
\hspace{3mm}20\,km Observatory &~ \CostContingencyTwenty  \\ 
\rowcolor{AntiqueWhite1}
\hspace{3mm}40\,km Observatory &~ \CostContingencyForty    \\ 
\rowcolor{AntiqueWhite1}
\hspace{3mm}Project Level &~ \CostContingencyProject  \\
\\[-2mm]

  \end{tabular}%
} %
  
\begin{minipage}{1.0\linewidth}
	\centering
	\begin{tabular}{p{25em} S}
		\rowcolor{Thistle1}
		\textbf{Grand Total for Reference Concept (2 Observatories)} & \bftab \CostTotal \\
	\end{tabular}
\end{minipage}  
  
  \caption{Reproduction of the cost estimate presented in \tref{costs}, but here in millions of 2021 US dollars (i.e., with no attempt to estimate future inflation). The cost estimate includes design, materials, construction, installation and project management for the civil engineering (buildings, roads, etc.), the vacuum system, and the detector.
The cost of alternate configurations can be estimated by adding the associated observatory costs to the project-level costs (e.g., \$\CostTwoTwenty\,M 2021 USD for two \twentykm\ observatories,
 or \$\CostJustForty\,M 2021 USD for a single \fortykm\ observatory).
  \tlabel{costs_2021}}
\end{table}

\vspace{2em}

\section[]{Inflation Estimates for Cosmic Explorer Project}

The cost tables in \sref{project} are given in ``2030 USD''.
This is intended to give the reader access to dollar values representative of what the Cosmic Explorer
 Project will cost in then-year dollars assuming the timeline show in \fref{timeline}.
The spending profile used for this computation is drawn from experience with other large projects,
 including LIGO construction and the Advanced LIGO upgrade.
In our estimation, spending on the CE project begins with the Conceptual Design phase of the MREFC process in 2024, peaks in the years around 2030 with observatory construction, and ramps-down to zero in 2035 as the Project enters the Operations phase.
Inflation in the years from 2022 to 2035 is assumed to be \SI{2.3}{\percent} per year.
Since spending peaks around 2030, the total project cost is inflated by an amount with is numerically equivalent
 to the estimated inflation from 2021 to 2030: roughly \SI{25}{\percent}.
Rather than presenting cost estimates in \sref{project} in then-year dollars,
 which would add significant complexity and indicate a level of precision that we feel is unwarranted at this stage,
 we simply inflate all values by \SI{25}{\percent} and label these inflated estimates as 2030 USD.
In recognition of the uncertainty in this estimate, especially in light of the COVID-19 pandemic,
 the previous section presents values in 2021 USD.

\chapter{Abbreviations}

\newcommand{\acro}[2]{\noindent\textbf{\textsf{#1}} \quad #2}

\acro{2G}{Second generation of gravitational-wave detectors}

\acro{3G}{Third generation of gravitational-wave detectors}

\acro{A+}{LIGO A+ upgrade}

\acro{AAS}{American Astronomical Society}

\acro{AISES}{American Indian Science and Engineering Society}

\acro{APS}{American Physical Society}

\acro{BBH}{Binary black hole}

\acro{BNS}{Binary neutron star}

\acro{BRDF}{Bidirectional reflectance distribution function}

\acro{CBO}{Compact-binary-optimized detector configuration}

\acro{CE}{Cosmic Explorer}

\acro{CERN}{European Organization for Nuclear Research}

\acro{DECIGO}{Decihertz Gravitational-Wave Observatory}

\acro{DOE}{Department of Energy}

\acro{ET}{Einstein Telescope}

\acro{EOS}{Equation of state}

\acro{GWAC}{Gravitational-Wave Agencies Correspondents}

\acro{GWADW}{Gravitational-Wave Advanced Detector Workshop}

\acro{GWIC}{Gravitational-Wave International Committee}

\acro{GWPAW}{Gravitational-Wave Physics and Astronomy Workshop}

\acro{IMBH}{Intermediate-mass black hole}

\acro{KAGRA}{Kamioka Gravitational-Wave Detector}

\acro{LIGO}{Laser Interferometer Gravitational-Wave Observatory}

\acro{LISA}{Laser Interferometer Space Antenna}

\acro{LSC}{LIGO Scientific Collaboration}

\acro{LVK}{LIGO--Virgo--KAGRA Collaboration}

\acro{MREFC}{NSF's Major Research Equipment and Facilities Construction}

\acro{NASA}{National Aeronautics and Space Administration} 

\acro{NEMO}{Neutron-Star Extreme Matter Observatory}

\acro{NIST}{National Institute of Standards and Technology}

\acro{NSF}{National Science Foundation}

\acro{PMO}{Postmerger-optimized detector configuration}

\acro{QCD}{Quantum chromodynamics}

\acro{SACNAS}{Society for Advancement of Chicanos/Hispanics and Native Americans in Science}

\acro{SMBH}{Supermassive black hole}

\acro{SNR}{Signal-to-noise ratio}

\acro{UHV}{Ultrahigh vacuum}

\acro{USD}{US dollars}

\AtNextBibliography{\small}
\printbibliography[heading=bibnumbered,title=References \hlight{[Evan]}]

\newpage
\pagecolor{black}
\thispagestyle{empty}
\begin{tikzpicture}[remember picture,overlay]
    \node[anchor=center] at (current page.center)
        {\includegraphics[width=5in]{CEsplash_white_solid_lp4.pdf}};
\end{tikzpicture}

\end{document}